\documentclass[oneside,12pt]{memoir}
\DoubleSpacing         
\usepackage{pwasu}     
\usepackage{cite}
\usepackage{pdfpages}
\usepackage{arydshln}
\usepackage{mathdots}
\usepackage{multirow}
\usepackage{appendix,nameref}
\usepackage{rotating}
\usepackage{pstricks, pst-node, pst-plot, pst-circ}
\usepackage[square, comma, sort&compress,numbers]{natbib}
\usepackage{url,epsfig}
\usepackage{bigints}
\usepackage{bbm}
\usepackage{chngcntr}
\usepackage{longtable}
\usepackage{tikz}
\usepackage{graphicx}  
\usepackage[T1]{fontenc}
\usepackage{mathtools}
\usepackage[cmintegrals]{newtxmath}
\usepackage{bm} 
\usepackage{array}
\usepackage[caption=false,font=footnotesize]{subfig}
\usepackage[linesnumbered, ruled, norelsize, algoruled, boxed, lined]{algorithm2e}
\usepackage{fancyvrb}

\usepackage{tgcursor}
\usepackage{setspace}

\SetKwRepeat{Do}{do}{while}%
\newcommand{\removelatexerror}{\let\@latex@error\@gobble}
\newcommand{\nosemic}{\renewcommand{\@endalgocfline}{\relax}}
\newcommand{\dosemic}{\renewcommand{\@endalgocfline}{\algocf@endline}}
\let\oldnl\nl
\newcommand{\nonl}{\renewcommand{\nl}{\let\nl\oldnl}}
\makeatother

\counterwithin{figure}{chapter}
\counterwithin{table}{chapter}

\usepgflibrary{shapes}


\newcommand*\xbar[1]{%
  \hbox{%
    \vbox{%
      \hrule height 0.45pt 
      \kern0.5ex
      \hbox{%
        \kern-0.1em
        \ensuremath{#1}%
        \kern-0.1em
      }%
    }%
  }%
}

\DoubleSpacing

\begin{document}

\maxsecnumdepth{subsubsection}
\setcounter{tocdepth}{2} 
\pagestyle{plain}  
\frontmatter
\thetitlepage

\setcounter{page}{1}
\setlength{\parindent}{.5in}
\footskip  .202in

\asuabstract
	Small wireless cells have the potential to overcome bottlenecks in
	wireless access through the sharing of spectrum resources. 
	A novel access backhaul network architecture based 
	on a Smart Gateway (Sm-GW) between the small cell base stations, e.g., 
	LTE eNBs, and the conventional backhaul gateways, 
	e.g., LTE Servicing/Packet Gateways (S/P-GWs) has been introduced to 
	address the bottleneck.  
	The Sm-GW flexibly schedules uplink transmissions for the eNBs.  
	Based on software defined networking (SDN) a
	management mechanism that allows multiple operator 
	to flexibly inter-operate via multiple Sm-GWs with a multitude of small cells
	has been proposed. This dissertation also comprehensively 
	survey the studies that examine the SDN paradigm in optical networks.
	Along with the PHY functional split improvements, the 
	performance of Distributed Converged Cable Access Platform (DCCAP) in the 
	cable architectures especially for the Remote-PHY and Remote-MACPHY nodes 
	has been evaluated. 
	In the PHY functional split, in addition to the re-use of 
	infrastructure with  a common FFT module for multiple technologies, 
	a novel cross functional split interaction to cache the 
	repetitive QAM symbols across time 
	at the remote node to reduce the transmission rate requirement 
	of the fronthaul link has been proposed. 
	 
\setdedication{\emph{I dedicate this dissertation 
		to my lovely mother Shobharani D., father Shivanna T. S. and younger
		brother Dr. Harshith Thyagaturu, who always stood by me and supported all my decisions,
		 to my grandmother and in her memory \mbox{Puttamma P.}, who would have been very 
		 proud to see me graduate with a Ph.D., finally to my close and very special friends 
		 Deepak Muckatira, Praveen Janarthanan, Ramya Ramasubramanian and 
		 Swathi Balakrishna, who selflessly showed unconditional love and support.}}
\asudedication

\asuacknowledgements
	My deepest regards and thanks to my advisor Prof. Martin Reisslein,
	without whom this would have not been possible.  
    
    I am immensely thankful to the committee members, Prof. Yanchao Zhang, 
	Prof. Patrick Seeling, and Prof. Cihan Tepedelenlioglu. 
	for their advice and continued support.
   
    My heartfelt thanks to my friends, lab-mates, colleagues and mentors, 
	Anu Mercian, Yousef Dashti, Ziyad Alharbi, Ahmed Nasrallah 
	Po-Yen Chen, Pranav Bhatkal and Yu Liu
	for their sincere love and support. 
	
    I am grateful to Prof. Daniel Bliss and Prof. Joseph Palais 
    for their generous trust in me, which made my Ph.D. possible through 
    assistantships and financial support. 
    
    In addition, I am also very grateful
    to Hesham ElBakoury, Chief Architect, Huawei for his insightful discussion
    as a part of our collaboration project with Huawei and 
    for generous funding support and which made my research possible.
          
    My greatest
    gratitude to the companies I worked for, as an Engineer at 
    \mbox{Qualcomm Inc.}, San Diego, CA (2013-2015), and as an
    Intern at Intel, San Diego, CA (2016-2017) for the inspiration to do great things.
    
    I am largely indebted to my friends who were my former colleagues 
    at Qualcomm, \mbox{Ambuj Agrawal} who was always available 
    for the technical discussions and \mbox{Mahati Godavarthi} 
    for her kindness in encouraging me throughout. 
    
    Lastly, but certainly not the least, 
    I would like to thank Stefan M{\"a}entele, \mbox{General Manager} at Intel, San Diego,
    for the job offer at Intel, San Diego, CA, USA, well before my graduation.    
\newpage

\setcounter{page}{3}   

\topmargin -0.028in
\headheight  .16in
\headsep  .25in
\footskip  .26in
\textheight  8.352 in 
\tableofcontents*
\textheight  8.352 in
\listoftables   
\listoffigures  

\addtocontents{toc}{\noindent CHAPTER\par}
\newpage
\topmargin -0.057in
\headheight  0in
\headsep  0in
\footskip  .45in
\textheight  8.6 in
\mainmatter
\pagestyle{asu}
\pagestyle{plain}

\newpage
\setcounter{page}{1}
\pagenumbering{arabic}

						
				    	\chapter{\MakeUppercase{SDN Based Smart Gateways (Sm-GWs) for Multi-Operator Small Cell
	Network Management}}

\section{\MakeUppercase{Introduction}}
\subsection{Motivation: Small Cells}
Recent wireless communications research has examined the
benefits of splitting the conventional cells in wireless cellular
communications into small cells for supporting the growing
wireless network traffic.
Small cells can coexist with
neighboring small cells while sharing the same spectrum resources \cite{khan2016cognitive}, and are thus
an important potential strategy for accommodating wireless network traffic
growth~\cite{jaf2015sma}.
Small cells are also sometimes referred to as ``femto'' cells in
the context of the Third Generation Partnership Project (3GPP)
Long Term Evolution (LTE) wireless standard; 
we use the general terminology ``small'' cells throughout.
However, small cells pose new challenges, including
interference coordination~\cite{Cong2014},
backhaul complexity \cite{Wei2013,Siddique2015},
and increased network infrastructure cost~\cite{Mugume2015}.
In this article~\cite{thyagaturu2016sdn} we propose a solution to
reduce the infrastructure cost and complexity of backhaul access
networks supporting small cells.

Small cell networks are expected to be
privatively owned~\cite{Haider2016}.
Therefore it is important
to enable usage flexibility and the freedom of investment in the new
network entities (e.g., gateways and servers) and the network infrastructures
(e.g., switches and optical fiber)
by the private owners of small cells.
While a plethora of studies has examined advanced enhanced Node~B (eNB)
resource
management, e.g.,~\cite{chen2015virtual,liu2015qos,samdanis2016td},
the implications of small cell deployments for backhaul
gateways have largely remained unexplored~\cite{chuang2015resource}.
Generally, backhaul access networks that interconnect
  small cell deployments with LTE gateways
   can employ a wide variety of link layer (L2) technologies, including
SONET/SDH, native Ethernet, 
and Ethernet over MPLS~\cite{Briggs2010,CiscoDeployment,men2013hie}.
In order to accommodate these heterogeneous L2 technologies, 
cellular LTE network interfaces, such as S1 and X2 interfaces,
are purposefully made independent of the L2 technology between
small cell deployments and gateways. Due to the independent nature 
of L2 technologies, a dedicated link with prescribed 
QoS, which can support the fundamental operations of cellular protocols,
must be established for each interface connection~\cite{Ghebretensae2010}. 
Statistical multiplexing is then limited 
by the aggregate of the prescribed QoS \cite{guck2016function, fitzek2002providing} requirements
and only long-term re-configurations, e.g.,
in response to deployment changes, can optimize the
backhaul transmissions~\cite{Mathew2015}.
Present wireless network deployments based on the 3GPP LTE standard do 
not provide feedback from the eNBs to a
central decision entity, e.g., an SDN orchestrator, 
which could flexibly allocate network resources based on eNB traffic demands.
Thus, present wireless backhaul architectures are characterized by
$(i)$ essentially static network
resource allocations between eNBs and operator gateways,
e.g., LTE Servicing/Packet Gateways (S/P-GWs), and
$(ii)$ lack of coordination between the eNBs and the operator gateways
in allocating these network resources,
resulting in under-utilization of the backhaul transmission resources.
Additionally, exhaustion of available ports at the operator gateways can
limit the eNB deployment in practice.

The static resource allocations and lack of eNB-gateway cooperation
are highly problematic since the aggregate uplink transmission bitrate of the
small cells within a small geographic area, e.g., in a building,
is typically much higher than the
uplink transmission bitrate available from the cellular operators.
Thus, small cell deployments create a bottleneck between the eNBs and the
operator gateways.  For instance, consider the deployment of
100 small cells in a building, whereby each small cell supports
1~Gbps uplink transmission bitrate.
Either each small cell can be allocated only one hundredth of the
operator bitrate for this building or the operator would need to
install 100~Gbps uplink transmission bitrate for this
single building, which would require cost-prohibitive operator gateway
installations for an organization with several buildings in a small
geographical area.
However, the uplink transmissions from the widespread data communication
applications consist typically of short high-bitrate bursts, e.g.,
100 Mbps bursts.
If typically no more than ten small cells burst simultaneously, then
the eNBs can dynamically share a 1~Gbps operator uplink
transmission bitrate.
An additional problem is that with the typically limited port counts on
operator gateways, connections to many new small cells may require
new operator gateway installations.
An intermediate Sm-GW can aggregate the small cell connections
and thus keep the required port count at operator gateways low.

\subsection{Overview of Network Management with SDN-based Sm-GW}
\begin{figure}
    \centering
    \includegraphics[width=4.5in]{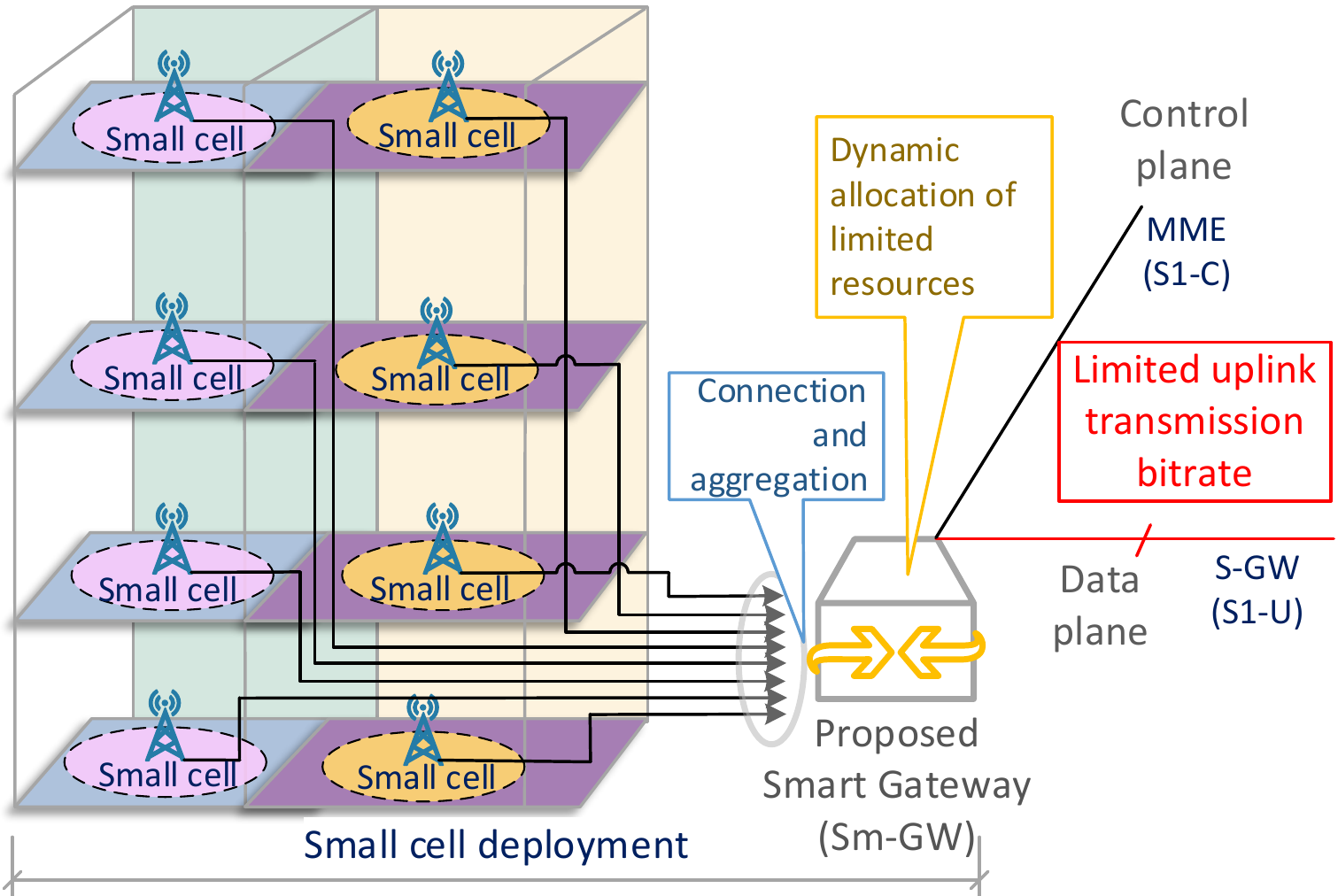}
    \caption{The proposed smart gateway (Sm-GW).} 
    \label{fig_enterprise}
\end{figure}
\begin{figure}[t]
    \centering
    \includegraphics[width=0.9\textwidth]{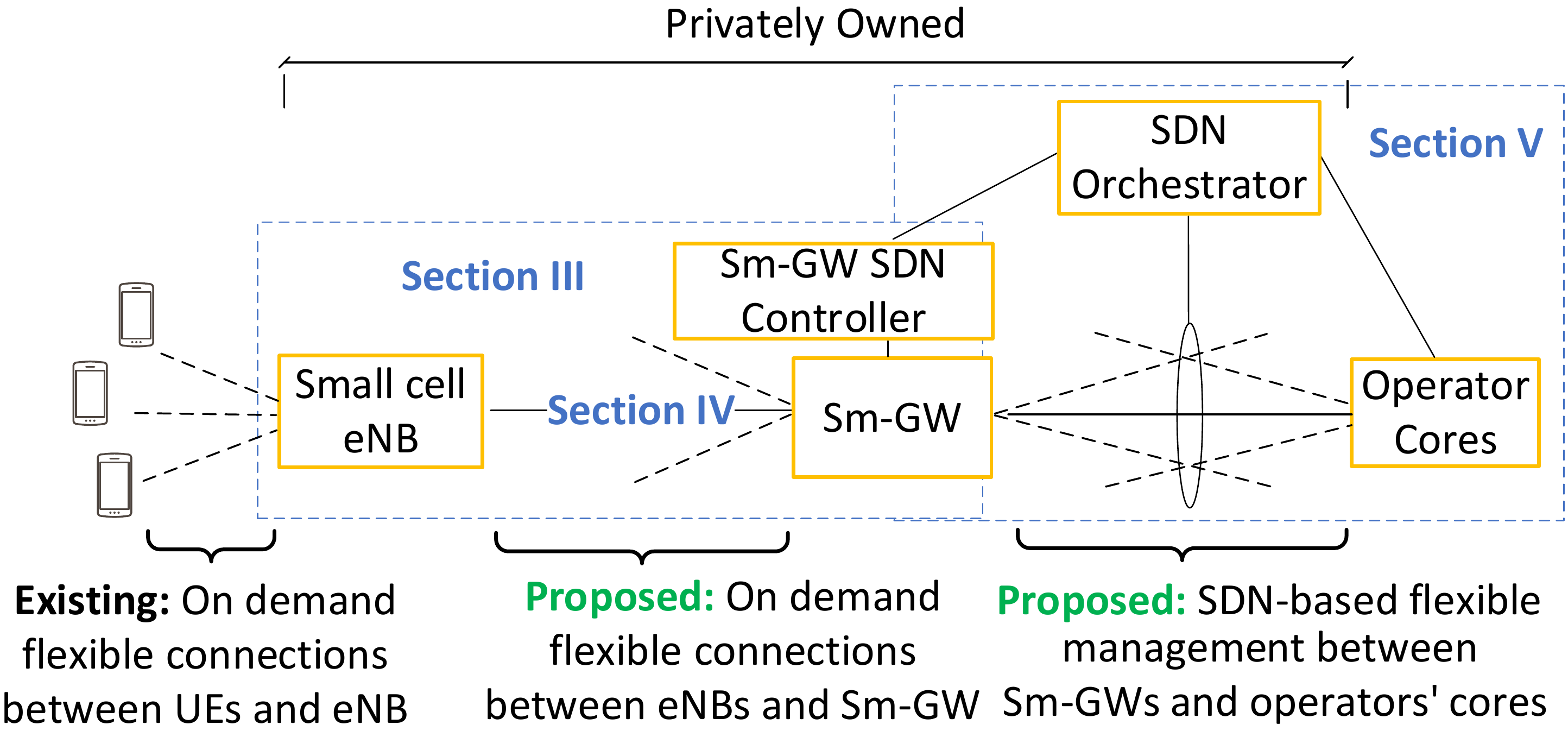}
    \caption{The Smart Gateway (Sm-GW) architecture.} 
    \label{fig_SMGW_overall}
\end{figure}

We present a new backhaul network framework for supporting small cell
deployments based on a new network entity, the Smart GateWay (Sm-GW).
Consider an exemplary small cell deployment throughout multiple
buildings of a university.
Each building has hundreds of small cells that are flexibly
connected to an Sm-GW, as illustrated in Fig.~\ref{fig_enterprise}.
Multiple Sm-GWs are then connected to core networks, i.e., the S-GWs
and P-GWs, of multiple cellular operators via
physical links (e.g., optical or microwave links)
\cite{mcgarry2012investigation, mcgarry2008ethernet, aurzada2008delay, mcgarry2006wdm, mcgarry2004ethernet, reisslein6metropolitan, mercian2013offline, aurzada2014delay},
as illustrated for a single Sm-GW in Fig.~\ref{fig_SMGW_overall}.
An SDN orchestrator owned by the university
manages the cellular infrastructure of the entire university.
The SDN orchestrator coordinates the resource allocations from
the operators to the Sm-GWs.

The main original contributions of this article are:
\begin{enumerate}
\item A novel comprehensive Smart Gateway (Sm-GW) architecture
  and protocol framework that accommodates a flexible number of eNBs
  while reducing the requirements at the operator's core, e.g.,
  at LTE S-GW and MME. The Sm-GW physically and logically aggregates
 the eNB connections so that a set of eNBs appears as a single virtual
 eNB to the operator gateways, see Section~\ref{arch:sec}.
    \item A Sm-GW scheduling framework to flexibly share the limited
     uplink transmission bitrate among all the small cell eNBs connected
      to an Sm-GW,  see Section~\ref{gwsch:sec}.
    \item An adaptive SDN-based multi-operator management framework that
      dynamically shares the uplink transmission bitrates of multiple
      operators among the Sm-GWs. An SDN orchestrator dynamically coordinates
      the sharing among the Sm-GWs, the transport network connecting the
      Sm-GWs to the operator gateways, and the operator gateways,
  see Section~\ref{MOM:sec}.
\end{enumerate}

\subsection{Related Work}
Recently proposed SDN based backhaul architectures, such as
CROWD~\cite{AliAhmad,seb2015dyn}, iJOIN~\cite{Dongyao},
U-WN~\cite{Shengli}, Xhaul~\cite{oli2015xha},
the multi-tiered SDN based backhaul architecture~\cite{elg2015thr},
and similar
architectures~\cite{cos2015sof,cos2015,dra2015dyn,hur2015sdn,
  Jungnickel,liu2013cas},
are revolutionary designs proposing new cellular infrastructure
  installations.
In contrast, our proposed SDN-based Sm-GW enables the softwarization of
\textit{existing}
cellular infrastructures consisting of eNBs and conventional operator gateways,
such as the S/P-GW in LTE core networks.
The proposed Sm-GW is inserted in the existing backhaul
infrastructure to inter-network and co-exist with the existing
LTE network core entities, such as the S/P-GW.

SDN based backhaul architectures with centralized
interference coordination have been proposed
in~\cite{akh2016syn, gop2016gen, nie2015wir}.
The SDN controller in these architectures maintains a global database
of spectrum resources \cite{akhtar2016white} and dynamically assigns the resources to base stations so
as to minimize the mutual interference among base stations in dense
deployments.
We note that wireless interference is a localized phenomenon.
Therefore, a base station is most affected by its neighboring base stations
in dense deployments.
The centralized interference coordination techniques
in~\cite{akh2016syn, gop2016gen, nie2015wir} are complementary
to our proposed Sm-GW architecture in that they can be
implemented at the Sm-GW instead of the SDN controller/orchestrator.

Schedulers at the eNB allow multiple user equipment (UE) devices to
share the wireless resources at the eNB.  For example, the LTE standard medium
access control (MAC) protocol~\cite{3gppLTEMAC} coordinates the
scheduling of wireless resources between an eNB and multiple UEs.
Generally, most wireless resource scheduling studies to date have focused on the
sharing of the wireless resources at a given single eNB.
For instance, quality of service
(QoS) aware uplink scheduling and
resource allocation at a given single small cell eNB in an LTE network have been
examined in~\cite{Chaudhuri2015}.
In contrast, we propose a novel
scheduling framework at the Sm-GW based on uplink transmission bitrate requests
from \textit{multiple eNBs}, i.e., we propose the sharing of
the backhaul network resources among multiple eNBs.

A similar sharing of network resources among small cell base
stations has been studied in~\cite{lak2013h}.
Specifically, the H-infinity scheduler for limited capacity
backhaul links~\cite{lak2013h} schedules the traffic
in the downlink. The centralized H-infinity scheduler
focused on buffer size requirements at
the base stations in the small cell networks.
In contrast, we focus on the \textit{uplink} traffic from the eNBs to the Sm-GW.
To the best of our knowledge, we propose the first
network protocol framework for the uplink transmissions
from multiple eNBs to the operator gateways in the context of LTE small cells.
We note that our Sm-GW framework is complementary to several recently studied
resource allocation mechanisms in cellular networks.
For instance, D2D resource allocation through traffic
offloading to small cell networks has been studied in~\cite{sem2015con}; this
D2D approach can be readily supported by our proposed Sm-GW.
Coordinated scheduling \cite{mer2016ups, wei2014dyc, che2015sim, das2016gro} in the context of small cells
with dynamic cell muting to mitigate the interference has
been discussed in~\cite{wan2015dyn}. The cell muting technique can be
further extended based on our approach of traffic scheduling to eNBs.
A flexible wireless resource allocation mechanism
based on the SDN programmability of traffic flows
from a single UE device to multiple
base stations in dense small cell networks
has been examined in~\cite{gas2015pro}.
The offloading of UE traffic for efficient traffic management in
  small cell networks has been examined in~\cite{tal2015eff}.
In contrast, we propose an SDN-based multi-operator resource allocation
mechanism that allocates limited backhaul link capacities to
multiple Sm-GWs (which in turn can flexibly allocate the capacities
to multiple eNBs).
The UE to eNB communication approach from~\cite{gas2015pro}
and UE traffic offloading~\cite{tal2015eff} are thus
complementary to our eNB to Sm-GW and Sm-GW to S/P GW
network management approaches.

\section{\MakeUppercase{Background: Conventional LTE Small Cell Backhaul}}
In this section we describe the conventional architectural model for
Home-eNodeB (HeNB) access  networks~\cite{HeNB3GPP} and the network
sharing mechanism in 3GPP LTE.
HeNBs are the small cell base stations of the LTE standard.
  We use the general terminology ``eNB'' to denote all types of
small cell base stations.
\begin{figure}[t]
    \centering
    \includegraphics[width=4.5in]{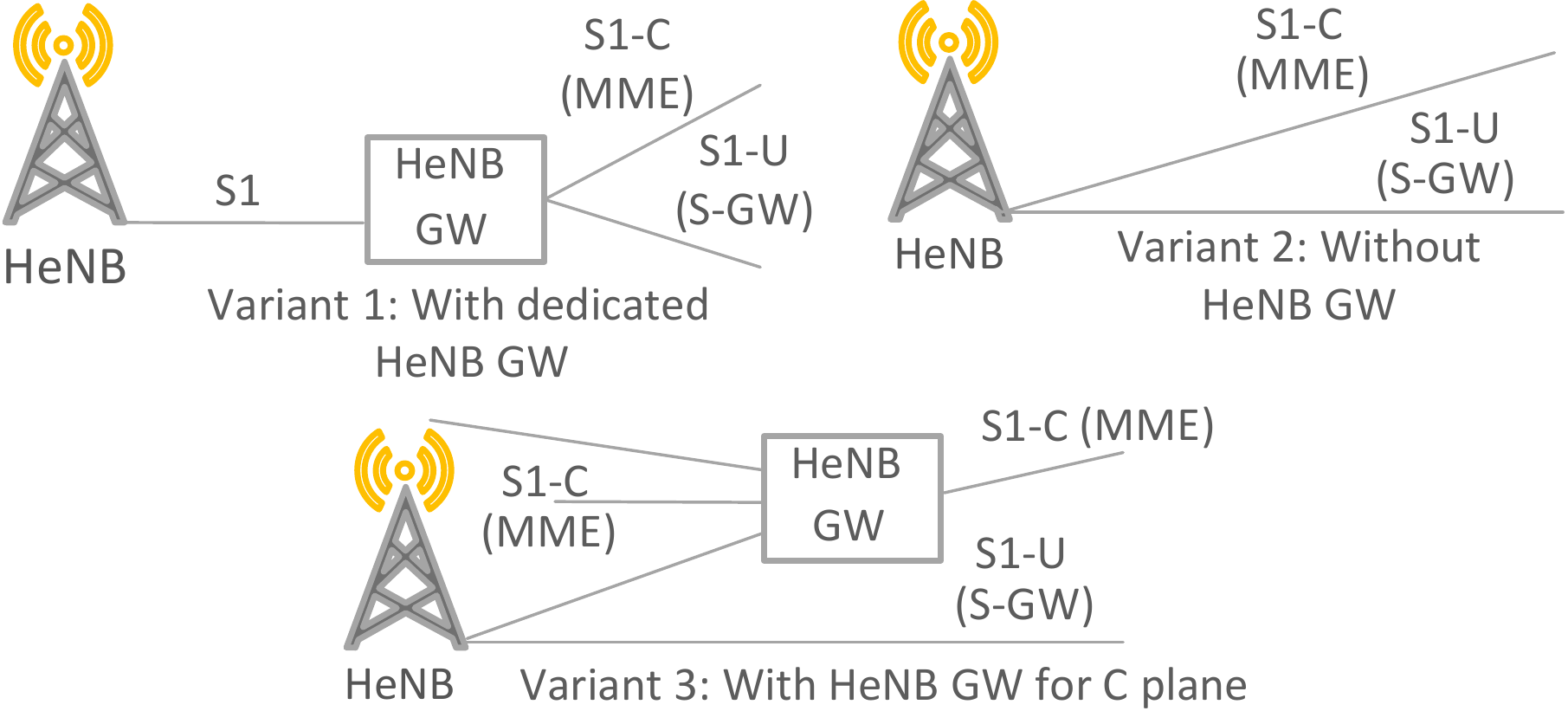}
    \caption{HeNB architectural models in 3GPP LTE.}
    \label{fig_HeNB}
\end{figure}

\subsection{HeNB Architectural Models in 3GPP LTE}
In Figure \ref{fig_HeNB}
we show the 3GPP HeNB architectural models:
1) with dedicated HeNB-GateWay (HeNB-GW), 2) without HeNB-GW,
and 3) with HeNB-GW for the control plane.

\subsubsection{With Dedicated HeNB-GW}
With a dedicated HeNB-GW, communication between
the HeNB and the HeNB-GW is secured by a mandatory security gateway
(Se-GW) network function. The HeNB-GW aggregates the
control plane connections (S1-MME) and user plane connections (S1-U) of all
HeNBs connected to the HeNB-GW to a single control and user plane
connection. The HeNB gateway appears as a single eNB to the outside
entities, such as S-GW and MME.
In a similar way, the HeNB-GW appears as both an
S-GW and an MME to the eNBs connected to the HeNB-GW.  The numbers of
ports required at the MME and S-GW are reduced through the
aggregation at the HeNB-GW.
Our proposed Sm-GW architecture is similar to the
  dedicated HeNB-GW architecture, in that the Sm-GW aggregates the eNBs
  connections both physically and logically.
In addition, our Sm-GW flexibly allocates uplink transmission
bitrates to small cell eNBs (see Section~\ref{gwsch:sec})
and allows for the adaptive allocation of
operator uplink transmission bitrates to the Sm-GW by the SDN orchestrator
(see Section~\ref{MOM:sec}).

\subsubsection{Without HeNB-GW}
Deployments of HeNBs without the HeNB-GWs increase the requirements on
the S-GW and MME to support large numbers of connections.  Large
deployments of small cells without gateway aggregation at the HeNBs
would greatly increase the total network infrastructure cost.

\subsubsection{With HeNB-GW for the Control Plane}
HeNB control plane connections are terminated at the HeNB-GW and a
single control plane connection is established from the
HeNB gateway to the MME. Although
the number of connections required at the MME is reduced due to the
control plane aggregation at the HeNB-GW, data plane connections are still
terminated directly at the S-GW, increasing requirements at the
S-GW. The Se-GW typically secures the communication to and from the
HeNB. In contrast, our proposed Sm-GW terminates all
the control and data connections from HeNBs.

\subsection{3GPP Network Sharing} 
Network sharing was introduced by 3GPP in Technical Specification
  TS~23.951~\cite{3GPPNetSharing}
  with the main motivation to share expensive radio spectrum
  resources among multiple operators.
 For instance, an operator without available spectrum 
in a particular geographic area can offer cellular
services in the area through sharing the spectrum of another operator. 
In addition to spectrum sharing, 3GPP specifies
core network sharing among multiple operators
through a gateway core network (GWCN) configuration~\cite{3GPPNetSharing}.
GWCN configurations are statically pre-configured at deployment
for fixed pre-planned core network sharing.
Thus, GWCN sharing can achieve only limited statistical
multiplexing gain as the sharing is based on the pre-configured 
QoS requirements of the eNB interface connections and not on the 
varying eNB traffic demands. 
Also, the GWCN configuration lacks a central entity for 
optimization of the resource allocations with global knowledge of the 
eNB traffic demands.
In contrast, our Sm-GW framework includes a central SDN orchestrator for
optimized allocations of backhaul transmission resources  
according to the varying eNB traffic demands (see Section~\ref{MOM:sec}).
 
\section{\MakeUppercase{Proposed Smart Gateway (Sm-GW)}} \label{arch:sec}
In this section we introduce the proposed Smart Gateway (Sm-GW) network
architecture for existing LTE deployments.
We describe the fundamental protocol mechanisms and interfaces
that integrate the proposed Sm-GW into the conventional LTE
protocols.

\subsection{LTE Protocol Modifications}  \label{prot:sec}
\begin{figure*}[!t]
    \centering
    \includegraphics[width=6in]{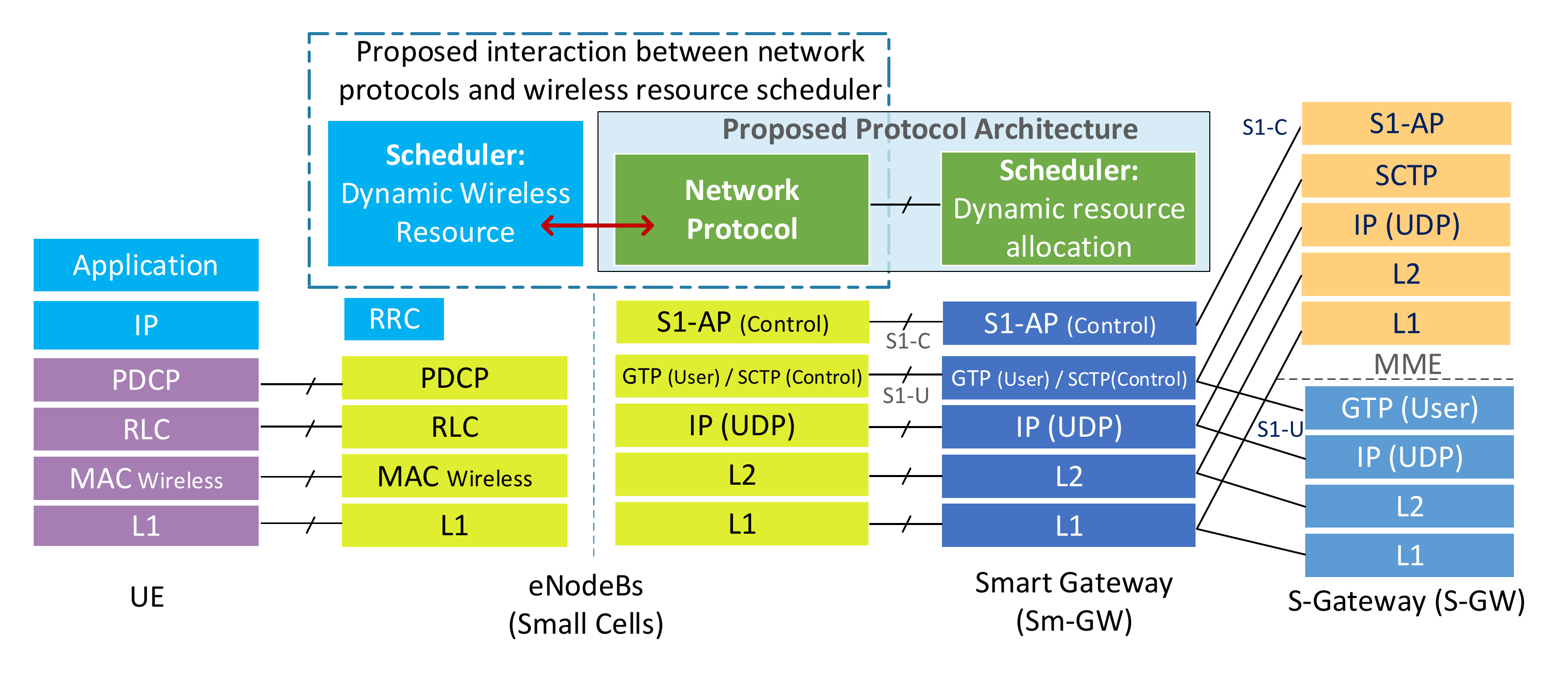}
    \caption{Illustration of proposed protocol mechanisms at eNB,
Sm-GW, and S-GW.} 
    \label{fig_protocol}
\end{figure*}

Fig.~\ref{fig_protocol} illustrates the proposed protocol
mechanisms between a set of $N_s$ eNBs and a given Sm-GW $s$.

\subsubsection{eNB} \label{prot:sec:eNB}
At the eNB, we newly introduce the eNB-to-Sm-GW
reporting protocol, which
operates on top of the GPRS tunneling protocol (GTP)~\cite{GTP}
and stream control transmission protocol (SCTP).
The reporting protocol $(i)$ evaluates the required uplink transmission
bitrate, and $(ii)$ sends the bitrate request messages to the Sm-GW.
The reporting protocol formulates the operator specific uplink
transmission bitrate requests
based on the requests of the UEs that are connected
via the eNB to multiple operators $o,\ o = 1, 2, \ldots, O$.

The eNB wireless resource scheduler is responsible for the sharing
of wireless resources between the eNB and the UEs.
The eNB wireless resource scheduler ensures
that only the resources available at the eNB are granted to the UEs.
UEs periodically send buffer status reports
(BSRs) to the eNB which they are connected to.
Therefore, the eNB-to-Sm-GW reporting protocol can estimate the UE traffic
requirements by interacting with the wireless resource scheduler.

\subsubsection{Smart Gateway (Sm-GW)} \label{prot:sec:Sm-GW}
The protocol stack at the Sm-GW is similar to the HeNB-GW protocol stack.
However, in the Sm-GW, an additional eNB coordination protocol,
a scheduler for the dynamic resource
allocation, and SDN capabilities are introduced.

The eNB coordination protocol collects request messages from eNBs.
The eNB uplink transmission grants are sized based on the
eNB requests and the available Sm-GW resources
according to the Sm-GW scheduling described in Section~\ref{gwsch:sec}.
The eNB coordination protocol sends grant messages
to all eNBs within a reasonable processing delay.

S1 based handovers for the downlink transmissions are typically
anchored at the S-GW. (For the uplink transmissions, an anchoring,
or buffering of packets, at a network entity, e.g., eNBs or S-GW, is
not required.) We emphasize that the
Sm-GW will be transparent to all the downlink packets from the S-GW
and hence not be limited by the network protocol scheduler.
This ensures that the S1 based handover mechanisms
at the S-GW and eNBs continue to function normally.

\subsubsection{SDN Operations of Sm-GW}
\begin{figure}[!t]
	\centering
	\includegraphics[width=3.5in]{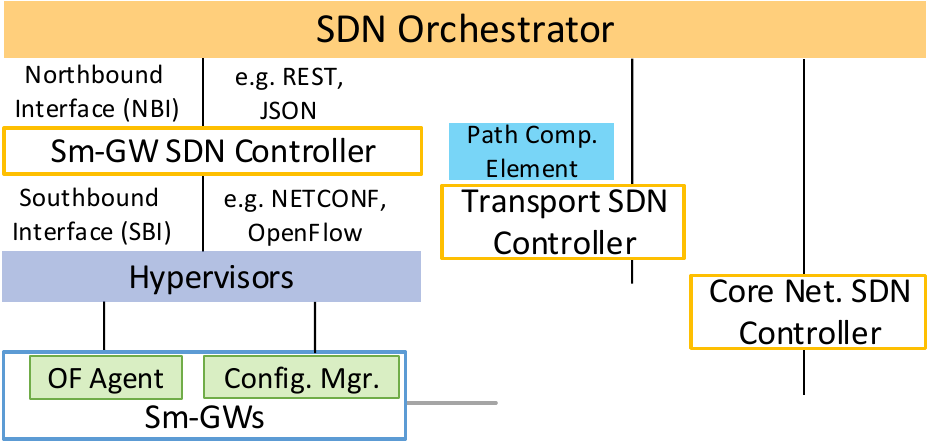}
        \caption{Illustration of Sm-GW embedding in the SDN ecosystem.}
	\label{fig_Sm-GW_SDN}
\end{figure}

\paragraph{SDN Infrastructure}
The Sm-GW SDN capabilities can be provided by an
OpenFlow (OF) agent and/or
a configuration manager at each Sm-GW, as illustrated
in Fig.~\ref{fig_Sm-GW_SDN}.
OpenFlow is a popular protocol for the southbound
interface (SBI) and can be employed on the SBI
between the Sm-GW SDN controller and Sm-GW.
The OpenFlow agent supports OpenFlow SDN functionalities at the Sm-GW,
making the Sm-GW configurable through the OpenFlow protocol.
The Sm-GW configuration manager can be controlled by the Sm-GW SDN controller,
e.g., through the NETCONF (or OpenFlow) SBI,
to dynamically reconfigure the Sm-GW.

The Sm-GW SDN controller configures the Sm-GWs to enable the
internal LTE X2 tunnel interfaces among all connected small cell eNBs,
as elaborated in Section~\ref{X2int:sec}. Also, the Sm-GW SDN
controller manages the external LTE X2 and S1 interfaces at the Sm-GW through
tunnel establishments to the external LTE network core entities,
i.e., MMEs and S/P-GWs.

Whereas the conventional LTE transport network between the eNBs and
S/P-GWs is configured with static MPLS/IP paths~\cite{Ghebretensae2010},
the flexible Sm-GW operation requires a flexible transport
network, that is controlled by a transport SDN controller, as illustrated in
Fig.~\ref{fig_Sm-GW_SDN}.
The flexible transport network can, for instance, be implemented through a
Software Defined Elastic Optical Network 
(SD-EON)~\cite{thy2016sof,Yoshida2015}.

\paragraph{Sm-GW Virtualization}   \label{smgwvirt:sec}
The SM-GW can support a variety of virtualization strategies, e.g., to
provide independent virtual networks for different operators.
One example virtualization strategy could let the
Sm-GWs abstract the connected eNBs. Sm-GWs could then be abstracted by
a hypervisor~\cite{ble2015sur,mij2016ne,nak2015wif,rig2015pro}
that intercepts the SBI, as illustrated in Fig.~\ref{fig_Sm-GW_SDN}.
Based on the operator configurations that are sent via the SDN orchestrator
to the Sm-GW SDN controller, resources at the Sm-GWs and the small cell eNBs
(which are privately owned by an organization~\cite{Haider2016}) can
be sliced to form operator-specific virtual networks of Sm-GWs and
eNBs. The configuration manger at each Sm-GW can allocate resources to each of
these virtual networks.

From the operator perspective,
  the Sm-GW virtualization \cite{ble2016con} essentially allows multiple operators to share the
physical small cell infrastructure of Sm-GWs and eNBs.
Thus, simultaneous services can be enabled to large UE populations
that belong to multiple operators, i.e., that have contracts with multiple
operators, while using the same physical small cell infrastructure.
Existing conventional cellular deployment structures do not support
the infrastructure sharing among multiple operators.

\paragraph{SDN Orchestration}
The SDN orchestrator coordinates the network management across multiple
domains of Sm-GWs (whereby each Sm-GW domain is controlled by its own
Sm-GW SDN controller), transport networks, and core networks.
The SDN orchestrator implements the multi-operator management
introduced in Section~\ref{MOM:sec} and configures the
Sm-GWs and transport networks based on
global multi-operator network optimizations.
For example, the SDN orchestrator
communicates with the path computation element (PCE) SDN application
on the transport SDN controller.
The PCE dynamically evaluates the label switched paths,
such as MPLS/IP paths, so as to flexibly enable and reconfigure
the transport network~\cite{thy2016sof,Yoshida2015}.

\subsubsection{LTE X2 Interfaces of eNBs with Sm-GW}  \label{X2int:sec}
X2 interfaces enable critical functionalities in LTE small cells,
such as X2-based handover as well as interference coordination and mitigation.
Typically, each eNB connected to a given Sm-GW
pertaining to an operator shares the same
MME; thus, each eNB needs an X2 interface to all other eNBs
within the same MME coverage area, the so-called tracking area.
Hence, eNBs connected to an Sm-GW must be
interconnected with X2 interfaces.

\paragraph{To External Macro-eNBs}
X2 traffic flows destined to eNBs located outside the scope of an
Sm-GW (typically to a macro-eNB) are not be limited by the scheduler
at the Sm-GW. X2 packets flow out of Sm-GW into the backhaul (i.e., to an
S-GW) as they originate at the eNBs. The Sm-GW appears as an
external router (or gateway) to the X2 external interfaces.

\paragraph{To Internal Small-eNBs} \label{prot:X2:conn}
The Sm-GW appears as a simple bridge or a router to the internal X2
interfaces, routing the internal X2 packets within.
Therefore, the scheduler at the Sm-GW does not limit any X2 packets.
For small cell deployments, an eNB can have multiple neighboring eNBs
in the tracking area;
these neighboring eNBs need to be interconnected with each other
with X2 connections.
On the order of $O(N(N-1))$ dedicated links would be required
  to interconnect the X2 interfaces of $N$ eNBs in the tracking area in a full
  mesh topology.
In contrast, a star
topology with the Sm-GW at the center requires only $O(N)$
connections to connect each eNB to the Sm-GW.
In summary, in our Sm-GW architecture, the Sm-GW manages the X2 interfaces
of all the internal small cell eNBs, thus eliminating
the distributed management of X2 interfaces at each eNB.

\subsubsection{Authentication of Sm-GW with EPC Core}
Typically HeNBs use IPSec tunneling for their
security and encryption, which creates overhead.
If Sm-GWs are authenticated, the HeNBs would no longer need IPsec tunneling.
Specifically, upon boot-up, the Sm-GW is authenticated with
an LTE Evolved Packet Core (EPC)
so as to eliminate the need for a network security-gateway
(Se-GW) function or IPsec tunneling between the eNBs and the P-GWs.
Critical cellular network functions, such as
security, authentication, and reliability,
require additional effort to be enabled in WiFi networks.
WiFi Passpoint~\cite{Passpoint} (Hotspot 2.0)
aims at providing an experience similar to
cellular connectivity in
WiFi networks by providing the cellular authentication mechanisms.
With the authentication of Sm-GWs,
the simplicity of WiFi networks
can be achieved by the small cell cellular networks.

\subsection{Downlink vs.~Uplink Communication} \label{updown:sec}
\subsubsection{Downlink Packets at the Sm-GW}
Traffic flows in the conventional downlink path from
an S/P-GW to an eNB are typically sent at rates that do not
exceed the wireless transmission rates from the eNB to the UE devices.
Thus, as long as the link rates from the S/P-GW to the
inserted Sm-GW and from the Sm-GW to the eNB are at least as high
as the conventional S/P-GW to eNB links, the Sm-GW can be
transparent to the downlink packets from the S/P-GW.

\subsubsection{Uplink Packets at Sm-GW}
In contrast to the downlink data traffic, the uplink data
traffic from the eNBs to an Sm-GW needs to be
regulated as the traffic flows from all the eNBs terminating at the
Sm-GW can overwhelm the outgoing link towards the operator S-GW.
Enforcing QoS strategies and fairness among eNBs requires scheduling
of the uplink packet traffic arriving from the eNBs at an Sm-GW.
Therefore, our focus is on frameworks for the uplink transmission scheduling of
the communication $(i)$ from eNBs to an Sm-GW (Section~\ref{gwsch:sec}),
and $(ii)$ from Sm-GWs to S-GWs (Section~\ref{MOM:sec}).

\section{\MakeUppercase{Proposed Sm-GW Scheduling Framework}} \label{gwsch:sec}
\subsection{Purpose}
The main purpose of the Sm-GW scheduling framework
is to maximize the utilization of
the network resources, and to ensure fair uplink transmission service for
 all eNBs connected to an Sm-GW.
Without scheduling, highly
loaded eNBs can impair the service for lightly loaded
eNBs connected to the same Sm-GW.
When many eNBs are
flexibly connected to an Sm-GW, traffic bursts from heavily loaded eNBs can
overwhelm the queue of an Sm-GW, resulting in excessive packet
drops and high delays, even for lightly loaded eNBs.
On the other hand, with scheduling, a large number of
eNBs can be flexibly connected to the Sm-GW while ensuring
prescribed QoS and fairness levels.
Each eNB can possibly have a different service level agreement.
The Sm-GW allows for the flexible deployment of a wide variety of
scheduling algorithms.
We outline two classes of Sm-GW scheduling algorithms, and illustrate
an elementary algorithm for each class.

\begin{table}[t]
\caption{Summary of Notation of Sm-GW Network Management}
     	\centering
\begin{tabular}{|p{0.4cm} p{7.2cm}|}
\hline
\multicolumn{2}{|c|}{\textbf{\rule{0pt}{1\normalbaselineskip}
    Sm-GW Sched. Framework (Sm-GW $\leftrightarrow$ eNBs),
           Sec.~\ref{gwsch:sec}}} \\   [.5ex]
$N_s$ &  Number of small cell eNBs at Sm-GW $s$ \\
$G_{so}$ &  Available uplink transm. bitrate [bit/s] from Sm-GW $s$ to operator $o$  \\
$W$ &  Duration [s] of scheduling cycle   \\
$\Gamma_{so}$  &  $= G_{so} W / N_s$, Max. eNB uplink transm. data amount [bit] per cycle with equal sharing \\
$\rho_{son}$ & Data amount [bit] that eNB $n$ at Sm-GW $s$ wants to
  transmit to operator $o$ in a cycle, i.e., request by eNB $n$ \\
$\gamma_{son}$ & Data amount [bit] that eNB $n$ at Sm-GW $s$ is allowed
to transmit to operator $o$ in a cycle, i.e., grant by Sm-GW $s$ \\
\hline
\multicolumn{2}{|c|}{\textbf{\rule{0pt}{1\normalbaselineskip}
   SDN Based Multi-Operator Managm. Framework, Sec.~\ref{MOM:sec}}} \\
            \multicolumn{2}{|c|}{\textbf{
    (Sm-GWs $\leftrightarrow$ Operator Gateways)}} \\[.5ex]
$o$  &  Index of operators, $o = 1, 2, \ldots, O$\\
$s$   & Index of Sm-GWs, $s = 1, 2, \ldots, S$ \\
$R_{so}$   &  Smoothed uplink transmission bitrate [bit/s] request
     from Sm-GW $s$ to operator $o$ \\
$K_o$ &  Max. available uplink transm. bitrate through operator $o$\\ 
$G_{so}$ & Granted uplink transm. bitrate from Sm-GW $s$ to operator $o$\\
$X_{so}$ & Actual uplink traffic bitrate from Sm-GW $s$ to operator $o$\\ [0.5ex]
		\hline
		\end{tabular}
		\end{table}

\subsection{Configuration Adaptive Scheduling} \label{gwsch:sec:equal}
Configuration adaptive scheduling adapts the scheduling,
i.e., the allocation of uplink transmission bitrates, according to
the number of eNBs connected to a given Sm-GW.
The Sm-GW tracks the number of connected eNBs and sends
a configuration message to all eNBs in the event of a change in
connectivity at the Sm-GW, i.e., addition of new eNB or
disconnection of existing eNB.
More specifically, consider $N_s$ eNBs at a given Sm-GW $s$ that has been
allocated the uplink transmission bitrate $G_{so}$ [bit/s] toward a
given operator $o$ (through the coordination techniques in
Section~\ref{MOM:sec}).

An elementary equal share scheduling shares the available uplink
transmission bitrate at the Sm-GW toward a given operator $o$
equally among all eNBs connected to the Sm-GW.
Each eNB $n,\ n = 1, 2, \ldots, N_s$, can then transmit at most
$\Gamma_{so} =  G_{so} W / N_s$ [Byte] of traffic during a cycle
of duration $W$ [seconds].
The traffic amount limit $\Gamma_{so}$ and cycle duration $W$
are sent to the eNBs as a part of the initial configuration message.
Each eNB schedules the uplink transmissions
such that no more than $\Gamma_{so}$ [Byte] of traffic are
send in a cycle of duration $W$ [seconds].

The simple equal share scheduler can flexibly accommodate
large numbers $N_s$ of eNBs.
However, the equal bandwidth assignments by the elementary
equal share scheduler to
the eNBs under-utilize the network resources when some eNBs have very
little traffic while other eNBs have high traffic loads.

\subsection{Traffic Adaptive Scheduling} \label{gwsch:sec:excess}
\begin{figure}[t!]
    \centering
    \includegraphics[width=3.5in]{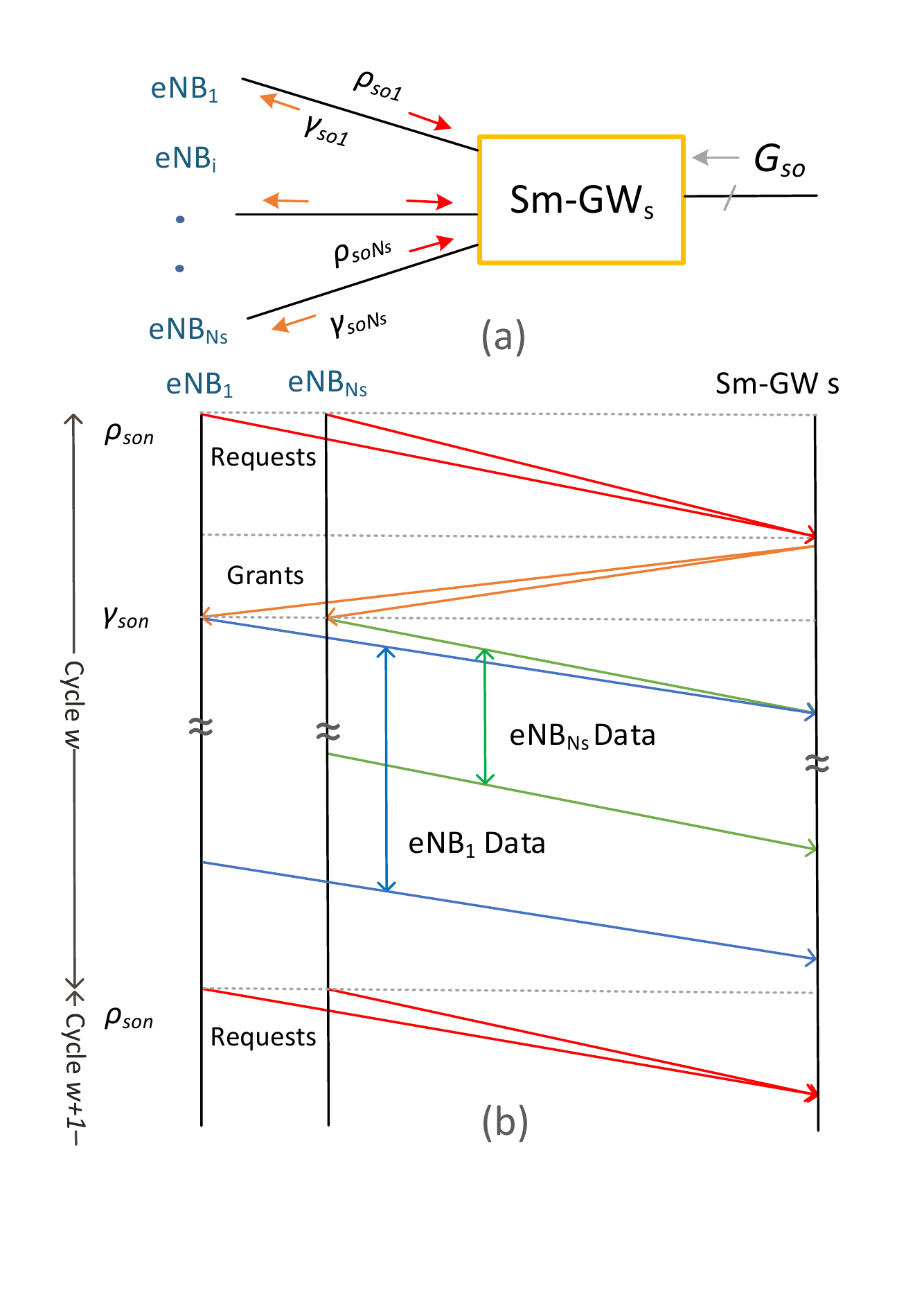}
\caption{Illustration of traffic adaptive Smart Gateway (Sm-GW)
    scheduling.} 
\label{fig_ExcessTiming}
\end{figure}
With traffic adaptive scheduling, the Sm-GW collects uplink transmission
requests from the eNBs. The Sm-GW then adaptively allocates portions of
the uplink transmission bitrate $G_{so}$ to the individual eNBs
according to their requests.
Traffic adaptive scheduling operates with a
request-allocate-transmit cycle of duration $W$ [seconds]
illustrated in Fig.~\ref{fig_ExcessTiming}. At the start of the
cycle, each eNB $n,\ n = 1, 2, \ldots, N_s$, sends an
uplink transmission bitrate request to Sm-GW~$s$.
We let $\rho_{son}$ denote the amount of traffic [in Byte] that
eNB $n$ wants to transmit to operator $o$ over the next cycle of duration $W$.
Once all requests have been received, i.e., following the
principles of the offline scheduling
framework~\cite{zhe2009sur}, portions of $G_{so}$ can be
allocated to the eNBs according to some scheduling policy.

An elementary excess share scheduling
policy~\cite{bai2006fai}
allocates the eNB grants as follows. Lightly loaded eNBs with
$\rho_{son} < \Gamma_{so}$ are granted their full request, i.e.,
receive the grant size $\gamma_{son} = \rho_{son}$, while their
unused (excess) portion of the equal share allocation is accumulated
in an excess pool: 
\begin{equation}
\xi = \sum_{\forall \rho_{son} \leq \Gamma_{so}} \Gamma_{so} - \rho_{son}.
\label{ExcessAccum}
\end{equation}
Following the principles of controlled equitable excess
allocation~\cite{bai2006fai}, highly loaded eNBs are allocated
an equal share of the excess up to their request. That is,
with $|\mathcal{H}|$ highly loaded eNBs, the grants are
\begin{equation}
\gamma_{son} = \min \left(\rho_{son},\; \Gamma_{so} +
       \frac{\xi}{|\mathcal{H}|}\right).
\label{HighGrant}
\end{equation}

\subsection{Scheduling Fairness} \label{sec:fariness}
Within the context of our proposed Sm-GW scheduling framework, fairness is
the measure of network accessibility of all $N_s$ eNBs
connected to an Sm-GW $s$ based on individual eNB uplink throughput level
requirements.
We denote $T_{son}$ for the long-run average throughput level [bit/s] of
uplink traffic generated at eNB $n,\ n = 1, 2, \ldots, N_s$,
at Sm-GW $s$ for operator $o$.
The throughput level $T_{son}$ can for instance be obtained through
smoothing of the requests $\rho_{son}$ over successive cycles $w$.
In order to avoid clutter, we omit the subscripts $s$ and $o$ in the remainder
of this fairness evaluation.
We define the following fair target throughput levels $\Omega_n$ [bit/s]:
Lightly loaded eNBs $l \in \mathcal{L}$
with throughput levels $T_l < \Gamma/W$,
should be able to transmit their full traffic load, i.e., $\Omega_l = T_l$.
Next, consider highly loaded eNBs $h \in \mathcal{H}$ with throughput
levels $T_h > \Gamma/W$.
If the total throughput requirement of all eNBs
$\sum_{l \in \mathcal{L}} T_{l} +\sum_{h \in \mathcal{H}}T_h$
is less than or equal to the
uplink transmission bitrate $G$, then the highly loaded
eNBs should be able to transmit their full traffic load, i.e.,
$\Omega_h = T_h$.
On the other hand, if the total traffic load exceeds the uplink transmission
bitrate, i.e., if $\sum_{l \in \mathcal{L}} T_{l} +\sum_{h \in \mathcal{H}}T_h > G$, then
the highly loaded eNBs should be able to transmit traffic up to
an equitable share of the uplink
transmission bitrate not used by the lightly loaded eNBs.
Thus, overall:
$\Omega_h = \min \{T_h,\ (G - \sum\limits_{l \in \mathcal{L}} T_{l} )  /
	|\mathcal{H}| \}$. We define the normalized distance $\mathcal{E}_n$
of the actually achieved (observed) throughput $\tau_n$ and the
target throughput $\Omega_n$, i.e.,
$\mathcal{E}_n = \tau_n - \Omega_n$.

Based on the preceding target throughput definitions, we obtain
the normalized distance throughput fairness index~\cite{JainIndex}
\begin{equation}
\mathcal{F}_T =
\frac{ \sqrt{ \sum_{n = 1}^N \mathcal{E}^2_n}}
{\sqrt{\sum_{n =1}^N\Omega^2_n}},
\label{FairIndex}
\end{equation}
whereby $\mathcal{F}_T$ close to zero indicates
fair Sm-GW scheduling.

\subsection{Sm-GW Scheduling Overhead}
In configuration adaptive Sm-GW scheduling, a reconfiguration event,
i.e., an eNB connect or disconnect event, triggers the re-evaluation
of the grant size limit $\Gamma_{so}$, see Section \ref{gwsch:sec:equal}.
The new $\Gamma_{so}$ value is sent to all eNBs.
Since reconfiguration events occur typically only rarely, e.g., on
the time scale of minutes or hours, the overhead for
configuration adaptive scheduling is negligible.

Traffic adaptive Sm-GW scheduling requires each eNB $n$ to send a request
every cycle of duration $W$ seconds.
Upon reception of the requests from all $N_{s}$ eNBs,
the Sm-GW evaluates and sends the grants to the respective eNBs,
as illustrated in Fig.~\ref{fig_ExcessTiming}(a).
The requests and grants can typically be sent simultaneously, i.e., in
parallel, over the individual eNB-to-Sm-GW links.
Thus, within a cycle duration $W$,
the overhead amounts to the transmission delays of the request and
grant messages, the maximum round-trip propagation delay between eNBs and Sm-GW,
and the schedule processing delay at the Sm-GW.
For typical parameter settings, such as
70~Byte messages transmitted at 1~Gbps, up to 500~m eNB-to-Sm-GW
propagation distance, $W=1$~ms cycle duration, and schedule processing
delay on the order of microseconds, the overhead is less than half a
percent.

\begin{figure*}[t]
	\centering
	\centering
	\includegraphics[width=3.5in]{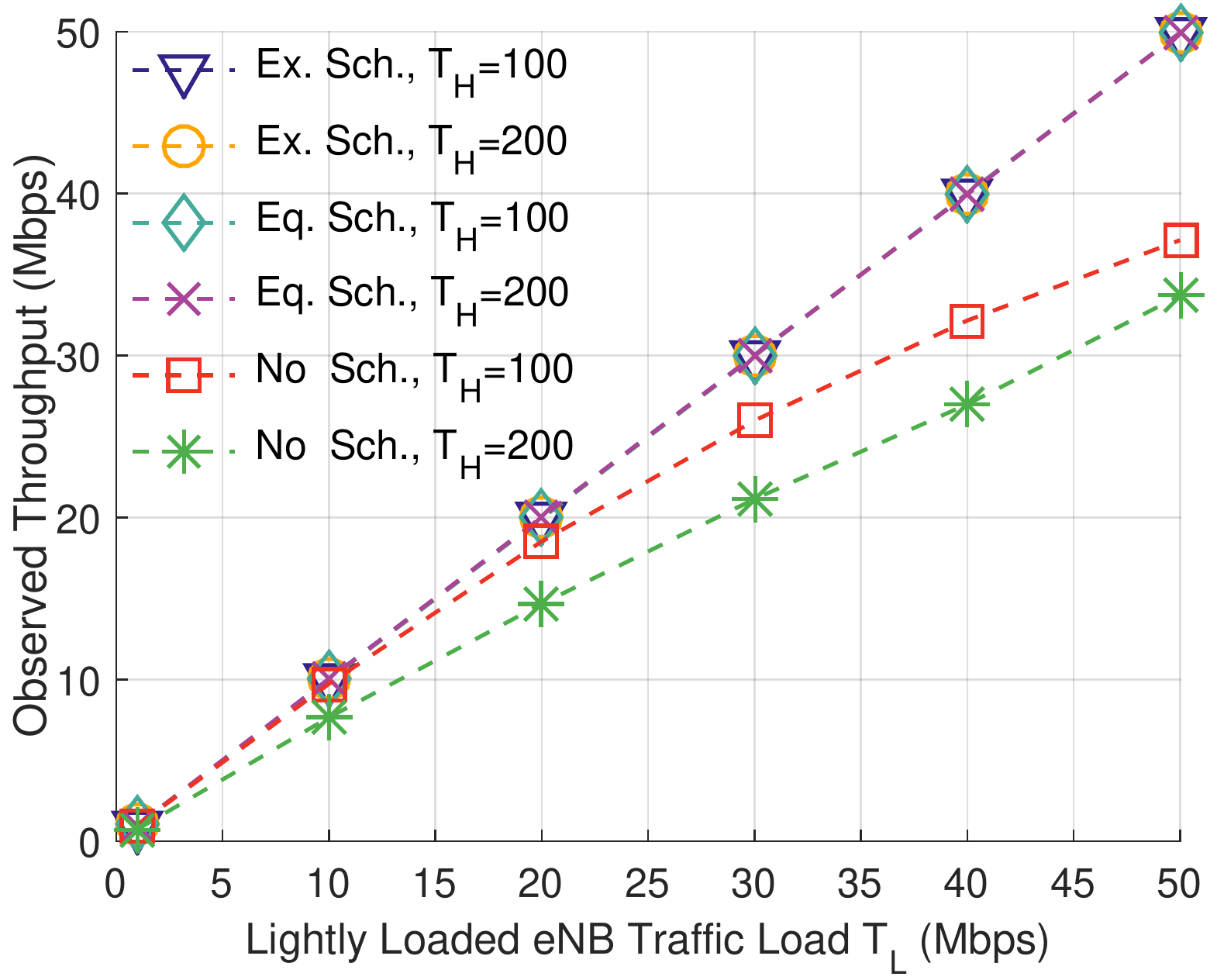}\\
	\footnotesize (a) Observed avg. throughput
	of lightly loaded eNBs  \\ \footnotesize
	\includegraphics[width=3.5in]{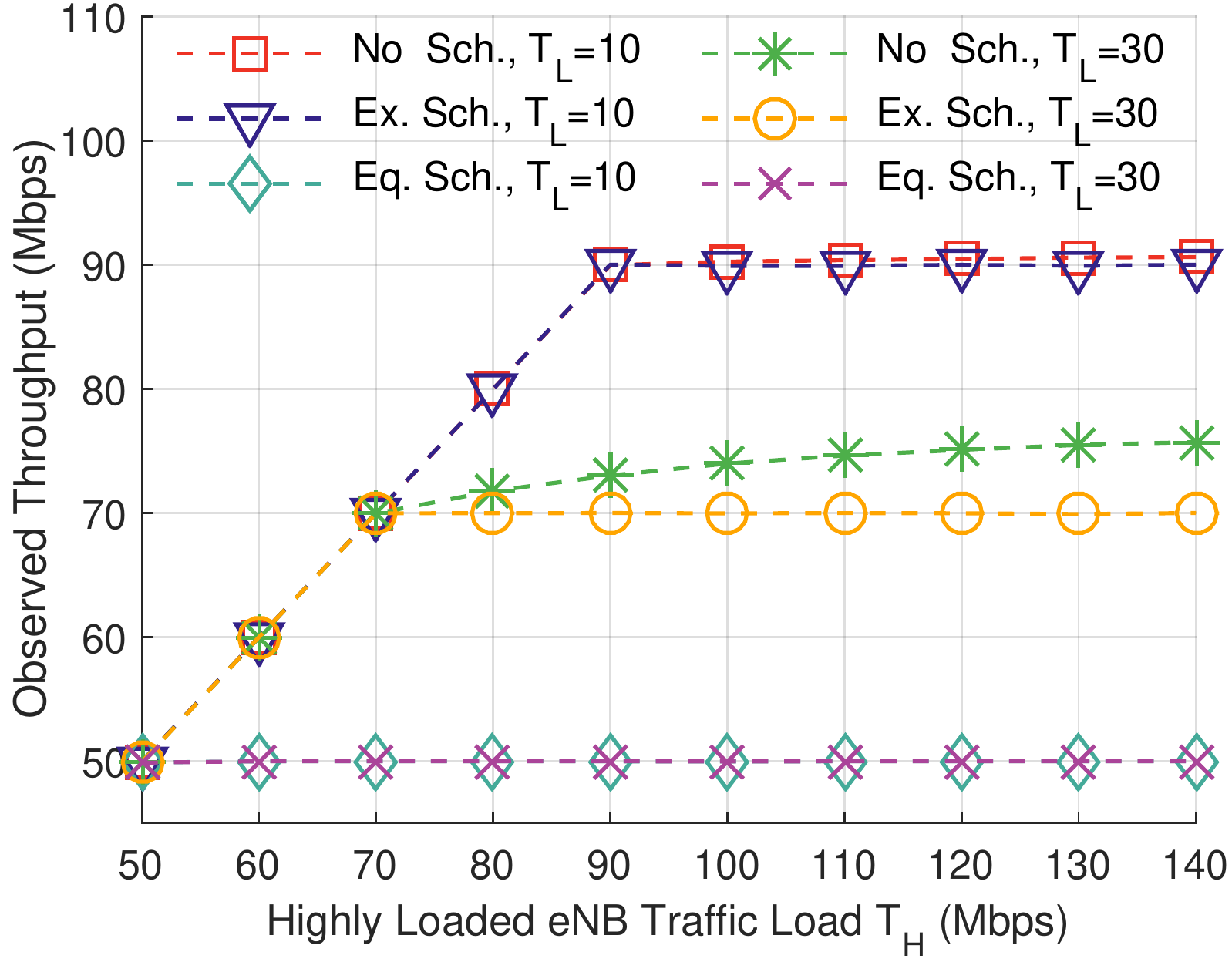} \\
	(b) Observed avg. throughput of highly loaded eNBs\\
	\label{fig_SmGW_sch2}
\end{figure*}

\begin{figure*}[t]
      \centering
      \centering
      \includegraphics[width=3.5in]{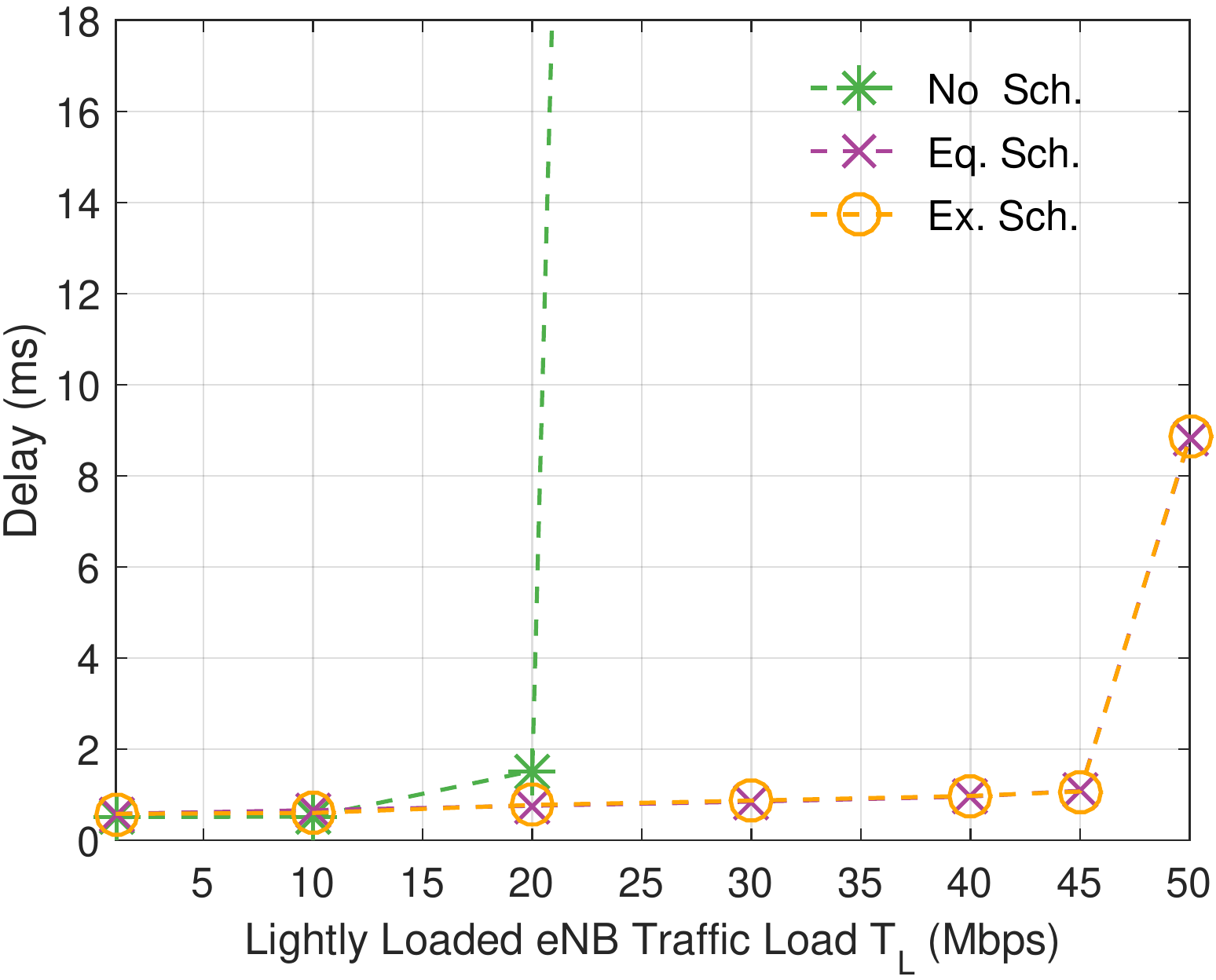}  \\
      \footnotesize (c) Avg. delay of lightly loaded $T_L$ eNBs,
      when $T_H = 80$ Mbps \\
      \includegraphics[width=3.5in]{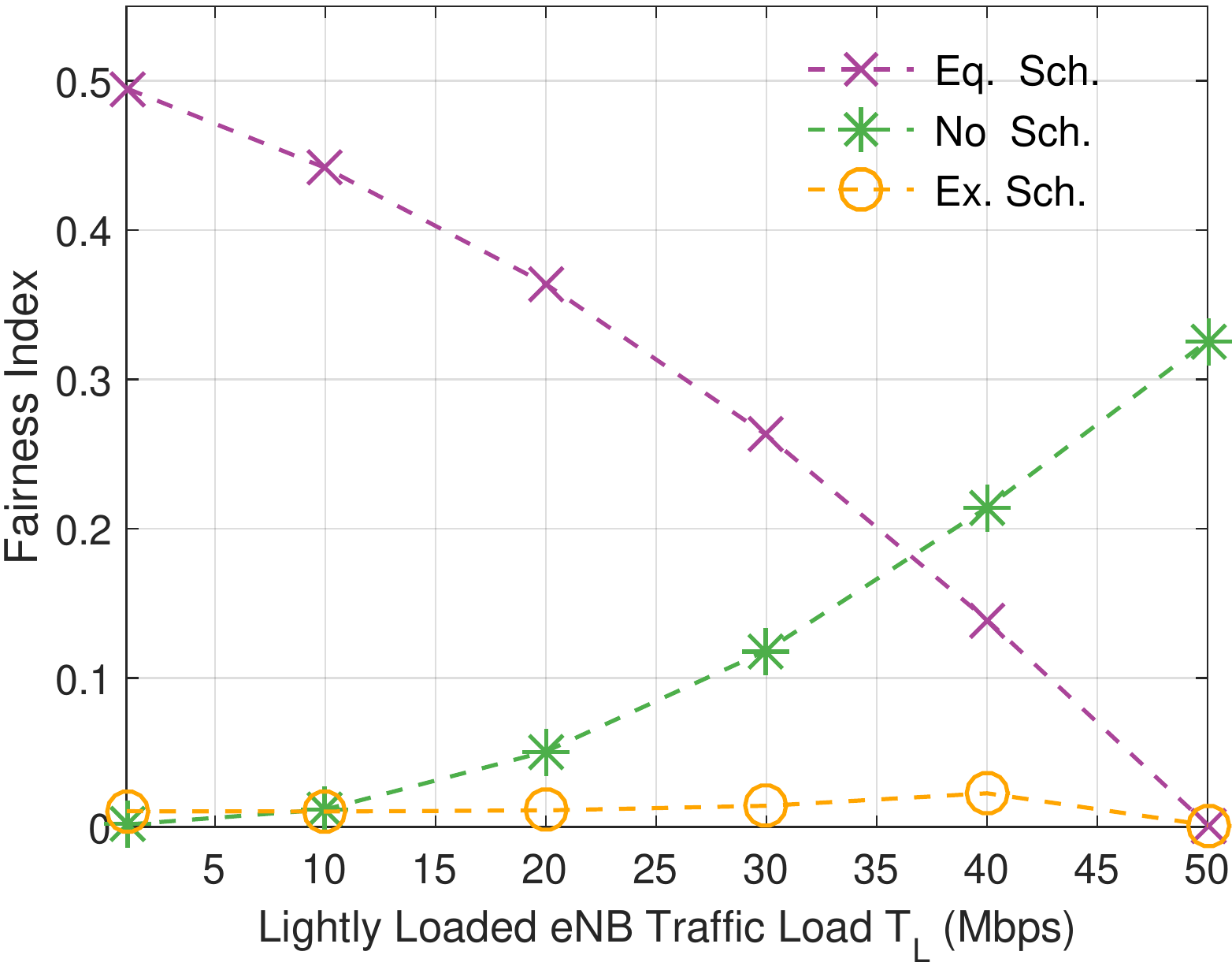} \\
      \footnotesize (d) Fairness Index $\mathcal{F}_T$,
      when $T_H$ = 200 Mbps.
      \caption{Simulation results for Sm-GW scheduling.}
      \label{fig_SmGW_sch}
\end{figure*}

\subsection{Evaluation of Sm-GW Scheduling}
\subsubsection{Simulation Setup}
We evaluate the performance of Sm-GW scheduling
with the discrete event simulator OMNET++.
We consider a given Sm-GW $s$ with an uplink
transmission bitrate to a given operator $o$ of
$G_{so} = 1$~Gbps. We omit the subscripts $s$ and $o$ in the remainder
of this evaluation section to avoid notational clutter.
The LTE access network typically requires the
packet delay to be less than 50~ms~\cite{3GPPQoS}.
Therefore, we set the Sm-GW queue size to
20~MBytes, which is equivalent to a maximum queuing delay of 20~ms
over the $G = 1$~Gbps link. Without any specific scheduling,
the Sm-GW operates in first-come-first-served mode with taildrop.

We simulate the typical bursty eNB traffic
generation pattern, with two eNB traffic rate states: low and heavy.
The sojourn time in
a given traffic rate state is randomly drawn from a
uniform distribution over 1~ms to 4~ms. At the end of the sojourn time,
a switch to another state occurs with a probability of 70~\%
in the low traffic state
and 30~\% in the heavy traffic state. The traffic bitrate
ratio between the heavy and low traffic states is $4:1$.
Within a given traffic rate state, data packets are randomly
generated according to independent Poisson processes.

We consider $|\mathcal{L}| = 10$ lightly
loaded eNBs and $|\mathcal{H}| = 10$ highly loaded eNBs connected to
the considered Sm-GW.
Each eNB, irrespective of whether it is lightly or highly loaded, generates
traffic according to the two traffic rate state (low and heavy) model.
The low and heavy traffic rates are set such that the long-run
average generated traffic rate corresponds to a prescribed required
throughput (load) level
$T_L < G/N = 50$~Mbps for a lightly loaded eNB and a prescribed required
throughput (load)
level $T_H > G/N$ for a highly loaded eNB.
For all simulations, the 95~\% confidence intervals are less
than 5~\% of the corresponding sample mean.

\subsubsection{Simulation Results}
\paragraph{Without Sm-GW Scheduling}
In Fig.~\ref{fig_SmGW_sch}, we show representative evaluation results
comparing configuration adaptive equal-share Sm-GW scheduling
and traffic adaptive excess-share Sm-GW scheduling with the
conventional backhaul without Sm-GW scheduling.
Figs.~\ref{fig_SmGW_sch}(a) and (b) show the actual (achieved, observed)
throughput
$\tau$ of lightly loaded and highly loaded eNBs, respectively, as a function
of the generated lightly loaded ($T_L$) and highly loaded ($T_H$)
throughput levels.
We observe from Figs.~\ref{fig_SmGW_sch}(a) that without scheduling,
the lightly loaded eNB suffer reductions in the achieved throughput, that
are especially pronounced (over 30~\%) for the high $T_H = 200$~Mbps
load of the highly loaded eNBs.
At the same time, we observe from Figure~\ref{fig_SmGW_sch}(b)
that without scheduling, the highly loaded eNBs achieve more than
their fair throughput share.
For instance, for the highly loaded eNB throughput requirement (load)
$T_H = 140$~Mbps, and $T_L = 30$~Mbps, the observed throughout of the highly
loaded eNBs is $\tau_H = 76$~Mbps, which is significantly higher than
the fair share of $(G - |\mathcal{L}| T_L) /|\mathcal{H}| = 70$~Mbps.
The unfairness arising without scheduling is further illustrated in
Fig.~\ref{fig_SmGW_sch}(c), where we observe a sharp delay increase
at $T_L = 20$~Mbps, when the total traffic load
$|\mathcal{L}| T_L + |\mathcal{H}| T_H$ approaches the uplink transmission
bitrate $G$.
Moreover, from Fig.~\ref{fig_SmGW_sch}(d), we observe an increasing
fairness index $\mathcal{F}_T$ as the lightly loaded eNBs generate more
traffic, i.e., as $T_L$ increases.
That is, as the lightly loaded eNBs try to transmit more traffic,
their achieved throughput falls more and more below their
fair share [see growing divergence between
the no scheduling curves and straight lines for
scheduling in Fig.~\ref{fig_SmGW_sch}(a)], leading to increasingly
unfair treatment of the lightly loaded eNBs.

\paragraph{Equal-share Sm-GW Scheduling}
We observe from Fig.~\ref{fig_SmGW_sch}(a) and (c) that
lightly loaded eNBs benefit from equal-share scheduling in that they
get the full share of their fair target throughput and experience
low delay.
However, we observe from Fig.~\ref{fig_SmGW_sch}(b) that
highly loaded eNBs achieve only a throughput of
$G/(|\mathcal{L}| + |\mathcal{H}|) = 50$~Mbps as
equal-share Sm-GW scheduling assigns a configuration adaptive allocation of
equal shares of the limited uplink transmission bitrate $G$ to all
eNBs irrespective of their traffic generation rates.
Correspondingly, we observe from Fig.~\ref{fig_SmGW_sch}(d), a high
fairness index $\mathcal{F}_T$ for low traffic loads of the
lightly loaded eNBs, as the highly loaded eNBs receive only
unfairly small shares of the uplink transmission bitrate $G$.

\paragraph{Excess-share Sm-GW Scheduling}
We observe from Fig.~\ref{fig_SmGW_sch}(a) and (b)
that with excess-share Sm-GW scheduling, both
lightly loaded eNBs and highly loaded eNBs
achieve their fair target throughput.
We further observe from Figs.~\ref{fig_SmGW_sch}(c) and (d)
that excess-share Sm-GW scheduling gives also
favorable delay and fairness index performance.

\paragraph{Summary}
We conclude that scheduling
of the Sm-GW uplink transmission bitrate $G$ is necessary to prevent backhaul
bandwidth starvation of lightly loaded eNBs due to the overwhelming
traffic rates of highly loaded eNBs. On the other hand,
simple configuration adaptive
allocation of equal uplink transmission bitrate
shares to each eNB wastes bandwidth.
Flexible traffic adaptive scheduling according to the
traffic loads of the eNBs, e.g., through excess-share
scheduling, can ensure fairness while efficiently utilizing the
uplink transmission bitrate.

\section{\MakeUppercase{SDN Based Multi-Operator Management}}  \label{MOM:sec}
\subsection{Overview}
In this section we introduce a novel SDN based network management
framework for flexible sharing of the backhaul resources of
multiple operators.
In particular, the framework
introduced in Sections~\ref{rap:sec}--\ref{optex:sec}
allows a set of Sm-GWs to flexibly share the
uplink transmission bitrate of a given single operator.
The inter-operator sharing introduced in Section~\ref{roam:sec}
allows the Sm-GWs to flexibly share the uplink transmission bitrates
of multiple operators.
Our proposed multi-operator management framework accommodates
dynamic changes of the traffic requirements of the small cells, such as
changes of the generated uplink traffic bitrates, as well as dynamic changes of
the operator characteristics, such as changes of the available
uplink traffic bitrates.
In the proposed multi-operator management framework,
an SDN orchestrator dynamically
configures the Sm-GWs and the transport network connecting
the Sm-GWs to the operator gateways to flexibly adapt to changes in
small cell traffic loads and the operator characteristics.

\subsection{Request and Allocation Procedures}  \label{rap:sec}
In a small cell deployment environment, such as a large organization,
multiple Sm-GWs can serve multiple buildings.
For example, in a university setting,
a library can be equipped with an Sm-GW
and the administration building can be equipped with another Sm-GW.
The throughput requirements and priorities
of these buildings typically vary widely over time.
For instance, the administration building experiences a large
visitor influx during graduation and student admission periods, while
many students visit the library during exam week.
Moreover, services from multiple operators may need to be
shared among the buildings in a given organization,
i.e., among multiple Sm-GWs.
Hence, there is a need for highly flexible traffic management
within the large organization based on time-varying priorities
and throughput requirements.

\begin{figure*}[t!]
    \centering
    \includegraphics[width=1\textwidth]{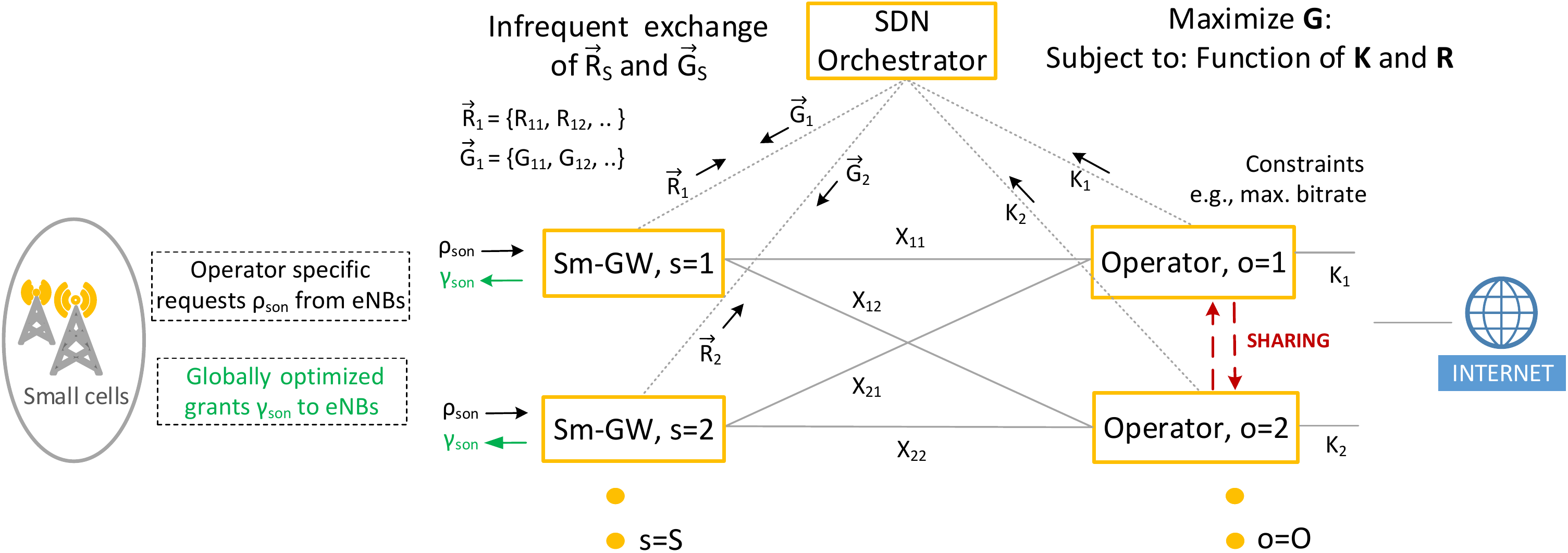}  
    \caption{Illustration of SDN based multiple-operator management
serving multiple Smart Gateways (Sm-GWs).}
    \label{fig_OPmanagement}
\end{figure*}

Suppose, the Sm-GWs $s,\ s = 1, 2, \ldots, S$, and
operators $o,\ o = 1, 2, \ldots, O$, are interconnected
in a full mesh transport network, as illustrated in Fig.~\ref{fig_OPmanagement}.
As described in Section~\ref{gwsch:sec:excess}, with traffic adaptive
Sm-GW scheduling, each eNB $n$ sends its operator $o$ specific uplink
transmission bitrate request to Sm-GW $s$ in every cycle.
The requested uplink transmission data amounts $\rho_{son}$
will typically vary over time scales that are long enough to
reasonably permit adaptations of the Sm-GW configurations.
For instance, the requests will typically change
 on the time scales of several seconds or minutes, or possibly even
longer, such as the seasonal variations in the visitor volume to
university buildings.
In order to obtain the variational characteristics of the
eNB requirements, the operator specific requests at the Sm-GWs can be
aggregated over the eNBs and appropriately smoothed, i.e., averaged over time,
to obtain an aggregate smoothed uplink transmission bitrate request
$R_{so}$ [bit/s] from Sm-GW $s$ to operator $o$.

Ideally, the backhaul network should adapt to varying requirements at
the Sm-GWs to maximize the network utilization.
We exploit the centralized control property of SDN to
adaptively configure the network for variable requirements.
More specifically, the SDN orchestrator in Fig.~\ref{fig_OPmanagement}
optimizes the allocations $G_{so}$ of operator $o$
uplink transmission bitrate [bit/s] to the individual Sm-GWs $s$.
The SDN orchestrator thus ensures that the grants to
the eNBs are globally maximized (subject to the operators' constraints
and requirements).  When the optimized allocations $G_{so}$ are used at the
Sm-GW scheduler (Section~\ref{gwsch:sec}),
the maximum allowed traffic flow is sent from the Sm-GWs
to each operator core.

\subsection{Optimization Decision Variables and Constraints for Multi-Operator Management with Sm-GWs}
In this section,
  we define a general optimization model for the multi-operator
management framework. Specifically, we define the constraints and
decision variables for optimizing the multi-operator management.
The defined decision variables and constraints are employed
for the operation of the SDN orchestrator,
as detailed in Section~\ref{sdnorchpro:sec}.
The SDN orchestrator can employ arbitrary objective functions and
constraint allocation strategies for the optimization, as illustrated
for an elementary example in Section~\ref{optex:sec}.

\subsubsection{Constraints}
Requests for the uplink transmission of $\rho_{son}$ [bits] from
eNBs $n,\ n = 1, 2, \ldots, N_s$, arrive at Sm-GW $s,\ s = 1, 2, \ldots, S$,
every cycle of duration $W$ seconds, i.e., on the order of milliseconds,
requesting uplink transmission bitrates from operator $o,\ o = 1, 2, \ldots, O$.
The Sm-GW aggregates the requests over the eNBs $n$ and smoothes the
aggregated requests to obtain the smoothed aggregated requests $R_{so}$.
Denoting $w$ for the cycle index, an elementary
weighted sampling smoothing computes
\begin{equation}
  R_{so}(w) = \alpha\; \left( \frac{1}{N_s}
  \sum_{n=1}^{N_s}  \frac{\rho_{son}(w)}{W}  \right)
   + (1-\alpha)\; R_{so}(w-1),
\label{filteredReports}
\end{equation}
where $\alpha$ denotes the weight for the most recent request sample.
A wide variety of other smoothing mechanism can be employed and optimized
according to the specific deployment settings.
The smoothed requests $R_{so}$ are periodically (with a period typically
much longer than the eNB reporting window) sent to the
SDN orchestrator.
In particular, each Sm-GW $s$, sends a vector of smoothed requests
$\overrightarrow{R_{s}} = [R_{s1} \; R_{s2} \; \cdots \;R_{sO}]$
containing the aggregated and smoothed requests for each operator $o$
to the SDN orchestrator.
The SDN orchestrator combines the request vectors $\overrightarrow{R_{s}}$ to
form the request matrix
\begin{equation}  \label{reportMatrix}
\mathbf{R} = [ R_{so} ],\ s = 1, 2, \ldots, S;\ o = 1, 2, \ldots, O.
\end{equation}

Each operator $o,\ o = 1, 2, \ldots, O$, can enforce
a set of constraints $K_{oc},\ c = 1, 2, \ldots, C$,
represented by a constraint vector
 $\overrightarrow{K_{o}} = [K_{o1} \; K_{o2} \; \cdots \;K_{oC}]$
that is sent to the SDN orchestrator.
Each constraint $c$ may be associated with a particular specification from
operator $o$, e.g., for traffic shaping of the flows or for the aggregate
maximum bitrate.
In order to avoid clutter and not to obscure the main ideas of our
overall multi-operator management framework, we consider in this study
a single constraint for each operator $o$.
That is, in place of the constraint vector $\overrightarrow{K_{o}}$
we consider a single (scalar) constraint $K_o$.
The SDN orchestrator combines the scalar constraints from the
various operators $o$ to form the constraint vector
\vspace{0cm}
\begin{equation}
\mathbf{K} = [K_{1} \; K_{2} \; \cdots \;K_{O}].
\label{constrantsVec}
\end{equation}

\subsubsection{Decision Variables}
The Sm-GW $s$ scheduler uses the operator $o$ specific grant size limits
$\Gamma_{so}$ to schedule/assign uplink transmission grants to eNBs
(see Sections~\ref{gwsch:sec:equal} and~\ref{gwsch:sec:excess}). By
controlling the variable $\Gamma_{so}$ specific to operator $o$ we
can control the flow of traffic outward from the Sm-GW, i.e.,
towards the respective operator $o$. The long-term average traffic
flow rates $X_{so}$ [bit/s] from the Sm-GW $s,\ s = 1, 2, \ldots, S$, to the
operators $o,\ o = 1, 2, \ldots, O$, can be expressed as matrix
\begin{equation}   \label{flowMatrix}
\mathbf{X} = [X_{so}],\ s = 1, 2, \ldots, S; o = 1, 2, \ldots, O.
\end{equation}

The operator $o$ specific uplink transmission bitrates
$G_{so}$ granted to the Sm-GWs are evaluated at the
SDN orchestrator, based on the
request matrix $\mathbf{R}$ and the constraint vector $\mathbf{K}$.
The orchestrator
responds to the request vector $\overrightarrow{R_s}$ from each
Sm-GW $s$ with a grant vector $\overrightarrow{G_s}$.
At the SDN orchestrator, the grant vectors $\overrightarrow{G_{s}}$
can be combined to form the orchestrator grant matrix
\begin{equation}   \label{GmaxMatrix}
\mathbf{G} = [G_{so}],\ s = 1, 2, \ldots, S; o = 1, 2, \ldots, O.
\end{equation}
$\mathbf{G}$ is a positive (non-negative) matrix, since the
matrix elements $G_{so},\ G_{so} \geq 0$, correspond granted
uplink transmission bitrates.

Our objective is to maximize the traffic flow rates $X_{so}$ from the
Sm-GWs $s$ to the operators $o$ subject to the operator
constraints $\mathbf{K}$.
In particular, the aggregated traffic
sent from the Sm-GWs $s,\ s = 1, 2, \ldots, S$,
to the operator $o$ core should satisfy the operator
constraint $K_o$, i.e.,
\begin{equation}
\sum_{s = 1}^S X_{so} \leq K_o,\ \ \ \forall o,\ o = 1, 2, \ldots, O.
\label{Eq:OpCnstGen}
\end{equation}

Using the grant vector $\overrightarrow{G_{s}}$ at Sm-GW $s$
to assign, i.e., to schedule, uplink traffic grants to the eNBs
(see Section~\ref{gwsch:sec}) ensures that the
traffic flow rates $X_{so}$ from Sm-GW $s$ to operator $o$ are bounded
by $G_{so}$, i.e., \begin{equation}
X_{so} \leq G_{so},\ \ \forall (s,o).
\label{eq:geqx}
\end{equation}
Thus, in order to ensure that the traffic flows
$X_{so}$ satisfy the operator constraints $\mathbf{K}$,
the grants $G_{so}$ must satisfy the operator constraints, i.e.,
\begin{equation}
\sum_{s = 1}^S G_{so} \leq K_o, \;\;\;\;\; \forall o,\ o = 1, 2, \ldots, O.
\label{Eq:OpCnstGenGrant}
\end{equation}
In order to maximize the traffic flows $X_{so}$ to each operator $o$,
the SDN orchestrator needs to grant each Sm-GW $s$
the maximum permissible uplink transmission bitrate $G_{so}$.

\begin{algorithm}[t!]
    \caption{SDN Orchestrator procedure}
    \label{algo:SDN}
    \SetKwInOut{Input}{input}
    \SetKwInOut{Output}{output}
    \nonl {\bfseries{1. Sm-GWs}} \newline
    \nl (a) Evaluate aggregate smoothed requests $R_{so}$ from
    eNB requests $\rho_{son}$,  Eqn~(\ref{filteredReports}). \newline
    (b) Periodically send request vector $\overrightarrow{R_s}$
     to SDN orchestrator
\newline
    \nonl \If{Grant vector $\overrightarrow{G_s}$ is received}{
     \nonl   Update SM-GW (to eNBs) grant size limits $\Gamma_{so}$,
    \nonl }

    \nonl   {\bfseries{2. Operators}} \newline
    (a) Send constraint $K_o$ to SDN orchestrator

    \nonl   {\bfseries{3. SDN Orchestrator}}   \newline
    \nonl     \If{Request vector $\overrightarrow{R_s}$ is received
   \textbf{OR} constraint $K_o$ is received}{
     \nonl   Re-optimize orchestrator (to Sm-GW) grants $\mathbf{G}$ \;
      \nonl  Send grant vector $\overrightarrow{G_s}$ to Sm-GW $s$;
    \nonl }

\end{algorithm}
\subsection{SDN Orchestrator Operation}  \label{sdnorchpro:sec}
The operational procedures for evaluating the
SDN orchestrator grant matrix $\mathbf{G}$ (\ref{GmaxMatrix})
are executed in parallel in the Sm-GWs, operators, and the SDN orchestrator,
as summarized in Algorithm~\ref{algo:SDN}.
The Sm-GWs aggregate and smooth the eNB requests and periodically
  send the request vector $\overrightarrow{R_s}$ to the SDN orchestrator.
The SDN orchestrator optimizes the grant matrix $\mathbf{G}$
upon the arrival of a new Sm-GW request vector $\overrightarrow{R_s}$ or
 a change in an operator constraint $K_o$.
The orchestrator updates the Sm-GWs with the newly
evaluated orchestrator grant vectors $\overrightarrow{G_s}$,
which update their grant size limits $\Gamma_{so}$.

Our SDN based multi-operator management framework
  allows for a wide variety of
resource (uplink transmission bitrate) allocations from the multiple operators
to the Sm-GWs.
In order to illustrate the introduced framework, we consider
next an elementary specific optimization problem formulation with
a linear objective function and a proportional constraint allocation
strategy that allocates the uplink transmission bitrate constraints
proportional to the requests.
More complex objective functions and allocation strategies,
e.g., objective functions
that prioritize specific grants are an interesting direction for
future research.
We note that this illustrative example does not exploit inter-operator
sharing, which is examined in Section~\ref{roam:sec}.

\subsection{Illustrative Optimization Example with Linear Objective
  Function and Request-Proportional Constraint Allocations}  \label{optex:sec}
Since the grants $G_{so}$ are non-negative, an elementary objective
function can linearly sum the grants $G_{so}$, i.e., as
$\sum_{s = 1}^S \sum_{o = 1}^O G_{so}$.
For the constraint allocation,
we consider the aggregate over all Sm-GWs $s$ of the
aggregated smoothed requests $R_{so}$
for a specific operator $o$, i.e., we consider
the unit norm of the request vector
$\| \overrightarrow{R_o} \|_1 = \sum_{s = 1}^S R_{so}$.
If $\| \overrightarrow{R_o} \|_1$ is less than the operator constraint $K_o$,
then the corresponding grants $G_{so}$ are set to the requests,
i.e., $G_{so} = R_{so}$.
On the other hand, if $\| \overrightarrow{R_o} \|_1 > K_o$,
then we proportionally assign the operator $o$ backhaul bandwidth $K_o$, i.e.,
we assign the proportion $R_{so} / \| \overrightarrow{R_o} \|_1$
of the constraint $K_o$. Thus,
\begin{equation}
G_{so} = \min \left( R_{so},\
    \frac{R_{so}}{\|\overrightarrow{R_o}\|_1} K_o\right).
\label{Eq:propGrantThresh}
\end{equation}
The resulting elementary optimization problem can be summarized as:
\vspace{0cm}
\begin{equation}
\begin{split}
\text{Maximize} \;\;\; \sum_{s = 1}^S \sum_{o = 1}^O G_{so}\;\;\;\;
\;\;\;\;\;\;\;\;\;\;\;\;\;\;\;\;\;\;\\
\text{Subject to:} \;\;\;\;\;\;\;\;\;\;\;\:\:\;\;\;\;\;\;\;\;
\;\;\;\;\;\;\;\;\;\;\:\:\;\;\;\;\\
\forall s \;\in\; \{ 1, 2,\ldots,S\}\;\;\; \text{and}
\;\;\;\forall o \; \in \; \{ 1, 2,\ldots,O\},\\
-G_{so} \leq 0,\;\;\;\;\;\;\;\:\;\;\;\;\;\;\;\;
\;\;\;\:\;\;\;\;\;\;\;\;\;\\
\;\;\; \;G_{so} \leq R_{so}, \;\;\;\;\;\;\;\;\;\;\;
\:\:\;\;\;\;\;\;\;\;\;\;\;\;\\
G_{so} \leq K_{o} \frac{R_{so}}{\| \overrightarrow{R_o}\|_1 }.
\;\;\;\;\;\;\;\;\;\:\;\;\;\;\;\:
\end{split}
\label{optimization}  \end{equation}

\subsection{Inter-Operator Sharing}
\label{roam:sec}
When the aggregate backhaul bandwidth $\|R_o\|_1$ requested from
an operator $o$ exceeds its constraint $K_o$, inter-operator sharing
can be employed to route the additional traffic through the
 network managed by another operator.
Our proposed Sm-GW multi-operator management
provides a distinctive advantage in
maintaining active connections with multiple operators to
easily route the excess traffic to a sharing operator.
We denote $o = m$ for the operator
that accepts the sharing traffic from an other operator $o = e$ whose
traffic constraint has been exceeded.
In this study, we focus on one operator accepting sharing traffic and
one operation with excess traffic.
The extension to sets of multiple operators accepting
sharing traffic and multiple operators with excess traffic
is left for future research. An operator in sharing $m$ should have
low incoming traffic as compared to the constraints $K_m$ in order to accept
the traffic from the operator in excess $e$. Therefore, for
the sharing ($o = m$) and excess ($o = e$) operators the requests
$R_{so}$ need to satisfy,
\begin{equation}
\sum_{s = 1}^S R_{sm} < K_m,\ \ \mbox{and}\ \
\sum_{s = 1}^S R_{se} > K_e.    \label{Eq:op:class}
\end{equation}
The traffic rate from excess operator~$e$ that
can be carried by sharing operator $m$ depends on the
unutilized slack uplink transmission bitrate of operator~$m$:
\begin{equation}
\begin{split}
\zeta = K_m - \sum_{s = 1}^S R_{sm}.
\end{split}
\label{eq:op:ex}
\end{equation}
If $\zeta > 0$, the last constraint in optimization problem
(\ref{optimization}) for the excess operator $e$ is replaced by the constraint
\begin{equation}
G_{se} \leq \left(K_{e} + \zeta\right)\frac{R_{se}}
   {\| \overrightarrow{R_e}\|_1 }\ \ \ \forall s.
\end{equation}

\begin{figure}[t!]  \centering
    \includegraphics[width=3.2in]{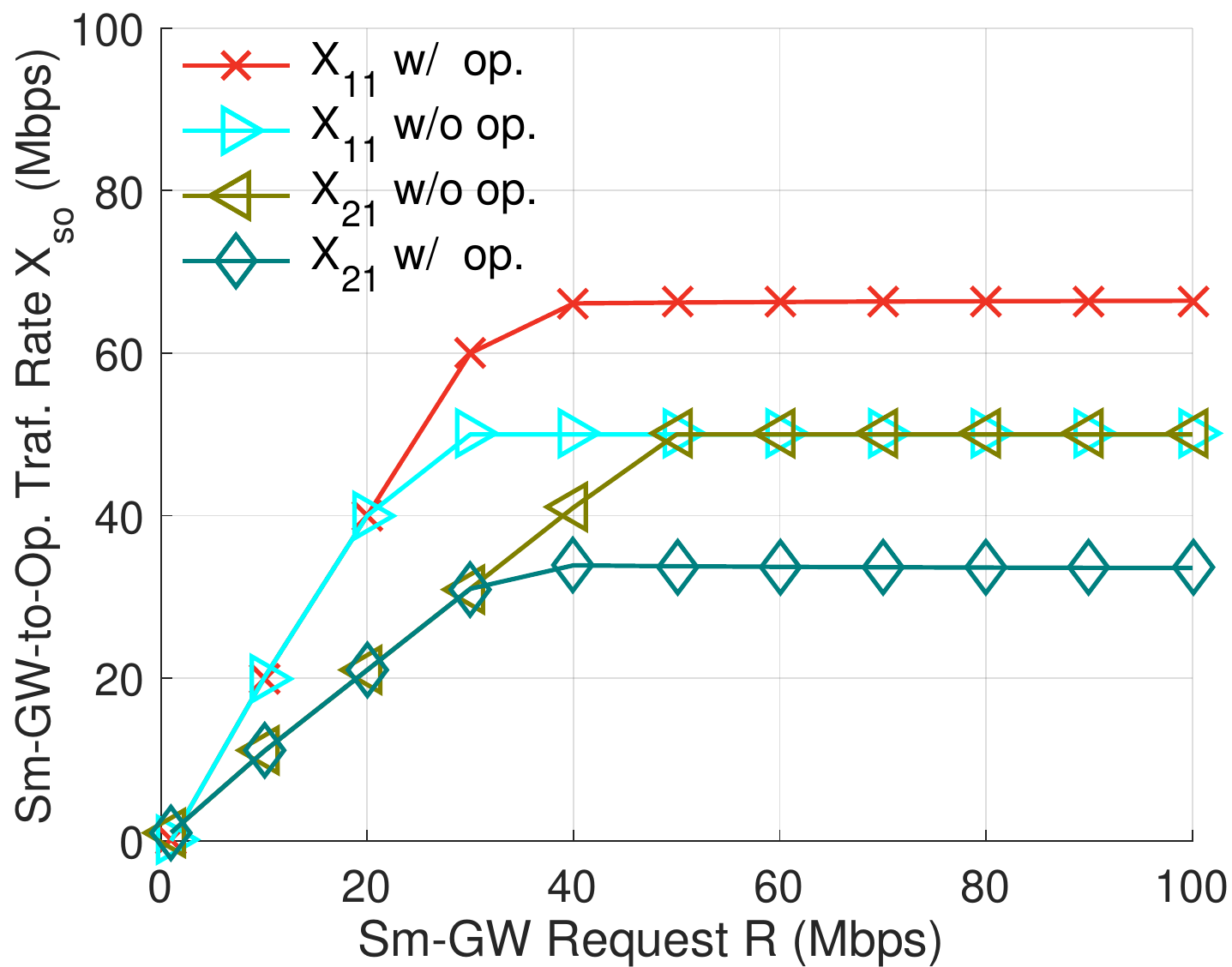}  
    \caption{Traffic rates simulation.}
    \label{fig_opti1}  \end{figure}

\begin{figure*}[t!]
	\centering
    \includegraphics[width=0.9\textwidth]{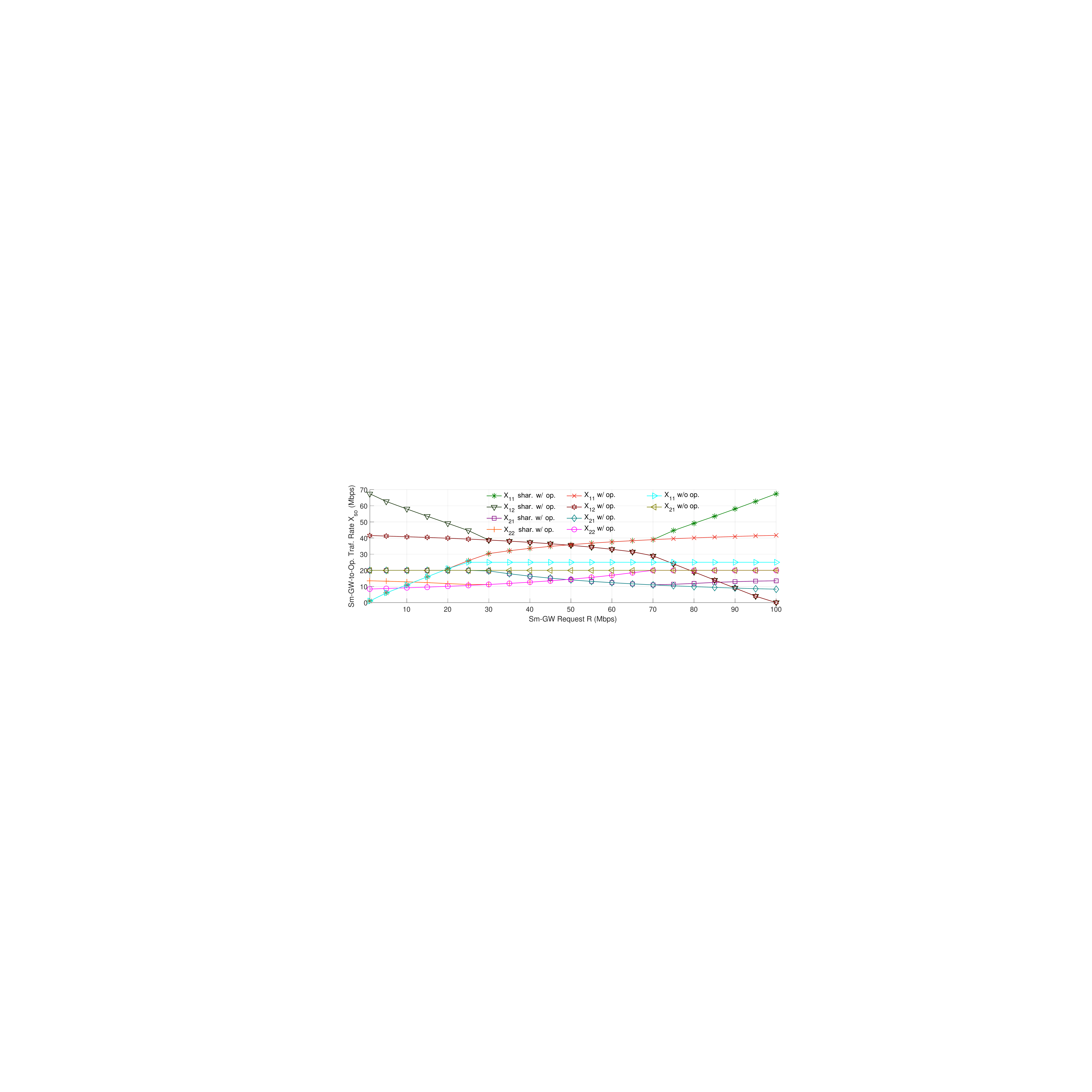}
	\caption{Inter-operator sharing evaluation.}
	\label{fig_roam2}
\end{figure*}

\subsection{Evaluation of Multi-Operator Management} \label{OMres:sec}
In order to showcase the effectiveness of the SDN based
multi-operator management framework,
we conducted simulations for the elementary optimization with linear
objective function and proportional constraint sharing
(see Section~\ref{optex:sec}).
We consider $S=2$ Sm-GWs and $O=2$ operators.
As comparison benchmark, we consider a static equal allocation of
operator uplink transmission bitrate $K_o$ to the $S$ Sm-GWs, i.e.,
each Sm-GW $s,\ s = 1, 2$, is allocated $K_o / S$ of the operator $o$
uplink transmission bitrate.

\subsubsection{Without Inter-Operator Sharing}
In Fig.~\ref{fig_opti1} we plot the Sm-GW $s$ to operator $o$
traffic flow rates $X_{so}$ resulting from the optimized
SDN orchestrator grants $G_{so}$ as a function of
the uplink transmission bitrate requested by Sm-GWs $s = 1$ and $s = 2$
from operator $o = 1$. Specifically, Sm-GW $s = 1$ requests bitrate
$R_{11} = 2R$ and Sm-GW $s = 2$ requests bitrate $R_{21} = R$ from operator
$o = 1$. The bitrate requests from operator $o = 2$ are fixed at 50~Mbps.
Each operator $o,\ o = 1, 2$,
has uplink transmission bitrate constraint $K_o = 100$~Mbps.

We observe from Fig.~\ref{fig_opti1} that for requests for
 operator $o = 1$ bitrate up to $R = 25$~Mbps,
the traffic rates $X_{11}$ and $X_{21}$ are equal to the requests, irrespective
of whether SDN orchestrated optimization is employed for not.
In contrast, as the requested bitrate increases above $R = 25$~Mbps,
i.e., the bitrate $R_{11} = 2R$ requested by Sm-GW $s = 1$ from operator
$o = 1$ increases above $K_1 / S = 50$~Mbps, the granted bitrate $G_{11}$
with SDN orchestration and the corresponding traffic flow $X_{11}$
continue to increase. On the other hand, the granted bitrate $G_{11}$ and traffic
flow $X_{11}$ without SDN orchestration stay limited at the static
equal share $X_{11} = G_{11} = K_1 / S = 50$~Mbps.

As the requested bitrate $R$ increases above 33.3~Mbps, i.e., a total of
$3R = 100$~Mbps requested from operator $o = 1$,
we observe from Fig.~\ref{fig_opti1} that without orchestration,
the traffic flow $X_{21} $ from Sm-GW $s = 2$ to operator $o = 1$
grows to and then remains at the static equal share $K_1 / S = 50$~Mbps.
That is, the conventional static uplink transmission bitrate allocation
results in unfair disproportional backhaul service.
In contrast, our dynamic multi-operator management with
SDN orchestrated optimization based on proportional sharing adaptively
assigns the operator $o = 1$ bitrate to Sm-GWs $s=1$ and $s = 2$
proportional to their requests.

\subsubsection{With Inter-Operator Sharing} \label{sec:withroaming}
In Fig.~\ref{fig_roam2}, we plot the Sm-GW $s$ to operator $o$
traffic flow rates $X_{so}$
as a function of the uplink transmission bitrate $R_{11} = R$ requested by
Sm-GW $s = 1$ from operator $o = 1$ when inter-operator sharing is employed.
Sm-GW
$s = 1$ requests bitrate $R_{12} = 100 - R$~Mbps from operator $o = 2$.
Also, Sm-GW $s = 2$ requests fixed bitrates
$R_{21} = R_{22} = 20$~Mbps from each operator. Each operator $o$ has a fixed
uplink transmission bitrate constraint of $K_o = 50$~Mbps. Note that operator
$o = (m) = 1$ has slack uplink transmission bitrate
when $R \leq 30$~Mbps and can thus serve
as roaming operator for the excess traffic to operator $e = 2$.
As $R$ increases and starts to exceed 70~Mbps, the roles are reversed, so
that operator $e = 1$ can send some of its excess traffic to roaming
operator $m = 2$.

Focusing initially on the case $R = 100$~Mbps, i.e., the right edge of
Fig.~\ref{fig_roam2}, we observe that without SDN orchestrated optimization,
Sm-GW $s = 1$ can only transmit at its fixed static allocation of
$X_{11} = K_1 / S = 25$~Mbps to operator $o = 1$, even though
Sm-GW $s = 1$ has a traffic load demanding $R_{11} = R = 100$~Mbps.
At the same time, Sm-GW $s = 2$ transmits at its requested rate
$X_{21} = R_{21}  = 20~\mbox{Mbps} < K_1/S$. Thus, the operator $o = 1$
uplink transmission bitrate $K_1$ is underutilized,
even though Sm-GW $s = 1$ has
more traffic to send, but cannot due to the inflexible static
uplink transmission bitrate allocations.

With SDN orchestrated optimization with proportional sharing,
the overloaded uplink transmission bitrate
$K_1 = 50$~Mbps of operator $o = 1$ is shared between
the two Sm-GWs, allowing Sm-GW $s = 1$ to transmit
$X_{11} = R_{11} / (R_{11} + R_{21}) = 41.7$~Mbps, while Sm-GW $s = 2$
transmits $X_{21} = R_{21} / (R_{11} + R_{21}) = 8.3$~Mbps.
However, the uplink transmission bitrate
$K_2$ of operator $o = 2$ is underutilized
with only Sm-GW $s = 2$ transmitting $X_{22} = 20$~Mbps.

With inter-operator sharing, the unutilized uplink transmission bitrate
$\zeta = K_2 - X_{22} = 30$~Mbps
of operator $o = 2$, is used to carry excess traffic from
operator $o = 1$.
In particular, the aggregate of the regular operator $o = 1$ uplink transmission
bitrate $K_1$ and the uplink transmission bitrate available to operator $o = 1$ through
traffic sharing to operator $o = 2$ ($\zeta = 30$~Mbps),
i.e., $K_1 + \zeta = 80$~Mbps is available to operator $o = 1$.
With proportional sharing, Sm-GW $s = 1$ can transmit
$X_{11} = (K_1 + \zeta) R_{11} / (R_{11} + R_{21}) = 66.7$~Mbps, while Sm-GW $s = 2$
can correspondingly transmit $X_{21} = 13.3$~Mbps, fully utilizing
the backhaul capacities of both operators.

Overall, we observe from Fig.~\ref{fig_roam2} that across the
entire range of traffic loads $R$ from Sm-GW $s = 1$ for operator $o = 1$,
our SDN based multi-operator orchestration with sharing is able to fully
utilize the uplink transmission bitrates of both operators. Note in
particular, that depending on the traffic load, the roles
of the two operators (excess or sharing) are dynamically adapted.

\section{\MakeUppercase{Conclusions}}
We have introduced a new backhaul architecture by inserting a novel
Smart Gateway (Sm-GW) between the wireless base stations (eNBs) and
the conventional operator gateways, e.g., LTE S/P-GWs. The Sm-GW enables
flexible support for large numbers of small cell base stations. In
particular, the Sm-GW adaptively schedules uplink
backhaul transmission grants to the individual eNBs on a
fast (typically millisecond) timescale. In addition, an SDN
orchestrator adapts the allocation of the uplink transmission bitrate of the
conventional gateways of multiple operators to the Sm-GWs on a slow
(typically minutes or hours) time scale. Simulation results have
demonstrated that the scheduling of eNB grants by the Sm-GW can
greatly improve the backhaul service over conventional
static backhaul uplink transmission bitrate allocations. Moreover, the
SDN orchestrator substantially improves the utilization of the
backhaul bandwidth, especially when inter-operator sharing is permitted.

There are several important directions for future research on the
Sm-GW architecture and small cell backhaul in general.
One direction is to examine a variety of scheduling algorithms
in the context of the Sm-GW.
Another direction
is to examine different specific optimization objective functions
within the general SDN orchestrator optimization introduced in this
article.
Moreover, it is of interest to investigate
QoS strategies for different traffic types, such as
data, voice, and video traffic \cite{li2009ene, see2015wvs, pul2013tra,van2008tra, rei2002tra}.

				    	\chapter{\MakeUppercase{Software Defined Optical Networks (SDONs): A Comprehensive Survey}}

\section{\MakeUppercase{Introduction}} \label{Sec1_Intro}
At least a decade ago \cite{ComerNetMgmtBook} it was recognized that new
network abstraction layers for network control functions needed to be
developed to both simplify and automate network management. Software
Defined Networking (SDN)~\cite{HuHB14,jai2013b4,Vau11} is the design
principle that emerged to structure the development of those new
abstraction layers.
Fundamentally, SDN is defined by three architectural
principles~\cite{SDNarch11,KreRVR15}:
$(i)$ the separation of control plane functions
and data plane functions, $(ii)$ the logical centralization of control, and
$(iii)$ programmability of network functions. The first two
architectural principles are related in that they combine to allow for network
control functions to have a wider perspective on the network. The idea is that
networks can be made easier to manage (i.e., control and monitor) with a
move away from significantly distributed control. A tradeoff is then
considered that balances ease of management arising from control
centralization and scalability issues that naturally arise from that
centralization.

\begin{figure*}[t!]	\centering
	\includegraphics[width=6in]{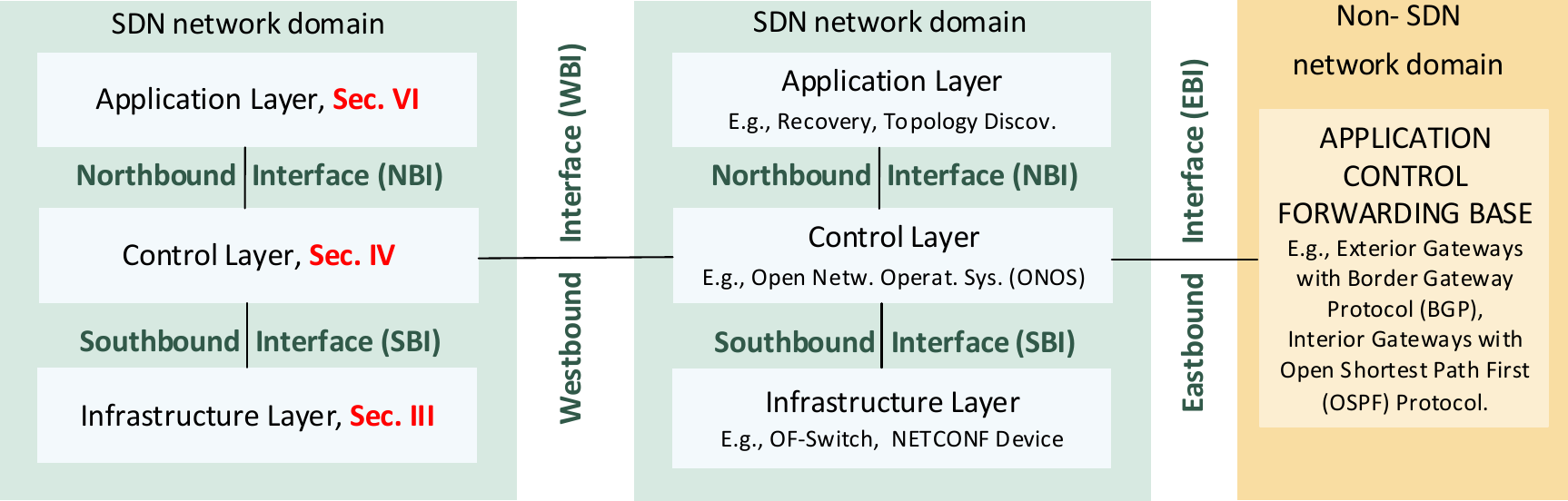}
    \caption{Illustration of Software Defined Networking (SDN) abstraction
     layers.}
	\label{fig:sdnlayers}
\end{figure*}

The SDN abstraction layering consists of three generally accepted
layers~\cite{SDNarch11} inspired by computing systems,
from the bottom layer to the top layer:
$(i)$ the \textit{infrastructure} layer, $(ii)$ the \textit{control} layer, and
($iii$) the \textit{application} layer,
as illustrated in Fig.~\ref{fig:sdnlayers}.
The interface between the
application layer and the control layer is referred to as the
NorthBound Interface (NBI), while the interface between the control layer
and the infrastructure layer is referred to as the SouthBound
Interface (SBI). There are a variety of standards emerging for these
interfaces, e.g., the OpenFlow protocol~\cite{LaKR14} for the SBI.

The \textit{application} layer is modeled after software applications
that utilize computing resources to complete tasks.
The \textit{control} layer is modeled after a computer's Operating System
(OS) that manages computer resources (e.g., processors and memory),
provides an abstraction layer to simplify interfacing with the
computer's devices, and provides a common set of services that all
applications can leverage. Device drivers in a computer's OS hide the
details of interfacing with many different devices from the
applications by offering a simple and unified interface for various
device types. In the SDN model both the unified SBI
as well as the control layer functionality provide
the equivalent of a device driver for interfacing with devices in the
\textit{infrastructure} layer, e.g., packet switches.

Optical networks play an important role in our modern information
technology due to their high transmission capacities.
At the same time, the specific optical (photonic) transmission
and switching characteristics, such as circuit, burst, and packet
switching on wavelength channels, pose challenges for controlling
optical networks.
This article presents a comprehensive survey of Software
Defined Optical Networks (SDONs).
SDONs seek to leverage the flexibility of
SDN control for supporting networking applications with an
underlying optical network infrastructure.
This survey comprehensively covers SDN related mechanisms that have
been studied to date for optical networks.

\subsection{Related Work}
The general principles of SDN have been extensively covered in several
surveys, see for instance,~\cite{aky2014roa,Chen2015a,cui2016big,far2015soft, FeRZ14, HuHB14, JaP13,
  jar2014sur, JaZH14, kha2015jons, KreRVR15, LaKR14, li2014sof, Lopes2015,MiCK13,
  NuMNO14,Racherla,Trois2016,van2014sca, wic2015sof, XiWF14}.
SDN security has been surveyed in~\cite{ahm2015sec,sco2015sur}, while
management of SDN networks has been surveyed in~\cite{wic2015sof} and
SDN-based satellite networking is considered in~\cite{ber2015sof}.

To date, there have been relatively few overview and survey articles
on SDONs.
Zhang et al.~\cite{zha2013surflexi} have presented a thorough survey
on flexible optical networking based on Orthogonal Frequency
Division Multiplexing (OFDM) in core (backbone) networks. The survey
briefly notes how OFDM-based elastic networking can facilitate
network virtualization and surveys a few studies on OFDM-based
network virtualization in core networks.

Bhaumik et al.~\cite{BhZCS14} have presented an overview of
SDN and network virtualization concepts and outlined principles
for extending SDN and network virtualization concepts to
the field of optical networking.
Their focus has been mainly on industry efforts, reviewing white papers
on SDN strategies from leading networking companies, such as
Cisco, Juniper, Hewlett-Packard, Alcatel-Lucent, and Huawei.
A few selected academic research projects on general SDN optical
networks, namely projects reported in the journal
articles~\cite{ChNS13,liu2013field} and a few related conference papers,
have also been reviewed by Bhaumik et al.~\cite{BhZCS14}.
In contrast to Bhaumik et al.~\cite{BhZCS14}, we provide a comprehensive
up-to-date review of academic research on
SDONs. Whereas Bhaumik et al.~\cite{BhZCS14}
presented a small sampling of SDON research organized by research projects,
we present a comprehensive SDON survey that is organized according
to the SDN infrastructure, control, and application layer architecture.

For the SDON sub-domain of access networks,
Cvijetic~\cite{Cvi14} has given an overview of access network challenges
that can be addressed with SDN.
These challenges include lack of support for on-demand modifications of
traffic transmission policies and rules and limitations to
vendor-proprietary policies, rules, and software.
Cvijetic~\cite{Cvi14} also offers a very brief overview of
research progress for SDN-based optical access networks, mainly focusing on
studies on the physical (photonics) infrastructure layer.
Cvijetic~\cite{CviSept14} has further expanded the overview of
SDON challenges by considering the incorporation of 5G wireless
systems. Cvijetic~\cite{CviSept14} has noted that
SDN access networks are highly promising for low-latency and high-bandwidth
back-hauling from 5G cell base stations and briefly surveyed
the requirements and areas of future research required for
integrating 5G with SDON access networks.
A related overview of general software defined access networks
based on a variety of physical transmission media, including copper
Digital Subscriber Line (DSL)~\cite{dslbook}
and Passive Optical Networks (PONs), has been
presented by Kerpez et al.~\cite{KeCGG14}.

Bitar~\cite{Bi14} has surveyed use cases for SDN controlled broadband access,
such as on-demand bandwidth boost, dynamic service
re-provisioning, as well as value-add services and service protection.
Bitar~\cite{Bi14} has discussed the commercial perspective of the
access networks that are enhanced with SDN to add
cost-value to the network operation.
Almeida Amazonas et al.~\cite{AmSS14} have surveyed the key issues
of incorporating SDN in optical and wireless access networks.
They briefly outlined the obstacles posed by
 the different specific physical characteristics of
optical and wireless access networks.

Although our focus is on optical networks, for completeness we note that
for the field of wireless and mobile networks, SDN based networking mechanisms
have been surveyed
in~\cite{ArSR15,BeDSB14,haq2016wir,jag2014sof,SaCKA15,soo2015sof,yan2014sof}
while network virtualization has been surveyed in~\cite{LiY15} for
general wireless networks and in~\cite{kha2015wir,lia2015wir} for
wireless sensor networks.
SDN and virtualization strategies for LTE wireless cellular networks
have been surveyed in~\cite{ngu2015sdn}. SDN-based 5G wireless
network developments for mobile networks have been outlined
in~\cite{Huawei14,PeLZW14,TrGVS15,YaKO14}.

\subsection{Survey Organization}
We have mainly organize our survey according to the three-layer SDN architecture
illustrated in Fig.~\ref{fig:sdnlayers}.
In particular, we have organized the survey in a bottom-up manner,
surveying first SDON studies focused on the
infrastructure layer in Section~\ref{sdninfra:sec}.
Subsequently, we survey SDON studies focused on the control layer
in Section~\ref{sdnctl:sec}.
The virtualization of optical networks is commonly closely related to
the SDN control layer. Therefore, we survey SDON studies
focused on virtualization in Section~\ref{virt:sec},
right after the SDON control layer section.
Resuming the journey up the layers in Fig.~\ref{fig:sdnlayers},
we survey SDON studies focused on the application layer in
Section~\ref{sdnapp:sec}.
We survey mechanisms for the overarching orchestration of the application
layer and lower layers, possibly across multiple network domains
(see Fig.~\ref{fig_control_orch}),
in Section~\ref{orch:sec}.
Finally, we outline open challenges and future research directions in
Section~\ref{sec:open} and conclude the survey in Section~\ref{sec:conclusion}.

\section{\MakeUppercase{Background}}
\label{bg:sec}
This section first provides background on
Software Defined Networking (SDN), followed by background on virtualization
and optical networking.
SDN, as defined by the Internet Engineering
Task Force (IETF)~\cite{rfc7426},
is a networking paradigm enabling the programmability of networks.
SDN abstracts and separates the data forwarding
plane from the control plane, allowing faster
technological development both in data and control planes.
We provide background on the SDN architecture, including its architectural
layers in Subsection~\ref{sdnarch:sec}.
The network programmability provides the flexibility to dynamically initialize,
control, manipulate, and manage the end-to-end network behavior via open
interfaces, which are reviewed in Subsection~\ref{sdnint:sec}.
Subsequently, we provide background on network virtualization in
Subsection~\ref{bgnetvit:sec}
and on optical networking in Subsection~\ref{bg_access:sec}.

\subsection{Software Defined Networking (SDN) Architectural Layers}
\label{sdnarch:sec}
SDN offers a simplified view of the underlying network infrastructure
for the network control and monitoring applications
through the abstraction of each independent network layer.
Fig.~\ref{fig:sdnlayers} illustrates the
three-layer SDN architecture model consisting of application,
control, and infrastructure layers as defined by the
Open Networking Foundation (ONF)~\cite{SDNarch11}. The ONF is
the organization that is responsible for
the publication of specifications for the OpenFlow protocol.
The OpenFlow protocol~\cite{HuHao2014,keo2008of,LaKR14}
has been the first protocol for the SouthBound Interface (SBI,
also referred to as Data-Controller Plane Interface (D-CPI))
between the control and infrastructure layers.
Each layer operates independently, allowing multiple solutions
to coexist within each layer, e.g., the
infrastructure layer can be built from any
programmable devices, which are commonly referred to as
network elements~\cite{SDNarch11_521} or network devices~\cite{rfc7426}
(or sometimes as forwarding elements~\cite{rfc3746}).
We will use the terminology network element throughout this survey.
The SouthBound Interface (SBI) and the NorthBound
Interface (NBI, also referred to as Application-Controller Plane
Interface (A-CPI)) are defined as the primary interfaces
interconnecting the SDN layers through abstractions.
An SDN network architecture can coexist with both
concurrent SDN architectures and non-SDN legacy network architectures.
Additional interfaces are defined namely the
EastBound Interface (EBI) and the WestBound
Interface (WBI)~\cite{JaZH14} to interconnect the SDN architecture
with external network architectures
(the EBI and WBI are also collectively
referred to as Intermediate-Controller Plane Interfaces (I-CPIs)).
Generally, EBIs establish communication links to
legacy network architectures (i.e., non-SDN networks); whereas,
links to concurrent (side-by-side) SDN architectures are facilitated
by the WBIs.

\subsubsection{Infrastructure Layer}
The infrastructure layer includes an environment for
(payload) data traffic forwarding (data plane)
either in virtual or actual hardware.
The data plane comprises a network of network elements,
which expose their capabilities through the SBI
to the control plane. In traditional networking, control mechanisms are
embedded within an infrastructure,
i.e., decision making capabilities are embedded within the
infrastructure to perform network actions, such as switching or routing.
Additionally, these forwarding actions in the traditional network elements
are autonomously
established based on self-evaluated topology information that is often
obtained through proprietary vendor-specific algorithms.
Therefore, the configuration setups of traditional network elements
are generally not reconfigurable without a service disruption,
limiting the network flexibility.
In contrast, SDN decouples the autonomous control functions, such as
forwarding algorithms and neighbor discovery of
the network nodes, and moves these control functions out of the infrastructure
to a centrally controlled logical node, the controller.
In doing so, the network elements act only as dumb switches which
act upon the instructions of the controller. This decoupling reduces the
network element complexity and improves reconfigurability.

In addition to decoupling the
control and data planes, packet modification capabilities at the line-rates
of network elements have been significantly improved with SDN.
P4~\cite{bosshart2014p4} is a programmable
protocol-independent packet processor, that can arbitrarily
match the fields within any formatted packet
and is capable of applying any arbitrary actions (as programmed)
on the packet before forwarding. A similar forwarding mechanism,
Protocol-oblivious Forwarding (PoF) has been
proposed by Huawei Technologies~\cite{Song2013}.

\subsubsection{Control Layer}
The control layer is responsible for programming (configuring)
the network elements (switches) via the SBIs.
The SDN controller is a logical entity that identifies the south bound
instructions to configure the network infrastructure
based on application layer requirements.
To efficiently manage the network, SDN controllers can
request information from the SDN infrastructures,
such as flow statistics, topology information, neighbor relations,
and link status from the network elements (nodes).
The software entity that implements the SDN
controller is often referred to as \textit{Network Operating System (NOS)}.
Generally, a NOS can be implemented independently of
SDN, i.e., without supporting SDN.
On the other hand, in addition to supporting SDN operations,
a NOS can provide advanced capabilities, such as virtualization,
application scheduling, and database management.
The Open Network Operating System (ONOS)~\cite{onos}
is an example of an SDN based NOS
with a distributed control architecture designed to operate over
Wide Area Networks (WANs).
Furthermore, Cisco has recently developed the one Platform Kit
(onePK)~\cite{onepk}, which consists of a set of
Application Program Interfaces (APIs)
that allow the network applications to control Cisco network devices
without a command line interface.
The onePK libraries act as an SBI for Cisco ONE controllers and
are based on C and Java compilers.

\subsubsection{Application Layer}
The application layer comprises network applications and services
that utilize the control plane to realize
network functions over the physical or virtual infrastructure.
Examples of network applications include
network topology discovery, provisioning, and fault restoration.
The SDN controller presents an abstracted view of the
network to the SDN applications to facilitate the realization of
application functionalities.
The applications can also include higher levels of network management,
such as network data analytics, or specialized functions requiring processing
in large data centers. For instance, the Central Office Re-architected as a
Data center (CORD)~\cite{cord} is an SDN application based
on ONOS~\cite{onos},
that implements the typical central office network functions, such as
optical line termination, as well as BaseBand Unit (BBU) and
Data Over Cable Interface (DOCSIS)~\cite{fel2001doc} processing as
virtualized software entities, i.e., as SDN applications.

\begin{figure}[t!]
	\centering 	\vspace{0cm}
	\includegraphics[width=3.4in]{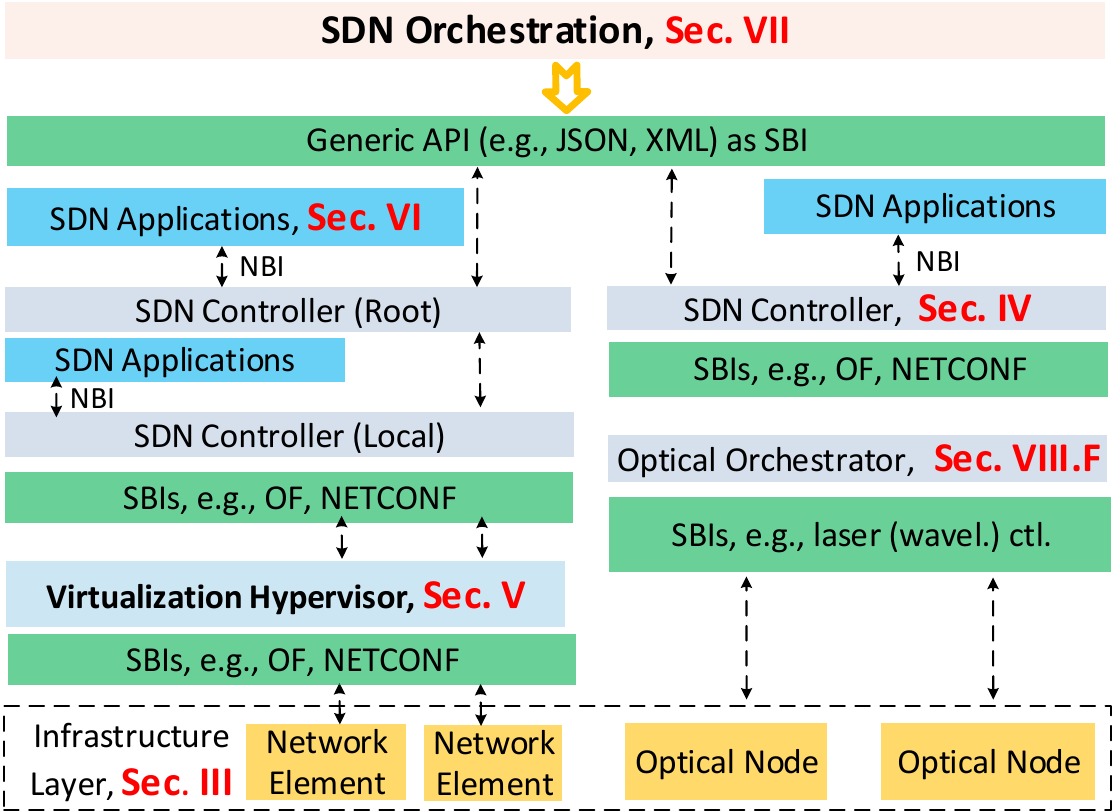}
    \caption{Overview of SDN orchestrator and SDN controllers.}
	\label{fig_control_orch}
\end{figure}
\subsubsection{Orchestration Layer}  \label{intro:orch:sec}
Although the orchestration layer is commonly not considered one of the
main SDN architectural layers illustrated in Fig.~\ref{fig:sdnlayers},
as SDN systems become more complex, orchestration becomes increasingly
important. We introduce therefore the orchestration layer as
an important SDN architectural layer in this background section.
Typically, an SDN orchestrator is the entity that coordinates
software modules within a single SDN controller,
a hierarchical structure of multiple SDN controllers, or a set of
multiple SDN controllers in a ``flat'' arrangement (i.e., without
a hierarchy) as illustrated in Fig.~\ref{fig_control_orch}.
An SDN controller in contrast
can be viewed as a logically centralized single
control entity. This logically centralized single control entity appears as the
directly controlling entity to the network elements.
The SDN controller is responsible for signaling the control actions or rules
that are typically predefined (e.g., through OpenFlow) to the network elements.
In contrast, the SDN orchestrator makes control
decisions that are generally not predefined.
More specifically, the SDN orchestrator could make an automated decision
with the help of SDN applications or seek a manual
recommendation from user inputs; therefore,
results are generally not predefined. These
orchestrator decisions (actions/configurations)
are then delegated via the SDN controllers
and the SBIs to the network elements.

Intuitively speaking, SDN orchestration can be viewed as a distinct
abstracted (higher) layer for coordination and management that is
positioned above the SDN control and application layers.  Therefore,
we generalize the term SDN orchestrator as an entity that realizes a
wider, more general (more encompassing) network functionality as
compared to the SDN controllers. For instance, a cloud SDN
orchestrator can instantiate and tear down Virtual Machines (VMs)
according to the cloud
workload, i.e., make decisions that span across multiple network
domains and layers.  In contrast, SDN controllers realize more
specific network functions, such as routing and path computation.

\subsection{SDN Interfaces}  \label{sdnint:sec}
\subsubsection{Northbound Interfaces (NBIs)}
A logical interface that interconnects the SDN controller and a
software entity operating at the application layer is commonly
referred to as a NorthBound Interface (NBI), or as Application-Controller Plane
Interface (A-CPI).

\paragraph{REST}
REpresentational State Transfer (REST)~\cite{REST15} is generally defined as
a software architectural style that supports flexibility, interoperability,
and scalability.
In the context of the SDN NBI, REST is commonly defined as
an API that meets the REST architectural style~\cite{LiChou2016},
i.e., is a so-called RESTful API:
\begin{itemize}
\item Client-Sever: Two software entities should follow the
  client-server model.  In SDN, a controller can be a server and the
  application can be the client.  This allows multiple
  heterogeneous SDN applications to coexist and operate over a common
  SDN controller.
\item Stateless: The client is responsible for managing all the states and
the server acts upon the client's request.
In SDN, the applications collect and maintain the states of the network, while
the controller follows the instructions from the applications.
\item Caching: The client has to support the temporary local storage
of information such that interactions between the
client and server are reduced so as to improve performance and scalability.
\item Uniform/Interface Contract:  An overarching technical
interface must be followed across all services using the REST API.
For example, the same data format, such as Java Script Object Notation (JSON)
or eXtended Markup Language (XML), has to be followed for all
interactions sharing the common interface.
\item Layered System: In a multilayered architectural solution,
  the interface should only be concerned with the next immediate node
  and not beyond. Thus, allowing more layers to be inserted, modified,
  or removed without affecting the rest of the system.
\end{itemize}

\subsubsection{Southbound Interfaces (SBIs)}
A logical interface that interconnects the SDN controller and the network
element operating on the infrastructure layer (data plane) is commonly
referred to as a SouthBound Interface (SBI), or as the
Data-Controller Plane Interface (D-CPI).
Although a higher level connection, such as a UDP or TCP connection, is
sufficient for enabling the communication between two entities of the SDN
architecture, e.g., the controller and the network elements,
specific SBI protocols have been proposed.
These SBI protocols are typically not interoperable
and thus are limited to work with SBI protocol-specific network elements
(e.g., an OpenFlow switch does not work with the NETCONF protocol).

\paragraph{OpenFlow Protocol}
The interaction between an OpenFlow switching element
(data plane) and an OpenFlow controller (control plane) is carried
out through the OpenFlow protocol~\cite{keo2008of,LaKR14}. This SBI (or D-CPI)
is therefore also sometimes referred to as the OpenFlow control channel.
SDN mainly operates through packet flows that are identified through
matches on prescribed packet fields that are specified in the
OpenFlow protocol specification. For matched packets, SDN switches
then take prescribed actions, e.g., process the flow's packets in a
particular way, such as dropping the packet, duplicating it on a different
port or modifying the header information.

\paragraph{Path Computation Element Protocol (PCEP)}
The PCEP enables communication between the Path Computation Client (PCC) of
the network elements and the Path Computation Element (PCE) residing within
the controller. The PCE centrally computes the paths based on constraints
received from the network elements. Computed paths are then forwarded to the
individual network elements through the PCEP protocol~\cite{rfc4655,rfc5440}.

\paragraph{Network Configuration Protocol (NETCONF) Protocol}
The NETCONF protocol~\cite{rfc6241}
provides mechanisms to configure, modify, and delete configurations
on a network device.
Configuration of the data and protocol messages are encoded in the NETCONF
protocol using an eXtensible Markup Language (XML).
Remote procedure calls are used to realize the NETCONF protocol
operations. Therefore, only devices that are enabled
with required remote procedure calls allow the NETCONF protocol to remotely
modify device configurations.

\paragraph{Border Gateway Protocol Link State Distribution (BGP-LS) Protocol}
The central controller needs a topology information database,
also known as Traffic Engineering Database (TED), for
optimized end-to-end path computation.
The controller has to request the information for building the TED,
such as topology and bandwidth utilization, via the SBIs
from the network elements.
This information can be gathered by a BGP extension,
which is referred to as BGP-LS.

\subsection{Network Virtualization} \label{bgnetvit:sec}
Analogously to the virtualization of computing
resources~\cite{gold1974sur,Douglis2013a},
network virtualization abstracts the underlying physical network
infrastructure so that one or multiple virtual networks
can operate on a given physical
network~\cite{bel2012res,duan2012sur,fis2013vir,han2015net,leo2003vir,mij2015net,pen2015gue,RajJain2013,wan2013net}.
Virtual networks can span over a
single or multiple physical infrastructures (e.g., geographically
separated WAN segments).
Network Virtualization (NV) can flexibly create independent virtual
networks (slices) for distinct
users over a given physical infrastructure.  Each network slice
can be created with prescribed resource allocations.
When no longer required, a
slice can be deleted, freeing up the reserved physical resources.

Network hypervisors~\cite{Khan2012,she2009flo}
are the network elements that abstract the
physical network infrastructure (including network elements,
communication links, and control functions) into logically isolated
virtual network slices.
In particular, in the case of an underlying physical SDN network,
an SDN hypervisor can create multiple isolated virtual SDN
networks~\cite{ble2015sur,dru2013sca}.
Through hypervisors, NV supports the implementation of a wide
range of network services belonging to the link and network protocol
layers (L2 and L3), such as switching and routing.
Additionally, virtualized infrastructures can also support
higher layer services, such as load-balancing of servers and firewalls.
The implementation of such higher layer services in a virtualized environment
is commonly referred to as Network Function Virtualization
(NFV)~\cite{haw2014nfv,LiChen2015,lin2016dem,mat2015tow,ye2016joi}.
NFV can be viewed as a special case of NV in which network
functions, such as address translation and intrusion detection functions,
are implemented in a virtualized environment. That is,
the virtualized functions are implemented in the form of software
entities (modules) running on a data center (DC) or
the cloud~\cite{Mijumbi2016}.
In contrast, the term NV emphasizes the virtualization of
the network resources, such as communication links and network nodes.

\subsection{Optical Networking Background}   \label{bg_access:sec}

\subsubsection{Optical Switching Paradigms}
Optical networks are networks that either maintain signals in the
optical domain or at least utilize transmission channels that carry
signals in the optical domain. In optical networks that
maintain signals in the optical domain, switching can be
performed at the \textit{circuit, packet, or burst} granularities.

\paragraph{Circuit Switching}
Optical \textit{circuit} switching can be performed in space,
waveband, wavelength, or time. The optical spectrum is divided into
wavelengths either on a fixed wavelength grid or on a flexible
wavelength grid. Spectrally adjacent wavelengths can be coalesced into
wavebands. The fixed wavelength grid standard (ITU-T G.694.1)
specifies specific center frequencies that are either 12.5~GHz, 25~GHz,
50~GHz, or 100~GHz apart. The flexible DWDM grid (flexi-grid) standard (ITU-T
G.694.1)~\cite{gon2015opt,jue2014sof,tom2014tut,zha2013surflexi}
allows the center frequency to be any multiple of 6.25~GHz
away from 193.1~THz and the spectral width to be any multiple of
12.5~GHz. Elastic Optical Networks (EONs)~\cite{cha2015rou,tal2014spe,yu2014spe}
that take advantage of the flexible
grid can make more efficient use of the optical spectrum but can cause
spectral fragmentation, as lightpaths are set up and torn down, the spectral
fragmentation counteracts the more efficient spectrum
utilization~\cite{Ger2012}.

\paragraph{Packet Switching}
Optical \textit{packet} switching performs packet-by-packet switching
using header fields in the optical domain as much as possible. An
all-optical packet switch requires~\cite{ram2009opt}:
\begin{itemize}
  \item Optical synchronization, demultiplexing, and multiplexing
  \item Optical packet forwarding table computation
  \item Optical packet forwarding table lookup
  \item Optical switch fabric
  \item Optical buffering
\end{itemize}
Optical packet switches typically relegate some of these design
elements to the electrical domain. Most commonly the packet forwarding
table computation and lookup is performed electrically. When there is
contention for a destination port, a packet needs to be buffered
optically, this buffering can be accomplished with rather impractical
fiber delay lines. Fiber delay lines are fiber optic cables whose lengths are
configured to provide a certain time delay of the optical signal; e.g.,
100 meters of fiber provides 500~ns of delay. An alternative to
buffering is to either drop the packet or to use deflection routing,
whereby a packet is routed to a different output that may or may not
lead to the desired destination.

\paragraph{Burst Switching}
Optical \textit{burst} switching alleviates the requirements of
optical packet forwarding table computation, forwarding table lookup,
as well as buffering while accommodating bursty traffic that would
lead to poor utilization of optical circuits. In essence, it permits
the rapid establishment of short-lived optical circuits to support the
transfer of one or more packets coalesced into a burst. A control
packet is sent through the network that establishes the lightpath for
the burst and then the burst is transmitted on the short-lived circuit
with no packet lookup or buffering required along the path~\cite{ram2009opt}.
Since the circuit is only established for the length of the burst, network
resources are not wasted during idle periods. To avoid any buffering
of the burst in the optical network, the burst transmission can begin
once the lightpath establishment has been confirmed (tell-and-wait) or
a short time period after the control packet is sent
(just-enough-time). \textit{Note}: Sending the burst immediately after
the control packet (tell-and-go) would require some buffering of the
optical burst at the switching nodes.

\subsubsection{Optical Network Structure}  \label{optnetstruct:sec}
Optical networks are typically structured into three main tiers, namely
access networks, metropolitan (metro) area networks,
and backbone (core) networks~\cite{sim2014opt}.

\paragraph{Access Networks}
In the area of optical access networks~\cite{for2015nex}, so-called Passive
Optical Networks (PONs), in particular, Ethernet PONs (EPONs) and
Gigabit PONs (GPONs)~\cite{haj2006epo,sku2009com},
have been widely studied.
A PON has typically an inverse tree structure with a central
Optical Line Terminal (OLT) connecting multiple distributed Optical
Network Units (ONUs; also referred to as Optical Network Terminals, ONTs)
to metro networks.
In the downstream (OLT to ONUs) direction, the OLT broadcasts transmissions.
However, in the upstream (ONUs to OLT) direction, the transmissions of the
distributed ONUs need to be coordinated to avoid collisions on the
shared upstream wavelength channel.
Typically, a cyclic polling based Medium Access Control (MAC) protocol,
e.g., based on the MultiPoint Control Protocol (MPCP, IEEE 802.3ah),
is employed.
The ONUs report their bandwidth demands to the OLT and the OLT then
assigns upstream transmission windows according to a Dynamic
Bandwidth Allocation (DBA)
algorithm~\cite{kan2012ban,mcg2010sho,mcg2012inv,zhe2009sur}.
Conventional PONs cover distances up to 20~km, while so-called
Long-Reach (LR) PONs cover distances up to
around 100~km~\cite{mer2013off,nag2016n,son2010lon}.

Recently, hybrid access networks that combine multiple transmission
media, such as Fiber-Wireless (FiWi)
networks~\cite{aur2014fiw,gha2011fib,liu2016new,sar2015arc,tsa2011sur} and
PON-DSL networks~\cite{gur2014pon}, have been explored to take
advantage of the respective strengths of the different transmission
media.

\paragraph{Networks Connected to Access Networks}
Optical access networks provide Internet connectivity for a wide range
of peripheral networks. Residential (home) wired or wireless
local area networks~\cite{che2014sur}
typically interconnect individual end devices (hosts) in a home or small
business and may connect directly with an optical access network.
Cellular wireless networks provide Internet access to a wide range of
mobile devices~\cite{cap2013dow,dam2011sur,sch2013pus}.
Specialized cellular backhaul
networks~\cite{LiPCYW14,LiZZW13,park2014fro,PeWLP15,raz2013bri,tip2011evo,YaLJS13} relay the traffic
to/from base stations of wireless cellular
networks to either wireless access
networks~\cite{aky2005wir,alo2012sur,ben2012wir,kur2007sur,pat2011sur,vij2013dis} or optical access networks.
Moreover, optical access networks are often employed to connect
Data Center (DC) networks to the Internet. DC networks interconnect highly
specialized server units
  that process and store large data amounts with specialized
networking technologies~\cite{cai2013sur,kac2012sur,sam2016sof,yan2016sud,zha2013sur}.
Data centers are
typically employed to provide the so-called  ``cloud'' services for
 commercial and social media applications.

\paragraph{Metropolitan Area Networks}
Optical Metropolitan (metro) Area Networks (MANs) interconnect the optical
access networks in a metropolitan area with each other and with
wide-area (backbone, core) networks.
MANs have typically a ring or star
topology~\cite{bia2013cos,bia2015sho,cha2013tow,mai2003hyb,rot2013rou,sch2003wav}
and commonly employ optical networking technologies.

\paragraph{Backbone Networks}
Optical backbone (wide area) networks interconnect the individual MANs
on a national
or international scale. Backbone networks have typically a mesh structure
and employ very high speed optical transmission links.
\begin{figure*}[t!]
\footnotesize
\setlength{\unitlength}{0.10in} 
\centering
\begin{picture}(40,35)
\put(-4,33){\textbf{SDN Controlled Photonic Communication Infrastructure Layer, Sec.~\ref{sdninfra:sec}}}
\put(-7,30){\line(1,0){50}}
\put(19,30){\line(0,1){2}}
\put(-7,30){\vector(0,-1){2}}

\put(-9,27){\makebox(0,0)[lt]{\shortstack[l]{			
	\textbf{Bandwidth Variable Transceivers (BVTs)},\\ Sec.~\ref{transc:sec}	\\	\\ 		
	Single-Flow BVTs~\cite{aut2013eva,ElA12,gri2010fle}, Sec.~\ref{SF-BVT:sec}\\
	Mach-Zehnder Modulator Based~\cite{Choi2013,Liu2013c}\\
	BVTs for PONs~\cite{LaAVKS14,IiSSK12,yeh2010usi,yu2008cen,VaBPF13,bol2014dig,Bolea2015a}\\
	BVTs for DC Netw.~\cite{Malacarne2014} \\ \\
	Sliceable Multi-Flow BVTs~\cite{jin2012mul}, \\ Sec.~\ref{MF-BVT:sec}	\\				
	Encoder Based~\cite{Sambo2014a,sam2015nex,sam2014sli} \\					
	DSP Based~\cite{Moreolo2016} \\					
	Subcar. $+$ Mod. Pool Based~\cite{Ou2016} \\					
	HYDRA~\cite{mat2015hyd}  
}}}

\put(18,30){\vector(0,-1){26}}

\put(9,3){\makebox(0,0)[lt]{\shortstack[l]{			
			\textbf{Space Div. Multipl.} \\
			\textbf{(SDM)-SDN}, \\ Sec.~\ref{SDM_SDN:sec}
			\cite{ama2013ful,ama2014sof,Galve2016}
}}}

\put(25,30){\vector(0,-1){2}}
\put(20,27){\makebox(0,0)[lt]{\shortstack[l]{						
			\textbf{Switching}, Sec.~\ref{infra_sw:sec} \\ \\ 				
			Switching Elements, Sec.~\ref{sw_elem:sec}	\\
			ROADMs~\cite{Co13,ama2013int,rof2013all,you2013eng,Way2013wav,Garrich2015}	\\			
			Open Transport Switch (OTS)~\cite{SaSPL13}	\\					
			Logical xBar~\cite{PaSMK13}	\\					
			Optical White Box~\cite{Nejabati2015}	\\					
			GPON Virt. Sw.~\cite{Lee2016,gu2014sof,Amokrane2014,amo2015dyn,Yeh2015}	\\				
			Flexi Access Netw. Node~\cite{FoG13,kon2015sdn} \\ \\
			Switching Paradigm, Sec.~\ref{ws_para:sec}	\\			
			Converged Pkt-Cir. Sw.~\cite{DaPM09,DaPMSG10,VeBB13,AzNEJ11,ShJKG12,KaT12,CeLR13}	\\					
			R-LR-UFAN~\cite{yin2013ult,ShYD14,Yin2015}	\\
			Flexi-grid~\cite{Cv13,CvTJSM14,OlSCH13,ZhZYY13}	
}}}	
\put(43,30){\vector(0,-1){26}}
\put(25, 3){\makebox(0,0)[lt]{\shortstack[l]{					
			\textbf{Opt. Perf. Monitoring}, Sec.~\ref{opm:sec} \\ \\
			Cognitive Netw. Infra.~\cite{MiDJF13,Oliveira2015,Giglio2015} \\
			Wavelength Selective Switch/Amplifier \\ \ \ Control~\cite{Moura2015,pao2015sup,Carvalho2015,Wang2015f}
}}}	
\end{picture}
\vspace{1cm}
\caption{Classification of physical infrastructure layer SDON studies.}
\label{infra_class:fig}
\end{figure*}
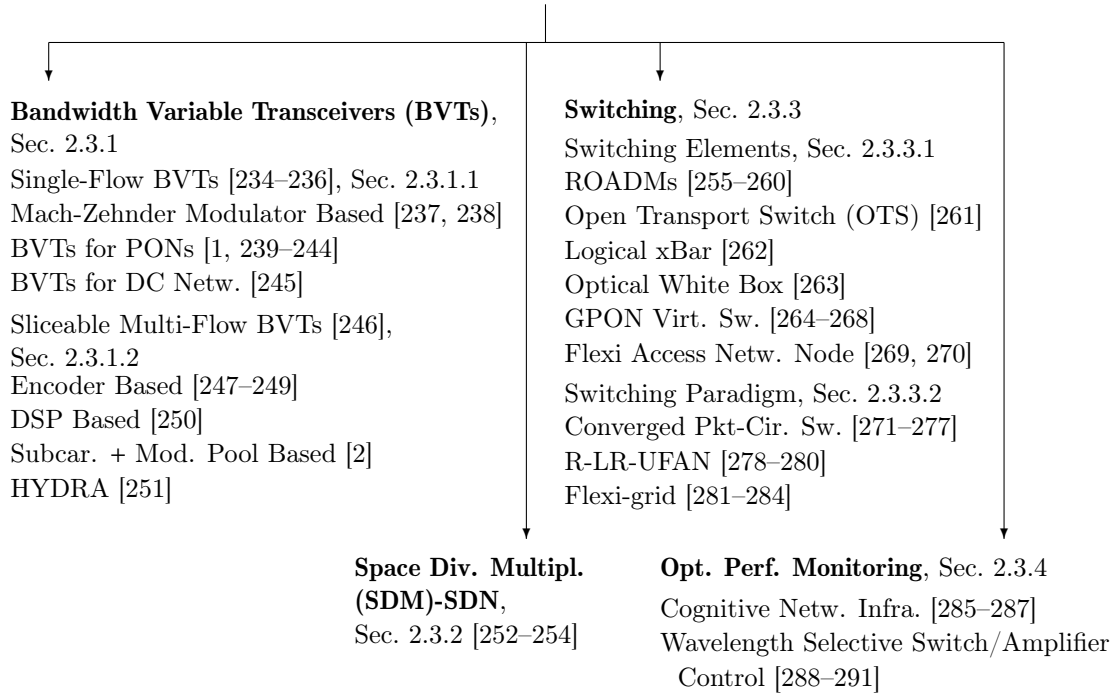
\section{\MakeUppercase{SDN Controlled Photonic Communication Infrastructure Layer}}
\label{sdninfra:sec}
This section surveys mechanisms for controlling physical layer
aspects of the optical (photonic) communication infrastructure through SDN.
Enabling the SDN control down to the photonic level operation of
optical communications allows for flexible adaptation of the
photonic components supporting optical networking
functionalities~\cite{ChNS13,gri2013ext,jin2013vir,RoD08}.
As illustrated in Fig.~\ref{infra_class:fig},
this section first surveys transmitters and receivers (collectively
referred to as transceivers or transponders) that permit SDN control of the
optical signal transmission characteristics, such as modulation format.
We also survey SDN controlled space division
multiplexing (SDM), which provides an emerging avenue for highly
efficient optical transmissions.
Then, we survey SDN controlled optical switching, covering first
switching elements and then overall switching paradigms, such as converged
packet and circuit switching.
Finally, we survey cognitive photonic communication infrastructures that
monitor the optical signal quality. The optical signal quality
information can be used to dynamically control the transceivers
as well as the filters in switching elements.

\subsection{Transceivers}  \label{transc:sec}
Software defined optical transceivers are optical transmitters and
receivers that can be flexibly configured by SDN to transmit or
receive a wide range of optical signals~\cite{Hillerkuss2016}.
Generally, software defined optical transceivers vary the modulation
format~\cite{win2006adv} of the transmitted optical signal by
adjusting the transmitter and receiver operation through Digital
Signal Processing (DSP)
techniques~\cite{cha2014adv,sch2010rea,yos2013dsp}. These
transceivers have evolved in recent years from Bandwidth Variable
Transceivers (BVTs) generating a single signal flow to sliceable
multi-flow BVTs. Single-flow BVTs permit SDN control to adjust the
transmission bandwidth of the single generated signal flow. In
contrast, sliceable multi-flow BVTs allow for the independent SDN
control of multiple communication traffic flows generated by a
single BVT.

\subsubsection{Single-Flow Bandwidth Variable Transceivers (BVTs)}
\label{SF-BVT:sec} Software defined optical transceivers have
initially been examined in the context of adjusting a single optical
signal flow for flexible WDM
networking~\cite{aut2013eva,ElA12,gri2010fle}. The goal has been to
make the photonic transmission characteristics of a given
transmitter fully programmable. We proceed to review a
representative single-flow BVT design for general optical mesh
networks in detail and then summarize related single-flow BVTs for
PONs and data center networks.

\paragraph{Mach-Zehnder Modulator Based Flexible Transmitter}
Choi and Liu et al.~\cite{Choi2013,Liu2013c} have demonstrated a
flexible transmitter based on Mach-Zehnder Modulators
(MZMs)~\cite{bar2003wid} and a corresponding flexible receiver for
SDN control in a general mesh network. The flexible transceiver
employs a single dual-drive MZM that is fed by two binary electric
signals as well as a parallel arrangement of two MZMs which are fed
by two additional electrical signals. Through adjusting the direct
current bias voltages and amplitudes of drive signals the
combination of MZMs can vary the amplitude and phase of the
generated optical signal~\cite{cho2012ber}. Thus, modulation formats
ranging from Binary Phase Shift Keying (BPSK) to Quadrature Phase
Shift Keying (QPSK) as well as 8 and 16 quadrature amplitude
modulation~\cite{win2006adv} can be generated. The amplitudes and
bias voltages of the drive signals can be signaled through an SDN OpenFlow
control plane to achieve the different modulation formats. The
corresponding flexible receiver consists of a polarization filter
that feeds four parallel photodetectors, each followed by an
Analog-to-Digital Converter (ADC). The outputs of the four parallel
ADCs are then processed with DSP techniques to automatically
(without SDN control) detect the modulation format. Experiments
in~\cite{Choi2013,Liu2013c} have evaluated the bit error rates and
transmission capacities of the different modulation formats and have
demonstrated the SDN control.

\paragraph{Single-Flow BVTs for PONs}
Flexible optical networking with real-time bandwidth adjustments
is also highly desirable for PON access and metro networks,
albeit the BVT technologies for access and metro networks should
have low cost and complexity~\cite{LaAVKS14}.
Iiyama et al.~\cite{IiSSK12} have developed a DSP based approach
that employs SDN to coordinate the downstream PON transmission of
On-Off Keying (OOK) modulation~\cite{yeh2010usi} and
Quadrature Amplitude Modulation (QAM)~\cite{yu2008cen} signals.
The OOK-QAM-SDN scheme involves a
novel multiplexing method, wherein all the data
are simultaneously sent from the OLT to the ONUs and the ONUs filter
the data they need.
The experimental setup in~\cite{IiSSK12} also demonstrated
digital software ONUs that concurrently transmit data by exploiting
the coexistence of OOK and QAM.
The OOK-QAM-SDN evaluations demonstrated the control of the receiving
sensitivity which is very useful for a wide range of transmission environments.

In a related study, Vacondio et al.~\cite{VaBPF13} have examined
Software-Defined Coherent Transponders (SDCT)
for TDMA PON access networks.
The proposed SDCT digitally processes the burst transmissions to
achieve improved burst mode transmissions according to the distance of a
user from the OLT.
The performance results indicate that the proposed flexible
approach more than doubles the average transmission capacity
per user compared to  a static approach.

\begin{figure*}[t]
    \centering
    \vspace{0cm}
    \includegraphics[width=6in]{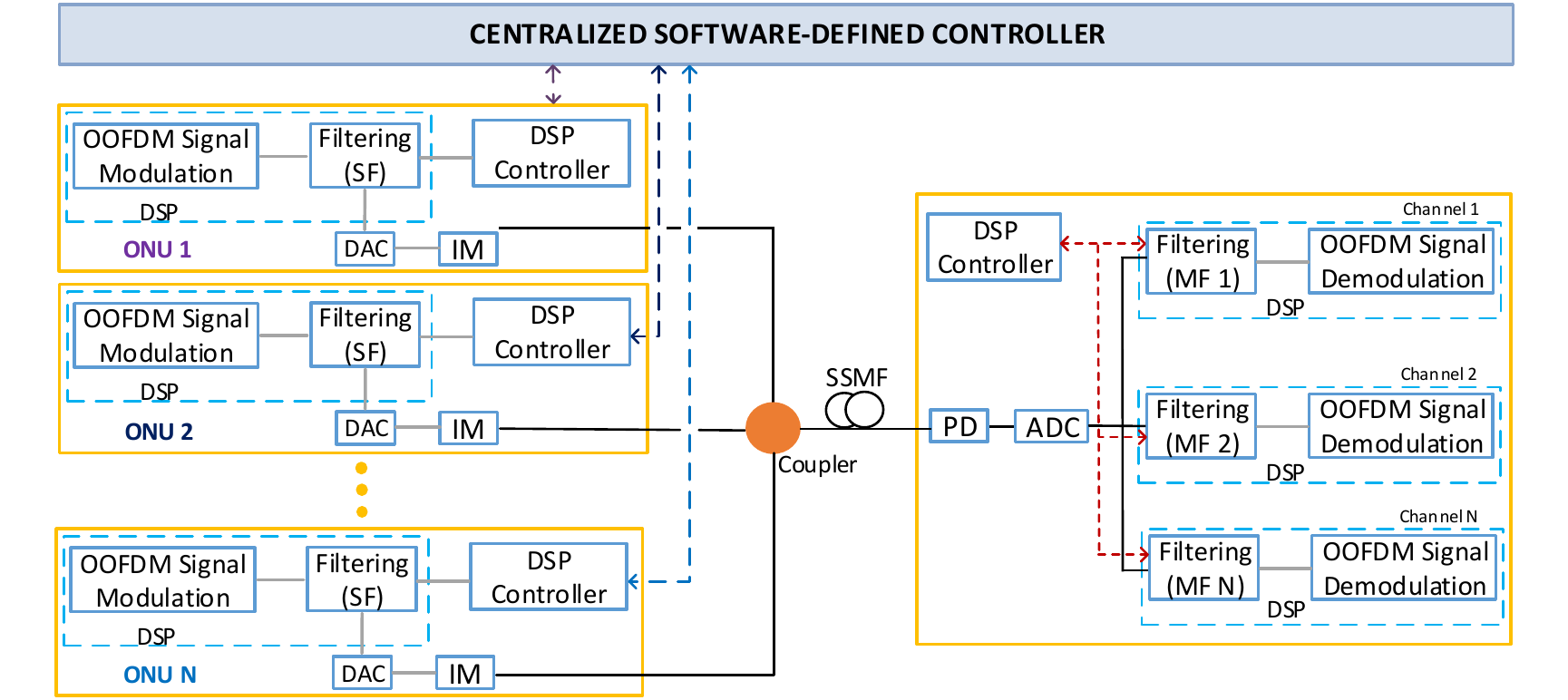}
    \vspace{0cm}
    \caption{Illustration of DSP reconfigurable ONU and OLT
    designs~\cite{Bolea2015a}.}
    \label{fig:bol2015}
\end{figure*}
Bolea et al.~\cite{bol2014dig,Bolea2015a} have recently
developed low-complexity DSP reconfigurable ONU and OLT designs for
SDN-controlled PON communication.
The proposed communication is based on carrierless amplitude and phase
modulation~\cite{rod2011car} enhanced with optical Orthogonal
frequency Division Multiplexing (OFDM)~\cite{bol2014dig}.
The different OFDM channels are manipulated through DSP filtering.
As illustrated in Fig.~\ref{fig:bol2015}, the ONU consists of a DSP controller
that controls the filter coefficients of the shaping filter.
The filter output is then passed through a Digital-to-Analog Converter (DAC)
and intensity modulator for electric-optical conversion.
At the OLT, a photo diode converts the optical signal to an electrical signal,
which then passes through an Analog-to-Digital Converter (ADC).
The SDN controlled OLT DSP controller
sets the filter coefficients in the matching filter to
correspond to the filtering in the sending ONU.
The OLT DSP controller is also responsible for ensuring the
orthogonality of all the ONU filters in the PON.
The performance evaluations in~\cite{Bolea2015a} indicate that
the proposed DSP reconfigurable ONU and OLT system
achieves ONU signal bitrates around 3.7~Gb/s for eight ONUs transmitting
upstream over a 25~km PON.
The performance evaluations also illustrate that long DSP filter lengths,
which increase the filter complexity, improve performance.

\paragraph{Single-Flow BVTs for Data Center Networks}
Malacarne et al.~\cite{Malacarne2014} have developed a low-complexity
and low-cost bandwidth adaptable transmitter for data center
networking.
The transmitter can multiplex Amplitude Shift Keying (ASK),
specifically On-Off Keying (OOK), and Phase Shift Keying (PSK)
on the same optical carrier signal without any
special synchronization or temporal alignment mechanism.
In particular, the transmitter design~\cite{Malacarne2014}
uses the OOK electronic signal to drive a Mach-Zehnder Modulator (MZM)
that is fed by the optical pulse modulated signal.
SDN control can activate (or de-activate) the OOK signal stream, i.e.,
adapt from transmitting only the PSK signal to transmitting
both the PSK and OOK signal and thus providing a higher transmission bit rate.

\subsubsection{Sliceable Multi-Flow Bandwidth Variable Transceivers}
\label{MF-BVT:sec} Whereas the single-flow transceivers surveyed in
Section~\ref{SF-BVT:sec} generate a single optical signal flow,
parallelization efforts have resulted in multi-flow transceivers
(transponders)~\cite{jin2012mul}. Multi-flow transceivers can
generate multiple parallel optical signal flows and thus form the
infrastructure basis for network virtualization.

\paragraph{Encoder Based Programmable Transponder}
Sambo et al.~\cite{Sambo2014a,sam2015nex} have developed an
SDN-programmable bandwidth-variable multi-flow transmitter and
corresponding SDN-programmable multi-flow bandwidth variable
receiver, referred to jointly as programmable bandwidth-variable
transponder. The transmitter mainly consists of a programmable
encoder and multiple parallel Polarization-Multiplexing Quadrature
Phase Shift Keying (PM-QPSK~\cite{win2006adv}) laser transmitters,
whose signals are multiplexed by a coupler. The encoder is
SDN-controlled to implement Low-Density Parity-Check (LDPC)
coding~\cite{bon2011low} with different code rates. At the receiver,
the SDN control sets the local oscillators and LDPC decoder. The
developed transponder allows the setting of the number of
subcarriers, the subcarrier bitrate, and the LDPC coding rate
through SDN. Related frequency conversion and defragmentation issues
have been examined in \cite{Sambo2015}. In~\cite{sam2014sli}, a
low-cost version of the SDN programmable transponder with a
multiwavelength source has been developed. The multiwavelength
source is based on a micro-ring resonator~\cite{ras2009dem} that
generates multiple signal carriers with only a single laser.
Automated configuration procedures for the comprehensive set of
transmission parameters, including modulation format, coding configuration,
and carriers have been explored in~\cite{cug2016tow}.

\paragraph{DSP Based Sliceable BVT}
Moreolo et al.~\cite{Moreolo2016} have developed an SDN controlled
sliceable BVT based on adaptive Digital Signal Processing (DSP) of
multiple parallel signal subcarriers. Each subcarrier is fed by a
DSP module that configures the modulation format, including the bit
rate setting, and the power level of the carrier by adapting a gain
coefficient. The output of the DSP module is then passed through
digital to analog conversion that drives laser sources. The parallel
flows can be combined with a wavelength selective switch; the
combined flow can be sliced into multiple distinct sub-flows for
distinct destinations. The functionality of the developed DSP based
BVT has been verified for a metropolitan area network with links
reaching up to 150~km.

\begin{figure}[t]
    \centering
    \vspace{0cm}
    \includegraphics[width=3.2in]{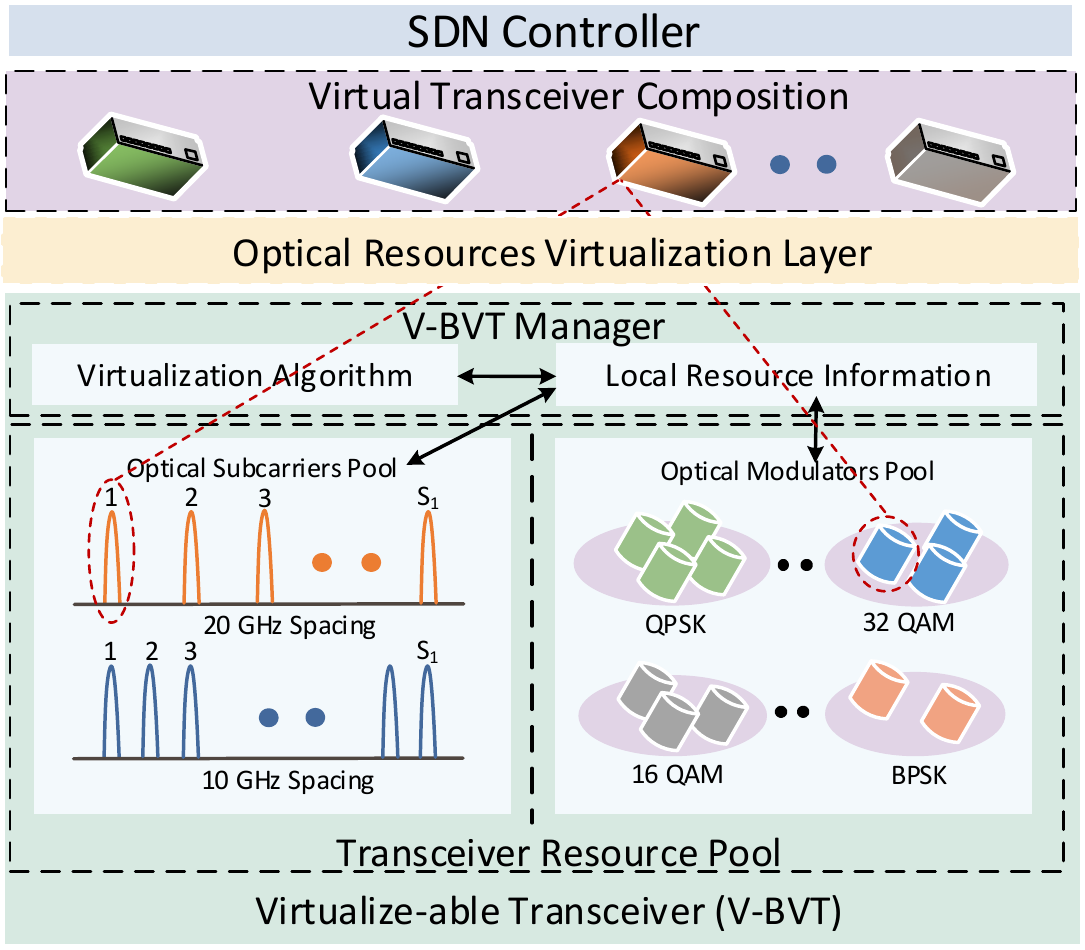}
    \vspace{0cm}
    \caption{Illustration of Subcarrier and Modulator Pool Based
    Virtualizable Bandwidth Variable Transceiver (V-BVT)~\cite{Ou2016}.}
    \label{fig:ou2016}
\end{figure}
\paragraph{Subcarrier and Modulator Pool Based Virtualizable BVT}
Ou et al.~\cite{Ou2016,ou2015onl} have developed a Virtualizable BVT
(V-BVT) based on a combination of an optical subcarriers pool
with an independent optical modulators pool, as illustrated in
Fig.~\ref{fig:ou2016}.
The emphasis of the design is on implementing
Virtual Optical Networks (VONs) at the transceiver level.
The optical subcarriers pool contains multiple optical carriers,
whereby channel spacing and central frequency (wavelength channel)
can be selected.
The optical modulators pool contains optical modulators that can
generate a wide variety of modulation formats.
The SDN control interacts with a V-BVT Manager that implements a
virtualization algorithm. The virtualization algorithm
generates a transceiver slice by combining a particular
set of subcarriers (with specific number of subcarriers, channel
spacing, and central frequencies) from the optical subcarriers pool
with a particular modulation (with specific number of modulators and
modulation formats) from the optical modulators pool.
The evaluations in~\cite{Ou2016} have evaluated the proposed V-BVT in
a network testbed with path lengths up to 200~km with 20~GHz channel
spacing and a variety of modulation formats, including
BPSK as well as 16QAM and 32QAM.

\paragraph{S-BVT Based Hybrid Long-Reach Fiber Access Network (HYDRA)}
HYDRA~\cite{mat2015hyd} is a novel hybrid long-reach fiber access
network architecture based on sliceable BVTs. HYDRA supports
low-cost end-user ONUs through an Active Remote Node (ARN) that
directly connects via a distribution fiber segment, a passive remote
node, and a trunk fiber segment to the core (backbone) network,
bypassing the conventional metro network. The ARN is based on an SDN
controlled S-BVT to optimize the modulation format. With the
modulation format optimization, the ARN can optimize the
transmission capacity for the given distance (via the distribution
and trunk fiber segments) to the core network. The evaluations
in~\cite{mat2015hyd} demonstrate good bit error rate performance of
representative HYDRA scenarios with a 200~km trunk fiber segment and
distribution fiber lengths up to 100~km. In particular, distribution
fiber lengths up to around 70~km can be supported without Forward
Error Correction (FEC), whereas distribution fiber lengths above
70~km would require standard FEC. The consolidation of the access
and metro network infrastructure~\cite{wan2016mig} achieved through
the optimized S-BVT transmissions can significantly reduce the
network cost and power consumption.

\subsection{Space Division Multiplexing (SDM)-SDN}  \label{SDM_SDN:sec}
Amaya et al.~\cite{ama2013ful,ama2014sof} have
demonstrated SDN control of
Space Division Multiplexing (SDM)~\cite{ric2013spa} in optical networks.
More specifically, Amaya et al.~employ SDN to control the
physical layer so as to achieve a bandwidth-flexible and programmable
SDM optical network. The SDN control can perform network slicing,
resulting in sliceable superchannels.
A superchannel consists of multiple spatial carriers to support
dynamic bandwidth and QoS provisioning.

Galve et al.~\cite{Galve2016} have built on the flexible SDN controlled
SDM communication principles to develop a reconfigurable Radio
Access Network (RAN). The RAN connects the BaseBand processing Units (BBUs)
in a shared central office with the corresponding distributed
Remote Radio Heads (RRHs) located at Base Stations (BSs).
A multicore fiber operated with SDM~\cite{ric2013spa} connects the
RRHs to the BBUs in the central office.
Galve et al.~introduce a radio over fiber operation mode where
SDN controlled switching maps the subcarriers dynamically to
spatial output ports.
A complementary digitized radio over fiber operating mode maintains a
BBU pool. Virtual BBUs are dynamically allocated to the cores of
the SDM operated multicore fiber.

\subsection{SDN-Controlled Switching} \label{infra_sw:sec}
\subsubsection{Switching Elements}   \label{sw_elem:sec}
\paragraph{ROADM} \label{ROADMs:par}
The Reconfigurable Optical Add-Drop Multiplexer (ROADM) is an
important photonic switching device for optical networks. Through
wavelength selective optical switches, a ROADM can drop (or add) one
or multiple wavelength channels carrying optical data signals from
(to) a fiber without requiring the conversion of the optical signal
to electric signals~\cite{he2014sur}. The ROADM thus provides an
elementary switching functionality in the optical wavelength domain.
Initial ROADM based node architectures for cost-effectively
supporting flexible SDN networks have been presented
in~\cite{Co13}. Conventional ROADM networks have
typically statically configured wavelength channels that transport
traffic along a pre-configured route. Changes of wavelength channels
or routes in the statically configured networks incur presently high
operational costs due to required physical interventions and are
therefore typically avoided. New ROADM node designs allow changes of
wavelength channels and routes through a management control plane.
Due to these two flexibility dimensions (wavelength and route),
these new ROADM nodes are referred to as ``colorless'' and
``directionless''. First designs for such colorless and
directionless ROADM nodes have been outlined in~\cite{Co13} and
further elaborated in~\cite{ama2013int,rof2013all}.
In addition to the colorless and directionless properties, the
contentionless property has emerged for ROADMs~\cite{gri2010fle}.
Contentionless ROADM operation means that any port can be
routed on any wavelength (color) in any direction without causing
resource contention.
Designs for such Colorless-Directionless-Contentionless (CDC)
ROADMs have been proposed in~\cite{you2013eng,Way2013wav}.
In general, the ROADM designs consist of an express bank that
interconnects the input and output ports coming from/leading to other ROADMs,
and an add-drop bank that connects the express bank with the local
receivers for dropped wavelength channels or transmitters for added
wavelength channels. The recent designs have focused on
the add-drop bank and explored different arrangements
of wavelength selective switches and multicast switches to
provide add-drop bank functionality with the CDC
property~\cite{you2013eng,Way2013wav}.

Garrich et al.~\cite{Garrich2015} have recently designed
and demonstrated a CDC ROADM with an add-drop bank based on
an Optical Cross-Connect (OXC) backplane~\cite{wan2012mul}.
The OXC backplane allows for highly flexible add/drop configurations
implemented through SDN control.
The backplane based ROADM has been analytically
compared with prior designs based on
wavelength selective and multicast switches and has been
shown to achieve higher flexibility and lower losses.
An experimental evaluation has tested the backplane based ROADM for
a metropolitan area mesh network extending over 100~km
with an aggregate traffic load of  close to 9~Tb/s.

\paragraph{Open Transport Switch (OTS)}
The Open Transport Switch (OTS) \cite{SaSPL13} is an OpenFlow enabled
optical virtual switch design. The OTS design abstracts the details
of the underlying physical switching layer (which could be packet
switching or circuit switching) to a virtual switch element. The OTS
design introduces three agent modules (discovery, control, and data
plane) to interface with the physical switching hardware. These
agent modules are controlled from an SDN controller through extended
OpenFlow messages. Performance measurements for an example testbed
network setup indicate that the circuit path computation latencies
on the order of 2--3 s that can be reduced through faster processing
in the controller.

\paragraph{Logical xBar}
The logical xBar~\cite{PaSMK13} has been defined to represent a
programmable switch. An elementary (small) xBar could consist of a
single OpenFlow switch. Multiple small xBars can be recursively
merged to form a single large xBar with a single forwarding table.
The xBar concept envisions that xBars are the building blocks for
forming large networks. Moreover, labels based on SDN and MPLS are
envisioned for managing the xBar data plane forwarding.
The xBar concepts have been further advanced in the Orion
study~\cite{fu2014ori} to achieve low computational complexity of
the SDN control plane.

\paragraph{Optical White Box}
Nejabati et al.~\cite{Nejabati2015} have proposed an optical white
box switch design as a building block for a completely softwarized
optical network. The optical white box design combines a
programmable backplane with programmable switching node elements.
More specifically, the backplane consists of two slivers, namely an
optical backplane sliver and an electronic backplane sliver. These
slivers are set up to allow for flexible arbitrary connections
between the switch node elements. The switch node elements include
programmable interfaces that build on SDN-controlled BVTs (see
Section~\ref{transc:sec}), protocol agnostic switching, and DSP elements.
The protocol agnostic switching
element is envisioned to support both wavelength channel and time
slot switching in the optical backplane as well as programmable
switching with a high-speed packet processor in the electronic
backplane. The DSP elements support both the network processing and
the signal processing for executing a wide range of network
functions. A prototype of the optical white box has been built with
only a optical backplane sliver consisting of a $192 \times 192$
optical space switch. Experiments have indicated that the creation of
a virtual switching node with the OpenDayLight SDN controller takes
roughly 400~ms.

\paragraph{GPON Virtual Switch}
Lee et al.~\cite{Lee2016} have developed a GPON virtual switch
design that makes the GPON fully programmable similar to a
conventional OpenFlow switch. Preliminary steps towards the GPON
virtual switch design have been taken by Gu et al.~\cite{gu2014sof}
who developed components for SDN control of a PON in a data center
and Amokrane et al.~\cite{Amokrane2014,amo2015dyn} who developed a
module for mapping OpenFlow flow control requests into PON configuration
commands. Lee et al.~\cite{Lee2016} have expanded on this groundwork
to abstract the entire GPON into a virtual OpenFlow switch. More
specifically, Lee et al. have comprehensively designed a hardware
architecture and a software architecture to allow SDN control to
interface with the virtual GPON as if it were a standard OpenFlow switch.
The experimental performance evaluation of the designed GPON virtual
switch measured response times for flow entry modifications from an
ONU port (where a subscriber connects to the virtual GPON switch) to
an SDN external port around 0.6~ms, which compares to 0.2~ms for a
corresponding flow entry modification in a conventional OFsoftswitch
and 1.7~ms in a EdgeCore AS4600 switch. In a related study on SDN
controlled switching in a PON, Yeh et al.~\cite{Yeh2015} have
designed an ONU with an optical switch that selects OFDM subchannels
in a TWDM-PON. The switch in the ONU allows for flexible dynamic
adaption of the downstream bandwidth through SDN.
Gu et al.~\cite{gu2016eff} have examined the flexible SDN controlled
re-arrangement of ONUs to OLTs so as to efficiently support PON
service with network coding~\cite{bas2013net}.

\paragraph{Flexi Access Network Node}
A flexi-node for an access network that flexibly aggregates
traffic flows from a wide range of networks,
such as local area networks and base stations of wireless
networks has been proposed in~\cite{FoG13}.
The flexi-node design is motivated by the shortcomings of the
currently deployed core/metro
network architectures that attempt to consolidate the access and
metro networks.
This consolidation forces all traffic in the access network to
traverse the metro network, even if the traffic is destined to
destination nodes in the coverage area of an access network.
In contrast, the proposed flexi-node encompasses
electrical and optical forwarding capabilities that can be
controlled through SDN. The flexi-node can thus serve as an
effective aggregation node in access-metro networks.
Traffic that is destined to other nodes in the coverage area of
an access network can be sent directly to the access network.

Kondepu et al. have similarly presented an SDN based PON aggregation
node~\cite{kon2015sdn}.
In their architecture, multiple ONUs
communicate with the SDN controller within the aggregation node
to request the scheduling of upstream transmission resources.
ONUs are then serviced by multiple Optical Service Units (OSUs)
which exist
within the aggregation node alongside with the SDN controller.
OSUs are then configured by the controller based on Time and Wavelength
Division Multiplexed (TWDM) PON. The OSUs step between normal and sleep-mode
depending on the traffic loads, thus saving power.

\subsubsection{Switching Paradigms}  \label{ws_para:sec}
\paragraph{Converged Packet-Circuit Switching}
Hybrid packet-circuit optical network infrastructures controlled by SDN
have been explored in a few studies.
Das et al.~\cite{DaPM09} have described how to unify the control and
management of circuit- and packet-switched networks using OpenFlow.
Since packet- and circuit-switched networking are extensively employed
in optical networks, examining their integration is an important research
direction.
Das et al.~have given a high-level overview of a flow abstraction for
each type of switched network and a common control paradigm.
In their follow-up work, Das et al.~\cite{DaPMSG10}
have described how a packet and circuit switching
network can be implemented in the context of an OpenFlow-protocol based testbed. The testbed is a standard Ethernet network that could generally
be employed in any access network with Time Division Multiplexing (TDM).
Veisllari et al.~\cite{VeBB13} studied packet/circuit hybrid optical
long-haul metro access networks.
Although Veisllari et al.~indicated that SDN can be used for load balancing in
the proposed packet/circuit network, no detailed study of
such an SDN-based load balancing has been conducted in~\cite{VeBB13}.
Related switching paradigms that integrate SDN with Generalized
Multiple Protocol Label Switching (GMPLS)
have been examined in~\cite{AzNEJ11,ShJKG12},
while data center specific aspects have been surveyed in~\cite{KaT12}.

Cerroni et al.~\cite{CeLR13} have further developed the concept of
unifying circuit- and packet-switching networks with OpenFlow,
which was initiated by Das et al.~\cite{DaPM09,DaPMSG10}.
The unification is accomplished with SDN on the network layer and
can be used in core networks. Specifically, Cerroni et
al.~\cite{CeLR13} have described an extension of the OpenFlow flow
concept to support hybrid networks. OpenFlow message format
extensions to include matching rules and flow entries have also been
provided. The matching rules can represent different transport
functions, such as a channel on which a packet is received in
optical circuit-switched WDM networks, time slots in TDM networks,
or transport class services (such as guaranteed circuit service or
best effort packet service). Cerroni et al.~\cite{CeLR13} have
presented a testbed setup and reported performance results for
throughput (in bit/s and packets/s) to demonstrate the feasibility
of the proposed unified OpenFlow switching network.

\paragraph{R-LR-UFAN}  \label{R_LR_UFAN:sec}
The Reconfigurable Long-Reach UltraFlow Access Network
(R-LR-UFAN)~\cite{yin2013ult,ShYD14} provides flexible dual-mode transport
service based on either the Internet Protocol (IP) or
Optical Flow Switching (OFS).
OFS~\cite{chan2012opt} provides dedicated end-to-end network paths
through purely optical switching, i.e., there is no electronic
processing or buffering at intermediate network nodes. The R-LR-UFAN
architecture employs multiple feeder fibers to form subnets within
the network.
UltraFlow coexists alongside the conventional PON OLT and ONUs. The
R-LR-UFAN introduces new entities, namely the Optical Flow Network
Unit (OFNU) and the SDN-controlled Optical Flow Line Terminal
(OFLT). A Quasi-PAssive Reconfigurable (QPAR) node~\cite{Yin2015} is
introduced between the OFNU and OFLT. The QPAR node can re-route
intra PON traffic between OFNUs without having to pass through the
OLFTs. The optically rerouted intra-PON channels can be used for
communication between wireless base stations supporting inter cell
device-to-device communication. The testbed evaluations indicate
that for an intra-PON traffic ratio of 0.3, the QPAR strategy
achieves power savings up to 24\%.

\paragraph{Flexi-grid}
The principle of flexi-grid (elastic) optical
networking~\cite{cha2015rou,gon2015opt,jue2014sof,she2016sur,tal2014spe,tom2014tut,yu2014spe,zha2013surflexi}
has been explored in several SDN infrastructure studies.
Generally, flexi-grid networking strives to enhance the efficiency
of the optical transmissions by adapting physical (photonic)
transmission parameters, such as modulation format, symbol rate,
number and spacing of subcarrier wavelength channels, as well as the
ratio of forward error correction to payload.
Flexi-grid transmissions have become feasible with
high-capacity flexible transceivers.
Flexi-grid transmissions use narrower
frequency slots (e.g., 12.5~GHz) than classical Wavelength Division
Multiplexing (WDM, with typically 50~GHz frequency slots for WDM)
and can flexibly form optical transmission channels that span multiple
contiguous frequency slots.

Cvijetic~\cite{Cv13} has proposed a hierarchical flexi-grid infrastructure
for multiservice broadband optical access utilizing
centralized software-reconfigurable resource management and digital signal
processing. The proposed flexi-grid infrastructure incorporates mobile
backhaul, as well as SDN controlled transceivers~\ref{transc:sec}.
In follow-up work, Cvijetic et al.~\cite{CvTJSM14} have designed a
dynamic flexi-grid optical access and aggregation network.
They employ SDN to control tunable lasers in the OLT for flexible
downstream transmissions.
Flexi-grid wavelength selective switches are controlled through SDN
to dynamically tune the passband for the upstream transmissions
arriving at the OLT.
Cvijetic et al.~\cite{CvTJSM14} obtained good results for the upstream and
downstream bit error rate
and were able to provide 150~Mb/s per wireless network cell.

Oliveira et al.~\cite{OlSCH13} have demonstrated a testbed for a
Reconfigurable Flexible Optical Network (RFON), which was one of the
first physical layer SDN-based testbeds. The RFON testbed is
comprised of 4 ROADMs with flexi-grid Wavelength Selective Switching
(WSS) modules, optical amplifiers, optical channel monitors and
supervisor boards. The controller daemon implements a node
abstraction layer and provides configuration details for an overall
view of the network. Also, virtualization of the GMPLS control plane
with topology discovery and Traffic Engineering (TE)-link
instantiation have been incorporated.
Instead of using OpenFlow, the
RFON testbed uses the controller language YANG~\cite{Schonwalder2010}
to obtain the topology information and collect monitoring data  for the
lightpaths.

Zhao et al.~\cite{ZhZYY13} have presented an architecture with
OpenFlow-based optical interconnects for intra-data center networking
and OpenFlow-based flexi-grid optical networks for inter-data center
networking.
Zhao et al.~focus on the SDN benefits for inter-data center networking
with heterogeneous networks.
The proposed architecture includes a service controller, an IP controller,
an and optical controller based on the Father Network Operating System
(F-NOX)~\cite{gud2008nox,zha2013uni}.
The performance evaluations in~\cite{ZhZYY13} include results for
blocking probability, release latency, and bandwidth spectrum characteristics.

\subsection{Optical Performance Monitoring}  \label{opm:sec}
\label{sec:sdmonitoring}
\subsubsection{Cognitive Network Infrastructure}
A Cognitive Heterogeneous Reconfigurable Dynamic Optical Network (CHRON)
architecture has been outlined in~\cite{MiDJF13,cab2014cog,dur2016exp}.
CHRON senses the current network
conditions and adapts the network operation accordingly.
The three main components of CHRON are monitoring elements, software
adaptable elements, and cognitive processes. The monitoring elements
observe two main types of optical transmission impairments, namely
non-catastrophic impairments and catastrophic impairments.
Non-catastrophic impairments include the photonic impairments that
degrade the Optical Signal to Noise Ratio (OSNR), such as the
various forms of dispersion, cross-talk, and non-linear propagation
effects, but do not completely disrupt the communication. In
contrast, a catastrophic impairment, such as a fiber cut or
malfunctioning switch, can completely disrupt the communication.
Advances in optical performance monitoring allow for in-band OSNR
monitoring~\cite{dah2011opt,sui2010osn,sch2011osn,sai2012ban} at
midpoints in the communication path, e.g., at optical amplifiers and
ROADMs.

The cognitive processes involve the collection of the monitoring
information in the controller, executing control algorithms, and
instructing the software adaptable components to implement the
control decisions. SDN can provide the framework for implementing
these cognitive processes.
Two main types of software adaptable components have been considered
so far~\cite{Oliveira2015,Giglio2015}, namely control of transceivers and
control of wavelength selective switches/amplifiers.
For transceiver control,
the cognitive control adjusts the transmission parameters.
For instance, transmission bit
rates can be adjusted through varying the modulation format or the
number of signal carriers in multicarrier communication
(see Section~\ref{transc:sec}).

\subsubsection{Wavelength Selective Switch/Amplifier Control}
In general, ROADMs (see Section~\ref{ROADMs:par}) employ wavelength
selective switches based on filters to add or drop wavelength
channels for routing through an optical network. Detrimental
non-ideal filtering effects accumulate and impair the
OSNR~\cite{pao2015sup}. At the same time, Erbium Doped Fiber
Amplifiers (EDFAs)~\cite{zim2004amp} are widely deployed in optical
networks to boost optical signal power that has been depleted
through attenuation in fibers and ROADMs. However, depending on
their operating points, EDFAs can introduce significant noise. Moura
et al.~\cite{Moura2015,mou2016cog} have explored SDN based adaptation
strategies for EDFA operating points to increase the OSNR. In a
complementary study, Paolucci et al.~\cite{pao2015sup} have
exploited SDN control to reduce the detrimental filtering effects.
Paolucci group wavelength channels that jointly traverse a sequence
of filters at successive switching nodes. Instead of passing these
wavelength channels through individual (per-wavelength channel)
filters, the group of wavelength channels is jointly passed through
a superfilter that encompasses all grouped wavelength channels. This
joint filtering significantly improves the OSNR.

While the studies~\cite{Moura2015,mou2016cog,pao2015sup} have focused on
either the EDFA or the filters, Carvalho et al.~\cite{Carvalho2015} and
Wang et al.~\cite{Wang2015f} have jointly considered
the EDFA and filter control.
More specifically, the EDFA gain and the filter attenuation (and
signal equalization) profile were adapted to improve the OSNR.
Carvalho et al.~\cite{Carvalho2015} propose and evaluate a specific
joint EDFA and filter optimization approach that exploits the global
perspective of the SDN controller. The global optimization achieves
ONSR improvements close to 5~dB for a testbed consisting of four ROADMs
with 100~km fiber links.
Wang et al.~\cite{Wang2015f} explore different combinations of
EDFA gain control strategies and filter equalization strategies for a
simulated network with 14 nodes and 100~km fiber links.
They find mutual interactions between the EDFA gain control and the
filter equalization control as well as an additional wavelength assignment
module.
They conclude that global SDN control is highly useful for synchronizing
the EDFA gain and filter equalization in conjunction with wavelength
assignments so as to achieve improved OSNR.

\subsection{Infrastructure Layer: Summary and Discussion}
The research to date on the SDN controlled infrastructure layer has
resulted in a variety of SDN controlled transceivers as well as a few designs
of SDN controlled switching elements.
Moreover, the SDN control of switching paradigms and optical
performance monitoring have been examined.
The SDN infrastructure studies have paid close attention to the physical
(photonic) communication aspects. Principles of isolation of control plane and
data plane with the goals of simplifying network management and
making the networks more flexible have been explored.
The completed SDN infrastructure layer studies have indicated
that the SDN control of the infrastructure layer can reduce costs, facilitate
flexible reconfigurable resource management, increase utilizations,
and lower latency.
However, detailed comprehensive optimizations of the infrastructure components
and paradigms
that minimize capital and operational expenditures are an important
area for future research. Also, further refinements of the optical
components and switching paradigms are needed to ease the
deployment of SDONs and make the networks operating on
the SDON infrastructures more efficient.
Moreover, the cost reduction of implementations, easy adoption by network
providers, flexible upgrades to adopt new technologies, and reduced
complexity require thorough future research.

Most SDON infrastructure studies have
focused on a particular network component or networking aspect, e.g.,
a transceiver or the hybrid
packet-circuit switching paradigm, or a particular application context, e.g.,
data center networking. Future research should comprehensively
examine SDON infrastructure components and
paradigms to optimize their interactions
for a wide set of networking scenarios and application contexts.

The SDON infrastructure studies to date
have primarily focused on the optical transmission medium.
Future research should explore complementary infrastructure
components and paradigms to
support transmissions in hybrid fiber-wireless and other hybrid fiber-$X$
networks, such as
fiber-Digital Subscriber Line (DSL) or fiber-coax cable
networks~\cite{Fuentes2014,gur2014pon,luo2013act}.
Generally, the flexible SDN control can
be very advantageous for hybrid networks composed of heterogeneous
network segments. The OpenFlow protocol can facilitate the topology
abstraction of the heterogeneous physical transmission media, which
in turn facilitates control and optimization at the higher network
protocol layers.

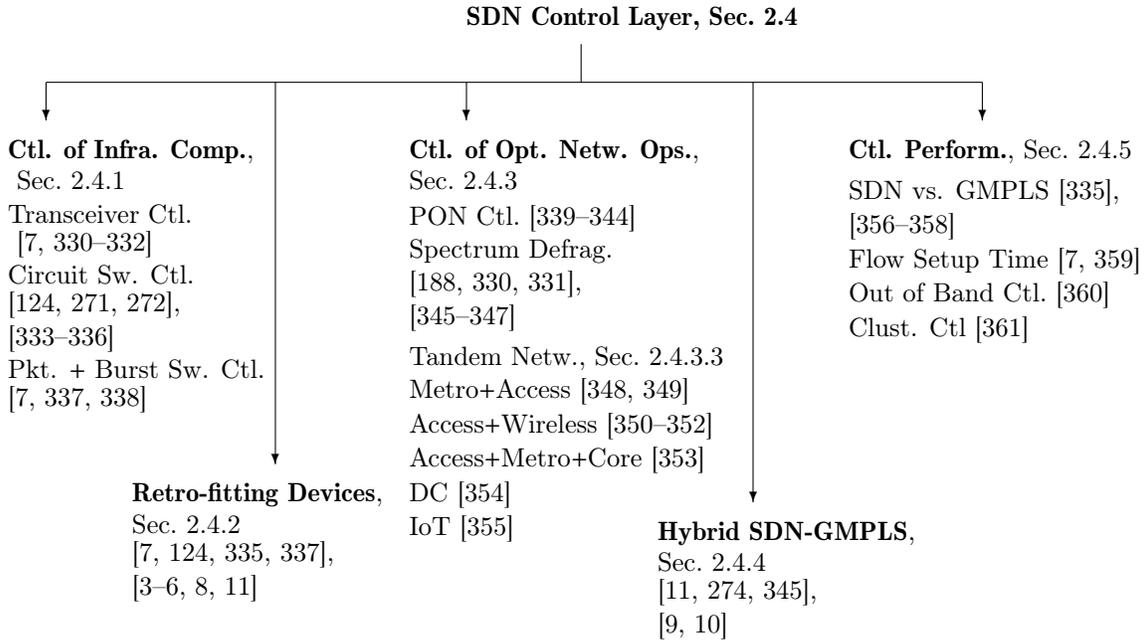
\begin{figure*}[t!]
\footnotesize
\setlength{\unitlength}{0.10in} 
\centering
\begin{picture}(40,33)
\put(15,33){\textbf{SDN Control Layer, Sec.~\ref{sdnctl:sec}}}
\put(-7,30){\line(1,0){49}}
\put(21,30){\line(0,1){2}}
\put(-7,30){\vector(0,-1){2}}
\put(-9,27){\makebox(0,0)[lt]{\shortstack[l]{			
\textbf{Ctl. of Infra. Comp.}, \\ \ Sec.~\ref{ctlinfra:sec}  \\ \\
		Transceiver Ctl.\\~\cite{YuZZG14,Chen2014,JiXia2014,liu2013field}    \\
		Circuit Sw. Ctl. \\ \cite{ChNS13,DaPMSG10,DaPM09}, \\
			\cite{DaPM09Jul,OFCircuitSwitch,ChNF13,Baik2014} \\
		Pkt. $+$ Burst Sw. Ctl.~\\ \cite{Cao2015,Harai2014,liu2013field}
		}}}
	\put(5,30){\vector(0,-1){20}}
	\put(-2.5,9){\makebox(0,0)[lt]{\shortstack[l]{			
              \textbf{Retro-fitting Devices},\\
                 Sec.~\ref{retfitctl:sec} \\
              \cite{ChNF13,ChNS13,Cao2015,liu2013field},\\
		\cite{Alvizu2014,LiTMG11,LiuTMGW11,LiCTM12,LiZTV12,ClSLT14}
	}}}
	\put(15,30){\vector(0,-1){2}}
	\put(12,27){\makebox(0,0)[lt]{\shortstack[l]{			
              \textbf{Ctl. of Opt. Netw. Ops.}, \\
                  Sec.~\ref{ctlops:sec} \\ \\
		PON Ctl.~\cite{KhRS14,LeK15,PaP13,par2014fut,YaZWZ14,KaCTW13} \\
		Spectrum Defrag.~\\\cite{Ger2012,YuZZG14,Chen2014},\\
		\cite{Munoz2014a,Zhu2015b,Meloni2016} \\ \\
		Tandem Netw., Sec.~\ref{ctltan:sec} \\
		Metro$+$Access~\cite{wu2014glo,zha2015sof}	\\	
		Access$+$Wireless~\cite{boj2013adv,TaC14,Costa2015}	\\	
		Access$+$Metro$+$Core~\cite{SlR14} \\		
		DC~\cite{MaHSC14}	   \\	
		IoT~\cite{wan2015nov}
	}}}
	\put(30,30){\vector(0,-1){22}}		
	\put(25,7){\makebox(0,0)[lt]{\shortstack[l]{			
	      \textbf{Hybrid SDN-GMPLS}, \\
               Sec.~\ref{hybsdnctl:sec}  \\
		\cite{AzNEJ11,Alvizu2014,Munoz2014a}, \\
		\cite{LiCTMMM12,Casellas2013b}
	}}}
	\put(42,30){\vector(0,-1){2}}
	\put(35,27){\makebox(0,0)[lt]{\shortstack[l]{			
\textbf{Ctl. Perform.}, Sec.~\ref{cltperf:sec} \\ \\
		SDN vs. GMPLS~\cite{ChNF13}, \\
		\cite{LiTM12,ZhaoZYY13,CviAPJT13} \\
		Flow Setup Time~\cite{VeSBR14,liu2013field} \\
		Out of Band Ctl.~\cite{Sanchez2013} \\
		Clust. Ctl~\cite{Penna2014}
	}}}		
	\end{picture}	
\caption{Classification of SDON control layer studies.}
                  \label{ctl_class:fig}
\end{figure*}
\section{\MakeUppercase{SDN Control Layer}}
\label{sdnctl:sec}
This section surveys the SDON studies that are focused on applying the
SDN principles at the SDN control layer to control the various optical network
elements and operational aspects.
The main challenges of SDON control include extensions of the OpenFlow
protocol for specifically controlling the optical transmission
and switching components surveyed in Section~\ref{sdninfra:sec}
and for controlling the optical
spectrum as well as for controlling optical networks spanning
multiple optical network tiers (see Section~\ref{optnetstruct:sec}).
As illustrated in Fig.~\ref{ctl_class:fig},
we first survey SDN control mechanisms and frameworks for controlling
infrastructure layer components, namely transceivers as well as
optical circuit, packet, and burst switches.
More specifically, we survey OpenFlow extensions for controlling the
optical infrastructure components.
We then survey mechanisms for retro-fitting non-SDN optical network elements
so that they can be controlled by OpenFlow.
The retro-fitting typically involves the insertion of an abstraction layer
into the network elements. The abstraction layer
makes the optical hardware controllable by OpenFlow.
The retro-fitting studies would also fit into Section~\ref{sdninfra:sec} as
the abstraction layer is inserted into the network elements; however,
the abstraction mechanisms closely relate to the OpenFlow extensions for
optical networking and we include the retro-fitting studies therefore in this
control layer section.
We then survey the various SDN control mechanisms for operational aspects of
optical networks, including the control
of tandem networks that include optical segments. Lastly, we survey
SDON controller performance analysis studies.

\subsection{SDN Control of Optical Infrastructure Components}
\label{ctlinfra:sec}

\subsubsection{Controlling Optical Transceivers with OpenFlow}
Recent generations of optical transceivers utilize digital signal
processing techniques that allow many parameters of the transceiver to
be software controlled (see Sections~\ref{SF-BVT:sec} and~\ref{MF-BVT:sec}).
These parameters include modulation scheme, symbol rate, and wavelength.
Yu et al.~\cite{YuZZG14} and Chen et al.~\cite{Chen2014} proposed adding a ``modulation format'' field to the OpenFlow cross-connect table entries to support this
programmable feature of some software defined optical transceivers.

Ji et al.~\cite{JiXia2014} created a testbed that places super-channel optical transponders and optical
amplifiers under SDN control. An OpenFlow extension is proposed to control these devices. The modulation
technique and FEC code for each optical subcarrier of the super-channel transponder
and the optical amplifier power level can be controlled via OpenFlow. Ji et al. do not discuss this explicitly
but the transponder subcarriers can be treated as OpenFlow switch ports that can be configured through the
OpenFlow protocol via port modification messages. It is unclear in \cite{JiXia2014} how the amplifiers would be
controlled via OpenFlow. However, doing so would allow the SDN controller to adaptively modify amplifiers to
compensate for channel impairments while minimizing energy consumption. Ji et al.~\cite{JiXia2014}
have established a testbed demonstrating the placement of transponders and EDFA optical amplifiers under SDN control.

Liu et al.~\cite{liu2013field} propose configuring optical transponder operation via flow table entries with
new transponder specific fields (without providing details). They also propose capturing failure alarms from optical
transponders and sending them to the SDN controller via OpenFlow Packet-In messages. These messages are normally
meant to establish new flow connections. Alternatively, a new OpenFlow message type could be created for the purpose
of capturing failure alarms~\cite{liu2013field}. With failure alarm information, the SDN controller can implement
protection switching services.

\subsubsection{Controlling Optical Circuit Switches with OpenFlow}
Circuit switching can be enabled by OpenFlow by adding new circuit switching flow table
entries~\cite{DaPM09Jul,DaPM09,DaPMSG10,Baik2014}. The OpenFlow circuit switching addendum~\cite{OFCircuitSwitch}
discusses the addition of cross-connect tables for this purpose.
These cross-connect tables are configured via OpenFlow messages inside the circuit switches. According to the addendum, a cross-connect table entry consists of the following
fields to identify the input:
\begin{itemize}
  \item Input Port
  \item Input Wavelength
  \item Input Time Slot
  \item Virtual Concatenation Group
\end{itemize}
and the following fields to identify the output:
\begin{itemize}
  \item Output Port
  \item Output Wavelength
  \item Output Time Slot
  \item Virtual Concatenation Group
\end{itemize}
These cross-connect tables cover circuit switching in space, fixed-grid wavelength, and time.

Channegowda et al.~\cite{ChNF13,ChNS13} extend the capabilities of the OpenFlow circuit switching addendum to
support flexible wavelength grid optical switching. Specifically, the wavelength identifier specified in the
circuit switching addendum to OpenFlow is replaced with two fields: \textit{center frequency}, and
\textit{slot width}. The \textit{center frequency} is an integer specifying the multiple of 6.25~GHz the center
frequency is away from 193.1~Thz and the \textit{slot width} is a positive integer specifying the spectral
width in multiples of 12.5~GHz.

An SDN controlled optical network testbed at the University of Bristol has been established to
demonstrate the OpenFlow extensions for flexible grid DWDM~\cite{ChNS13}. The testbed consists of
both fixed-grid and flexible-grid optical switching devices.
South Korea Telekom has also
built an SDN controlled optical network testbed~\cite{Shin2014}.

\subsubsection{Controlling Optical Packet and Burst Switches with OpenFlow}
OpenFlow flow tables can be utilized in
optical packet switches for expressing the forwarding table and its computation can be offloaded to an SDN
controller. This offloading can simplify the design of highly complex optical packet switches~\cite{Cao2015}.

Cao et al.~\cite{Cao2015} extend the OpenFlow protocol to work with Optical Packet Switching (OPS) devices by creating:
$(i)$ an abstraction layer that converts OpenFlow configuration messages to the native OPS configuration, $(ii)$ a process that
converts optical packets that do not match a flow table entry to the electrical domain for forwarding to the SDN
controller, and $(iii)$ a wavelength identifier extension to the flow table entries. To compensate for either the
lack of any optical buffering or limited optical buffering, an SDN controller, with its global view, can provide more
effective means to resolve contention that would lead to packet loss in optical packet switches.
Specifically, Cao et al. suggest to select the path with the most available resources among multiple available paths between two nodes~\cite{Cao2015}.
Paths can be re-computed periodically or on-demand to account for changes
in traffic conditions. Monitoring messages can be defined to keep the SDN controller updated of network traffic
conditions.

Engineers with Japan's National Institute of Information and Communications Technology~\cite{Harai2014}
have created an optical circuit and packet switched demonstration system in which the packet portion is SDN
controlled. The optical circuit switching is implemented with Wavelength Selective Switches (WSSs) and the
optical packet switching is implemented with an Semiconductor Optical Amplifier (SOA) switch.

OpenFlow flow tables can also be used to configure optical burst switching devices~\cite{liu2013field}. When there is
no flow table entry for a burst of packets, the optical burst switching device can send the Burst Header Packet (BHP) to
the SDN controller to process the addition of the new flow to the network~\cite{liu2013field} rather than the first
packet in the burst.

\subsection{Retro-fitting Devices to Support OpenFlow}  \label{retfitctl:sec}
\begin{figure}[t!]
    \centering
    \includegraphics[width=3in]{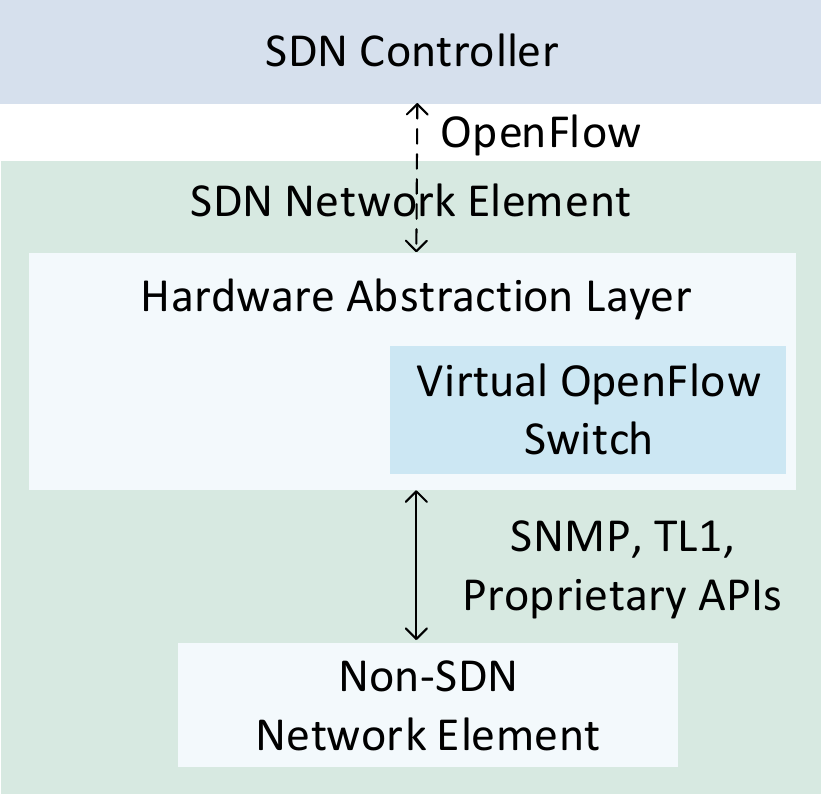}
    \caption{Traditional non-SDN network elements can be retro-fitted for control by an SDN controller using OpenFlow using
    a hardware abstraction layer~\cite{LiTMG11,LiuTMGW11,LiCTM12,LiZTV12,liu2013field}.}
    \label{fig:retrofit}
\end{figure}

An abstraction layer can be used to turn non-SDN optical switching
devices into OpenFlow controllable switching
devices~\cite{ChNF13,ChNS13,liu2013field,Alvizu2014,Cao2015}.
As illustrated in Fig.~\ref{fig:retrofit}, the
abstraction layer provides a conversion layer between OpenFlow
configuration messages and the optical switching devices' native
management interface, e.g., the Simple Network Management Protocol
(SNMP), the Transaction Language 1 (TL1) protocol, or a proprietary
(vendor-specific) API. Additionally, a virtual OpenFlow switch with
virtual interfaces that correspond to physical switching ports on the
non-SDN switching device completes the abstraction
layer~\cite{LiTMG11,LiuTMGW11,LiCTM12,LiZTV12,liu2013field}. When a
flow entry is added between two virtual ports in the virtual OpenFlow
switch, the abstraction layer uses the switching devices' native
management interface to add the flow entry between the two
corresponding physical ports.

\begin{figure}[t!]
    \centering
    \includegraphics[width=3.5in]{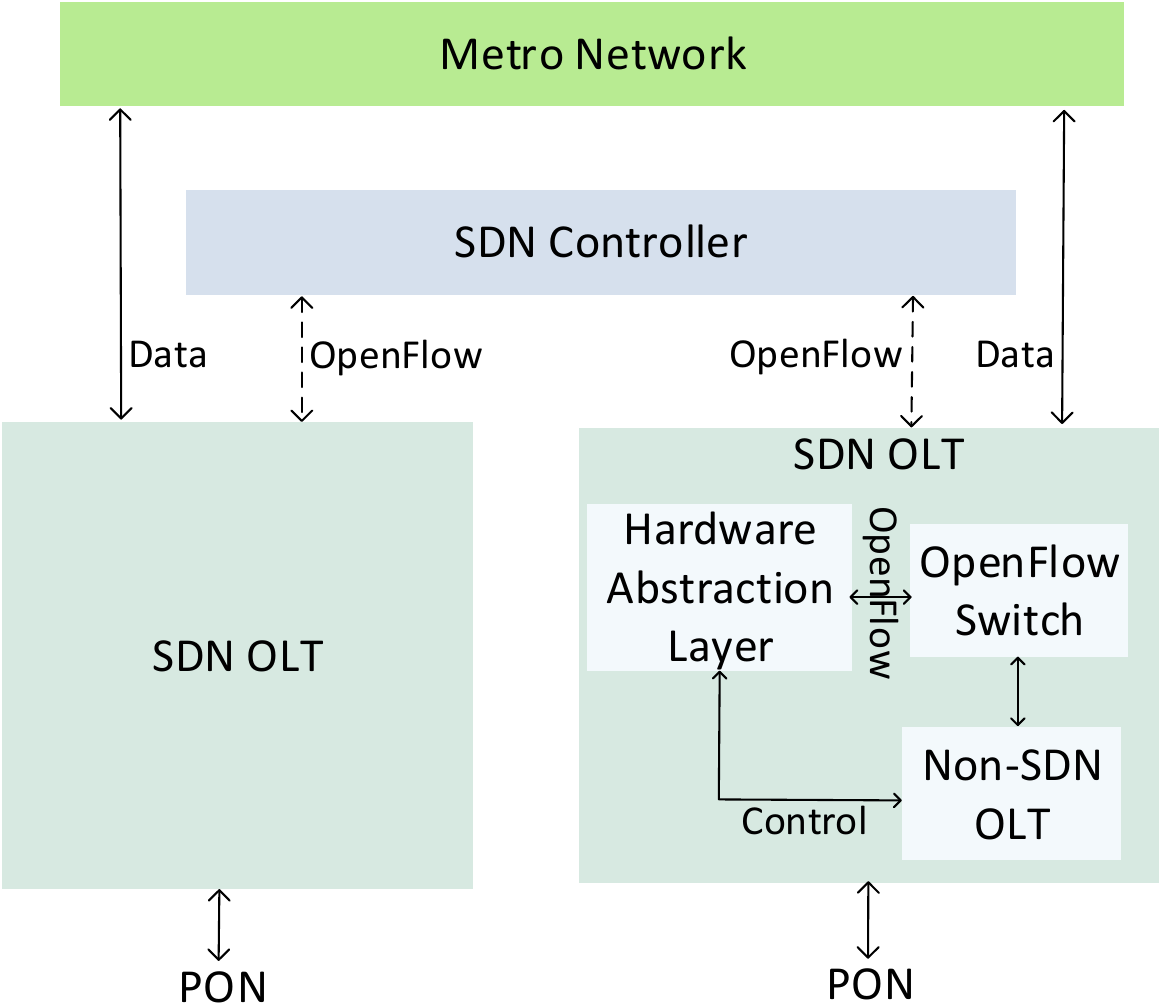}
    \caption{Non-SDN OLTs can be retro-fitted for control by an SDN controller using OpenFlow~\cite{ClSLT14}.}
    \label{fig:oltretrofit}
\end{figure}

A non-SDN PON OLT can be supplemented with a two-port OpenFlow switch and a hardware abstraction layer that
converts OpenFlow forwarding rules to control messages understood by the non-SDN OLT~\cite{ClSLT14}. Fig.~\ref{fig:oltretrofit} illustrates this OLT retro-fit for SDN control via OpenFlow. In this way the
PON has its switching functions controlled by OpenFlow.

\subsection{SDN Control of Optical Network Operation}  \label{ctlops:sec}

\subsubsection{Controlling Passive Optical Networks with OpenFlow}
\label{ponctl:sec}
An SDN controlled PON can be created by upgrading OLTs to SDN-OLTs that can be
controlled using a Southbound Interface, such as OpenFlow~\cite{KhRS14,LeK15}. A centralized PON controller,
potentially executing in a data center, controls one or more SDN-OLTs. The advantage of using SDN is the
broadened perspective of the PON controller as well as the potentially reduced cost of the SDN-OLT
compared to a non-SDN OLT.

Parol and Pawlowski~\cite{PaP13,par2014fut} define OpenFlowPLUS to extend the OpenFlow SBI for GPON. OpenFlowPLUS extends SDN programmability to both OLT and ONU devices whereby each act
as an OpenFlow switch through a programmable flow table. Non-switching functions (e.g., ONU registration,
dynamic bandwidth allocation) are outside the scope of OpenFlowPLUS. OpenFlowPLUS extends OpenFlow
by channeling OpenFlow messages through the GPON ONU Management and Control Interface (OMCI) control channel and adding PON specific action
instructions to flow table entries. The PON specific action instructions defined in OpenFlowPLUS are:
\begin{itemize}
  \item (new \textit{gpon} action type): map matching packets to PON specific traffic identifiers, e.g.,
  GPON Encapsulation Method (GEM) ports and GPON Traffic CONTainers (T-CONTs)
  \item (\textit{output} action type): activate PON specific framing of matching packets
\end{itemize}

Many of the OLT functions operate at timescales that are problematic for the controller due
to the latency between the controller and OLTs. However, Khalili et al.~\cite{KhRS14} identify ONU
registration policy and coarse timescale DBA policy as functions that operate at timescales that allow
effective offloading to an SDN controller. Yan et al.~\cite{YaZWZ14} further identify OLT and ONU power
control for energy savings as a function that can be effectively offloaded to an SDN controller.

There is also a movement to use PONs in edge networks to provide
connectivity inside a multitenant building or on a campus with multiple
buildings~\cite{PaP13,par2014fut}. The use of PONs in this edge scenario
requires rapid re-provisioning from the OLT. A software controlled PON
can provide this needed rapid reprovisioning~\cite{PaP13,par2014fut}.

Kanonakis et al. \cite{KaCTW13} propose leveraging the
broad perspective that SDN can provide to perform dynamic bandwidth
allocation across several Virtual PONs (VPONs). The VPONs are
separated on a physical PON by the wavelength bands that they
utilize. Bandwidth allocation is performed at the granularity of OFDMA
subcarriers that compose the optical spectrum.

\subsubsection{SDN Control of Optical Spectrum Defragmentation}
In a departure from the fixed wavelength grid
(ITU-T G.694.1), elastic optical networking allows flexible use of the optical spectrum. This flexibility can permit higher spectral efficiency by avoiding consuming an entire fixed-grid
wavelength channel when unnecessary and avoiding unnecessary guard bands in certain circumstances \cite{Ger2012}.
However, this flexibility causes fragmentation of the optical spectrum as flexible grid lightpaths are established
and terminated over time.

Spectrum fragmentation leads to the circumstance in which there is enough spectral capacity to satisfy a demand but
that capacity is spread over several fragments rather than being consolidated in adjacent spectrum as required. If
the fragmentation is not counter-acted by a periodic defragmentation process than overall spectral utilization will
suffer. This resource fragmentation problem appears in computer systems in main memory and long term storage. In those
contexts the problem is typically solved by allowing the memory to be allocated using non-adjacent segments. Memory and
storage is partitioned into pages and blocks, respectively. The allocations of pages to a process or blocks to a file do
not need to be contiguous. With communication spectrum this would mean combining multiple small bandwidth channels
through inverse multiplexing to create a larger channel~\cite{Munoz2014a}.

An SDN controller can provide a broad network perspective to empower the periodic optical spectrum defragmentation
process to be more effective~\cite{Munoz2014a}. In general, optical spectrum defragmentation operations can reduce
lightpath blocking probabilities from 3\%~\cite{YuZZG14} up to as much as 75\%~\cite{Chen2014,Zhu2015b}. Multicore
fibers provide additional spectral resources through additional transmission cores to permit quasi-hitless
defragmentation~\cite{Meloni2016}.

\subsubsection{SDN Control of Tandem Networks}  \label{ctltan:sec}

\paragraph{Metro and Access}
Wu et al.~\cite{wu2014glo,zha2015sof} propose leveraging the broad perspective that SDN can provide to
improve bandwidth allocation. Two cooperating stages of SDN controllers: $(i)$ access stage that controls each SDN
OLT individually, and $(ii)$ metro stage that controls global bandwidth allocation strategy, can coordinate bandwidth
allocation across several physical PONs~\cite{wu2014glo,zha2015sof}. The bandwidth allocation is managed cooperatively
among the two stages of SDN controllers to optimize the utilization of the access and metro network bandwidth. Simulation
experiments indicate a 40\% increase in network bandwidth utilization as a result of the global coordination compared
to operating the bandwidth allocation only within the individual PONs~\cite{wu2014glo,zha2015sof}.

\paragraph{Access and Wireless}
Bojic et al. \cite{boj2013adv} expand on the concept of SDN controlled OFDMA enabled VPONs~\cite{KaCTW13} to
provide mobile backhaul service. The backhaul service can be provided for wireless small-cell sites (e.g., micro and femto cells)
that utilize millimeter wave frequencies. Each small-cell site contains an OFDMA-PON ONU that provides the backhaul service
through the access network over a VPON. An SDN controller is utilized to assign bandwidth to each small-cell site through
OFDMA subcarrier assignment in a VPON to the constituent ONU. The SDN controller leverages its broad view of the network
to provide solutions to the joint bandwidth allocation and routing across several network segments. With this broad
perspective of the network, the SDN controller can make globally rather than just locally optimal bandwidth allocation and
routing decisions. Efficient optimization algorithms, such as genetic algorithms, can be used to provide computationally
efficient competitive solutions, mitigating computational complexity issues associated with optimization for large networks.
Additionally, network partitioning with an SDN controller for each partition can be used to mitigate unreasonable
computational complexity that arises when scaling to large networks. Tanaka and Cvijetic~\cite{TaC14} presented one
such optimization formulation for maximizing throughput.

Costa-Requena et al.~\cite{Costa2015} described a proof-of-concept LTE testbed they have constructed
whereby the network consists of software defined base stations and various network functions
executing on cloud resources. The testbed is described in broad qualitative terms, no technical
details are provided. There was no mathematical or experimental analysis provided.

\paragraph{Access, Metro, and Core}
Slyne and Ruffini~\cite{SlR14} provide a use case for SDN switching control across network segments: use Layer 2 switching
across the access, metro, and core networks. Layer 2 (e.g., Ethernet) switching does not scale well due to a lack of hierarchy
in its addresses. That lack of hierarchy does not allow for switching rules on aggregates of addresses thereby limiting the
scaling of these networks. Slyne and Ruffini~\cite{SlR14} propose using SDN to create hierarchical pseudo-MAC addresses that
permit a small number of flow table entries to configure the switching of traffic using Layer 2 addresses across network
segments. The pseudo-MAC addresses encode information about the device location to permit simple switching rules. At the entry
of the network, flow table entries are set up to translate from real (non-hierarchical) MAC addresses to hierarchical pseudo-MAC
addresses. The reverse takes place at the exit point of the network.

\paragraph{DC Virtual Machine Migration}
Mandal et al.~\cite{MaHSC14} provided a cloud computing use case for SDN bandwidth allocation across network segments:
Virtual Machine (VM) migration between data centers. VM migrations require significant network bandwidth. Bandwidth
allocation that utilizes the broad perspective that SDN can provide is critical for reasonable VM migration latencies
without sacrificing network bandwidth utilization.

\paragraph{Internet of Things}
Wang et al.~\cite{wan2015nov} examine another use case for SDN bandwidth allocation across network segments: the Internet
of Things (IoT). Specifically, Wang et al. have developed a Dynamic Bandwidth Allocation (DBA) protocol that exploits
SDN control for multicasting and suspending flows. This DBA protocol is studied in the context of a virtualized
WDM optical access network that provides IoT services through the distributed ONUs to individual devices.
The SDN controller employs multicasting and flow suspension to efficiently prioritize the IoT service requests.
Multicasting allows multiple requests to share resources in the central nodes that are responsible for processing a
prescribed wavelength in the central office (OLT). Flow suspension allows high-priority requests (e.g., an emergency
call) to suspend ongoing low-priority traffic flows (e.g., routine meter readings). Performance results for a real-time
SDN controller implementation indicate that the proposed bandwidth (resource) allocation with multicast and flow
suspension can improve several key performance metrics, such as request serving ratio, revenue, and delays
by 30--50~\%~\cite{wan2015nov}.

\subsection{Hybrid SDN-GMPLS Control}  \label{hybsdnctl:sec}
\subsubsection{Generalized MultiProtocol Label Switching (GMPLS)}
Prior to SDN, MultiProtocol Label Switching (MPLS) offered a mechanism to separate the control and data planes
through label switching. With MPLS, packets are forwarded in a connection-oriented manner through Label Switched
Paths (LSPs) traversing Label Switching Routers (LSRs). An entity in the network establishes an LSP through a network
of LSRs for a particular class of packets and then signals the label-based forwarding table entries to the LSRs. At
each hop along an LSP, a packet is assigned a label that determines its forwarding rule at the next hop. At the next
hop, that label determines that packet's output port and label for the next hop; the process repeats until the packet
reaches the end of the LSP. Several signalling protocols for programming the label-based forwarding table entries
inside LSRs have been defined, e.g., through the Resource Reservation Protocol (RSVP). Generalized MPLS (GMPLS) extends MPLS to offer circuit switching
capability. Although never commercially deployed~\cite{liu2013field}, GMPLS and a centralized Path Computation Element
(PCE)~\cite{mun2014pce,oki2005dyn,pao2013sur,Casellas2013} have been considered for control of optical networks.

\subsubsection{Path Computation Element (PCE)}  \label{PCE:sec}
A PCE is a concept developed by the IETF (see RFC 4655) to refer to an entity that computes network paths given a topology
and some criteria. The PCE concept breaks the path computation action from the forwarding action in switching devices.
A PCE could be distributed in every switching element in a network domain or there could be a single centralized PCE
for an entire network domain. The network domain could be an area of an Autonomous System (AS), an AS, a conglomeration
of several ASes, or just a group of switching devices relying on one PCE. Some of an SDN controller's functionality
falls under the classification of a centralized PCE. However, the PCE concept does not include the external
configuration of forwarding tables. Thus, a centralized PCE device does not necessarily have a means to configure the
switching elements to provision a computed path.

When the entity requesting path computation is not co-located with the PCE, a PCE Communication Protocol (PCEP) is used
over TCP port 4189 to facilitate path computation requests and responses. The PCEP consists of the following message
types:
\begin{itemize}
  \item Session establishment messages (open, keepalive, close)
  \item PCReq -- Path computation request
  \item PCRep -- Path computation reply
  \item PCNtf -- event notification
  \item PCErr -- signal a protocol error
\end{itemize}

The path computation request message must include the end points of the path and can optionally include the requested
bandwidth, the metric to be optimized in the path computation, and a list of links to be included in the path. The Path
computation reply includes the computed path expressed in the Explicit Route Object format (see RFC 3209) or an
indication that there is no path. See RFC 5440 for more details on PCEP.

A PCE has been proposed as a central entity to manage a GMPLS-enabled optical circuit switched network. Specifically,
the PCE maintains the network topology in a structure called the Traffic Engineering Database (TED). The traffic
engineering modifier (see RFC 2702) signifies that the path computations are made to relieve congestion that is caused
by the sub-optimal allocation of network resources. This modifier is used extensively in discussions of MPLS/GMPLS
because their use case is for traffic engineering; in acronym form the modifier is TE (e.g., TE LSP, RSVP-TE).

If the PCE is stateful with complete control over its network domain, it will also maintain an LSP database recording the provisioned GMPLS lightpaths. A lightpath request can be sent to the PCE, it will use the topology and LSP
database to find the optimal path and then configure the GMPLS-controlled optical circuit switching nodes using NETCONF
(see RFC 6241) or proprietary command line interfaces (CLIs)~\cite{Munoz2014a}. This stateful PCE with instantiation
capabilities (capabilities to provision lightpaths) operates similarly to an SDN controller. For that reason, GMPLS with a
centralized stateful PCE with instantiation capabilities can provide a baseline for performance analysis of an SDN
controller as well as provide a mechanism to be blended with an SDN controller for hybrid control~\cite{ChNF13,ChNS13,Alvizu2014}.

\subsubsection{Approaches to Hybrid SDN-GMPLS Control}
\begin{figure}[t!]
	\centering
	\includegraphics[width=5in]{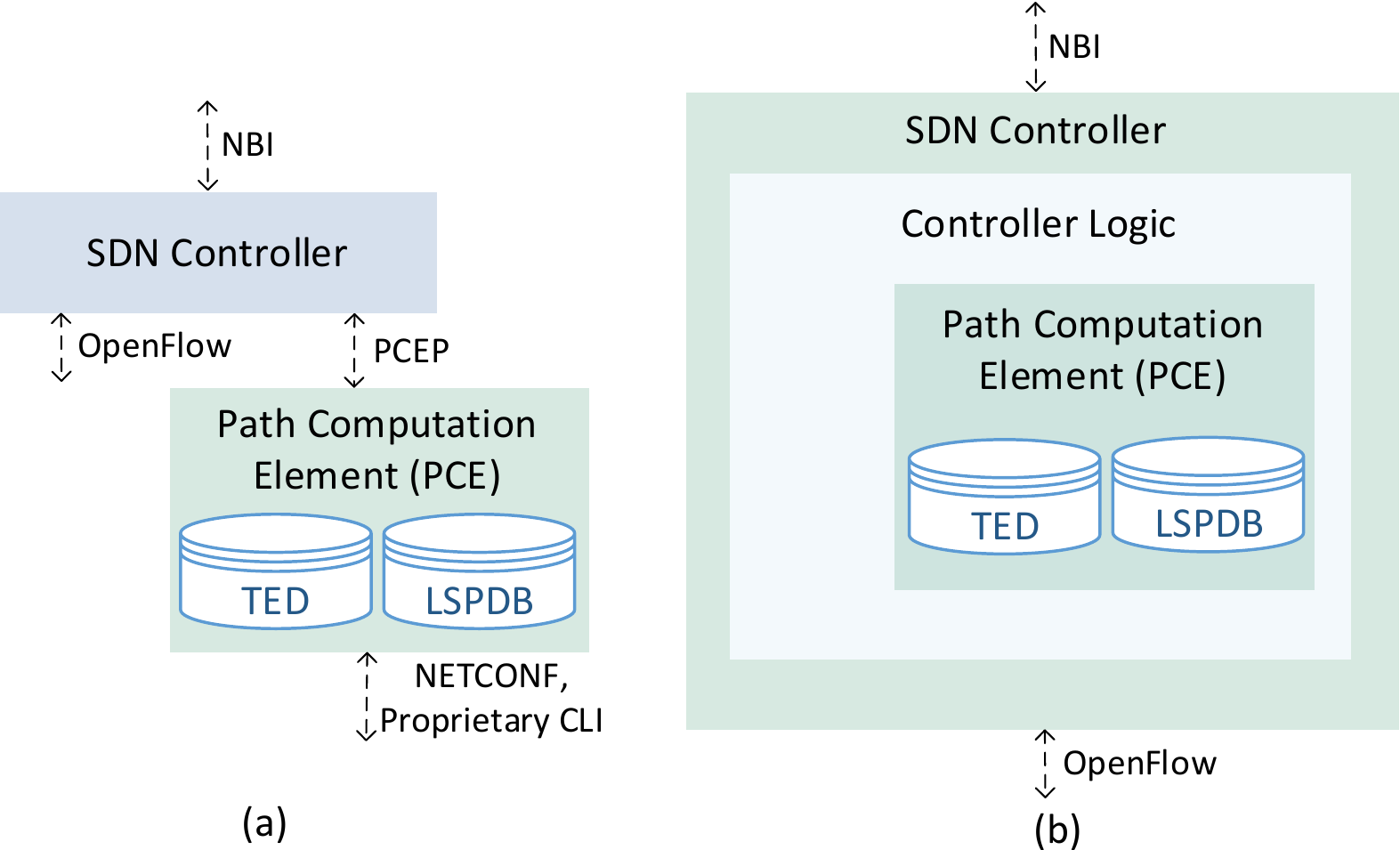}
    \caption{Hybrid GMPLS/PCE and SDN network control~\cite{LiCTMMM12,Casellas2013b,Alvizu2014}.}
	\label{fig:hybridctl}
\end{figure}

Hybrid GMPLS/PCE and SDN control can be formed by allowing an SDN controller to leverage a centralized PCE to control a
portion of the infrastructure using PCEP as the SBI~\cite{AzNEJ11,Munoz2014a}; see illustration a) in Fig.~\ref{fig:hybridctl}.
The SDN controller builds higher functionality above what the PCE provides and can possibly control a large network that
utilizes several PCEs as well as OpenFlow controlled network elements.

Alternatively, the SDN controller can leverage a PCE for its path computation abilities with the SDN controller handling
the configuration of the network elements to establish a path using an SBI protocol, such as
OpenFlow~\cite{LiCTMMM12,Casellas2013b,Alvizu2014};
see illustration b) in Fig.~\ref{fig:hybridctl}.

\subsection{SDN Performance Analysis}  \label{cltperf:sec}
\subsubsection{SDN vs. GMPLS}
Liu et al.~\cite{LiTM12} provided a qualitative comparison of GMPLS,
GMPLS/PCE, and SDN OpenFlow for control of wavelength switched optical
networks. Liu et al.~noted that there is an evolution of centralized
control from GMPLS to GMPLS/PCE to OpenFlow. Whereas GMPLS offers
distributed control, GMPLS/PCE is commonly regarded as having centralized path
computation but still distributed provisioning/configuration; while
OpenFlow centralizes all of the network control. In
our discussion in Section \ref{hybsdnctl:sec} we noted that a
stateful PCE with instantiation capabilities centralizes all
network control and is therefore very similar to SDN. Liu et al.~have
also pointed out that GMPLS/PCE is more technically mature compared to
OpenFlow with IETF RFCs for GMPLS (see RFC 3471) and PCE (see RFC
4655) that date back to 2003 and 2006, respectively. SDN has just
recently, in 2014, received standardization attention from the IETF
(see RFC 7149).

A comparison of GMPLS and OpenFlow has been conducted by Zhao et al.~\cite{ZhaoZYY13} for large-scale optical networks.
Two testbeds were built, based on GMPLS and on Openflow, respectively. Performance metrics, such as
blocking probability, wavelength utilization, and lightpath setup time were evaluated for a 1000 node topology. The
results indicated that GMPLS gives slightly lower blocking probability. However, OpenFlow gives higher wavelength
utilization and shorter average lightpath setup time. Thus, the results suggest that OpenFlow is overall advantageous
compared to GMPLS in large-scale optical networks.

Cvijetic et al.~\cite{CviAPJT13} conducted a numerical analysis to compare the computed shortest path
lengths for non-SDN, partial-SDN, and full-SDN optical networks. A full-SDN network enables path lengths
that are approximately a third of those computed on a non-SDN network. These path lengths can also translate
into an energy consumption measure, with shortest paths resulting in reduced energy consumption. An SDN
controlled network can result in smaller computed shortest paths that translates to smaller network latency
and energy consumption~\cite{CviAPJT13}.

Experiments conducted on the testbed described in \cite{ChNF13} show a 4~\% reduction in lightpath blocking probability
using SDN OpenFlow compared to GMPLS for lightpath provisioning. The same experiments show that lightpath setup times
can be reduced to nearly half using SDN OpenFlow compared to GMPLS. Finally, the experiments show that an Open vSwitch
based controller can process about three times the number of flows per second as a NOX~\cite{gud2008nox} based controller.

\subsubsection{SDN Controller Flow Setup}
Veisllari et al.~\cite{VeSBR14} evaluated the use of SDN to support both circuit and packet
switching in a metropolitan area ring network that interconnects access network segments
with a backbone network. This network is assumed to be controlled by a single SDN controller.
The objective of the study~\cite{VeSBR14} was to determine the effect of packet service flow
size on the required SDN controller flow service time to meet stability conditions at the
controller. Toward this end, Veisllari et al.~produced a mean arrival rate function of new packet and
circuit flows at that controller. This arrival rate function was visualized by varying the length
of short-lived (``mice'') flows, the fraction of long-lived (``elephant'') flows, and the volume
of traffic consumed by ``elephant'' flows. Veisllari et al.~discovered, through these visualizations,
that the length of ``mice'' flows is the dominating parameter in this model.

Veisllari et al.~translated the arrival rate function analysis to an analysis of the ring MAN network
dimensions that can be supported by a single SDN controller. The current state-of-the-art Beacon
controller can handle a flow request every 571~ns. Assuming mice flows sizes of 20~kB and average circuit
lifetimes of 1 second, as the fraction of packet traffic increases from 0.1 to 0.9, the network
dimension supported by a single Beacon SDN controller decreases from 14 nodes with 92 wavelengths per
node to 5 nodes with 10 wavelengths per node.

Liu et al.~\cite{liu2013field} use a multinational (Japan, China, Spain) NOX:OpenFlow controlled
four-wavelength optical circuit and burst switched network to study path setup/release times as well as path
restoration times. The optical transponders that can generate failure alarms were also under NOX:OpenFlow
control and these alarms were used to trigger protection switching. The single SDN controller was located
in the Japanese portion of the network. The experiments found the path setup time to vary from 250--600 ms and the path release times to vary from 130--450 ms. Path restoration times varied
from 250--500~ms. Liu et al.~noted that the major contributing factor to these times was the OpenFlow message delivery time~\cite{liu2013field}.

\subsubsection{Out of Band Control}
Sanchez et al.~\cite{Sanchez2013} have qualitatively compared four SDN controlled ring metropolitan network architectures. The architectures vary in whether the SDN control traffic is carried
in-band with the data traffic or out-of-band separately from the data traffic. In a single wavelength ring
network, out-of-band control would require a separate physical network that would come at a high cost,
but provide reliability of the network control under failure of the ring network. In a multiwavelength
ring network, a separate wavelength can be allocated to carry the control traffic. Sanchez et al.~\cite{Sanchez2013}
focused on a Tunable Transceiver Fixed Receiver (TTFR) WDM ring node architecture. In this architecture each
node receives data on a home wavelength channel and has the capability to transmit on any of the available
wavelengths to reach any other node. The addition of the out-of-band control channel on a separate wavelength
requires each node to have an additional fixed receiver, thereby increasing cost. Sanchez et al. identified a clear tradeoff between cost and reliability when comparing the four architectures.

\subsubsection{Clustered SDN Control}
Penna et al.~\cite{Penna2014} described partitioning a wavelength-switched optical network into administrative
domains or clusters for control by a single SDN controller. The clustering should meet certain
performance criteria for the SDN controller. To permit lightpath establishment across clusters, an inter-cluster
lightpath establishment protocol is established. Each SDN controller provides a lightpath establishment function
between any two points in its associated cluster. Each SDN controller also keeps a global view of the network
topology. When an SDN controller receives a lightpath establishment request whose computed path traverses
other clusters, the SDN controller requests lightpath establishment within those clusters via a WBI.

The formation of clusters can be performed such that for a specified number of clusters the average distance to
each SDN controller is minimized~\cite{Penna2014}. The lightpath establishment time decreases exponentially
as the number of clusters increases.

\subsection{Control Layer: Summary and Discussion}
A very large body of literature has explored how to expand the OpenFlow protocol to support various optical network
technologies (e.g., optical circuit switching, optical packet switching, passive optical networks). A significant
body of literature has investigated methodologies for retro-fitting non-SDN network elements for OpenFlow control
as well as integrating SDN/OpenFlow with the GMPLS/PCE control framework. A variety of SDN controller use cases have
been identified that motivate the benefits of the centralized network control made possible with SDN (e.g., bandwidth
allocation over large numbers of subscribers, controlling tandem networks).

However, analyzing the performance of SDN controllers for optical network applications is still in a state of infancy.
It will be important to understand the connection between the implementation of the SDN controller (e.g., processor
core architecture, number of threads, operating system) and the network it can effectively control (e.g., network
traffic volume, network size) to meet certain performance objectives (e.g., maximum flow setup time). At present there
are not enough relevant studies to gain an understanding of this connection. With this understanding network service
providers will be able to partition their networks into control domains in a manner that meets their performance
objectives.

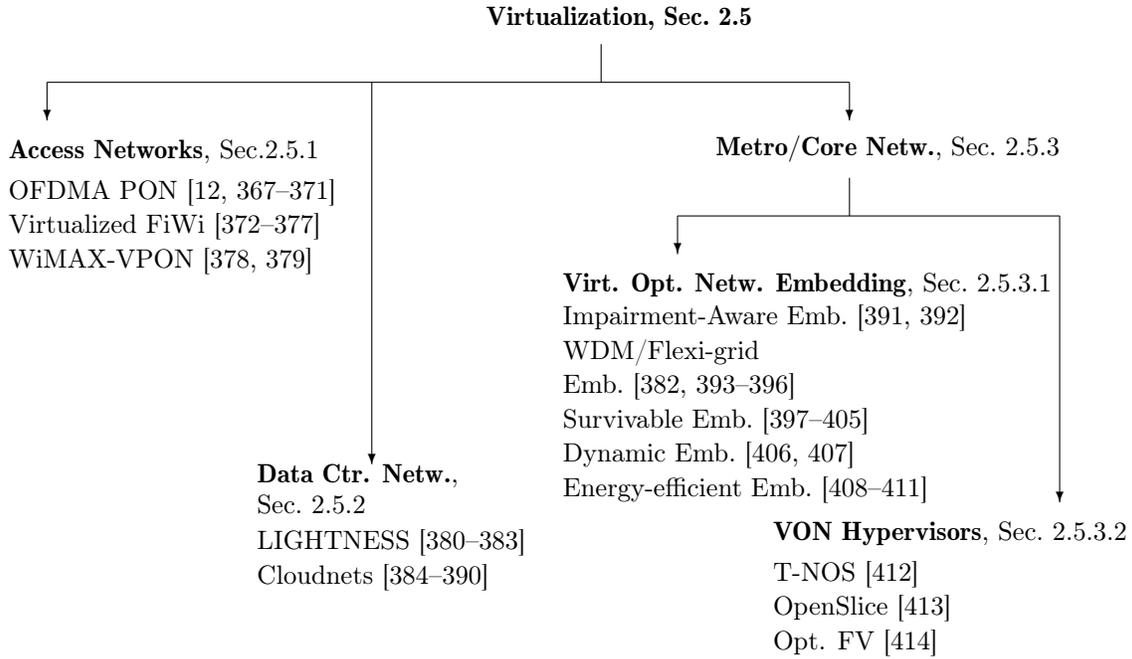
\begin{figure*}[t!]
\footnotesize
\setlength{\unitlength}{0.10in} 
\centering
\begin{picture}(40,33)
\put(15,33){\textbf{Virtualization, Sec.~\ref{virt:sec}}}
\put(-8,30){\line(1,0){42}}
\put(25,23){\line(1,0){20}}
\put(27,26.2){\textbf{Metro/Core Netw.}, Sec.~\ref{virt_core:sec}}
\put(34,25){\line(0,-1){2}}
\put(21,30){\line(0,1){2}}
\put(-8,30){\vector(0,-1){2}}
\put(-10,27){\makebox(0,0)[lt]{\shortstack[l]{			
\textbf{Access Networks}, Sec.\ref{virt_access:sec}	\\	\\
OFDMA PON~\cite{wei2009pon,wei2009pro,wei2010ada,li2016pro,jin2009vir,zhou2015dem}\\		
Virtualized FiWi~\cite{QiZSH14,QiGYZ13,ShGYZ13,he2013int,men2014eff,QiZHG14}\\		
WiMAX-VPON~\cite{dha2010wim,DhPX10}
}}}

\put(9,30){\vector(0,-1){20}}
\put(3,10){\makebox(0,0)[lt]{\shortstack[l]{			
\textbf{Data Ctr. Netw.}, \\
Sec.~\ref{virt_dc:sec} \\ \\
LIGHTNESS~\cite{miao2015sdn,pen2015mul,pag2015opt,Saridis2016} \\
Cloudnets~\cite{kan2015res,ahm2014enh,dov2015usi,xie2014dyn,vel2014tow,zha2015ren,tza2014con}
}}}				
\put(34,30){\vector(0,-1){2}}
\put(25,23){\vector(0,-1){2}}
\put(19,20){\makebox(0,0)[lt]{\shortstack[l]{
\textbf{Virt. Opt. Netw. Embedding}, Sec.~\ref{virt_emb:sec} \\
Impairment-Aware Emb.~\cite{pen2011imp,PeNS13} \\
WDM/Flexi-grid \\ Emb.~\cite{zha2013net,zha2013vir,gon2014vir,wan2015vir,pag2015opt} \\
Survivable Emb.~\cite{hu2013sur,ye2015sur,xie2014sur,che2016cos,jia2015ava,son2011ban,pao2014mul,ass2016net,Hong2015} \\
Dynamic Emb.~\cite{ye2014upg,zha2015dyn} \\
Energy-efficient Emb.~\cite{non2015ene,che2016pow,she2014fol,wan2014hie}
}}}							
\put(45,23){\vector(0,-1){15}}		
\put(30,7){\makebox(0,0)[lt]{\shortstack[l]{			
\textbf{VON Hypervisors}, Sec.~\ref{virt_hv:sec} \\ \\
T-NOS~\cite{siq2015pro} \\
OpenSlice~\cite{LiuMCTMM13} \\
Opt. FV~\cite{Azod2012}
}}}
\end{picture}	
\caption{Classification of SDON virtualization studies.}
\label{virt_class:fig}
\end{figure*}
\section{\MakeUppercase{Virtualization}}  \label{virt:sec}
This section surveys control layer mechanisms for virtualizing
SDONs.
As optical infrastructures have typically high costs, creating
multiple VONs over the optical network infrastructure is
especially important for access networks, where the costs need to
be amortized over relatively few users. Throughout, accounting
for the specific optical transmission and signal propagation characteristics
is a key challenge for SDON virtualization.
Following the classification structure illustrated in
Fig.~\ref{virt_class:fig},
we initially survey virtualization mechanisms for access networks
and data center networks, followed by virtualization mechanisms for
optical core networks.

\begin{figure*}[t!]
	\centering
	\vspace{0cm}
	\begin{tabular}{c}
		\includegraphics[width=3.5in]{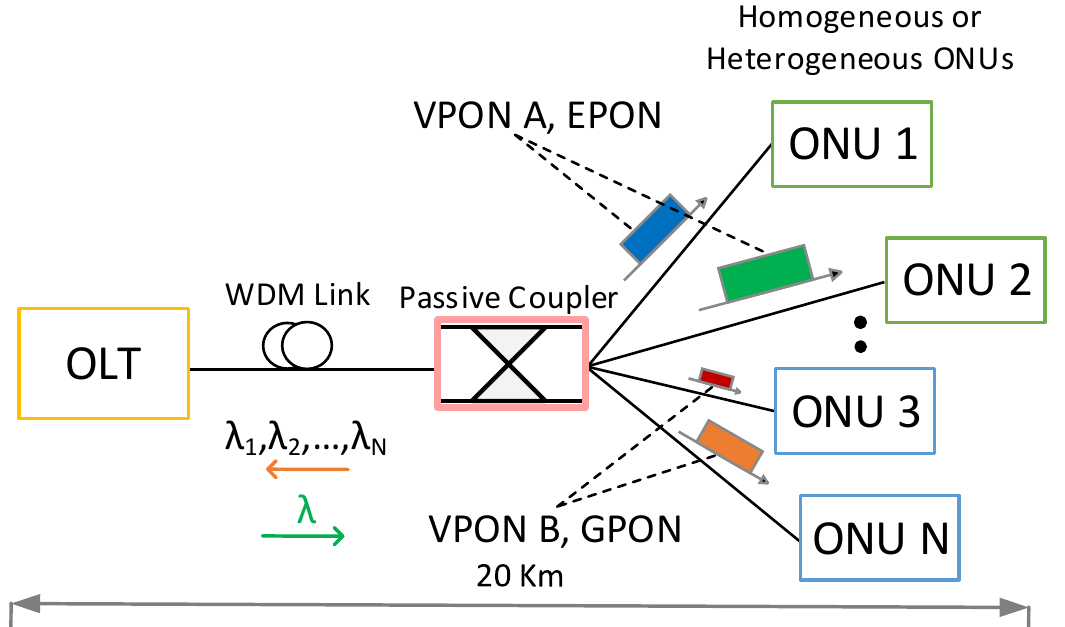} \\
		\footnotesize{(a) Overall virtualization structure} \\
		\includegraphics[width=4in]{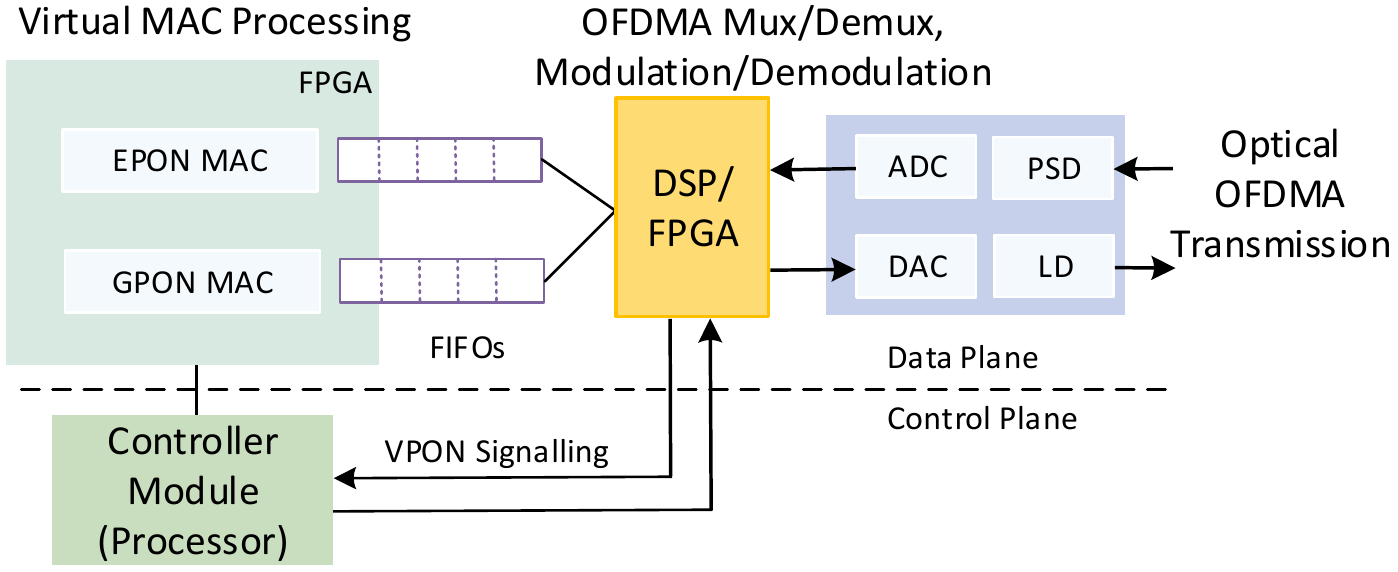} \\		
		\footnotesize{(b) Virtualization of OLT}
	\end{tabular}
	\vspace{0cm}
	\caption{Illustration of OFDMA based virtual access
    network~\cite{wei2009pon}.}
	\vspace{0cm}
	\label{fig_poniard}
\end{figure*}

\subsection{Access Networks}  \label{virt_access:sec}
\subsubsection{OFDMA Based PON Access Network Virtualization}
Wei et al.~\cite{wei2009pon,wei2009pro,wei2010ada} have developed
a link virtualization mechanism that can span from
optical access to backbone networks
based on Orthogonal Frequency Division Multiplexing Access (OFDMA).
Specifically, for access networks, a Virtual PON (VPON) approach
based on multicarrier OFDMA over WDM has been proposed. Distinct
network slices (VPONs) utilize distinct OFDMA subcarriers, which
provide a level of isolation between the VPONs. Thus, different
VPONs may operate with different MAC standards, e.g., as illustrated
in Fig.~\ref{fig_poniard}(a), VPON A may operate as an Ethernet PON
(EPON) while VPON~B operates as a Gigabit PON (GPON). In addition,
virtual MAC queues and processors are isolated to store
  and process the data from multiple VPONs, thus creating virtual
MAC protocols, as illustrated in Fig.~\ref{fig_poniard}(b).
The OFDMA transmissions and receptions are processed in a
DSP module that is controlled by a central SDN control module.
The central SDN control module also controls the different
virtual MAC processes in Fig.~\ref{fig_poniard}(b), which feed/receive
data to/from the DSP module.
Additional bandwidth partitioning between VPONs can be achieved through
Time Division Multiple Access (TDMA).
Simulation studies compared a static allocation of subcarriers to
VPONs with a dynamic allocation based on traffic demands.
The dynamic allocation achieved significantly higher numbers of
supported VPONs on a given network infrastructure as well as lower
packet delays than the static allocation.
A similar strategy for flexibly employing different dynamic
bandwidth allocation modules for different groups of ONU queues has been
examined in~\cite{li2016pro}.

Similar OFDMA based slicing strategies for supporting
cloud computing have been examined by Jinno et al.~\cite{jin2009vir}.
Zhou et al.~\cite{zhou2015dem} have explored a FlexPON with similar
virtualization capabilities. The FlexPON employs OFDM for adaptive
transmissions. The isolation of different VPONs is
mainly achieved through separate MAC processing.
The resulting VPONs allow for flexible port
assignments in ONUs and OLT, which have been demonstrated
in a testbed~\cite{zhou2015dem}.

\subsubsection{FiWi Access Network Virtualization}  \label{virt_fiwi:sec}
\paragraph{Virtualized FiWi Network}
Dai et al.~\cite{QiZSH14,QiGYZ13,ShGYZ13} have examined the
virtualization of FiWi networks~\cite{Bock2014,Effenberger2015}
to eliminate the differences between
the heterogeneous segments (fiber and wireless). The virtualization
provides a unified homogenous (virtual) view of the FiWi network.
The unified network view simplifies flow control and other
operational algorithms for traffic transmissions over the
heterogeneous network segments. In particular, a virtual resource
manager operates the heterogeneous segments. The resource manager
permits multiple routes from a given source node to a given
destination node. Load balancing across the multiple paths has been
examined in~\cite{he2013int,men2014eff}. Simulation results indicate
that the virtualized FiWi network with load balancing significantly
reduces packet delays compared to a conventional FiWi network. An
experimental OpenFlow switch testbed of the virtualized FiWi
network has been presented in~\cite{QiZHG14}. Testbed measurements
demonstrate the seamless networking across the heterogeneous fiber
and wireless networks segments. Measurements for nodal throughput,
link bandwidth utilization, and packet delay indicate performance
improvements due to the virtualized FiWi networking approach.
Moreover, the FiWi testbed performance is measured for a video
service scenario indicating that the virtualized FiWi networking approach
improves the Quality of Experience
(QoE)~\cite{che2014qos,seu2014sur} of the video streaming. A
mathematical performance model of the virtualized FiWi network has
been developed in~\cite{QiZHG14}.

\paragraph{WiMAX-VPON}
WiMAX-VPON~\cite{dha2010wim,DhPX10} is a Layer-2 Virtual Private Network
(VPN) design for FiWi access networks.
WiMAX-VPON executes a common MAC protocol across the
wireless and fiber network segments.
A VPN based admission control mechanism in conjunction with a VPN
bandwidth allocation ensures per-flow Quality of Service (QoS).
Results from discrete event simulations demonstrate that the
proposed WiMAX-VPON achieves favorable performance.
Also, Dhaini et al.~\cite{dha2010wim,DhPX10}
demonstrate how the WiMAX-VPON design can be extended
to different access network types with polling-based wireless and optical
medium access control.

\subsection{Data Centers}  \label{virt_dc:sec}

\subsubsection{LIGHTNESS}
LIGHTNESS~\cite{miao2015sdn,pen2015mul,pag2015opt,Saridis2016} is a
European research project examining an optical Data
Center Network (DCN) capable of providing dynamic,
programmable, and highly available DCN connectivity services.
Whereas conventional DCNs have rigid control and management
platforms, LIGHTNESS strives to introduce flexible control and
management through SDN control.
The LIGHTNESSS architecture comprises server racks that are
interconnected through optical packet
switches, optical circuit switches, and hybrid Top-of-the-Rack
(ToR) switches. The server racks and switches  are all
controlled and managed by an SDN controller.
LIGHTNESS control consists of an SDN controller above the optical
physical layer and OpenFlow agents that interact with the optical
network and server  elements. The SDN controller in cooperation with
the OpenFlow-agents provides a programmable data plane to the
virtualization modules.
The virtualization creates  multiple
Virtual Data Centers (VDCs), each with its own virtual computing
and memory resources, as well as virtual networking resources, based on
a given physical data center.
The virtualization is achieved through a VDC planner module and an
NFV application that directly interact with
the SDN controller.
The VDC planner composes the VDC slices through mapping of the VDC requests
to the physical SDN-controlled switches and server racks.
The VDC slices are monitored by the NFV application, which interfaces
with the VDC planner. Based on monitoring data, the NFV application
and VDC planner may revise the VDC composition, e.g., transition from
optical packet switches to optical circuit switches.

\subsubsection{Cloudnets}
Cloudnets~\cite{azo2013sdn,ban2013mer,bari2013data,FerLR13,WoRSV11,WoRVS10}
exploit network virtualization for
pooling resources among distributed data centers. Cloudnets support
the migration of virtual machines across networks to achieve
resource pooling. Cloudnet designs can be supported through
optical networks~\cite{ShNS13}.
Kantarci and Mouftah~\cite{kan2015res} have examined
designs for a virtual cloud backbone network that interconnects
distributed backbone nodes, whereby each backbone node is
associated with one data center. A network resource manager
periodically executes a virtualization algorithm to accommodate
traffic demands through appropriate resource provisioning. Kantarci
and Mouftah~\cite{kan2015res} have developed and evaluated algorithms for
three provisioning objectives: minimize the outage probability of
the cloud, minimize the resource provisioning, and minimize a
tradeoff between resource saving and cloud outage probability. The
range of performance characteristics for outage probability,
resource consumption, and delays of the provisioning approaches have
been evaluated through simulations. The outage probability of
optical cloud networks has been reduced in~\cite{ahm2014enh} through
optimized service re-locations.

Several complementary aspects of optical cloudnet networks have
recently been investigated.
A multilayer network architecture with an SDN based
network management structure for cloud services has been
developed in~\cite{dov2015usi}.
A dynamic variation of the sharing of optical network resources
for intra- and inter-data center networking has been examined
in~\cite{xie2014dyn}.
The dynamic sharing does not statically assign optical network resources
to virtual optical networks; instead, the network resources are
dynamically assigned according to the time-varying traffic demands.
An SDN based optical transport mode for data center traffic has been
explored in~\cite{vel2014tow}.
Virtual machine migration mechanisms that take the characteristics
of renewable energy into account have been examined in~\cite{zha2015ren}
while general energy efficiency mechanisms for optically networked
could computing resources have been examined in~\cite{tza2014con}.

\subsection{Metro/Core Networks}  \label{virt_core:sec}

\subsubsection{Virtual Optical Network Embedding}  \label{virt_emb:sec}
Virtual optical network embedding seeks to map requests for virtual
optical networks to a given physical optical network infrastructure (substrate).
A virtual optical network consists of both a set of virtual
nodes and a set of interconnecting links that need to
be mapped to the network substrate.
This mapping of virtual networks consisting of both
network nodes and links is fundamentally different from the
extensively studied virtual topology design for optical wavelength
routed networks~\cite{dut2000sur}, which only considered network links
(and did not map nodes).
Virtual network embedding of both nodes and link has already been
extensively studied in general network graphs~\cite{fis2013vir,rah2013svn}.
However, virtual optical network embedding requires additional
constraints to account for the special optical transmission characteristics,
such as the wavelength continuity constraint and the transmission reach
constraint.
Consequently, several studies have begun to examine virtual network
embedding algorithms specifically for optical networks.

\paragraph{Impairment-Aware Embedding}
Peng et al.~\cite{pen2011imp,PeNS13} have modeled the optical
transmission impairments to facilitate the embedding of isolated
VONs in a given
underlying physical network infrastructure.
Specifically, they model the physical (photonic)
layer impairments of both single-line
rate and mixed-line rates~\cite{nag2010opt}.
Peng et al.~\cite{PeNS13} consider intra-VON impairments
from Amplified Spontaneous Emission (ASE) and inter-VON impairments
from non-linear impairments and four wave mixing.
These impairments are captured in a $Q$-factor~\cite{azo2009sur,sar2009phy},
which is considered in the mapping of
virtual links to the underlying physical link resources,
such as wavelengths and wavebands.

\paragraph{Embedding on WDM and Flexi-grid Networks} \label{wdm_flexi_emb:sec}
Zhang et al.~\cite{zha2013net} have considered the embedding of
overall virtual networks encompassing both virtual nodes and virtual links.
Zhang et al. have considered both conventional WDM networks as well as
flexi-grid networks.
For each network type, they formulate the virtual node and virtual link
mapping as a mixed integer linear program.
Concluding that the mixed integer
linear program is NP-hard, heuristic solution approaches
are developed. Specifically, the overall embedding (mapping) problem
is divided into a node mapping problem and a link mapping problem.
The node mapping problem is heuristically solved through a greedy
MinMapping strategy that maps the largest computing resource demand to the
node with the minimum remaining computing capacity (a complementary
MaxMapping strategy that maps the largest demand to the
node with the maximum remaining capacity is also considered).
After the node mapping, the link mapping problem is solved
with an extended grooming graph~\cite{zhu2003nov}.
Comparisons for a small network indicate that
the MinMapping strategy approaches the optimal mixed integer linear program
solution quite closely;
whereas the MaxMapping strategy gives poor results.
The evaluations also indicate that the flexi-grid network requires only about
half the spectrum compared to an equivalent WDM network for several
evaluation scenarios.

The embedding of virtual optical networks in the context of
elastic flexi-grid optical networking has been further examined in several
studies.
For a flexi-grid network based on OFDM~\cite{zha2013surflexi},
Zhao et al.~\cite{zha2013vir} have compared a greedy heuristic that maps
requests in decreasing order of the required resources with an arbitrary
first-fit benchmark.
Gong et al.~\cite{gon2014vir} have considered flexi-grid networks with
a similar overall strategy of node mapping followed by link mapping
as Zhang et al.~\cite{zha2013net}.
Based on the local resource constraints at each node,
Gong et al.~have formed a layered auxiliary graph for the node mapping.
The link mapping is then solved with a shortest path routing approach.
Wang et al.~\cite{wan2015vir} have examined an embedding approach
based on candidate mapping patterns that could provide the requested
resources. The VON is then embedded according to a shortest path
routing.
Pages et al.~\cite{pag2015opt} have considered embeddings that
minimize the required optical transponders.

che2016cos

\paragraph{Survivable Embedding}  \label{surv_emb:sec}
Survivability of a virtual optical network, i.e., its continued
operation in the face of physical node or link failures, is
important for many applications that require dependable service. Hu
et al.~\cite{hu2013sur} developed an embedding that can survive the
failure of a single physical node. Ye et al.~\cite{ye2015sur} have
examined the embedding of virtual optical networks so as to survive
the failure of a single physical node or a physical link.
Specifically, Ye et al. ensure that each virtual
node request is mapped to a primary physical node as well as a
distinct backup physical node. Similarly, each virtual link is
mapped to a primary physical route as well as a node-disjoint backup
physical route. Ye et al. mathematically formulate an optimization
problem for the survivable embedding and then propose a Parallel
Virtual Infrastructure (VI) Mapping (PAR) algorithm. The PAR
algorithm finds distinct candidate physical nodes (with the highest
remaining resources) for each virtual node request. The candidate
physical nodes are then jointly examined with pairs of
 shortest node-disjoint paths.
The evaluations in~\cite{ye2015sur} indicate that the parallel
PAR algorithm reduces the blocking probabilities of virtual network requests
by 5--20~\% compared to a sequential algorithm benchmark.
A limitation of the survivable embedding~\cite{ye2015sur} is that it
protects only from a single link or node failure.
As the optical infrastructure is expected to penetrate
deeper in the access network deployments (e.g., mobile backhaul),
it will become necessary to consider multiple failure points.
Similar survivable network embedding algorithms that employ node-disjoint
shortest paths in conjunction with specific cost metrics for
node mappings have been investigated by Xie et al.~\cite{xie2014sur}
and Chen et al.~\cite{che2016cos}.
Jiang et al.~\cite{jia2015ava} have examined a solution variant based
on maximum-weight maximum clique formation.

The studies~\cite{son2011ban,pao2014mul,ass2016net} have examined
so-called bandwidth squeezed restoration for virtual topologies.
With bandwidth squeezing, the back-up path
bandwidths of the surviving virtual topologies are generally lower than
the bandwidths on the working paths.

Survivable virtual topology design in the context of multidomain optical
networks has been studied by Hong et al.~\cite{Hong2015}.
Hong et al.~focused on minimizing the total network link cost
for a given virtual traffic demand. A heuristic algorithm
for partition and contraction
mechanisms based on cut set theory has been proposed for the
mapping of virtual links onto multidomain optical networks.
A hierarchical SDN control plane is split between local controllers
that to manage individual domains and a
global controller for the overall management.
The partition and contraction
mechanisms abstract inter- and intra-domain information as a
method of contraction. Survivability conditions are ensured
individually for inter- and intra-domains such that survivability is
met for the entire network.
The evaluations in~\cite{Hong2015} demonstrate successful virtual
network mapping at the scale required by commercial Internet
service providers and infrastructure providers.

\paragraph{Dynamic Embedding}
The embedding approaches surveyed so far have mainly focused on the
offline embedding of a static set of virtual network requests.
However, in the ongoing network operation the dynamic embedding of
modifications (upgrades) of existing virtual networks, or the
addition of new virtual networks are important. Ye et
al.~\cite{ye2014upg} have examined a variety of strategies for
upgrading existing virtual topologies. Ye et al. have considered
both scenarios without advance planning (knowledge) of virtual
network upgrades and scenarios that plan ahead for possible
(anticipated) upgrades. For both scenarios, a divide-and-conquer
strategy and an integrate-and-cooperate strategy are
examined. The divide-and conquer strategy sequentially maps all the
virtual nodes
and then the virtual links. In contrast, the integrate-and-cooperate
strategy jointly
considers the virtual node and virtual link mappings. Without
advance planning, these strategies are applied sequentially, as the
virtual network requests arrive over time, whereas, with planning,
the initial and upgrade requests are jointly considered. Evaluation
results indicate that the integrate-and-cooperate strategy
slightly increases a revenue
measure and request acceptance ratio compared to the divide-and-conquer
strategy.
The results also indicate that planning has the potential to
substantially increase the revenue and acceptance ratio. In a
related study, Zhang et al.~\cite{zha2015dyn} have examined
embedding algorithms for virtual network requests that arrive
dynamically to a multilayer network consisting of electrical and
optical network substrates.

\paragraph{Energy-efficient Embedding}
Motivated by the growing importance of green networking and information
technology~\cite{BiPCR12}, a few studies have begun to consider
the energy efficiency of the embedded virtual optical networks.
Nonde et al.~\cite{non2015ene} have developed and evaluated
mechanisms for embedding virtual cloud networks so as to minimize
the overall power consumption, i.e., the aggregate of the
power consumption for communication and computing (in the data centers).
Nonde et al. have incorporated the power consumption of the
communication components, such as transponders and optical switches,
as well as the power consumption characteristics of data center servers
into a mathematical power minimization model.
Nonde et al. then develop a real-time heuristic for energy-optimized
virtual network embedding.
The heuristic strives to consolidate computing requests in the
physical nodes with the least residual computing capacity.
This consolidation strategy is motivated by the typical
power consumption characteristic of a compute server that
has a significant idle power consumption and then grows linearly with
increasing computing load; thus a fully loaded server is more
energy-efficient than a lightly loaded server.
The bandwidth demands are then routed between the nodes according to
 a minimum hop algorithm.
The energy optimized embedding is compared with a cost optimized
 embedding that only seeks to minimize the number of utilized wavelength
 channels.
 The evaluation results in~\cite{non2015ene} indicate
 that the energy optimized embedding significantly reduces the overall energy
 consumption for low to moderate loads on the physical infrastructure;
 for high loads, when all physical resources need to be utilized,
 there are no significant savings. Across the entire load range, the
 energy optimized embedding saves on average 20~\% energy compared to the
 benchmark minimizing the wavelength channels.

Chen~\cite{che2016pow} has examined a similar energy-efficient
virtual optical network embedding that considers primary and
link-disjoint backup paths, similar to the survivable embeddings in
Section~\ref{surv_emb:sec}. More specifically, virtual link requests
are mapped in decreasing order of their bandwidth requirements to
the shortest physical transmission distance paths, i.e., the highest
virtual bandwidth demands are allocated to the shortest physical
paths. Evaluations indicate that this link mapping approach roughly
halves the power consumption compared to a random node mapping
benchmark. Further studies focused on energy savings have examined
virtual link embeddings that maximize the usage of nodes with
renewable energy~\cite{she2014fol} and the traffic
grooming~\cite{wan2014hie} onto sliceable BVTs~\cite{zha2015ene}.

\subsubsection{Hypervisors for VONs} \label{virt_hv:sec}
The operation of VONs over a given
underlying physical (substrate) optical network requires an
intermediate hypervisor. The hypervisor presents the physical
network as multiple isolated VONs to the corresponding VON
controllers (with typically one VON controller per VON). In turn,
the hypervisor intercepts the control messages issued by a VON
controller and controls the physical network to effect the control
actions desired by the VON controller for the corresponding VON.

Towards the development of an optical network hypervisor, Siquera et
al.~\cite{siq2015pro} have developed a SDN-based controller for an
optical transport architecture. The controller implements a
virtualized GMPLS control plane with offloading to facilitate the
implementation of hypervisor functionalities, namely the creation
optical virtual private networks, optical network slicing, and
optical interface management. A major contribution of Siquera et
al.~\cite{siq2015pro} is a Transport Network Operating System
(T-NOS), which abstracts the physical layer for the controller and
could be utilized for hypervisor functionalities.

OpenSlice~\cite{LiuMCTMM13} is a comprehensive
OpenFlow-based hypervisor that creates VONs over underlying elastic
optical networks~\cite{cha2015rou,tal2014spe}. OpenSlice
dynamically provisions end-to-end paths and offloads IP traffic by
slicing the optical communications spectrum. The paths are set up
through a handshake protocol that fills in cross-connection table
entries. The control messages for slicing the optical communications
spectrum, such as slot width and modulation format, are carried in
extended OpenFlow protocol messages. OpenSlice relies on special
distributed network elements, namely bandwidth variable wavelength
cross-connects~\cite{jin2009spe} and multiflow optical
transponders~\cite{jin2012mul} that have been extended for control
through the extended OpenFlow messages. The OpenSlice evaluation
includes an experimental demonstration. The evaluation results
include path provisioning latency comparisons with a GMPLS-based
control plane and indicate that OpenFlow outperforms GMPLS for paths
with more than three hops.
OpenSlice extension and refinements to
multilayer and multidomain networks are surveyed in
Section~\ref{orch:sec}.
An alternate centralized Optical FlowVisor that does not require
extensions to the distributed network elements has been investigated
in~\cite{Azod2012}.

\subsection{Virtualization: Summary and Discussion}
The virtualization studies on access
networks~\cite{wei2009pon,wei2009pro,wei2010ada,jin2009vir,zhou2015dem,QiZSH14,QiGYZ13,ShGYZ13,he2013int,men2014eff,QiZHG14,dha2010wim,DhPX10} have primarily
focused on exploiting and manipulating the specific properties of the
optical physical layer (e.g., different OFDMA subcarriers)
and MAC layer (e.g., polling based MAC protocol) of the optical
access networks for virtualization.
In addition, to virtualization studies on
purely optical PON access networks, two sets of studies, namely
sets~\cite{QiZSH14,QiGYZ13,ShGYZ13,he2013int,men2014eff,QiZHG14}
and WiMAX-VPON~\cite{dha2010wim,DhPX10} have examined
virtualization for two forms of FiWi access networks.
Future research needs to consider virtualization of a wider set of
FiWi network technologies, i.e., FiWi networks that consider
optical access networks with a wider variety of wireless access
technologies, such as different forms of cellular access or combinations
of cellular with other forms of wireless access.
Also, virtualization of integrated access and metropolitan area
networks~\cite{ahm2012rpr,seg2007all,val2015exp,woe2013sdn} is
an important future research direction.

A set of studies has begun to explore optical networking support for
SDN-enabled cloudnets that exploit virtualization to
dynamically pool resources across distributed data centers.
One important direction for
future work on cloudnets is to examine moving data center resources
closer to the users and the subsequent resource pooling across edge
networks~\cite{man2013clo}. Also, the exploration of the benefits of
FiWi networks for decentralized
cloudlets~\cite{din2013sur,sat2009cas,ScSF12,ver2012clo} that
support mobile wireless network services is an important future
research direction~\cite{mai2015inv}.

\begin{figure*}[t!]
	\footnotesize
	\setlength{\unitlength}{0.10in} 
	\centering
	\begin{picture}(40,33)
	\put(15,33){\textbf{Applications, Sec.~\ref{sdnapp:sec}}}
	\put(-6,30){\line(1,0){51}}
	\put(21,30){\line(0,1){2}}
	
	\put(-6,30){\vector(0,-1){2}}
	\put(-9,27){\makebox(0,0)[lt]{\shortstack[l]{			
\textbf{QoS}, Sec.~\ref{app_qos:sec} \\ \\
				Long-term QoS~\cite{ZhZZ13,kho2016qua} \\
				Short-term QoS~\cite{Li2014,PatelJiWang2013} \\
				Virt. Top. Reconfig.~~\cite{WetteKarl2013} \\
				QoS Routing~\cite{Tariq2015,SgPCVC13,Ilchmann2015,ChangLi2015} \\
				QoS Management~\cite{RuBRK14,TeMAD14} \\
				Video Appl.~\cite{chi2015app,Chitimalla2015,Li2014video}
			}}}
			
		\put(10,30){\vector(0,-1){17}}
		\put(5,12){\makebox(0,0)[lt]{\shortstack[l]{		
\textbf{Access Control and} \\
\textbf{Security}, Sec.~\ref{app_secur:sec}\\ \\  	
				Flow-based Access \\ Ctl.~\cite{MaGMT14,nay2009res}	\\			
				Lightpath Hopping Sec.~\cite{Li2016c}	\\			
				Flow Timeout~\cite{ZhuFan2015}
			}}}
					
		\put(31,30){\vector(0,-1){2}}
		\put(23,27){\makebox(0,0)[lt]{\shortstack[l]{			
\textbf{Energy Eff.}, Sec.~\ref{ene_eff:sec}	\\ \\  			
				Appl. Ctl.~\cite{Ji2014,yan2013mul,yan2015per,yan2016per}			\\	
				Routing~\cite{Tego2014,Wang2015,Yevsieieva2015,Yoon2015,val2015exp}
			}}}
	
		\put(45,30){\vector(0,-1){17}}		
		\put(33,12){\makebox(0,0)[lt]{\shortstack[l]{			
\textbf{Failure Recov. $+$} \\
\textbf{Restoration}, Sec.~\ref{app_fail_rec:sec} \\ \\
				Netw. Reprov.~\cite{sav2015bac} \\
				Restoration~\cite{Giorgetti2015} \\
		Reconfig.~\cite{Aguado2016,aib2016sof,SlKMPR14,Kim2015} \\
				Hierarchical Surv.~\cite{ZhangSong2014} \\
				Robust Power Grid~\cite{Rastegarfar2016}
			}}}
									
\end{picture}	
\caption{Classification of application layer SDON studies.}
\label{app_class:fig}
\end{figure*}
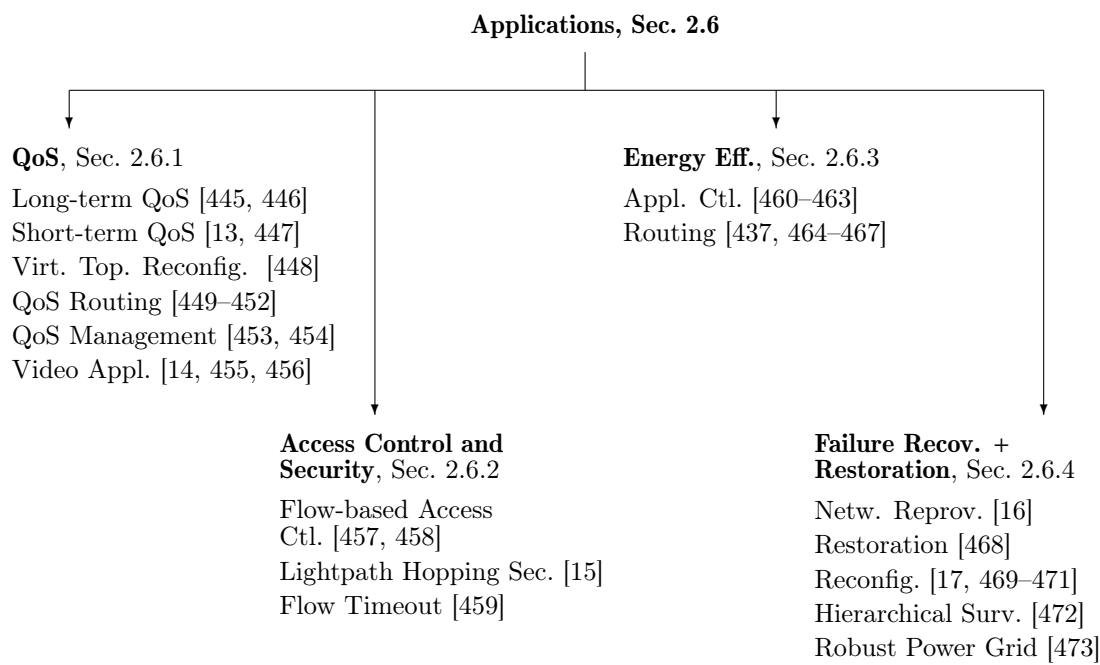
A fairly extensive set of studies has examined virtual network embedding
for metro/core networks.
The virtual network embedding studies have considered the
specific limitations and constraints of optical networks and have
begun to explore specialized embedding strategies that strive to
meet a specific optimization objective,
such as survivability, dynamic adaptability,
or energy efficiency.
Future research should seek to develop a comprehensive framework of
embedding algorithms that can be tuned with weights to achieve
prescribed degrees of the different optimization objectives.

A relatively smaller set of studies has developed and refined hypervisors
for creating VONs over metro/core optical networks.
Much of the SDON hypervisor research has centered on the OpenSlice
hypervisor concept~\cite{LiuMCTMM13}. While OpenSlice accounts
for the specific characteristics of the optical transmission medium,
it is relatively complex as it requires a distributed implementation
with specialized optical networking components.
Future research should seek to achieve the hypervisor functionalities
with a wider set of common optical components so as to reduce cost
and complexity.
Overall, SDON hypervisor research should examine the performance-complexity/cost
tradeoffs of distributed versus centralized approaches.
Within this context of examining the spectrum of distributed to
centralized hypervisors,
future hypervisor research should further refine and optimize the
virtualization mechanisms so as to achieve strict isolation between
virtual network slices, as well as low-complexity hypervisor
deployment, operation, and maintenance.

\section{\MakeUppercase{SDN Application Layer}}  \label{sdnapp:sec}
In the SDN paradigm, applications interact with the controllers to
implement network services. We organize the survey of the studies on
application layer aspects of SDONs according to the main application
categories of quality of service (QoS), access control and security,
energy efficiency, and failure recovery, as illustrated in
Fig.~\ref{app_class:fig}.

\subsection{QoS}  \label{app_qos:sec}
\subsubsection{Long-term QoS: Time-Aware SDN}
Data Center (DC) networks move data back and forth between
DCs to balance the computing load and the data storage usage
(for upload)~\cite{DeCusatis2014}.
These data movements between DCs can span large geographical areas and help
ensure DC service QoS for the end users.
Load balancing algorithms can exploit the characteristics
of the user requests.
One such request characteristic is the high degree of time-correlation over
various time scales
ranging from several hours of a day (e.g., due to a sporting event)
to several days in a year (e.g., due to a political event).
Zhao et al.~\cite{ZhZZ13} have proposed a time-aware SDN application using
OpenFlow extensions to dynamically balance the load across the
DC resources so as to improve the QoS.
Specifically, a time correlated PCE algorithm based on flexi-grid optical
transport (see Section~\ref{PCE:sec})
has been proposed. An SDN application monitors
the DC resources and applies network rules to preserve the QoS.
Evaluations of the algorithm indicate improvements
in terms of network blocking probability, global blocking probability, and
spectrum consumption ratio.
This study did not consider short time scale traffic bursts,
which can significantly affect the load conditions.

We believe that in order to avoid pitfalls in the operation of load balancing
through PCE algorithms implemented with SDN, a wide range of
traffic conditions needs to be considered.
The considered traffic range should include short and long term traffic
variations, which should be traded off with various QoS aspects,
such as type of application
and delay constraints, as well as the resulting costs and control overheads.
Khodakarami et al.~\cite{kho2016qua} have taken steps in this direction by
forming a traffic forecasting model for both long-term and short-term forecasts
in a wide-area mesh network.
Optical lightpaths are then configured based on the overall traffic forecast,
while electronic switching capacities are allocated based on short-term
forecasts.

\begin{figure}[t!]
    \centering
    \vspace{0cm}
    \includegraphics[width=4in]{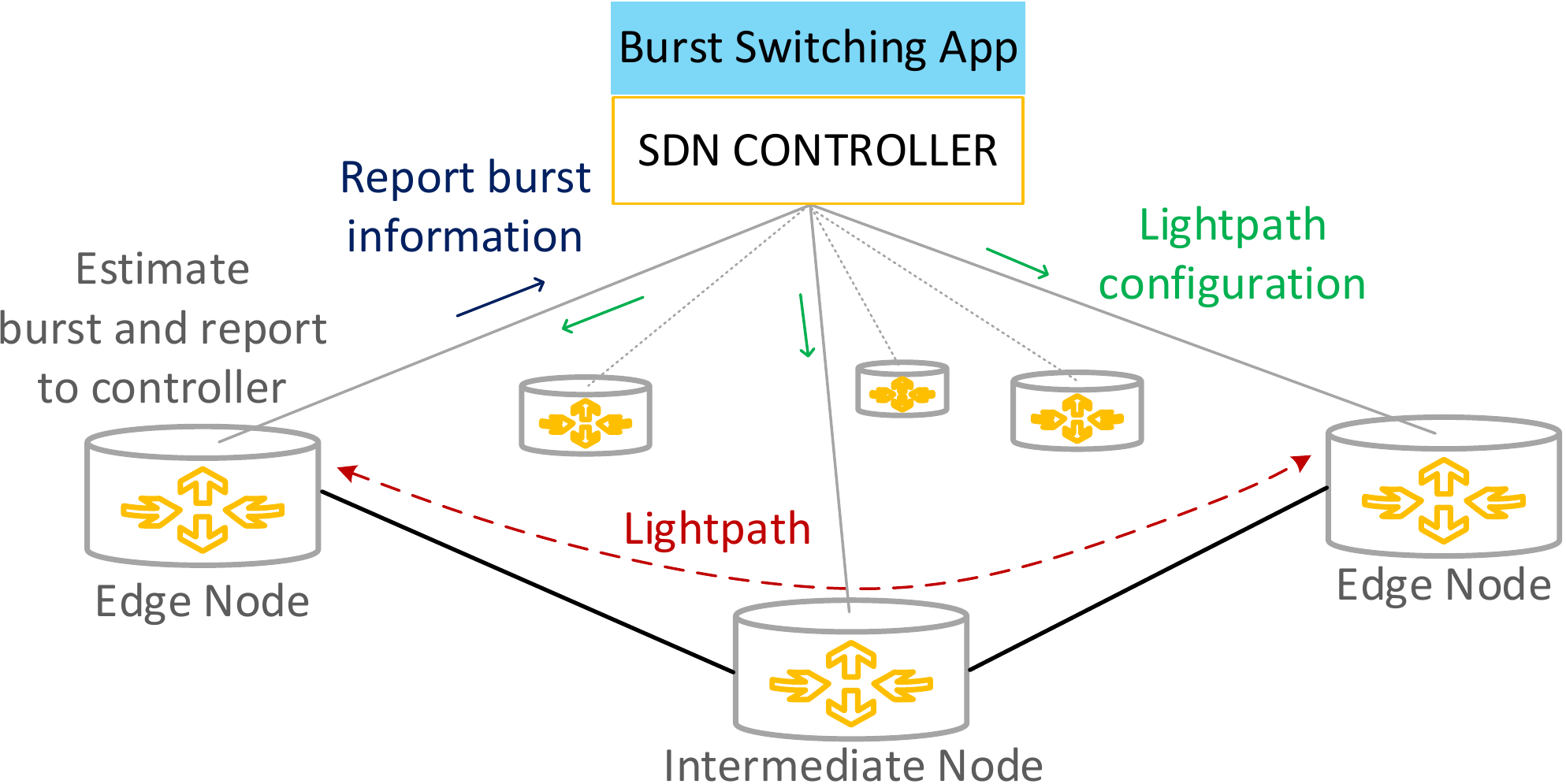}
    \caption{Optical SDN-based QoS-aware burst switching
      application~\cite{PatelJiWang2013}.}
    \label{fig_app_qos}
\end{figure}
\subsubsection{Short Term QoS}
Users of a high-speed FTTH access network may request
very large bandwidths
due to simultaneously running applications that require high data rates.
In such a scenario,
applications requiring very high data rates may affect each other.
For instance, a video conference running simultaneously with
the streaming of a sports video may result in
call drops in the video conference application and
in stalls of the sports video.
Li et al.~\cite{Li2014} proposed an SDN based bandwidth provisioning
application in the broadband remote access server~\cite{Dietz2015}
network. They defined and assigned the minimum bandwidth, which they named
``sweet point'', required
for each application to experience good QoE.
Li et al. showed that maintaining the ``sweet point'' bandwidth for each
application can significantly improve the QoE while other
applications are being served
according to their bandwidth requirements.

In a similar study, Patel et al.~\cite{PatelJiWang2013} proposed
a burst switching mechanism based on a software
defined optical network. Bursts typically originate
at the edge nodes and the aggregation
points due to statistical multiplexing of high speed
optical transmissions. To ensure QoS for multiple traffic classes,
bursts at the edge nodes have to be managed by deciding their end-to-end
path to meet their QoS requirements, such as minimum delay and data rate.
In non-SDN based mechanisms, complicated
distributed protocols, such as GMPLS~\cite{mun2014pce,pao2013sur},
are used to route the burst traffic.
In the proposed application,
the centralized unified control plane decides the routing
path for the burst based on latency and QoS requirements.
A simplified procedure involves $(i)$ burst evaluation at the edge node,
$(ii)$ reporting burst information to the SDN controller, and
$(iii)$ sending of configurations to the optical nodes by the controller
to set up a lightpath as illustrated in Fig.~\ref{fig_app_qos}.
Simulations indicate an increase of performance in terms of
throughput, network blocking probability, and latency
along with improved QoS when compared to non-SDN GMPLS methods.

\subsubsection{Virtual Topology Reconfigurations}
The QoS experienced by traffic flows greatly depends on their
route through a network.
Wette et al.~\cite{WetteKarl2013} have examined an application algorithm that
reconfigures WDM network virtual topologies
(see Section~\ref{wdm_flexi_emb:sec})
according to the traffic levels.
The algorithm considers the localized traffic information and
optical resource availability at the nodes.
The algorithm does not require synchronization,
thus reducing the overhead while simplifying the network design.
In the proposed architecture, optical switches are connected to ROADMs.
The reconfiguration application manages
and controls the optical switches through the SDN controller.
A new WDM controller is introduced to configure the lightpaths taking
 wavelength conversion and lightpath switching at the ROADMs into consideration.
The SDN controller operates on the optical network which appears as a
static network, while the WDM controller configures (and re-configures)
the ROADMs to create
multiple virtual optical networks according to the traffic levels.
Evaluation results indicate improved utilization and throughput.
The results indicate that virtual topologies reconfigurations can
significantly increase the flexibility of the network while achieving
the desired QoS.  However, the control overhead and the delay aspects
due to virtualization and separation of control and lightwave paths
needs to be carefully considered.

\subsubsection{End-to-End QoS Routing}
Interconnections between DCs involve typically
multiple data paths. All the interfaces existing
between DCs can be utilized by MultiPath TCP (MPTCP).
Ensuring QoS in such an MPTCP setting
while preserving throughput efficiency in a
reconfigurable underlying burst switching optical network is a
challenging task. Tariq et al.~\cite{Tariq2015} have
proposed QoS-aware bandwidth reservation for
MPTCP in an SDON.
The bandwidth reservation proceeds in two stages
$(i)$ path selection for MPTCP, and
$(ii)$ OBS wavelength reservation to assign the priorities for
latency-sensitive flows. Larger portions of a wavelength reservation are
assigned to high priority flows, resulting in reduced
burst blocking probability
while achieving the higher MPTCP throughput.
The simulation results in~\cite{Tariq2015} validate the two-stage algorithm
for QoS-aware  MPTCP over an SDON, indicating decreased dropping probabilities,
and increased throughputs.

\begin{figure}[t!]
    \centering
    \vspace{0cm}
    \includegraphics[width=3.5in]{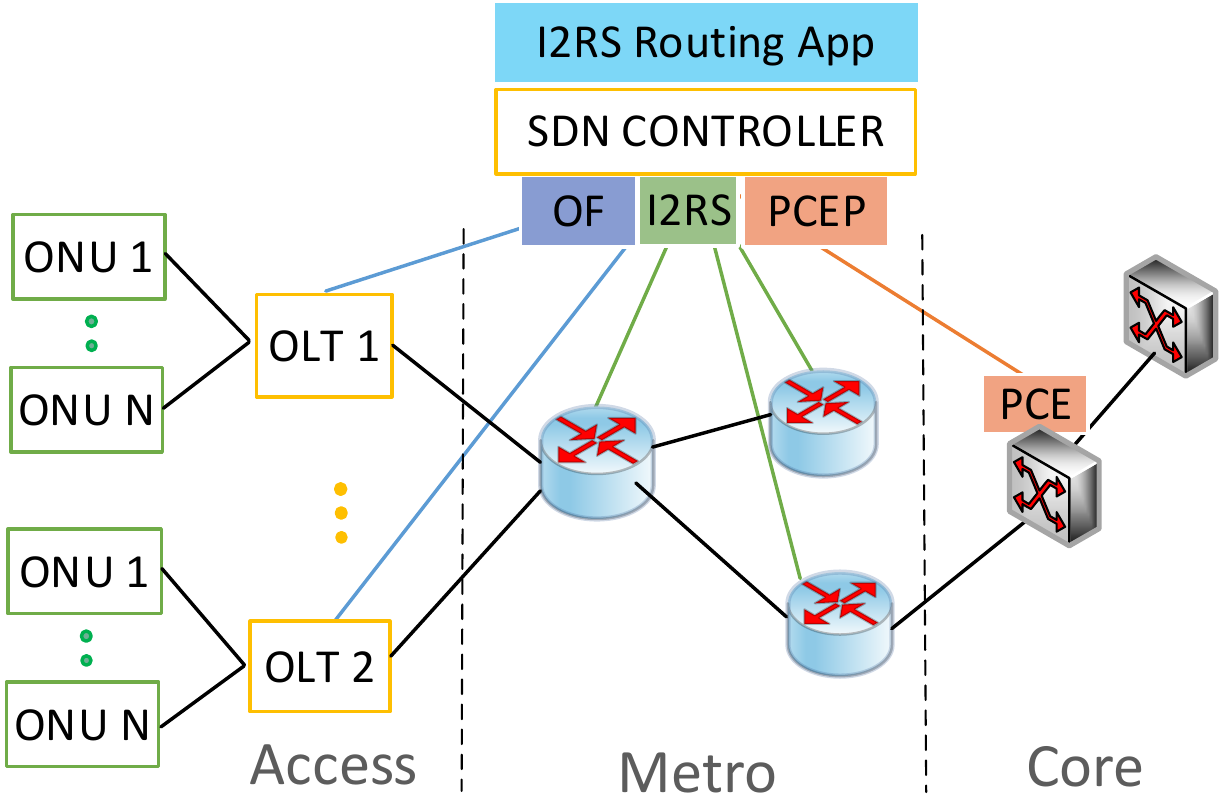}
    \vspace{0cm}
    \caption{Illustration of routing application.}  
   \vspace{0cm}
    \label{fig_app_i2rs}
\end{figure}
Information To the Routing System (I2RS)~\cite{I2RSietf} is a high-level
architecture for communicating and interacting with routing systems,
such as BGP routers.
A routing system may consists of several complex
functional entities, such as a Routing Information Base (RIB),
an RIB manager, topology and policy databases,
along with routing and signalling units.
The I2RS provides a programmability platform that enables
access and modifications of the configurations of the routing system elements.
The I2RS can be extended with SDN principles to achieve global network
management and reconfiguration~\cite{har2013sof}.
Sgambelluri et al.~\cite{SgPCVC13} presented an SDN based routing application
within the I2RS framework to integrate the control of the
access, metro, and core networks as illustrated in Fig.~\ref{fig_app_i2rs}.
The SDN controller communicates with the Path Computation
Elements (PCEs) of the core network
to create Label Switched Paths (LSPs)
based on the information received by the OLTs.
Experimental demonstrations validated the routing optimization
based on the current traffic status and previous load as well as
the unified control interface for access, metro, and core networks.

Ilchmann et al.~\cite{Ilchmann2015} developed an SDN application that communicates to an SDN controller via
an HTTP-based REST API. Over time, lightpaths in an optical network can become inefficient for a number of
reasons (e.g., optical spectrum fragmentation). For this reason, Ilchmann et al. developed an SDN application
that evaluates existing lightpaths in an optical network and offers an application user the option to
re-optimize the lightpath routing to improve various performance metrics (e.g., path length). The application
is user-interactive in that the user can see the number of proposed lightpath routing changes before they are made
and can potentially select a subset of the proposed changes to minimize network down-time.

At the ingress and egress routers of optical networks (e.g., the
edge routers between access and metro networks), buffers are highly
non-economical to implement, as they require large buffers sizes to
accommodate the channel rates of 40~Mb/s or more. To reduce the buffer
requirements at the edge routers, Chang et al.~\cite{ChangLi2015}
have proposed a backpressure application referred to as Refill and
SDN-based Random Early Detection (RS-RED).  RS-RED implements a
refill queue at the ingress device and a droptail
queue at the egress device, whereby both queues are centrally
managed by the RS-RED algorithm running on the SDN
controller. Simulation results showed that at the expense of small delay
increases, edge router buffer sizes can be significantly reduced.

\subsubsection{QoS Management}
Rukert et al.~\cite{RuBRK14} proposed SDN based controlled home-gateway
supporting heterogeneous wired technologies, such as DSL, and
wireless technologies, such as LTE and WiFi.
SDN controllers managed by the ISPs optimize the traffic flows
to each user while accommodating large numbers of users  and ensuring their
minimum QoS.
Additionally, Tego et al.~\cite{TeMAD14} demonstrated an experimental
SDN based QoS management setup to optimize the energy utilization.
GbE links are switched on and off based on the traffic levels.
The QoS management reroutes the traffic to avoid congestion and achieve
efficient throughput. SDN applications conduct active QoS
probing to monitor the network QoS characteristics.
Evaluations have indicated that the SDN based techniques
achieve significantly higher throughput than non-SDN techniques~\cite{TeMAD14}.

\subsubsection{Video Applications}
The application-aware SDN-enabled resource allocation
application has been introduced by Chitimalla et al.~\cite{chi2015app}
to improve the video QoE in a PON access network.
The resource allocation application
uses application level feedback to schedule the optical
resources. The video resolution is incrementally increased or
decreased based on the buffer utilization statistics that the client
sends to the controller. The scheduler at the OLT schedules the
packets based on weights calculated by the SDN controller, whereby
the video applications at the clients communicate with the
controller to determine the weights. If the network is congested,
then the SDN controller communicates to the clients to reduce the
video resolution so as to reduce the stalls and to improve the QoE.

\begin{figure}[t!]
    \centering
    \vspace{0cm}
    \includegraphics[width=5in]{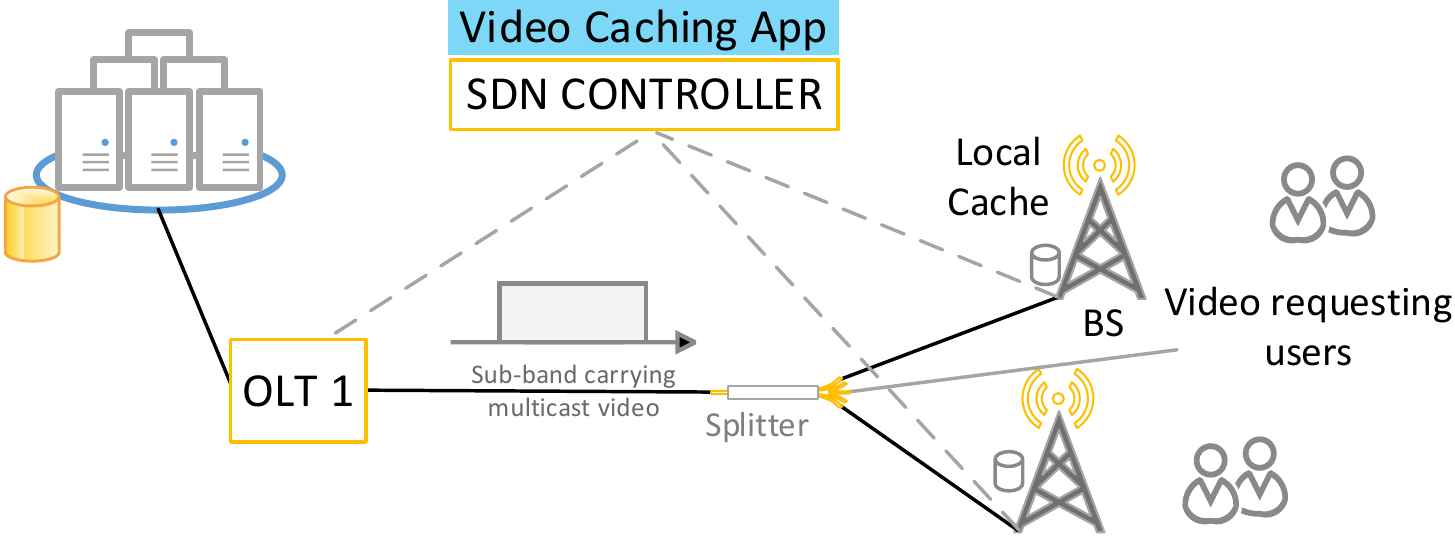}
    \caption{SDN based video caching application in PON
        for mobile users~\cite{Li2014video}.}
    \label{fig_app_video} 
\end{figure}
Caching of video data close the users is generally beneficial for
improving the QoE of video services~\cite{Ahlgren2012,Choi2012}.
Li et al.~\cite{Li2014video} have introduced
caching mechanisms for software-defined PONs.
In particular, Li et al.~have proposed
joint provisioning of the bandwidth to service the video and
the cache management, as illustrated in Fig.~\ref{fig_app_video}.
Based on the request frequency for specific
video content, the Base Station (BS) caches the content with the
assistance of the SDN controller.
The proposed \textit{push}-based mechanism delivers (pushes) the video
to the BS caches when the PON is not congested.
A specific PON transmission sub-band can be used to multicast video
content that needs to be cached at multiple BSs.
The simulation evaluation in~\cite{Li2014video} indicate
that up to 30\% additional videos can be serviced while the service
response delay is reduced to 50\%.

\subsection{Access Control and Security} \label{app_secur:sec}
\subsubsection{Flow-based Access Control}
Network Access Control (NAC) is a networking application that
regulates the access to network services~\cite{cas2007eth,par2014fut}.
A NAC based on traffic flows has been developed by Matias~\cite{MaGMT14}.
Flow-NAC exploits the forwarding rules of OpenFlow switches, which are set by a
central SDN controller, to control the access of traffic flows to
network services.
FlowNAC can implement the access control based on various flow identifiers,
such as MAC addresses or IP source and destination addresses.
Performance evaluations measured
the connections times for flows on a testbed and found average connection
times on the order of 100~ms for completing the flow access control.

In a related study, Nayak et al.~\cite{nay2009res} developed the
Resonance flow based access control system for an enterprise
network. In the Resonance system, the network elements, such as the
routers themselves, dynamically enforce access control policies. The
access control policies are implemented through real-time alerts and
flow based information that is exchanged with SDN principles. Nayak
et al. have demonstrated the Resonance system on a production
network at Georgia Tech. The Resonance design can be readily
implemented in SDON networks and can be readily extended to wide
area networks. Consider for example multiple heterogeneous DCs of
multiple organizations that are connected by an optical backbone
network. The Resonance system can be extended to provide access
control mechanisms, such as authentication and authorization,
through such a wide area SDON.

\subsubsection{Lightpath Hopping Security}
\begin{figure}[t!]
    \centering     \vspace{0cm}
    \includegraphics[width=3.5in]{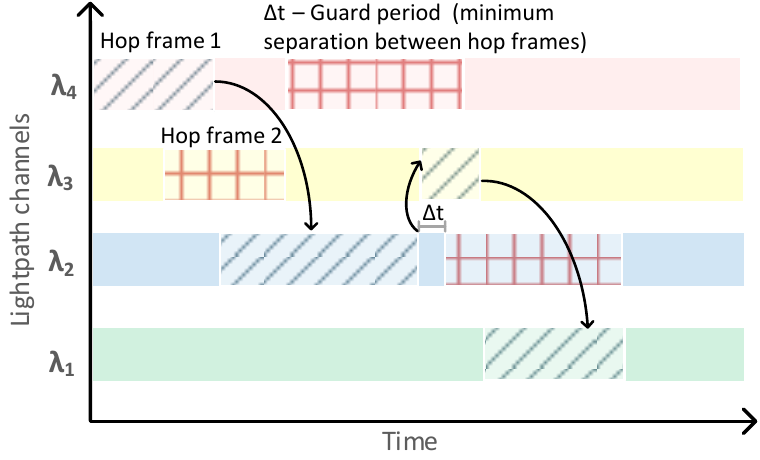} 
    \caption{Overview of optical light path hopping mechanism~\cite{Li2016c}.} 
    \label{fig_app_hop}
\end{figure}
The broad network perspective of SDN controllers facilitates
the implementation of security functions that require this broad
perspective~\cite{ahm2015sec, sco2015sur, ShJJJ14}.
However, SDN may also be vulnerable to a wide range of attacks and
vulnerabilities, including unauthorized access, data leakage,
data modification, and misconfiguration.
Eavesdropping and jamming are security threats on the physical layer
and are especially relevant for the optical layer of SDONs.
In order to prevent
eavesdropping and jamming in an optical lightpath, Li et al.~\cite{Li2016c}
have proposed an SDN based fast lightpath hopping mechanism.
As illustrated in Fig.~\ref{fig_app_hop}, the hopping mechanism operates
over multiple lightpath channels. Conventional optical
lightpath setup times range from several hundreds of milliseconds
to several seconds and would result in a very low hopping frequency.
To avoid the
optical setup times during each hopping period, an SDN based
high precision time synchronization has been proposed.
As a result, a fast hopping
mechanism can be implemented and executed in a coordinated manner. A hop frame
is defined and guard periods are added in between hop frames.
The experimental evaluations indicate that a maximum hopping frequency
of~1~MHz can be achieved with a BER of $1 \times 10^{-3}$.
However, shortcomings of such mechanisms are the secure exchange of hopping
sequences between the transmitter and
the receiver. Although, centralized SDN control provides authenticated
provisioning of the hopping sequence, additional mechanisms to secure
the hopping sequence from
being obtained through man-in-the-middle attacks should be investigated.

\subsubsection{Flow Timeout}
SDN flow actions on the forwarding and switching elements have generally a
validity period.
Upon expiration of the validity period, i.e., the flow action timeout,
the forwarding or switching
element drops the flow action from the forwarding information base or the flow
table. The switching
element  CPU must be able to access the flow action information with very
low latency so as to perform switching actions at the line rate.
Therefore, the flow actions are commonly stored in Ternary Content
Addressable Memories (TCAMs)~\cite{Pagiamtzis2006}, which are limited
to storing on the order of thousands of distinct entries.
In SDONs, the optical network elements perform
the actions set by the SDN controller. These actions have to be stored in
a finite memory space.  Therefore, it is important to utilize the finite
memory space as efficiently as
possible~\cite{Bull2015, Liang2015, ngu2016rul, Xie2014d, Zhang2015timeout}.
In the dynamic timeout approach~\cite{ZhuFan2015},
the SDN controller tracks the TCAM occupancy levels in the switches and adjusts
timeout durations accordingly.
However, a shortcoming of such techniques
is that the bookkeeping processes at the SDN controllers can become
cumbersome for a large network.  Therefore, autonomous
timeout management techniques that are implemented at the hypervisors
can reduce the controller processing load and are
an important future research direction.

\subsection{Energy Efficiency}  \label{ene_eff:sec}
The separation of the control plane from the data plane
and the global network perspective are unique advantages of SDN for
 improving the energy efficiency of networks,
which is an important goal~\cite{tuc2011gre,zha2010ene}.

\subsubsection{Power-saving Application Controller}
Ji et al.~\cite{Ji2014} have proposed an all optical
energy-efficient network centered around an application
controller~\cite{yan2013mul,yan2015per} that monitors power consumption
characteristics and enforces power savings policies.
Ji et al. first
introduce energy-efficient variations of Digital-to-Analog
Converters (DACs) and wavelength selective ROADMs as components for
their energy-efficient network. Second, Jie et al. introduce
an energy-efficient switch architecture that consists of multiple
parallel switching planes, whereby each plane consists of three
stages with optical burst switching employed in the second (central)
switching stage.
Third, Jie et al. detail a multilevel SDN based control architecture
for the network built from the introduced components and switch.
The control structure accommodates multiple networks domains,
whereby each network domain can involve multiple switching
technologies, such as time-based and frequency-based optical switching.
All controllers for the various domains and technologies are placed
under the control of an application controller.
Dedicated power monitors that are distributed throughout the network
update the SDN based application controller about the energy consumption
characteristics of each network node.
Based on the received energy consumption updates,
the application controller executes power-saving strategies.
The resulting control actions are signalled by the application
controller to the various controllers for the different network domains
and technologies.
An extension of this multi-level architecture to cloud-based
radio access networks has been examined in~\cite{yan2016per}.

\subsubsection{Energy-Saving Routing}
Tego et al.~\cite{Tego2014} have proposed an energy-saving
application that switches off under-utilized GbE network links.
Specifically, Tego et al. proposed two methods: Fixed Upper Fixed
Lower (FUFL) and Dynamic Upper and Fixed Lower (DLFU). In FUFL, the
IP routing and the connectivity of the logical topology are
\textit{fixed}. The utilization of physical GbE links (whereby
multiple parallel physical links form a logical link) is compared
with a threshold to determine whether to switch off or on individual
physical links (that support a given logical link). The traffic on a
physical link that is about to be switched off is rerouted on a
parallel physical GbE link (within the same logical link). In
contrast, in the DLFU approach, the energy saving application
monitors the load levels on the virtual links. If the load level on
a given virtual link falls below a threshold value, then the virtual
link topology is reconfigured to eliminate the virtual link with the
low load. A general pitfall of such link switch-off techniques is
that energy savings may be achieved at the expense of deteriorating
QoS. The QoS should therefore be closely monitored when switching
off links and re-routing flows.

A similar SDN based routing strategy that strives to save energy while
preserving the QoS has been examined in the context of a GMPLS
optical networks in~\cite{Wang2015}.
Multipath routing optimizing applications that strive to save
energy in an SDN based transport optical network have been presented
in~\cite{Yevsieieva2015}.
A similar SDN based optimization approach
for reducing the energy consumption in
data centers has been examined by Yoon et al.~\cite{Yoon2015}.
Yoon et al. formulated a mixed integer linear program that models the switches
and hosts as queues. Essentially, the optimization decides on the
switches and hosts that could be turned off.
As the problem is NP-hard,
annealing algorithms are examined.
Simulations indicate that energy savings of more than 80\% are possible for
low data center utilization rates, while the energy savings decrease to less
than 40\% for high data center utilization rates.
Traffic balancing in the metro optical access networks through the
SDN based reconfiguration of optical subscriber units
in a TWDM-PON systems for energy
savings has been additionally demonstrated in~\cite{val2015exp}.

\subsection{Failure Recovery and Restoration}  \label{app_fail_rec:sec}
\subsubsection{Network Reprovisioning}
\begin{figure}[t!]
    \centering
    \includegraphics[width=3.5in]{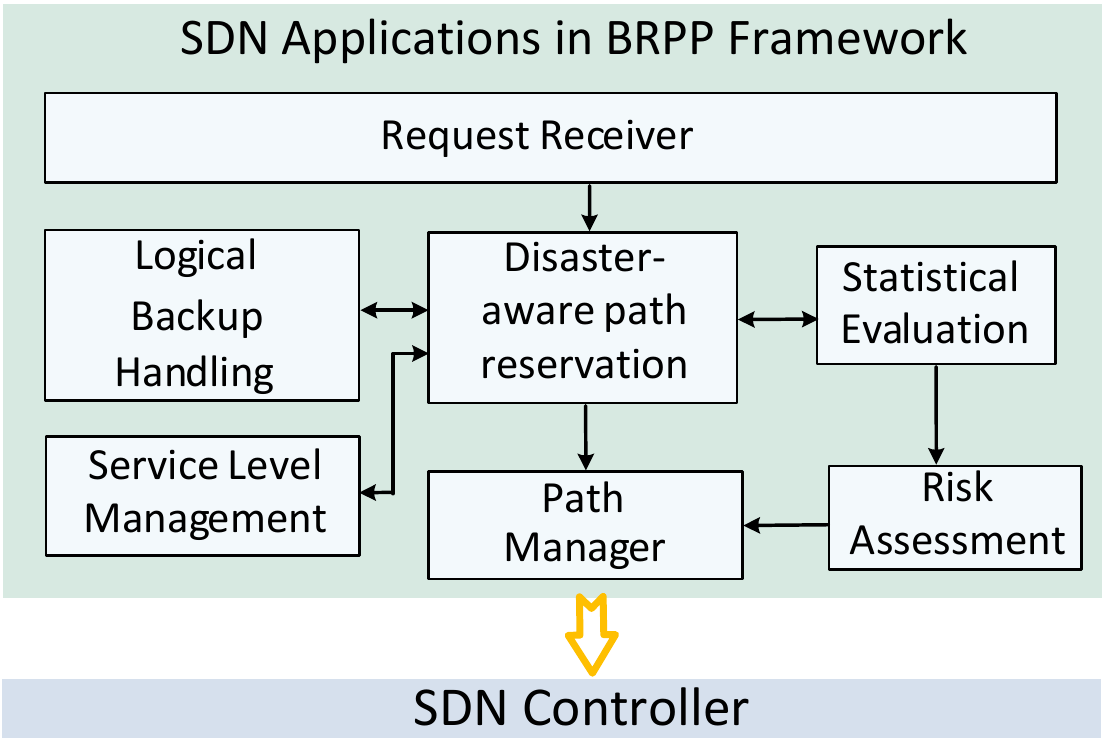}
    \caption{Illustration of application layer
        for disaster aware networking~\cite{sav2015bac}.}
    \label{fig_app_reprov}
\end{figure}
Network disruptions can occur due to various natural and/or man-made
factors.
Network resource reprovisioning is a process to
change the network configurations, e.g., the network topology and routes,
to recover from failures.
A Backup Reprovisioning with Path Protection (BRPP), based on SDN for
optical networks has been presented by Savas et al.~\cite{sav2015bac}.
An SDN application framework as illustrated in
Fig.~\ref{fig_app_reprov} was designed to support the
reprovisioning with services, such
as provisioning the new connections, risk assessment, as well as service level
and backup management.
When new requests are received by the BRPP application
framework, the statistics module evaluates the network state to find the
primary path and a link-disjoint backup path.  The computed backup
paths are stored as logical links without being
provisioned on the physical network. The logical backup module manages
and recalculates the logical links when a new backup path cannot be
accommodated or to optimize the existing backup paths (e.g., minimize the
backup path distance). Savas et al.~introduce a degraded backup path mechanism
that reserves not the full, but a lower (degraded) transmission capacity
on the backup paths, so as to accommodate more requests.
Emulations of the proposed mechanisms indicate improved
network utilization while effectively provisioning the backup paths
for restoring the network after network failures.

As a part of DARPA's core
optical networks CORONET project, a non-SDN based Robust Optical Layer
End-to-end X-connection (ROLEX)
protocol has been demonstrated and presented along with the lessons
learned~\cite{VonLehmen2015}.
ROLEX is a distributed protocol for failure recovery
which requires a considerable amount of signaling between nodes
for the distributed management.
Therefore to avoid the pitfall of excessive signalling,
it may be worthwhile to examine a ROLEX version with
centralized SDN control in future research
to reduce the recovery time and
signaling overhead, as well as the costs of restored
paths while ensuring the user QoS.

\subsubsection{Restoration Processing}
During a restoration, the network control plane simultaneously triggers backup
provisioning of all disrupted paths.
In GMPLS restoration, along with signal flooding,
there can be contention of signal messages at the network nodes.
Contentions may arise due to spectrum conflicts of the lightpath,
or node-configuration overrides, i.e., a new configuration request arrives
while a preceding reconfiguration is under way.
Giorgetti et al.~\cite{Giorgetti2015} have proposed
dynamic restoration in the elastic optical
network to avoid signaling contention in SDN (i.e., of OpenFlow messages).
Two SDN restoration mechanisms were presented:
$(i)$ the independent restoration scheme (SDN-ind),
and $(ii)$ the bundle restoration scheme (SDN-bund).
In SDN-ind, the controller triggers simultaneous independent
flow modification (Flow-Mod) messages for each backup path
to the switches involved in the reconfigurations.
During contention, switches enqueue the multiple received Flow-Mod messages
and process them sequentially. Although SDN-ind achieves reduced
recovery time as compared to non-SDN GMPLS, the waiting of messages in
the queue incurs a delay. In SDN-bund,
the backup path reconfigurations are bundled into a single message,
i.e., a Bundle Flow-Mod message, and sent to each involved switch.
Each switch then configures the flow modifications in
one reconfiguration, eliminating the delay incurred by the queuing of
Flow-Mod messages. A similar OpenFlow enabled restoration
in Elastic Optical Networks (EONs) has been studied in~\cite{Liu2015d}.

\subsubsection{Reconfiguration}
\begin{figure}[t!]
    \centering
    \vspace{0cm}
    \includegraphics[width=3in]{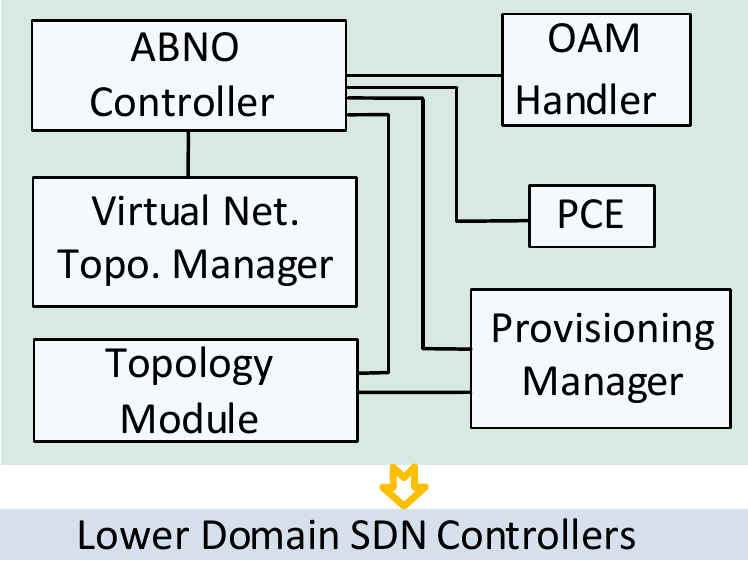}
    \vspace{0cm}
    \caption{Illustration of Application-Based Network Operation (ABNO)
         architecture~\cite{Aguado2016}.}
    \label{fig_app_abno}
\end{figure}
Aguado et al.~\cite{Aguado2016} have demonstrated
a failure recovery mechanism as part of the EU FP7 STRAUSS project with
dynamic virtual reconfigurations using SDN.
They considered multidomain hypervisors and
domain-specific controllers to virtualize the multidomain networks.
The Application-Based Network Operations (ABNO) framework
illustrated in Fig.~\ref{fig_app_abno} enables network automation
and programmability.
ABNO can compute end-to-end optical paths and delegate the
configurations to lower layer domain SDN controllers.
Requirements for fast recovery from network failures
would be in the order of tens of milliseconds, which is challenging to
achieve in large scale networks. ABNO reduces the recovery
times by pre-computing the backup connections after the first failure,
while the Operation, Administration and Maintenance (OAM)
module~\cite{Paolucci2015}
communicates with the ABNO controller to configure the new end-to-end
connections in response to a failure alarm. Failure alarms are triggered
by the domain SDN controllers monitoring the traffic via the optical
power meters when power is below $-20$~dBm.
In order to ensure survivability, an adaptive survivability scheme that
takes routing as well as spectrum assignment and modulation into
consideration has been explored in~\cite{aib2016sof}.

A similar design for end-to-end protection and failure
recovery has been demonstrated by Slyne et al.~\cite{SlKMPR14} for a long-reach
(LR) PON. LR-PON failures are highly
likely due to physical breaks in the long feeder fibers. Along with the
high impact of connectivity break down or degraded service,
physical restoration time can be very long. Therefore, 1:1 protection for
LR-PONs based on SDN has been proposed, where primary and secondary (backup)
OLTs are used without traffic duplication.
More specifically, Slyne et al. have
devised and demonstrated an OpenFlow-Relay located at the switching unit.
The OpenFlow-Relay
detects and reports a failure along with fast updating of forwarding rules.
Experimental demonstration show the backup OLT carrying protected traffic
within $7.2$ ms after a failure event.

An experimental demonstration utilizing multiple paths in optical
transport networks for failure recovery has been discussed by
Kim et al.~\cite{Kim2015}.
Kim et al. have used commercial grade IP WDM network equipment and
implemented multipath TCP
in an SDN framework to emulate inter-DC communication.
They developed an SDN application, consisting of an
cross-layer service manager module and a
cross-layer multipath transport module to
reconfigure the optical paths
for the recovery from connection impairments.
Their evaluations show increased bandwidth
utilization and reduced cost while being resilient to network impairments
as the cross-layer multipath transport module does not reserve the backup
path on the transport network.

\subsubsection{Hierarchical Survivability}
Networks can be made survivable by introducing resource redundancy.
However, the cost of the network increases with increased redundancy.
Zhang et al.~\cite{ZhangSong2014} have demonstrated a highly survivable
IP-Optical multilayered transport network.
Hierarchal controllers are placed for multilayer
resource provisioning. Optical nodes are controlled by
Transport Controllers (TCs), while higher
layers (IP) are controlled by unified controllers (UCs).
The UCs communicate with the TCs
to optimize the routes based on cross-layer information. If a fiber causes a
service disruption, TCs may directly set up alternate routes or
ask the UCs for optimized routes. A pitfall of such
hierarchical control techniques can be long restoration times. However,
the cross layer restorations
can recover from high degrees of failures, such as
multipoint and concurrent failures.

\subsubsection{Robust Power Grid}
The lack of a reliable communication infrastructure for power grid
management was one the many reasons for the widespread blackout in
the Northeastern U.S.A. in the year 2003, which affected the lives
of 50 million people~\cite{Parandehgheibi2014}. Since then building
a reliable communication infrastructure for the power grid has
become an important priority. Rastegarfar et
al.~\cite{Rastegarfar2016} have proposed a communication
infrastructure that is focused on monitoring and can react to and recover from
failures so as to reliably support power grid applications.
More specifically, their
architecture was built on SDN based optical networking for
implementing robust power grid control applications. Control and
infrastructure in the SDN based power grid management exhibits an
interdependency i.e., the physical fiber relies on the control plane
for its operations and the logical control plane relies on the
same physical fiber for its signalling communications. Therefore,
they only focus on optical protection switching instead of IP layer
protection, for the resilience of the SDN control. Cascaded failure
mechanisms were modeled and simulated for two geographical
topologies (U.S. and E.U.). In addition, the impacts of cascaded
failures were studied for two scenarios $(i)$ static optical layer
(static OL), and $(ii)$ dynamic optical layer (dynamic OL). Results
for a static OL illustrated that the failure cascades are persistent
and are closely dependent on the network topology. However, for a
dynamic OL (i.e., with reconfiguration of the physical layer),
failure cascades were suppressed by an average of 73\%.

\subsection{Application Layer: Summary and Discussion}
The SDON QoS application studies have mainly examined traffic and
network management mechanisms that are supported through the
OpenFlow protocol and the central SDN controller.
The studied SDON QoS applications are structurally very similar in
that the traffic conditions or
network states (e.g., congestion levels) are probed
or monitored by the central SDN controller.
The centralized knowledge of the traffic and network is then utilized
to allocate or configure resources, such as DC resources in~\cite{ZhZZ13},
application bandwidths in~\cite{Li2014}, and
topology configurations or routes
in~\cite{WetteKarl2013,Tariq2015,SgPCVC13,ChangLi2015}.
Future research
 on SDON QoS needs to further optimize the interactions of the
controller with the network applications and data plane
to quickly and correctly react to changing user demands and
network conditions, so as to assure consistent QoS.
The specific characteristics and requirements
of video streaming applications have
been considered in the few studies on video
QoS~\cite{chi2015app,Chitimalla2015,Li2014video}.
Future SDON QoS research should consider a wider range of
specific prominent application traffic types with
specific characteristics and requirements, e.g., Voice over IP (VoIP)
traffic has relatively low bit rate requirements, but requires low
end-to-end latency.

\begin{figure*}[t!]
	\centering
	\includegraphics[width=5in]{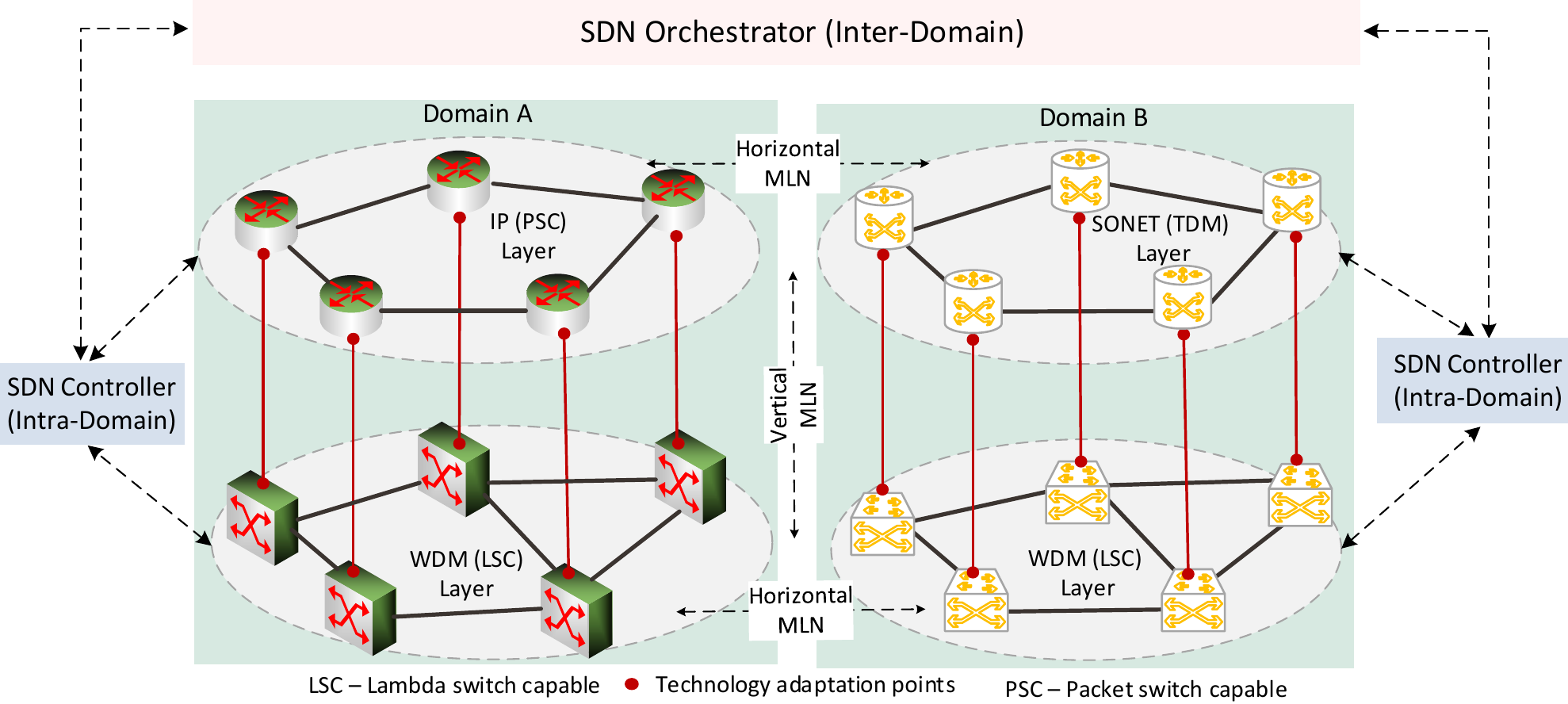}
	\caption{Illustration of SDN orchestration of multilayer networking.}
	\label{fig_control_multilay}
\end{figure*}
Very few studies have considered security and access control for SDONs.
The thorough study of the broad topic area of security and privacy is
an important future research direction in SDONs, as outlined in
Section~\ref{futworksec:sec}
Energy efficiency is similarly a highly important topic
within the SDON research area that has received relatively little
attention so far and presents overarching research challenges,
see Section~\ref{futworkene:sec}.

One common theme of the SDON application layer studies focused
on failure recovery and restoration has been to exploit
the global perspective of the SDN control.
The global perspective has been exploited for
for improved planning of the recovery and
restoration~\cite{sav2015bac,Aguado2016,ZhangSong2014} as well as
for improved coordination of the execution of the restoration
processes~\cite{Giorgetti2015,Liu2015d}.
Generally, the existing failure recovery and restoration studies
have focused on network (routing) domain that is owned by a particular
organizational entity. Future research should seek to examine the
tradeoffs when exploiting the
global perspective of orchestration of multiple routing domains, i.e.,
the failure recovery and restoration techniques surveyed in this section
could be combined with the multidomain orchestration techniques
surveyed in Section~\ref{orch:sec}.
One concrete example of multidomain orchestration could be to coordinate
the specific LR-PON access network protection and
failure recovery~\cite{SlKMPR14}
with protection and recovery techniques for metropolitan and core
network domains, e.g.,~\cite{sav2015bac,Aguado2016,Kim2015,ZhangSong2014},
for improved end-to-end protection and recovery.

\section{\MakeUppercase{Orchestration}}  \label{orch:sec}
As introduced in Section~\ref{intro:orch:sec}, orchestration
accomplishes higher layer abstract coordination of network services
and operations. In the context of SDONs, orchestration has mainly
been studied in support of multilayer networking.
Multilayer networking in the context of SDN and network virtualization
generally refers to networking across multiple network layers
and their respective technologies, such as
IP, MPLS, and WDM, in combination with networking across multiple
routing domains~\cite{LeZGF11,leo2003vir,rui2011sur,tou2001net,vig2005mul}.
The concept of multilayer networking is generally an abstraction of
providing network services with multiple networking layers (technologies) and
multiple routing domains.
The different network layers and their technologies are sometimes classified into
Layer~0 (e.g., fiber-switch capable), Layer~1 (e.g., lambda switching
capable), Layer~1.5 (e.g., TDM SONET/SDH), Layer~2 (e.g., Ethernet),
Layer~2.5 (e.g., packet switching capable using MPLS), and Layer~3
(e.g., packet switching capable using IP routing)~\cite{YoMKN14}.
Routing domains are also commonly referred to as
network domains, routing areas, or levels~\cite{LeZGF11}.

The recent multilayer
networking review article~\cite{LeZGF11} has introduced a range of capability
planes to represent the grouping of related functionalities for a
given networking technology. The capability planes include the data
plane for transmitting and switching data. The control plane and the
management plane directly interact with the data plane for
controlling and provisioning data plane services as well as for
trouble shooting and monitoring the data plane. Furthermore, an
authentication and authorization plane, a service plane, and an
application plane have been introduced for providing network
services to users.

Multilayer networking can involve vertical layering or
horizontal layering~\cite{LeZGF11}, as illustrated in
Fig.~\ref{fig_control_multilay}. In vertical layering, a given
layer, e.g., the routing layer, which may employ a
particular technology, e.g., the Internet Protocol (IP),
uses another (underlying) layer, e.g., the Wavelength
Division Multiplexing (WDM) circuit switching layer, to provide
services to higher layers. In horizontal layering, services are
provided by ``stitching'' together a service path across multiple
routing domains.

SDN provides a convenient control framework for these
flexible multilayer networks~\cite{LeZGF11}. Several research
networks, such as ESnet, Internet2, GEANT, Science DMZ
(Demilitarized Zone) have experimented with these multilayer
networking concepts~\cite{KiSTP13,RoMSS14}.
In particular, SDN based multilayer network
architectures, e.g.,~\cite{woe2013sdn, iiz2016mul, mun2016nee},
are formed by conjoining
the layered technology regions
$(i)$ in vertical fashion i.e., multiple technology layers internetwork
within a single domain, or
$(ii)$ in horizontal layering fashion across multiple domains,
i.e., technology layers internetwork across distinct domains.
Horizontal multilayer networking can be viewed as a generalization of
vertical multilayer networking in that the horizontal networking
may involve the same or different (or even multiple) layers in the
distinct domains. As illustrated in
Fig.~\ref{fig_control_multilay}, the formed SDN based multilayer network
architecture is controlled by an SDN orchestrator.
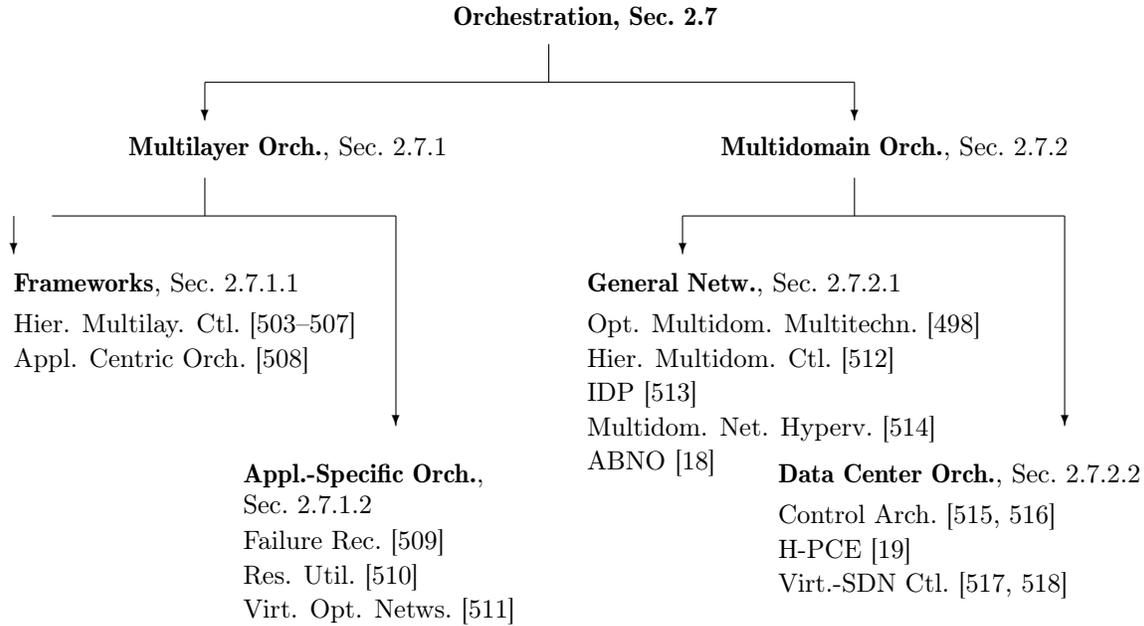
\begin{figure*}[t!]
\footnotesize
\setlength{\unitlength}{0.10in} 
\centering
\begin{picture}(40,33)
\put(13,33){\textbf{Orchestration, Sec.~\ref{orch:sec}}}
\put(18,30){\line(0,1){2}}
\put(0,30){\line(1,0){34}}
\put(0,30){\vector(0,-1){2}}
\put(-4,26.2){\textbf{Multilayer Orch.}, Sec.~\ref{mullayorch:sec} }
\put(0,25){\line(0,-1){2}}

\put(25,23){\line(1,0){20}}
\put(27,26.2){\textbf{Multidomain Orch.}, Sec.~\ref{muldomorch:sec}}
\put(34,25){\line(0,-1){2}}

\put(-8,23){\line(1,0){18}}
\put(-10,23){\vector(0,-1){2}}
\put(10,23){\vector(0,-1){11}}

\put(-10,20){\makebox(0,0)[lt]{\shortstack[l]{			
\textbf{Frameworks}, Sec.~\ref{mullayorchfr:sec} \\ \\
Hier. Multilay. Ctl.~\cite{Felix2014,ger2013dem,sal2013inf,sun2014des,ShZBLT12}\\
Appl. Centric Orch.~\cite{Gerstel2015}
}}}

\put(2,10){\makebox(0,0)[lt]{\shortstack[l]{			
\textbf{Appl.-Specific Orch.}, \\ Sec.~\ref{applorch:sec} \\ \\
Failure Rec.~\cite{Khaddam2015} \\
Res. Util.~\cite{Liu2015e} \\
Virt. Opt. Netws.~\cite{vil2015net}
}}}				
\put(34,30){\vector(0,-1){2}}
\put(25,23){\vector(0,-1){2}}
\put(20,20){\makebox(0,0)[lt]{\shortstack[l]{
\textbf{General Netw.}, Sec.~\ref{muldomorchgen:sec} \\ \\
Opt. Multidom. Multitechn.~\cite{YoMKN14} \\
Hier. Multidom. Ctl.~\cite{Jing2015} \\
IDP~\cite{Zhu2015} \\
Multidom. Net. Hyperv.~\cite{vil2015mul} \\
ABNO~\cite{Munoz2015}
}}}							
\put(45,23){\vector(0,-1){11}}		
\put(30,10){\makebox(0,0)[lt]{\shortstack[l]{			
\textbf{Data Center Orch.}, Sec.~\ref{muldomorchdc:sec} \\ \\
Control Arch.~\cite{Liu2015b, may2016sdn} \\
H-PCE~\cite{cas2015sdn} \\
Virt.-SDN Ctl.~\cite{mun2015int,Vilalta2016}
}}}
\end{picture}	
\caption{Classification of SDON orchestration studies.}
\label{orch_class:fig}
\end{figure*}
As illustrated in Fig.~\ref{orch_class:fig} we organize
the SDON orchestration studies according to their focus into studies
that primarily address the orchestration of vertical multilayer
(multitechnology) networking,
i.e., the vertical networking across multiple layers (that typically
implement different technologies) within a given domain,
and into studies that primarily
address the orchestration of horizontal multilayer (multidomain)
networking, i.e., the horizontal networking across multiple
routing domains (which may possibly involve different or multiple
vertical layers in the different domains).
We subclassify the vertical multilayer studies into general (vertical)
multilayer networking frameworks and studies focused on supporting specific
applications through vertical multilayer networking.
We subclassify the multidomain (horizontal multilayer) networking studies into
studies on general network domains and studies focused on internetworking
with Data Center (DC) network domains.

\subsection{Multilayer Orchestration}  \label{mullayorch:sec}

\subsubsection{Multilayer Orchestration Frameworks} \label{mullayorchfr:sec}

\paragraph{Hierarchical Multilayer Control}
Felix et al.~\cite{Felix2014} presented an hierarchical
SDN control mechanism for packet optical networks.
Multilayer optimization techniques
are employed at the SDN orchestrator
to integrate the optical transport technology
with packet services by provisioning end-to-end Ethernet services.
Two aspects are investigated, namely
$(i)$ bandwidth optimization for the optical transport services,
and $(ii)$ congestion control for packet network services
in an integrated packet optical network.
More specifically, the SDN controller initially allocates the
minimum available bandwidth required
for the services and then dynamically scales
allocations based on the availability.
Optical-Virtual Private Networks (O-VPNs) are created
over the physical transport network. Services are then mapped
to O-VPNs based on class of service requirements.
When congestion is detected for a service,
the SDN controller switches the service to another O-VPN, thus
balancing the traffic to maintain the required class of service.

Similar steps towards the orchestration of multilayer networks
have been taken within the OFELIA
project~\cite{ger2013dem,sal2013inf,sun2014des}. Specifically,
Shirazipour et al.~\cite{ShZBLT12} have explored
extensions to OpenFlow version 1.1 actions to enable
multitechnology transport layers, including Ethernet transport and
optical transport. The explorations of the extensions include
justifications of the use of SDN in circuit-based transport
networks.

\paragraph{Application Centric Orchestration}
Gerstel et al.~\cite{Gerstel2015} proposed an application centric network
service provisioning approach based on multilayer orchestration.
This approach enables the network applications
to directly interact with the physical layer
resource allocations to achieve the desired
service requirements.
Application requirements for
a network service may include
maximum end-to-end latency, connection setup and hold times,
failure protection, as well as security and encryption.
In traditional IP networking, packets from multiple applications
requiring heterogeneous services
are simply aggregated and sent over a common transport link (IP services).
As a result, network applications are typically
assigned to a single (common) transport service within an optical link.
Consider a failure recovery process with multiple available paths.
IP networking typically selects the single path
with the least end-to-end delay.
However, some applications may tolerate higher
latencies and therefore, the traffic can be split
over multiple restoration paths achieving better traffic management.
The orchestrator needs to interact with multiple
network controllers operating across multiple (vertical) layers
supported by north/south bound interfaces
to achieve the application centric control.
Dynamic additions of new IP links are demonstrated
to accommodate the requirements of multiple application services
with multiple IP links when
the load on the existing IP link was increased.

\subsubsection{Application-specific Orchestration}  \label{applorch:sec}
\paragraph{Failure Recovery}
Generally, network CapEx and OpEx increase as more protection against
network failures is added.
Khaddam et al.~\cite{Khaddam2015} propose
an SDN based integration of multiple layers, such as WDM and IP, in a failure
recovery mechanism to improve the utilization
(i.e., to eventually reduce CapEx and OpEx while maintaining
high protection levels).
An observation study was conducted
over a five year period to understand the impact of
network failures on the real deployment
of backbone networks.
Results showed $75$ distinct failures following a Pareto distribution,
in which, $48\%$ of the total deployed capacity was affected
by the top (i.e., the highest impact) $20\%$ of the failures.
And, $10\%$ of the total deployed
capacity was impacted by the top two failure instances.
These results emphasize the significance of backup
capacities in the optical links for restoration processes.
However, attaining the optimal protection capacities while
achieving a high utilization of the optical links is challenging.
A failure recovery mechanism is proposed based
on a ``hybrid'' (i.e., combination of optical transport and IP)
multilayer optimization.
The hybrid mechanism improved the optical link utilization up to 50~\%.
Specifically, 30~\% increase of the transport
capacity utilization is achieved by dynamically reusing the remainder
capacities in the optical links, i.e.,
the capacity reserved for failure recoveries.
The multilayer optimization technique was validated on an experimental
testbed utilizing central path-computation (PCE)~\cite{rfc5440}
within the SDN framework.
Experimental verification of failure
recovery mechanism resulted in recovery times
on the order of sub-seconds for MPLS restorations and
several seconds for optical WSON restorations.

\paragraph{Resource Utilization}
Liu et al.~\cite{Liu2015e} proposed a method to improve
resource utilization and to reduce transmission latencies
through the processes of virtualization and service abstraction.
A centralized SDN control implements the service abstraction
layer (to enable SDN orchestrations) in order to integrate the
network topology management (across both IP and WDM),
and the spectrum resource allocation in a single control platform.
The SDN orchestrator also achieves dynamic and simultaneous
connection establishment across both IP and OTN layers
reducing the transmission latencies. The control plane design is split
between local (child) and root (parent) controllers.
The local controller realizes the label switched paths on the optical nodes
while the root controller realizes the forwarding rules
for realizing the IP layer.
Experimental evaluation of average transfer time measurements
showed IP layer
latencies on the order of several milliseconds, and
several hundreds of milliseconds for the OTN latencies, validating
the feasibility of control plane unification for IP over
optical transport networks.

\paragraph{Virtual Optical Networks (VONs)}
Vilalta et al.~\cite{vil2015net} presented
controller orchestration to integrate multiple transport network technologies,
such as IP and GMPLS. The proposed architectural framework devises
VONs to enable the virtualization
of the physical resources within each domain.
VONs are managed by lower level physical controllers (PCs), which
are hierarchically managed by an SDN network orchestrator (NO). Network
Virtualization Controllers (NVC) are introduced (on top of the NO) to
abstract the virtualized multilayers across multiple domains.
End-to-end provisioning of VONs is facilitated through
hierarchical control interaction over three levels, the
customer controller, the NO\&NVCs, and the PCs.
An experimental evaluation demonstrated average VON
provisioning delays on the order of
several seconds (5~s and 10~s), validating the flexibility of
dynamic VON deployments over the optical transport networks.
Longer provisioning delays may impact the network
application requirements, such as failure recovery processes,
congestion control, and traffic engineering.
General pitfalls of such hierarchical structures are
increased control plane complexity, risk of controller failures, and
maintenance of reliable communication links between control plane entities.

\subsection{Multidomain Orchestration}  \label{muldomorch:sec}
Large scale network deployments typically involve multiple domains,
which have often heterogeneous layer technologies.
Achieve high utilization of the networking resources
while provisioning end-to-end network paths and services across multiple
domains and their respective layers and respective technologies is
highly challenging~\cite{Mayoral2015,Munoz2015b,Yu2015a}.
Multidomain SDN orchestration studies have sought to exploit
the unified SDN control plane to aid the resource-efficient
provisioning across the multiple domains.

\subsubsection{General Multidomain Networks}  \label{muldomorchgen:sec}

\paragraph{Optical Multitechnologies Across Multiple Domains}
Optical nodes are becoming
increasingly reconfigurable (e.g., through variable BVTs and
OFDM transceivers, see Section~\ref{sdninfra:sec}),
adding flexibility to the switching elements. When a single
end-to-end service
establishment is considered, it is more likely that a service is supported by
different optical technologies that operate across multiple domains.
Yoshida et al.~\cite{YoMKN14} have demonstrated SDN based orchestration with
emphasis on the physical interconnects between multiple domains and multiple
technology specific controllers so as to realize end-to-end services.
OpenFlow capabilities have been extended for fixed-length variable capacity
optical packet switching~\cite{Losada2015}.
That is, when an optical switch matches the label on an
incoming optical packet, if a rule
exists in the switch (flow entry in the table) for a specific label,
a defined action is performed on the optical packet by the switch.
Otherwise, the optical packet is dropped
and the controller is notified. Interconnects between optical packet
switching networks and elastic optical networks are enabled
through a novel OPS-EON interface card.
The OPS-EON interface is designed as an extension
to a reconfigurable, programmable
and flexi-grid EON supporting the OpenFlow protocol.
The testbed implementation of OPS-EON interface cards demonstrated
the orchestration of multiple domain controllers and the reconfigurability
of FL-VC OPS across multidomain, multilayer, multitechnology scenarios.

\paragraph{Hierarchical Multidomain Control}
Jing et al.~\cite{Jing2015} have also examined
the integration of multiple optical transport
technologies from to multiple
vendors across multiple domains, focusing on
the control mechanisms across multiple domains.
Jing et al.~proposed hierarchical SDN orchestration with parent and
domain controllers.
Domain controllers abstract the physical layer by
virtualizing the network resources.
A Parent Controller (PC) encompasses a Connection Controller (CC) and
a Routing Controller (RC) to process the abstracted virtual network.
When a new connection setup request is
received by the PC, the RC (within the PC) evaluates the end-to-end
routing mechanisms and forwards the
information to the CC.
The CC breaks the end-to-end routing information into shorter
link segments belonging to a domain.
Segmented routes are then sent to the respective domain
controllers for link provisioning over the physical infrastructures.
The proposed mechanism was experimentally verified on a testbed
built with the commercial OTN equipment.

\paragraph{Inter-Domain Protocol}
\begin{figure}[t!]
	\centering
	\vspace{0cm}
	\includegraphics[width=4.2in]{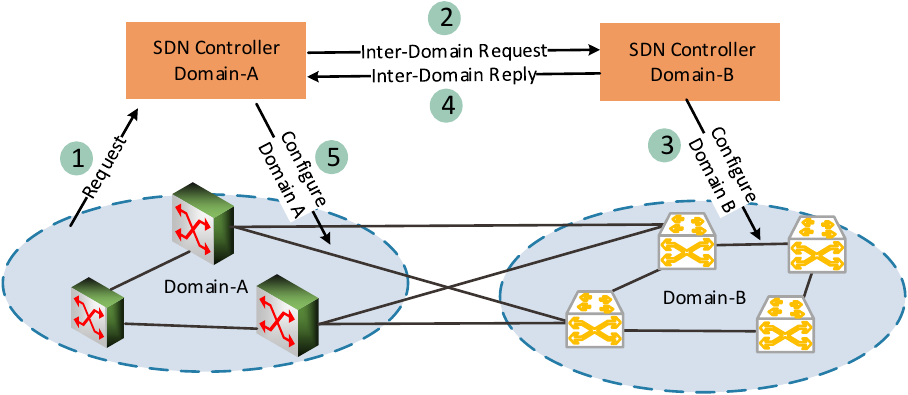}
	\vspace{0cm}
	\caption{Inter-domain lightpath provisioning mechanism.}
	\label{fig_multilay_idp}
\end{figure}
Zhu et al.~\cite{Zhu2015} followed a different approach for the
SDN multidomain control mechanisms by considering the
flat arrangement of controllers as shown in Fig.~\ref{fig_multilay_idp}.
Each domain is autonomously managed by
an SDN controller specific to the domain.
An Inter-Domain Protocol (IDP) was devised to establish the communication
between domain specific controllers
to coordinate the lightpath setup across multiple domains.
Zhu et al.~also proposed a Routing and
Spectrum Allocation (RSA) algorithm for the end-to-end provisioning
of services in the SD-EONs.
The distributed RSA algorithm operates on the domain specific controllers
using the IDP protocol. The RSA considers both transparent lightpath
connections, i.e., all-optical lightpath,
and translucent lightpath connections, i.e.,
optical-electrical-optical connections.
The benefit of such techniques is privacy, since
the domain specific policies and topology information are not
shared among other network entities.
Neighbor discovery is  independently conducted by the domain specific controller
or can initially be configured.
A domain appears as an abstracted virtual node to all other domain specific
controllers. Each controller then assigns the shortest path
routing within a domain between its border nodes.
An experimental setup validating the
proposed mechanism was demonstrated across
geographically-distributed domains in the USA and China.

\paragraph{Multidomain Network Hypervisors}
\begin{figure}[t!]
	\centering
	\includegraphics[width=3in]{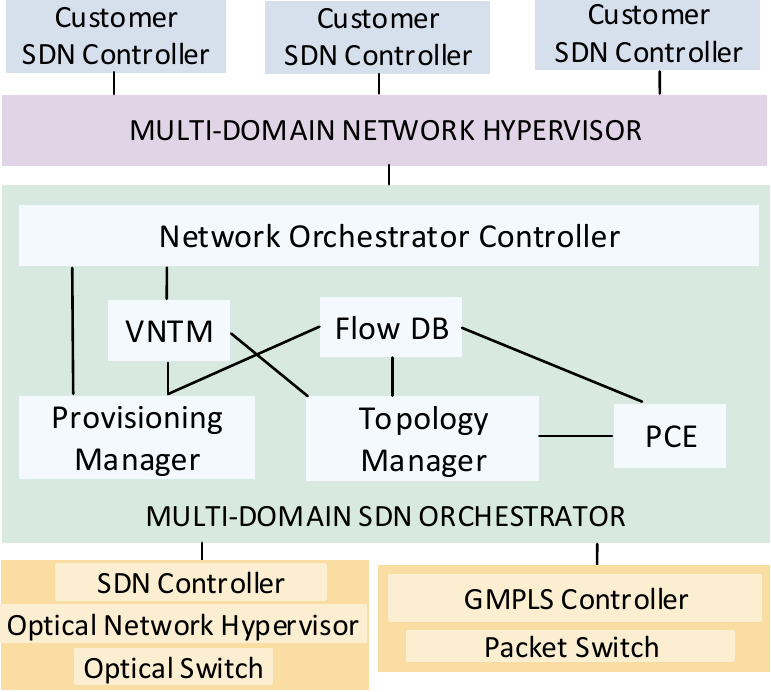}
	\caption{Illustration of multilevel virtualization.} 
	\label{fig_multilay_mnh}
\end{figure}
Vilalta et al.~\cite{vil2015mul} presented a mechanism for virtualizing
multitechnology optical, multitenant networks.
The Multidomain Network Hypervisor (MNH) creates customer specific
virtual network slices managed by the customer specific SDN controllers
(residing at the customers' locations) as illustrated
in Fig.~\ref{fig_multilay_mnh}.
Physical resources are managed by their domain specific
physical SDN controllers.
The MNH operates over the network orchestrator and
physical SDN controllers for provisioning VONs
on the physical infrastructures.
The MNHs abstracts both $(i)$ multiple optical transport technologies,
such as optical packet switching and Elastic Optical Networks (EONs),
and $(ii)$ multiple control domains, such as GMPLS and OpenFlow.
Experimental assessments on a testbed achieved VON provisioning
within a few seconds (5~s), and
control overhead delay on the order of several tens of milliseconds.
Related virtualization mechanisms for multidomain optical SDN networks
with end-to-end provisioning have been
investigated in~\cite{SzAEK14,vil2016hie}.

\paragraph{Application-Based Network Operations}
\begin{figure}[t!]
	\centering
	\includegraphics[width=3.5in]{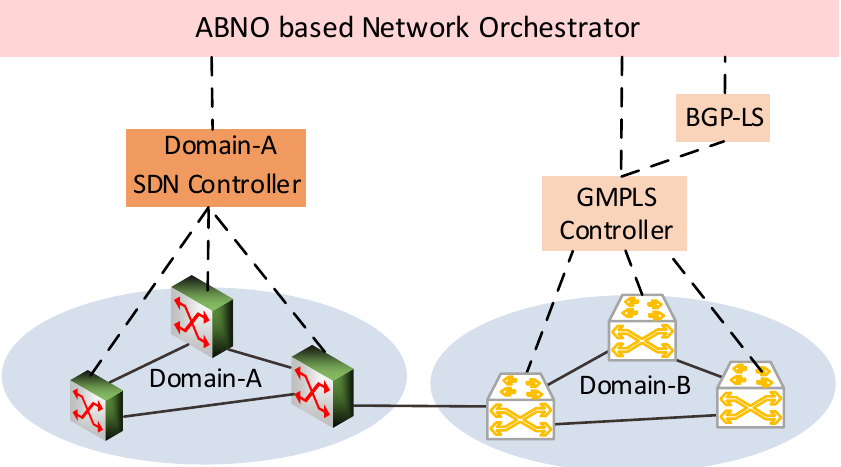}
	\caption{The application-based network operations (ABNO)~\cite{Munoz2015}}
	\label{fig_multilay_abno}
\end{figure}
Mu{\~{n}}oz et al.~\cite{Munoz2015}, have presented an SDN
orchestration mechanism based on the
application-based network operations (ABNO) framework, which
is being defined by the IETF~\cite{rfc7491}.
The ABNO based SDN orchestrator integrates
OpenFlow and GMPLS in transport networks.
Two SDN orchestration designs have been presented: $(i)$ with centralized
physical network topology aware path computation
(illustrated in Fig.~\ref{fig_multilay_abno}), and
$(ii)$ with topology abstraction and distributed path computation.
In the centralized design, OpenFlow and GMPLS controllers (lower level control)
expose the physical
topology information to the ABNO-orchestrator (higher level control).
The PCE in the ABNO-orchestrator has the global view of the network
and can compute end-to-end paths with complete knowledge of the network.
Computed paths are then provisioned through
the lower level controllers. The pitfalls of such centralized designs
are $(i)$ computationally intensive path computations,
$(ii)$ continuous updates of topology and traffic information,
and $(iii)$ sharing of confidential network information and policies
with other network elements.
To reduce the computational load at the orchestrator,
the second design implements distributed path computation
at the lower level controllers (instead of path computation at the
centralized orchestrator). However, such distributed mechanisms may
lead to suboptimal solutions due to the limited network knowledge.

\subsubsection{Multidomain Data Center Orchestration} \label{muldomorchdc:sec}
\paragraph{Control Architectures}
Geographically distributed DCs are typically
interconnected by links traversing multiple domains.
The traversed domains may be homogeneous i.e., have the same type of network
technology, e.g., OpenFlow based ROADMs,
or may be heterogeneous, i.e., have different types of network
technologies, e.g.,
OpenFlow based ROADMs and GMPLS based WSON. The SDN control structures
for a multidomain network can be broadly classified into the categories of
$(i)$ single SDN orchestrator/controller, $(ii)$ multiple mesh SDN
controllers, and
$(iii)$ multiple hierarchical SDN controllers~\cite{Liu2015b, may2016sdn}.
The single SDN orchestrator/controller has to support heterogeneous SBIs
in order to operate with multiple heterogeneous domains, e.g.,
the Path Computation Element Protocol (PCEP) for GMPLS network domains
and the OpenFlow protocol for OpenFlow supported ROADMs.
Also, domain specific details, such as topology, as well as
network statistics and configurations, have to be exposed to an external
entity, namely the single SDN orchestrator/controller,
raising privacy concerns. Furthermore, a single
controller may result in scalability issues.
Mesh SDN control connects the domain-specific controllers side-by-side
by extending the east/west bound interfaces.
Although mesh SDN control addresses the scalability and privacy issues,
the distributed nature of the control mechanisms may lead to sub-optimal
solutions.
With hierarchical SDN control, a logically centralized controller
(parent SDN controller) is placed above the domain-specific controllers
(child SDN controllers), extending the north/south bound interfaces.
Domain-specific controllers virtualize
the underlying networks inside their domains, exposing only the abstracted
view of the domains to the parent controller, which addresses the privacy
concerns. Centralized
path computation at the parent controller can achieve optimal solutions.
Multiple hierarchical levels can address the scalability issues.
These advantages of hierarchal SDN control are achieved at the expense of
an increased number of network entities,
resulting in the operational complexities.

\paragraph{Hierarchical PCE}
\begin{figure}[t!]
	\centering
	\includegraphics[width=3.5in]{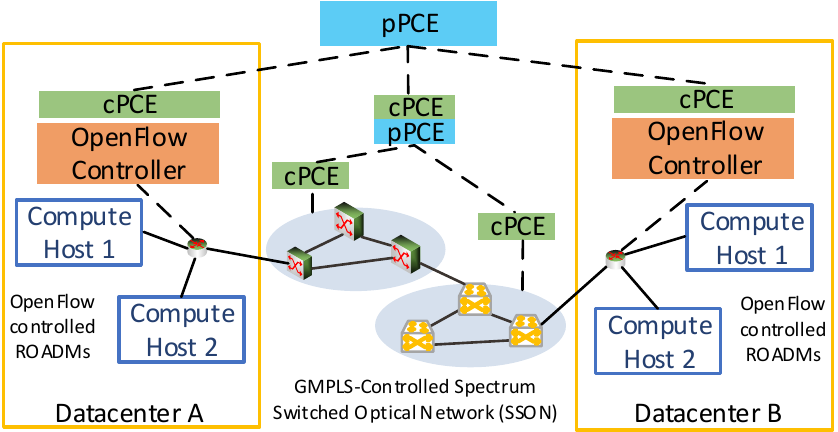}
  \caption{Illustration of SDN orchestration based on Hierarchical
  Path Computation Element (H-PCE)~\cite{cas2015sdn}.}
	\label{fig_multilay_HPCE}
\end{figure}
Casellas et al.~\cite{cas2015sdn} considered
DC connectivities involving both intra-DC and inter-DC communications.
Intra-DC communications enabled through OpenFlow networks
are supported by an OpenFlow controller.
The inter-DC communications
are enabled by optical transport
networks involving more complex control,
such as GMPLS, as illustrated in Fig.~\ref{fig_multilay_HPCE}. To
achieve the desired SDN benefits of flexibility and scalability,
a common centralized control platform spanning across heterogeneous
control domains is proposed.
More specifically, an Hierarchical PCE (H-PCE) aggregates PCE states
from multiple domains.
The end-to-end path setup between DCs is orchestrated by a parent-PCE (pPCE)
element, while the paths are provisioned
by the child-PCEs (cPCEs) on the physical resources, i.e.,
the OpenFlow and GMPLS domains.
The proposed mechanism utilizes existing protocol interfaces,
such as BGP-LS and PCEP, which are extended with OpenFlow to support the H-PCE.

\paragraph{Virtual-SDN Control}
Mu{\~n}oz et al.~\cite{mun2015int,Vilalta2016} proposed a
mechanism to virtualize the SDN control functions
in a DC/cloud by integrating SDN with Network Function Virtualization (NFV).
In the considered context, NFV refers to realizing network functions
by software modules running on
generic computing hardware inside a DC; these network functions were
conventionally implemented on specialized hardware modules.
The orchestration of Virtual Network Functions (VNFs)
is enabled by an integrated SDN and NFV management which
dynamically instantiates virtual SDN controllers.
The virtual SDN controllers
control the Virtual Tenant Networks (VTNs), i.e., virtual
multidomain and multitechnology networks.
Multiple VNFs running on a Virtual Machine (VM)
in a DC are managed by a VNF manger.
A virtual SDN controller is responsible for creating, managing, and tearing
down the VNF achieving the flexibility in the control plane management
of the multilayer and the multidomain networks.
Additionally, as an extension to the proposed mechanism,
the virtualization of the control functions of the
LTE Evolved Packet Core (EPC) has been discussed in~\cite{mar2016int}.

\subsection{Orchestration: Summary and Discussion}
Relatively few SDN orchestration studies to date have focused on
vertical multilayer networking within a given domain. The few studies
have developed two general orchestration frameworks and have examined
a few orchestration strategies for some specific applications.
More specifically, one orchestration framework has
focused on optimal bandwidth allocation based mainly on
congestion~\cite{Felix2014}, while the other
framework has focused on exploiting application traffic tolerances for
delays for efficiently routing traffic~\cite{Gerstel2015}.
SDN orchestration of vertical multilayer optical networking is thus
still a relatively little explored area.
Future research can develop orchestration frameworks that accommodate
the specific optical communication technologies in the various layers and
rigorously examine their performance-complexity tradeoffs.
Similarly, relatively few applications have been
examined to date in the application-specific orchestration
studies for vertical multilayer
networking~\cite{Khaddam2015,Liu2015e,vil2015net}.
The examination of the wide range of existing applications and
any newly emerging network application in the context of
SDN orchestrated vertical multilayer networking presents rich research
opportunities.
The cross-layer perspective of the SDN orchestrator over a given
domain could, for instance, be exploited for strengthening security and
privacy mechanisms or for accommodating demanding real-time multimedia.

Relatively more SDN orchestration studies to date have
examined multidomain networking than multilayer networking (within a single
domain). As the completed multidomain orchestration studies
have demonstrated, the SDN orchestration can help greatly in coordinating
complex network management decisions across multiple distributed
routing domains.
The completed studies have illustrated the fundamental
tradeoff between centralized decision making in a hierarchical
orchestration structure and distributed decision making in a flat
orchestration structure.
In particular, most studies have focused on hierarchical
structures~\cite{Jing2015,cas2015sdn,mun2015int}, while only one study
has mainly focused on a flat orchestration structure~\cite{Zhu2015}.
In the context of DC internetworking, the
studies~\cite{Liu2015b, may2016sdn} have sought to bring out
the tradeoffs between these two structures by examining
a range of structures from centralized to distributed.
While centralized orchestration can make decisions with
a wide knowledge horizon across the states in multiple domains,
distributed decision making preserves the privacy of network status
information, reduces control traffic, and can make fast localized
decisions.
Future research needs to shed further light on these complex
tradeoffs for a wide range of combinations of optical technologies employed
in the various domains.
Throughout, it will be critical to abstract and convey the key
characteristics of optical physical layer components and switching nodes to
the overall orchestration protocols. Optimizing each abstraction
step as well as the overall orchestration and examining the various
performance tradeoffs are important future research directions.

\section{\MakeUppercase{Open Challenges and Future SDON Research Directions}} \label{sec:open}
We have outlined open challenges and future
Software Defined Optical Network (SDON) research directions
for each sub-category of surveyed SDON studies in the Summary and
Discussion subsections in the preceding survey sections.  In this
section, we focus on the overall cross-cutting open challenges that
span across the preceding considered categories of SDON studies.
That is, we focus on open challenges and research directions
that span the vertical (inter-layer) and horizontal (inter-domain)
SDON aspects. The vertical SDON aspects encompass the
seamless integration of the various (vertical) layers of the
SDON architecture; especially the optical layer, which is not considered
in general SDN technology.
The horizontal SDON aspects include the integration of SDONs with
existing non-SDN optical networking elements, and the internetworking
with other domains, which may have  similar or different
SDN architectures.
A key challenge for SDON research is to enable the use of SDON concepts
in operational real-time network infrastructures.
Importantly, the SDON concepts need to demonstrate performance
gains and cost reductions to be considered by network and service providers.
Therefore, we cater some of the open challenges and
future directions towards enabling and demonstrating the successful use
of SDON in operational networks.

The SDON research and development effort to date
have resulted in insights for making the use of SDN in optical transport
networks feasible and have demonstrated advantages of SDN based optical network
management.
However, most network and service providers
depend on optical transport to integrate with multiple industries to
complete the network infrastructure. Often, network and service providers
struggle to integrate hardware components and to provide accessible software
management to customers. For example, companies that develop hardware
optical components do not always have a complete associated software stack
for the hardware components. Thus, network and service providers using the
hardware optical components often have to maintain a software development team
to integrate the various
hardware components through software based management into their network,
which is often a costly endeavor.
Thus, improving SDN technology so that it seamlessly integrates with
components of various industries
and helps the integration of components from various industries
is an essential underlying theme for future SDON research.

\subsection{Simplicity and Efficiency}   \label{simpl_fut:sec}
Optical network structures typically span heterogeneous devices ranging
from the end user nodes and local area networks via ONUs and OLTs in the
access networks to edge routers and metro network nodes and on
to backbone (core) network infrastructures.
These different devices often come from different vendors.
The heterogeneity of devices and their vendors often requires
manual configuration and maintenance of optical networks.
Moreover, different communication technologies typically
require the implementation of native functions that are specific
to the communication technology characteristics, e.g., the transmission and
propagation properties.
By centralizing the optical network control in an
SDN controller, the SDN networking paradigm creates a unified view
of the entire optical network.
The specific native functions for specific communication devices
can be migrated to the software layer and be implemented by a central node,
rather than through manual node-by-node configurations.
The central node would typically be readily accessible and could reduce
the required physical accesses to distributed devices at their
on-site locations.
This centralization can simplify the network management and
reduce operational expenditures.
An important challenge in this central management is the efficient
SDN control of components from multiple vendors.
Detailed vendor contract specifications of open-source middleware may be
needed to efficiently control components from different vendors.

The heterogeneity of devices may reduce the efficiency of
network infrastructures
due to the required multiple software and hardware modules for a
complete networking solution. Future research should
investigate efficient mechanisms for making complete networking solutions
available for specific use cases. For example, the use of
SDON for an access network provider may require multiple SDN controllers
co-located within the OLT to enable the control of the access network
infrastructure from one central location.  While the SDON studies reviewed in
this survey have led initial investigations of simple and dynamic
network management, future research needs to refine these management
strategies and optimize their operation across combinations of
network architecture structures and across various network
protocol layers.
Simplicity is an essential part of this
challenge, since overly complicated solutions are generally not deployed
due to the risk of high expenditures.

\subsection{North Bound Interface} \label{nbi:fut}
The NorthBound Interface (NBI) comprises the communication from the
controller to the applications. This is an important area of
future research as applications and their needs are generally
the driving force for deploying SDON infrastructures.
Any application, such as video on demand, VoIP, file
transfer, or peer-to-peer networking, is applied from the NBI to the
SDN controller which consequently conducts the necessary actions to implement
the service behaviors on the physical network infrastructure.
Applications often require specific service behaviors that
need to be implemented on the overall network infrastructure.
For example,
applications requiring high data rates and reliability, such as Netflix,
depend on data centers and the availability
of data from servers with highly resilient failure protection mechanisms.
The associated management network needs to stack redundant devices as
to safeguard against outages. Services are provided as
policies through the NBI to the SDN controller, which in turn generates flow
rules for the switching devices. These flow rules can be
prioritized based on the customer use cases.
An important challenge for future NBI research is to
provide a simple interface for a wide variety of
service deployments without vendor lock-in, as vendor lock-in
generally drives costs up.
Also, new forms of communication to the controller, in
addition to current techniques, such as
REpresentational State Transfer (REST)~\cite{REST15} and HTTP, should
be researched.
Moreover, future research should develop an NBI framework that spans
horizontally across multiple controllers, so that service customers are not
restricted to using only a single controller.

Future research should examine control mechanisms that optimally
exploit the central SDN control to provide simple and efficient
mechanisms for automatic network management and dynamic service
deployment~\cite{ZhYZ14}.  The NBI of SDONs is a
challenging facet of research and development because of the multitude
of interfaces that need to be managed on the physical layer and
transport layer. Optical physical layer components and infrastructures
require high capital and
operational expenditures and their management is generally not associated with
network or service providers but rather with optical component/infrastructure
vendors. Future research should develop novel Application Program Interfaces
(APIs) for optical layer components and infrastructures that facilitate
SDN control and are amenable to efficient NBI communication.
Essentially, the challenge of efficient NBI communication with the
SDN controller should be
considered when designing the APIs that interface with the
physical optical layer components and infrastructures.

One specific strategy for simplifying network management and operation
could be to explore the grouping of control policies of similar
service applications, e.g., applications with similar QoS requirements.
The grouping can reduce the number of control policies at the
expense of slightly coarser granularity of the service offerings.
The emerging Intent-Based Networking (IBN) paradigm, which
drafts intents for services and policies,
can provide a specific avenue for simplifying
dynamic automatic configuration and virtualization~\cite{BlDM08,CoBR13}.
Currently network applications are deployed based on how the network
should behave for a specific action. For example, for inter domain routing,
the Border Gateway Protocol (BGP) is used, and the network gateways
are configured to communicate with the BGP protocol.
This complicates the provisioning of
services that typically require multiple protocols and limits the
flexibility of service provisioning. With IBN, the application gives an intent,
for example, transferring video across multiple domains.
This intent is then
associated with automated dynamic configurations of the network elements
to communicate data over the domains using appropriate protocols.
The grouping of service policies, such as intents, can facilitate
easy and dynamic service provisioning.
Intent groups can be described in a graph to simplify the compilation of
service policies and to resolve conflicts~\cite{PrLT15}.

\subsection{Reliability, Security, and Privacy} \label{futworksec:sec}
The SDN paradigm is based on a centrally managed network.
Faulty behaviors, security infringements, or failures of the
control would likely result in extensive disruptions and
performance losses that are exacerbated by the centralized nature of the
SDN control. Instances of extensive disruptions and losses due to
SDN control failures or infringements would likely reduce the trust in
SDN deployments. Therefore, it is very important to ensure reliable network
operation~\cite{rak2016inf}
and to provision for security and privacy of the communication.
Hence, reliability, security, and privacy are prominent SDON research
challenges. Security in SDON techniques is a fairly open research
area, with only few published findings.
As a few reviewed studies (see Section~\ref{app_fail_rec:sec}) have explored,
the central SDN control can facilitate reliable network service through speeding
up failure recovery.
The central SDN control can continuously scan the network and
the status messages from the network devices.
Or, the SDN control can
redirect the status messages to a monitoring service that analyzes
the data network.  Security breaches can be controlled by broadcasting
messages from the controller to all affected devices to block traffic
in a specific direction. Future research should refine these
reliability functions to optimize automated fault and performance
diagnostics and reconfigurations for quick failure recovery.

Network failures can either occur within the physical layer
infrastructure, or as errors within the higher protocol
layers, e.g., in the classical data link (L2), network (L3), of
transport (L4) layers.
In the context of SDONs, physical
layer failures present important future research opportunities.
Physical layer devices need to be carefully monitored by sending
feedback from the devices to the controller.
The research and development on communication between
the SDN controller and the network devices has mainly focused on sending
flow rules to the network devices while feedback communicated from the devices
to the controller has received relatively little attention.
For example, there are three types of OpenFlow messages, namely Packet-In,
Packet-Out, and Flow-Mod. The Packet-In messages are
sent from the OpenFlow switches to the controller,
the Packet-Out message is sent from
the controller to the device, and the Flow-Mod message is used to
modify and monitor
the flow rules in the flow table. Future research should examine extensions of
the Packet-In message to send specific status updates in support of
network and device failure monitoring to the controller.
These status messages could be monitored
by a dedicated failure monitoring service.
The status update messages could be broadly defined to
cover a wide range of network management aspects, including
system health monitoring and network failure protection.

A related future research direction is to secure configuration and
operation of SDONs through trusted encryption and key management
systems~\cite{ahm2015sec}. Moreover, mechanisms to
ensure the privacy of the communication should be explored.
The security and privacy mechanisms should strive to exploit the
natural immunity of optical transmission segments to electro-magnetic
interferences.

In summary, security and privacy of SDON communication are largely open
research areas.
The optical physical layer infrastructure has traditionally not
been controlled remotely, which in general reduces the occurrences
of security breaches.
However, centralized SDN management and control increase the risk of
security breaches, requiring extensive research on SDON security, so
as to reap the benefits of centralized SDN management and control in a
secure manner.

\subsection{Scalability} \label{scalability:fut}
Optical networks are expensive and used for high-bandwidth
services, such as long-distance network access and data center interconnections.
Optical network infrastructures either
span long distances between multiple geographically distributed locations,
or could be short-distance incremental additions (interconnects)
of computing devices. Scalability in multiple dimensions
is therefore an important aspect for future SDON research.
For example, a
myriad of tiny end devices need to be provided with network access in
the emerging Internet of Things (IoT) paradigm~\cite{wan2015nov}.
The IoT requires access network architectures and
protocols to scale vertically (across protocol layers and technologies)
and horizontally (across network domains).
At the same time, the ongoing growth of
multimedia services requires data centers to scale up optical network bandwidths
to maintain the quality of experience of the multimedia services.
Broadly speaking, scalability includes in the vertical
dimension the support for multiple network devices and technologies.
Scalability in
the horizontal direction includes the communication between
a large number of different domains as well as support for existing non-SDON
infrastructures.

A specific scalability challenge arising with SDN infrastructure is that the
scalability of the control plane (OpenFlow protocol signalling)
communication
and the scalability of the data plane communication which transports
the data plane flows
need to be jointly considered. For example, the Openflow protocol~1.4
currently supports 34 Flow-Mod messages~\cite{OF2016}, which can
communicate between the network devices and the controller. This
number limits the functionality of the SBI communication. Recent studies
have explored a protocol-agnostic
approach \cite{HuLi2015, bosshart2014p4}, which is a data plane
protocol that extends the use of multiple protocols for
communication between the control plane and data plane.
The protocol-agnostic approach resolves
the challenges faced by OpenFlow and, in general, any particular protocol.
Exploring this novel protocol-agnostic approach presents many new SDON
research directions.

Scalability would also require SDN technology to overlay and scale
over existing non-SDN infrastructures. Vendors provide support
for known non-SDN devices, but this area is still a challenge. There
are no known protocols that could modify the flow tables of existing
popularly described ``non-OpenFlow'' switches. In the case of optical
networks, as SDN is still being incrementally deployed, the overlaying
with non-SDN infrastructure still requires significant attention.
Ideally, the overlay mechanisms should ensure seamless integration and should
scale with the growing deployment of SDN technologies while incurring only
low costs.
Overall, scalability poses highly important future SDON research directions
that require economical solutions.

\subsection{Standardization}   \label{std:sec}
Networking protocols have traditionally followed a uniform
standard system for all the
communication across multiple domains. Standardization has helped vendors
to provide products that work in and across different network infrastructures.
In order to ensure the compatible inter-operation of SDON components (both hardware and software) from a various vendors,
key aspects of the inter-operation protocols need to be standardized.
Towards the standardization goal,
communities, such as Open Networking Foundation (ONF), have created boards
and committees to standardize protocols, such as OpenFlow.
Standardization should ensure that SDON infrastructures can be flexibly
configured and operated with components from various vendors.
The use of open-source
software can further facilitate the inter-operation.
Proprietary hardware and software components generally create vendor lock-in,
which restricts the flexibility of network operation and reduces
the innovation of network and service providers.

As groundwork for standardization, it may be necessary to develop
and optimize a common (or a small set) of SDON architectures
and network protocol configurations that can serve as a basis
for standardization efforts.
The standardization process may involve a common platform that
is built thorough the cooperation of multiple manufacturers.
Another thrust of standardization groundwork could be the development of
open-source software that supports SDON architectures.
For example, Openstack is a cloud based management framework that has been
adopted and supported by multiple networking vendors. Such efforts should
be extended to SDONs in future work.

\subsection{Multilayer Networking} \label{multilayer:fut}
As discussed in Section~\ref{orch:sec}, multilayer networking
involves vertical multilayer networking
across the vertical layers as well as horizontal multilayer (multidomain)
networking across multiple domains.
We proceed to outline open challenges and
future research directions for vertical multilayer networking in the context of
SDON, which includes an optical physical layer, in this subsection.
Horizontal multilayer (multidomain) networking is considered in
Section~\ref{multidomain:fut}.

For the vertical multilayer networking in a single domain,
the optical physical layer is the key distinguishing feature of
SDONs compared to conventional SDN architectures for general IP networks.
Most of the higher layers in SDONs have similar multilayer
networking challenges as general IP networks.
However, the optical physical layer
requires the provisioning of specific optical
transmission parameters, such as wavelengths and signal strengths.
These parameters are managed by optical devices, such as the OLT in PON
networks. For SDON networks, so-called \textit{optical orchestrators},
which are commercially available, e.g., from ADVA Optical Networking,
provide a single interface to provision the optical layer parameters.
We illustrate this optical orchestrator layer in the context of an
SDON multilayer network in the rightmost branch of Fig.~\ref{fig_control_orch}.
The optical orchestrator resides
above the optical devices and below the SDN controller.
The optical orchestrator uses common SDN SBI interface protocols,
such as OpenFlow,
to communicate with the optical devices in the south-bound direction
and with the controller in the north-bound direction.

The SDN controller in the control plane is responsible for the management of the
SDN-enabled switches, potentially via an optical orchestrator.
Communicating over the SBI using different
protocols can be challenging for the controller.
This challenge can
be addressed by using south-bound renderers. South-bound renderers are
APIs that reside within the
controller and provide a communication channel to any desired
SBI protocol.  Most SDN controllers currently have an
OpenFlow renderer to be able to communicate to Openflow network switches.
But there are also SNMP and NETCONF-based renderers, which
communicate with traditional non-OpenFlow switches. This enables the
existence of hybrid networks with already existing switches.
The effective support of such hybrid networks, in conjunction with
appropriate south-bound renderers and optical orchestrators, is an important
direction for future research.

\subsection{Multidomain Networks} \label{multidomain:fut}
A network domain usually belongs to a single organization that
owns (i.e., financially supports and uses) the network domain.
The management of multidomain networking involves the
important aspects of configuring the access control as well as the
authentication, authorization, and accounting. Efficient
SDN control mechanisms for configuring these multidomain networking
aspects is an important direction for future research and development.

Multidomain SDONs may also need novel routing
algorithm that enhance the capabilities of the currently used BGP
protocol. Multidomain research \cite{phe2014dis} has now taken
interest in the Intent-Based Networking (NBI) paradigm for
SDN control, where Intent-APIs can solve the
problems of spanning across multiple domains. For instance, the intent of an
application to transfer information across multiple domains is
translated into service instances that access configurations between
domains that have been pre-configured based on contracts.
Currently, costly manual configurations between
domains are required for such applications.
Future research needs to develop concrete models for
NBI based multidomain networking in SDONs.

\subsection{Fiber-Wireless (FiWi) Networking} \label{wireless:fut}
The optical (fiber) and wireless network domains have
many differences.
At the physical layer, wireless networks are characterized
by varying channel qualities, potentially high losses, and generally lower
transmission bit rates than optical fiber.
Wireless end nodes are typically mobile and may connect dynamically
to wireless network domains.
The mobile wireless nodes are generally the end-nodes in a
FiWi network and connect via intermediate optical nodes to
the Internet.
Due to these different characteristics, the management of
wireless networks with mobile end nodes is very different from the
management of optical network nodes.
For example, wireless access points should maintain their own
routing table to accommodate access to dynamically connected mobile
devices. Combining the control of both wireless and
optical networks in a single SDN controller requires
concrete APIs that handle the respective control functions of
wireless and optical networks.
Currently, service providers maintain separate physical management services
without a unified logical control and management plane for
FiWi networks.
Developing integrated controls for FiWi networks
can be viewed as a special case of multilayer networking and integration.

Developing specialized multilayer networking strategies for
FiWi networks is an important future research directions as many aspects of
wireless networks have dramatically advanced in recent
years. For instance, the cell structure of wireless cellular
networks~\cite{Ohlen2016} has advanced to femtocell
networks~\cite{cha2008fem} as well as heterogeneous and multitier
cellular structures~\cite{els2013sto,lou2011tow}. At the same time,
machine-to-machine communication~\cite{has2013ran,lay2014ran} and
energy savings~\cite{ana2015opt,has2011gre} have drawn research attention.

\subsection{QoS and Energy Efficiency}  \label{futworkene:sec}
Different types of applications have vastly different
traffic bit rate characteristics and QoS requirements.
For instance, streaming high-definition video requires high bit rates,
but can tolerate
some delays with appropriate playout buffering. On the other hand,
VoIP (packet voice) or video conference applications have
typically low to moderate bit rates, but require low latencies.
Achieving these application-dependent QoS levels in an energy-efficient
manner~\cite{has2011gre,ShZL14,wan2015ene}
is an important future research direction.
A related future research direction is to
exploit SDN control for QoS adaptations of real-time media and
broadcasting services.
Broadcasting services involve typically data rates ranging from
3--48~Gb/s to deliver video at various resolutions to
the users within a reasonable time limit.
In addition to managing the QoS, the network
has to manage the multicast groups for efficient
routing of traffic to the users.
Recent studies \cite{Butler2015, Ellerton2015} discuss
the potential of SDN, NFV, and optical technologies
to achieve the growing demands of broadcasters and media.
Moreover, automated
provisioning strategies of QoS and the incorporation of quality of
protection and security with traditional QoS are important direction for
future QoS research in SDONs.

\subsection{Performance Evaluation}
Comprehensive performance evaluation methodologies and metrics need to
be developed to assess the SDON designs addressing the
preceding future research directions ranging from simplicity and
efficiency (Section~\ref{simpl_fut:sec}) to optical-wireless networks
(Section~\ref{wireless:fut}).
The performance evaluations need to encompass the data plane,
the control plane, as well as the overall data and control plane interactions
with the SDN interfaces and need to take virtualization
and orchestration mechanisms into
consideration. In the case of the SDON infrastructure, the performance
evaluations will need include the optical
physical layer~\cite{Azodolmolky2014}.
While there have been some efforts to develop evaluation frameworks for
general SDN switches~\cite{OFTest,rot2014ope}, such evaluation frameworks
need to be adapted to the specific characteristics of SDON architectures.
Similarly, some evaluation frameworks for general SDN controllers have
been explored~\cite{jar2012fle,jar2014ofc}; these need to be extended
to the specific SDON control mechanisms.

Generally, performance metrics obtained with SDN and virtualization
mechanisms should be benchmarked against the corresponding
conventional network without any SDN or virtualization components.
Thus, the performance tradeoffs and costs of the flexibility
gained through SDN and virtualization mechanism can be quantified.
This quantified data would then need to be assessed and compared
in the context of business needs. To identify some of the important aspects of
performance we analyze the sample architecture in Fig.~\ref{fig_app_i2rs}.
The SDN controller in the SDON architecture in Fig.~\ref{fig_app_i2rs}
spans across multiple elements, such as ONUs, OLTs,
routers/switches in the metro-section, as well as PCEs in the core section.
A meaningful performance evaluation of such a network
requires comprehensive analysis of data plane performance aspects and
related metrics,
including noise spectral analysis, bandwidth and link rate monitoring,
as well as evaluation of failure resilience.
Performance evaluation mechanisms need to be
developed to enable the SDON controller to obtain and analyze these
performance data. In addition, mechanisms for control layer
performance analysis are needed.
The control plane performance evaluation should, for instance
assess the controller efficiency and performance characteristics,
such as the OpenFlow message rates and the rates and delays of flow table
management actions.

\section{\MakeUppercase{Conclusion}}  \label{sec:conclusion}
We have presented a comprehensive survey of software defined
optical networking (SDON) studies to date.
We have mainly organized our survey according to the SDN
infrastructure, control, and application layer structure. In addition,
we have dedicated sections to SDON virtualization and orchestration
studies.
Our survey has found that SDON infrastructure studies
have examined optical (photonic) transmission and switching components
that are suitable for flexible SDN controlled operation.
Moreover, flexible SDN controlled switching paradigms and optical
performance monitoring frameworks have been investigated.

SDON control studies have developed and evaluated SDN control
frameworks for the wide range of optical network transmission
approaches and network structures. Virtualization allows for
flexible operation of multiple Virtual Optical Networks (VONs) over a given
installed physical optical network infrastructure.
The surveyed SDON virtualization studies have examined the provisioning
of VONs for access networks, exploiting the
specific physical and Medium Access Control (MAC) layer characteristics of
access networks. The virtualization studies have also examined
the provisioning of VONs in metro and
backbone networks, examining algorithms for embedding the VON topologies
on the physical network topology under consideration of the
optical transmission characteristics.

SDON application layer studies have developed mechanisms for achieving
Quality of Service (QoS), access control and security, as well as
energy efficiency and failure recovery.
SDON orchestration studies have examined coordination mechanisms
across multiple layers (in the vertical dimension of the network protocol
layer stack) as well as across multiple network domains (that may belong
to different organizations).

While the SDON studies to date have established basic principles
for incorporating and exploiting SDN control in optical networks,
there remain many open research challenges. We have
outlined open research challenges for each individual category of
studies as well as cross-cutting research challenges.

				    	\chapter{\MakeUppercase{R-FFT: Functional Split at IFFT/FFT in Unified LTE and Cable Access 
		Network}}

\section{\MakeUppercase{Introduction}} 
\subsection{Motivation}
Today the user Internet connectivity 
has been dominated by the cellular wireless and 
cable/DSL access network technologies 
For a long time, wireless technologies, such as LTE 
have enjoyed large growth opportunities due to the
proliferation of wireless smart phones and hand-held devices 
in contrast to the cable technologies~\cite{kuzlu2013assessment}.
With the recent development of DOCSIS~3.1 specifications for cable
technologies, Multi-System Operators (MSOs)
are able to offer the services which are equivalent 
to the advance wireless technologies supporting gigabit 
connectivity to the users in both upstream and downstream~\cite{Hamzeh2015}. 
The next generation 5G technology~\cite{agiwal2016next} 
focuses on the unification of heterogeneous platforms 
at the access, backhaul and core networks especially through the 
softwarization and virtualization of the network infrastructures.
Towards this end, integration of cable and wireless 
technologies could be the first step
in the process of unifying the heterogeneous access platform.
5G is also envisioned to support a wide range of applications 
ranging from lower data rate Internet of Things (IoT) to
ultra-reliable health monitoring, and 
larger data rates for 4K HD video streaming~\cite{chen2015virtual}.
Challenges in the 5G technologies~\cite{chin2014emerging} include backhaul complexity
resulting from densification of radio nodes, 
interference with the neighboring cells 
and network deployment costs~\cite{Chih-Lin2015}. 
Promising technologies, such as Software Defined 
Optical Networking (SDN)~\cite{thyagaturu2016software} and
Network Function Virtualization (NFV)~\cite{blenk2016survey} 
are not only addressing the challenges posed by 5G but also 
facilitating the faster integration of 
heterogeneous-access networks while supporting the larger date rate, 
lower latencies and high reliability.
Moreover, current network infrastructures for both wireless and cable
technologies can rely on common optical fiber technologies, such as 
optical digital Ethernet~\cite{gomes2016new}. 
Therefore, MSOs are showing large interests in
converging to a common access platform through generic 
protocols and standardization,  
such as CPRI~\cite{de2016overview} for integration of heterogeneous 
infrastructures.

\subsection{Overview of Softwarization}
\subsubsection{Cloud-Radio Access Networks (CRAN)}
Softwarization of network functions can fundamentally 
reduce implementation complexity while increasing the 
flexibility~\cite{galis2013softwarization}. Software implementation of 
the RAN functions
in a cloud environment is referred to as Cloud-RAN 
or CRAN~\cite{makhanbet2016overview}.
In the CRAN, the functional implementation of the 
Radio Access Network (RAN) is split 
between Remote Radio Units (RRUs) which corresponds to
antenna units with RF passband processing capabilities
for the physical transmission of the signal, 
and Base Band Units (BBUs) which 
implements the complementary baseband signal processing.
Communication between CRAN and RRU is typically enabled by
optical technologies, such as Passive Optical Networks (PON) and
Optical Wavelength Division Multiplexing.
Traditionally, the connection between RRU and BBU is established 
by a cable RF link. Cable RF contributes towards considerable signal degrade 
for the larger distances between RRU and BBU.
In contrast to downlink signals, 
signal degrade in the uplink direction has more negative consequences 
due to the lower signal levels.
The signal degrade between BBU and RRU can be avoided by digitizing the 
RF signal and transmitting the digitized symbols over an optical fiber.

As noted earlier, important aspect of the RRUs in the context of
CRAN is the reduced complexity in radio equipment hardware 
which results in lower CAPEX/OPEX for
the large scale deployment of small cells.
RRUs in a building environment are also
referred to as Distributed Antenna Systems (DAS)~\cite{heath2013current}. 
DAS play an important role in realizing the large number of 
small cell deployments in meeting the coverage and capacity demands for
indoor applications. 
Centralization of BBU can 
support multiple RRUs which can provide a common platform for 
the centralized management of resources. 
BBUs are typically implemented on a generic computing 
hardware, where by a specific virtual machine (VM) 
implements the base band processing operations 
on a virtualized hardware~\cite{dai2014uplink}.

\subsubsection{Distributed Converged Cable Access Platform (DCCAP) Architectures}
Connectivity to the Cable Modems (CM) in the Hybrid Fiber Coax (HFC) network 
is delivered in part by optical and in part by cable segment.  
The Cable Modem Termination System (CMTS)
connects the CMs to Central Office (CO)/headend
through the protocol defined by Data Over Cable Service 
Interface Specifications (DOCSIS).
In the traditional HFC network deployments, analog signals from the
CMTS are carried over an optical link to a remote analog fiber node where
the signal is converted to analog RF signal which can be 
transmitted over the cable link.
But, an analog signal over an optical fiber gradually degrades
as the optical propagation distance is increased 
limiting the effective HFC operational distance (i.e., from the headend to CM). 
Modular Headend Architecture version 2 (MHAv2) allows 
CMTS functions to be implemented in a modular fashion allowing 
some of the modular entities to be implemented at the newly 
designed digital remote nodes. 
In order to improve the signal quality over the cable segment,
the remote nodes are connected to headend through the digital 
optical link where the remote node can be deployed in close
proximity to the users reducing the effective cable length.
Distributed Converged Cable Access Platform (DCCAP) 
architecture implements the modular CMTS functions at the
remote nodes.
Remote node is similar to a RRU in the CRAN whereby the 
RAN implementation at the cloud is similar to 
CMTS implementation at the headend.
The Remote-PHY (R-PHY) and Remote-MACPHY (R-MACPHY) technology 
of the DCCAP 
extends the modular CMTS and 
modular headend architectures with a 
goal to reduce the complexity and increasing 
serviceability of the cable 
infrastructure. In particular, R-PHY implements
the DOCSIS PHY functions on a 
remote node, Remote-PHY Device (RPD), 
while centralizing the MAC and higher layer network functions at the headend. 
Whereas, the R-MAC implements the DOCSIS PHY and MAC 
functions on a remote node, Remote-MACPHY Device (PMD),
while centralizing the IP forwarding
and higher layer network
functions at the headend. 
Centralized DOCSIS functions for RPD and RMD 
can be implemented as a software on a 
virtualized entity either at the headend or cloud.

The critical aspect of CRAN and DCCAP in the time domain I/Q
transport are the lower latency and higher data rate requirements
on the optical fiber~\cite{Wubben2014}.
A constant high bit rate and low latency 
connection must be maintained constantly between 
baseband processing and radio node regardless of the user traffic.
The analog RF signals must be transmitted and received at all times even 
when there is no user activity. For e.g., the passband signal with the 
cell broadcast information, reference or pilot tones must be transmitted always. 
Thus, I/Q samples of the RF passband must always be transported at the 
constant rate at all the times. 
Moreover, the requirements of an optical fiber
increases linearly with increase in the 
number of nodes. Therefore, numerous  
techniques such as~\cite{nieman2013time, nanba2013new, guo2013lte, guo2012cpri} 
have been proposed to enable the dynamic compression of RF I/Q samples 
for the effective transmissions over the optical fiber.
However the compression techniques are lossy because of the quantization of 
RF signal reducing the sensitivity of the receiver in the upstream.
Nevertheless, the data rate requirements between cloud and radio unit
leads to the dedicated deployments of
optical fiber connections and static allocations of resources 
for the CRAN and DCCAP. Functional split architectures~\cite{Wubben2014,maeder2014towards} 
transport the upper layer
data to the radio unit such that the constraints on
the optical connection are relaxed. However, the additional processing at the 
radio unit increases the complexity at the node as well as the flexibility
of the infrastructure is reduced as compared to the traditional 
CRAN is due to the distributed implementation of split RAN functions. 

\subsection{Contributions and Organization}
In this article, we primarily focus on the trade offs and benefits of 
the functional-split in both LTE and cable networks. In particular, 
we focus on the split-PHY architecture which implements the IFFT/FFT at
the radio units of both LTE and cable networks. 
Towards this end, the main contributions of this article are as follows:
\begin{itemize}
	\item[i)] Overview and trade off discussions on the functional split 
	and split-PHY architectures for both LTE and cable networks. 
	Presented in Sec.~\ref{sec:backgroud}.
	\item[ii)] A novel cross-split interaction 
	mechanism for remote caching and prefetching of QAM symbols
	to reduce the data rate required for the 
	I/Q transmissions between cloud and 
	radio unit, especially in the \textit{downstream} direction 
	while implementing the IFFT/FFT at the remote nodes. 
	Presented in Sec.~\ref{sec:split:inter}.
	\item[iii)] A novel mechanism to commonly share the 
	infrastructure resources among LTE and cable networks 
	The discussion also includes the general timing analysis 
	for interleaving multiple IFFT/FFT computations 
	of LTE and cable on a single IFFT/FFT module.
	Presented in Sec.~\ref{sec:common:infra}.
	\item[v)] In the end, we present the evaluation of 
	FFT sharing mechanism and mean packet 
	delay performance evaluation of 
	the radio unit which implements IFFT/ FFT for the 
	simultaneous LTE and cable operation in comparison
	to existing R-PHY technology.
	Presented in Sec.~\ref{sec:perf:eval}.  
\end{itemize} 

\subsection{Related work}
Checko et al.~\cite{checko2015cloud} has conducted the
CRAN technology review which includes advantages, challenges, 
SDN/NFV applications, as well as the compression techniques 
for I/Q transport over the fronthaul network. 
A more detailed discussions on the internals of the 
BBU and RRU units has been discussed by Wu et al. in~\cite{wu2015cloud}.
They show that cooperative signal processing in the CRAN 
achieves better spectral efficiency and improves utilization 
as compared to traditional networks. 
CRAN also provides a platform to provide computing services 
to the users through the access network. 
A comprehensive survey on Mobile Cloud Computing (MCC) in the area
of CRAN has been conducted by 
Fernando et al.~\cite{fernando2013mobile}. 
Fundamentals of the functional and protocol split uplink as well as the 
downlink and uplink compression techniques has been 
presented in~\cite{peng2016recent}. The impact of overhead due to 
packatization on the data rate and latency 
of the fronthaul link while supporting 
various functional split in the CRAN
has been presented in~\cite{chang2016impact}. 

Miyamoto et al.in their 
wireless performance evaluation~\cite{miyamoto2016analysis}  
claim that their proposed split-PHY architecture reduces the 
fronthaul bandwidth by up to 97\% compared to traditional
CRAN with the penalty of 2~dB SNR. However, their proposed
split-PHY architecture requires specialized 
transceiver designs and optical transmission mechanisms
which may increase the cost of CRAN deployments.
In a typical RRH deployment scenario, 
several RRHs experience different
load due to the independent channel qualities of the UEs reducing
the overall load RRH. 
Therefore, especially when the higher functional splits are considered,
all the RRHs connected to the CRAN would likely
not result in the peak load.
User specific traffic variation can provide the  
multiplexing opportunity and the impact of such multiplexing 
for the RRH deployments has been studied by 
Chang et al.~\cite{chang2016impact}.
Further to split-PHY, Nishihara et al.~\cite{nishihara2016study} 
have evaluated the performance of 
MACPHY-split where both MAC and PHY are implemented at the RRH. 
Their simulation results show
that MACPHY-split approach reduces the bandwidth from
10~Gbps to 600~Mbps while supporting the multiplexing.

CRAN involve large number of RRH deployments 
which can potentially
be the source of large power consumptions. 
Efforts to reduce the carbon foot print over the
cellular networks supporting both CRAN as well as traditional
deployments has been presented in~\cite{hasan2011green, oh2013dynamic,wu2015energy,de2014enabling}. 
Whereas the energy saving mechanism in the wireless
networks has been presented in~\cite{suarez2012overview}.
Stephen et al.~\cite{stephen2016green} have designed an 
cache enabled OFDM resource allocation mechanism which saves
the transmissions over the fronthaul link. 
However, the content
has to be saved in parts marked with identifiers in the remote node
In contrast to all the previous CRAN studies, we have
considered the caching of
resource elements at the RRH on the split-PHY architecture 
to reduce the data rate over the fronthaul link potentially 
saving the energy consumption when 
user data is not present.

Complementary to the wireless cellular networks
HFC network provides the broadband access to the residential 
users. Bisdikian et al.~\cite{bisdikian1996cable}
has presented the overview of initial designs and protocol 
mechanisms for the cable networks. 
As compared to the earlier designs, current and 
future broadband access networks extends the
capabilities of connectivity to Gigabit 
speeds~\cite{effenberger2016future}. 
Even further, present cable networks can be designed to support cellular 
networks services as described in Gambini et al.~\cite{Gambini2012, gambini2010lte}. 
The economic benefits of infrastructure sharing between residential wired and 
cellular wireless networks has been identified in~\cite{pereira2012infrastructure}.
Additionally, economic benefits form the
integration of LTE and DOCSIS has also 
been discussed in~\cite{gruber2014broadband}.

Articles \cite{liu2015bandwidth,zeng2016demonstration,liu2016cpri} have
demonstrated efficient transmissions of CPRI equivalent 
 fronthaul data up  to 256 Gbps using the optical bandwidth of 10 GHz,
especially for supporting MIMO LTE applications. 
However, the proposed mechanism in the articles use specialized 
optical transmission with constant traffic over the optical link 
without allowing the possibility of multiplexing. In addition, 
we propose the novel remote caching and prefetching of QAM symbols, 
which saves the fiber transmissions. Especially, when there are no 
users, optical fiber connectivity can be effectively turned off to 
save the power or share the resources with other nodes. 
In comparison to most of the access networks, such as, 
macro base stations, DAS, femto cells, and HFC,
the traditional macro base 
station backhauled through a microwave link at low traffic
results in the most 
energy economical option~\cite{fiorani2016joint}.
As cable networks are one of the sources for growing
carbon footprint in the Internet connectivity,
there has been several studies proposed to address the issue
such as~\cite{zhu2011novel,Zhu2012,Lu2013}.

\section{\MakeUppercase{Background on Functional Split in LTE and Cable Networks}} 
\label{sec:backgroud}
\subsubsection{Downstream v/s Upstream}
\paragraph{Upstream} In the upstream direction the RRU
receives the RF signal transmitted from the users. This analog
passband signal is down converted to baseband and digitized 
for the transmission to BBU for baseband processing. Unlike the 
cable link in the traditional cellar network and antenna infrastructures,
the CRAN connects the BBU and RRU with an digital optical fiber. 
The cable link adds significant attenuation to the upstream signal especially
due to the low level of received signal from the devices. Whereas, the
digital fiber does not constitute towards the attenuation loss as it 
carries the signal in the digital form. To note, an extreme care 
is needed at the RRUs for digitizing the uplink signal from the users 
as an additional loss cannot be afforded due to the low signal 
level of uplink RF signal at the RRU. For example, if the cable link 
accounts for $2$~dB of loss and noise floor is $-120$dB, then the received
signal at the RRU connected a BBU over cable link must be $\leq-118$~dB 
for the successful detection, whereas the received signal can be $\leq 
-120$~dB if the RRU is connected to a BBU 
through the digital fronthaul link, thus increasing the dynamic range of the system by 2~dB. 
Although Single Carrier Orthogonal
Frequency Multiplexing (SC-OFDM) 
uplink modulation format is used in the typical deployments, 
the technology is advancing towards the uplink OFDM systems especially
for the MIMO applications~\cite{studer2016quantized, pitarokoilis2016performance}.
Therefore, in the scope of this article we focus on the symmetrical OFDM systems
in both upstream and downstream directions. In addition to the increased dynamic 
range, the processing of upstream 
for the detection and extraction of information from the RF 
uplink signals can be centrally processed at the cloud on the generic hardware, 
such as the general purpose processors.

\paragraph{Downstream} In the downstream direction BBU sends the 
information to RRUs for the generation of passband signal to
transmit over the physical antennas. The RRUs can easily set the 
transmit power level gain states for RF signals. As in the upstream direction,
there is no significant difference in terms of 
signal generation or the dynamic range of the systems 
between cable and digital fronthaul links.
Similar to the centralized processing of upstream at the cloud, 
the information is centrally processed on a generic hardware, such as the general 
purpose processors to generate the baseband downlink signals.

\subsection{Functional Split in LTE}
\begin{figure}[t!] \centering
	\includegraphics[width=5in]{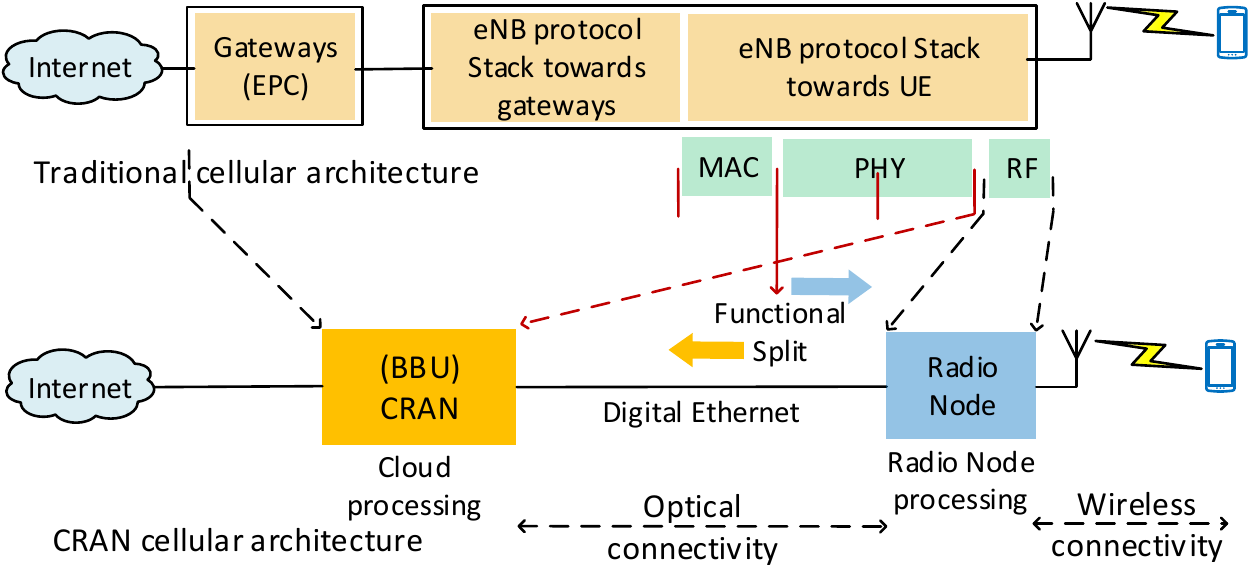}
	\caption{The Cloud-RAN (CRAN).} 
   \label{fig_CRAN}
\end{figure}
Figure~\ref{fig_CRAN} shows the
traditional CRAN deployment in comparison conventional cellular deployments.
A radio base station eNB of LTE protocol stack towards the UE 
can be functionally split and implemented flexibly over radio node and cloud
seprated by a fronthaul link which is typically an optical fiber connectivity
The traditional CRAN transports the baseband time domain I/Q samples
over the optical fiber to the RRUs. The number of RRUs that is
supported over a fiber deployment depends on the amount of traffic
that is required to support the operations of each RRU. Suppose if 
$C$ is the capacity of the fronthaul optical connectivity and $R_i$ 
is the data rate required by $i^th$ RRU,
then maximum number of RRUs $N$ that can be supported over that specific
fronthaul link is $\max_{N} \{\sum_{i = 1}^N R_{i} \}$ such at 
$\sum_{i = 1}^N R_{i} \leq C$. In the present deployments of CRAN
the resources of the fronthaul link are statically allocated with dedicated
connectivity. Therefore in symmetrical and homogeneous deployment, 
$R_1 = R_2 \ldots R_N = R$ resulting in the total number of RRUs 
that can be supported over the fronthaul link as $N = C/R$. The main
bottleneck for the CRAN deployments is delay and capacity of the fronthaul link $C$.
Heterogeneous fronthaul links where multiple RRUs requirements
of delay and data rate changing over time independent of each other 
would require an advance resource management mechanisms to avoid 
the capacity wastages as compared to the 
worst case tolerance deployment practices.  
\begin{table}[t]
	\caption{Typical LTE CRAN Parameters}
    \label{tab:CRANparams}
	\centering
	\begin{tabular}{|p{1cm}|p{5cm}|p{2cm}|} 
		\hline 
		\textbf{Param.} & \textbf{Description} &  \textbf{Value} \\ [.5ex]
		$N$ & Number of RRU per CRAN &  1 \\
		$B$ & LTE cell bandwidth &  20 MHz  \\
		$K$ & Bits per I/Q & 10 bits   \\
		$W$ & Number of Tx/Rx antennas & 2 \\
		$T_s$ &  OFDM symbol duration & 66.6 $\mu$s \\
		$f_s$ &  Sampling frequency & 30.72 MHz \\
		$f_c$ &  Carrier frequency & 2 GHz \\		
		\hline
	\end{tabular}
\end{table}
To understand the fronthaul requirements of an RRU, 
we estimate the data rate $R$ required by the traditional CRAN where 
the baseband I/Q is transported from the cloud BBU to RRU 
for the most common LTE deployment scenarios. 
Table~\ref{tab:CRANparams} summarizes the important list of parameters
used for evaluation in the context of LTE for the 
fronthaul optical link connecting RRU to CRAN. 
The data rate comparisons of various functional 
split in the LTE protocol stack has 
been conducted in~\cite{maeder2014towards, Wubben2014}. In contrast, 
we take a closer look on the data rate requirements based on the implementation 
specifics of the protocol stack. That is, we track the flow of information 
across multiple protocol stack layers of the LTE and 
identify the key characteristics that dictate the 
requirements of fronthaul link. 

Based on the computationally intensive operation of the FFT The data flow 
between BBU and RRU can be categorically divided into 
two types i) time domain samples and ii) frequency domain samples.  

\subsubsection{Time Domain I/Q Forwarding}
\label{sec:time:iq}
The time domain I/Q samples
represents the RF signal in the digital form either in the passband 
or baseband representation. Typically, the digital representation of the 
passband signal requires a very large data rate depending on the 
physical transmission frequency band. Thus, passband time domain I/Q forwarding
is non economical. For example, in an LTE system, the passband signal
is sampled at twice the carrier frequency $f_c$ with each sample requiring
10 bits for digital representation, the passband 
I/Q data rate $R_i^P$ required over the fronthaul link is
\begin{equation}
\begin{split}
R^P_i = N \times W  \times 2f_c \times 10 \;\;\;\;\;\;\;\;\;\;\;\;\;\;\;\;\;\;\;\;\;\;\;\;\\
= 1 \times 2 \times 2(2 \times 10^9) \times 10 = 80~\text{Gbps}.
\end{split}
\end{equation}

The baseband signal for an OFDM symbol in the 
time-domain consists time samples equal to the number of 
OFDM subcarriers because of the symmetric input and output samples \
from the IFFT/FFT structure.
The cyclic prefix is added to the OFDM signal to avoid the 
inter symbol interference. In order to reduce the constraints on the 
RF signal generation at the RRU, the baseband signal is sampled
at the frequency of 30.72 MHz with each sample requiring
10 bits for digital representation and an oversampling factor of 2, 
the baseband I/Q data rate $R_i^B$ 
required over the fronthaul link is
\begin{equation}
\begin{split}
R^B_i = N \times W  \times (2 \times f_s) \times (2 \times K) \;\;\;\;\;\;\;\;\;\;\;\\
= 1 \times 2 \times (2\times 30.72 \times 10^6) \times (2 \times 10)\\
 = 2.46~\text{Gbps}.\;\;\;\;\;\;\;\;\;\;\;\;\;\;\;\;\;\;\;\;\;\;\;
 \;\;\;\;\;\;\;\;\;\;\;\;\;\;\;\;\;
\end{split}
\end{equation}

Although the data rate is significantly reduced as compared to the passband 
I/Q forwarding, the baseband I/Q data rate scales linearly with the
number of antennas and bandwidth. Thus, for large number of 
antennas and larger aggregated bandwidth the data rate $R^B_i$ can be very large.
  
\subsubsection{Frequency Domain I/Q Forwarding}
\label{sec:freq:iq}
In the 20 MHz LTE system One OFDM symbol duration 
including the cyclic prefix is $71.3~\mu$s which 
corresponds to 2192 samples. The useful symbol duration in OFDM 
is $66.7~\mu$s of 2048 samples while the cyclic prefix duration 
is $4.7~\mu$s of 144 samples. 
Thus, each of 2048 samples in the OFDM symbol (excluding the cyclic prefix) 
corresponds to 2048 subcarriers ($B_{sub}$) when transformed by the FFT. 
However, only 1200 of these subcarriers are used for 
signal transmission which corresponds to 100 resource blocks (RBs) 
of 12 subcarriers each the rest is zero-padded 
and serves as guard carriers. This 
leads to $(2048-1200)/2048=0.41=41~\%$ of unused guard carriers.
Each subcarrier in the OFDM is modulated by a complex value mapped 
by a QAM alphabet. LTE QAM alphabet size is based on QAM bits 
such as, 64 QAM and 256 QAM. If only the subcarrier information which consists of
the complex values taken from the QAM alphabet i.e., only the frequency domain data
is used to transport between BBU to RRU. 
The resulting data rate $R^F_i$ is only dependent 
on the number of subcarriers $B_{sub}$. A vector of 
complex valued QAM alphabet symbols of size $B_{sub}$ needs to be sent 
once every OFDM symbol duration $T_s$ i.e., the data rate 
\begin{equation}
\begin{split}
R^F_i = N \times W \times B_{sub} \times T_s^{-1} \times (2 \times K)
\;\;\;\;\;\;\;\;\;\;\;\;\;\;\;\\
= 1 \times 2 \times 1200 \times (66.7\times 10^{-6})^{-1} \times (2 \times 10)\\
= 720~\text{Mbps}.\;\;\;\;\;\;\;\;\;\;\;\;\;\;\;\;\;\;\;\;\;\;\;\;\;\;\;
\;\;\;\;\;\;\;\;\;\;\;\;\;\;\;\;\;\;\;\;\;\;\;
\end{split}
\end{equation}
In comparison to time domain baseband I/Q data rate $R_i^B$, the $R_i^F$ is
approximately reduced by $(R_i^B - R_i^F)/R_i^B = (2460-720)/2460 = 70~\%$. 
Further attempts to reduce the data rate $R_i$ 
require moving complex functions, such as MAC to the RRU.
The MAC HARQ (Hybrid ARQ) 
processes requires complex operations and 
large buffers to be implemented 
at the RRU increasing the complexity of the RRU.
Thus, with the savings of 70~\%, keeping the simplicity at the RRU
is worthwhile to explore the transport mechanism of 
frequency domain I/Q symbols over the fronthaul link. 

\subsection{Functional Split in Cable Distributed Converged Cable Access Platform (DCCAP) Architectures}
\begin{figure}[t!] \centering
	\includegraphics[width=4.5in]{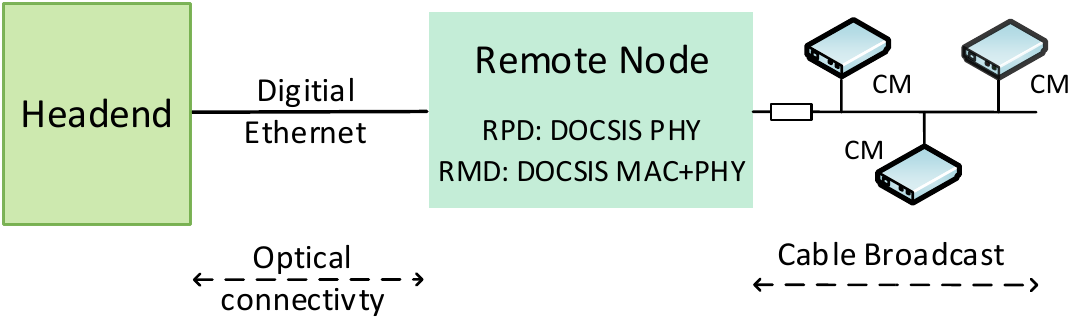}
	\caption{Distributed Converged Cable Access Platform (DCCAP).} 
	\label{fig_remote_node}
\end{figure}
Traditional CCAP architecture for the HFC network implements 
the CMTS at the headend and 
transports analog optical signal to an Optical Remote 
Node (ORN) over the optical fiber. 
ORN then converts the optical
analog signal to electrical RF signal to transmit over the cable 
segment. However, such architectures which rely on
analog optics suffer from attenuation of the analog signal in both 
optical fiber segment as well as cable segment of the HFC network because
of the placement of the ORN. If the ORN is deployed far from the headend,
the attenuation in optical signal will be dominant, similarly
if the ORN is deployed far from the CM (users) then the attenuation in 
the RF signal of the cable will be dominant. Modular Headend
Architecture (MHA) overcomes the downside of CCAP architectures by
enabling the functional splitting of CMTS allowing the modular 
implementation of the CMTS functions. Implementation of 
modular CMTS functions distributively across multiple nodes
are referred to as Distributed Converged Cable 
Access Platform (DCCAP) Architecture. As shown in the Fig.~\ref{fig_remote_node}
DCCAP architecture defines a new
Remote Node (RN) which is connected to the headend through digital
Ethernet fiber. Digital connection between the RN and headend eliminates the 
attenuation of the analog optical signal allowing the RN to be 
deployed more deeper into the users network 
reducing the cable segment length which in turn reduces 
the analog RF attenuation improving the overall Signal to Noise Ratio (SNR)
at the CM. The network which connects the RN to headend is referred 
to as Converged Interconnect Network (CIN). 
MHA version 2 (MHAv2)~\cite{MHAv2} architecture
defines two DCCAP architectures Remote-PHY and Remote-MACPHY.  
 
\subsubsection{Remote-PHY (R-PHY)}
\paragraph{Overview}
In the R-PHY architecture, the DOCSIS PHY function  
in the CMTS protocol stack is implemented at the 
RN referred to as Remote-PHY Device (RPD), while all the
higher layers in the CMTS protocol stack including 
the MAC as well as the upstream scheduler are implemented at the headend.
A virtual-MAC (vMAC) entity virtualizes the DOCSIS MAC on a 
generic hardware which can be flexibly implemented either at the headend 
or cloud/remote data center. Since the upstream scheduler is implemented
at the headend, the request from the CM is sent to the headend for processing
the grants to coordinate the upstream transmissions over the cable link. 
The extensive performance evaluation of R-PHY and R-MACPHY have been conducted
in~\cite{ziyad2017,thyagaturu2017r}.
Therefore, the CM request-to-grant delay depends on the distance of 
RN from the headend (i.e., the CIN distance)~\cite{Chapman2014, Chapman2015}.
\paragraph{Advantages}
R-PHY node is simple to implement and hence the CAPEX and OPEX of the
RPD can be reduced. At the headend, more computational resources 
can be dedicated to support the complex schedulers for 
mode advance coordination mechanism of CM transmissions over the 
broadcast cable medium.
\paragraph{Disadvantages}
As the distance of RPD from headend increase, the CM request-to-grant
delay increases due to increases RTT between RPD and headend.
Applications such as, virtual and augmented reality requires
Ultra Low Latency (ULL) over the cable link can be impacted
by the larger CM request-to-grant delays.

\subsubsection{Remote-MACPHY (R-MACPHY)}
\paragraph{Overview}
In the R-MACPHY architecture, the DOCSIS PHY and MAC functions along with
the upstream scheduler in the CMTS protocol stack are implemented at the 
RN referred to as Remote-MAC Device (RMD), while all the
higher layers in the CMTS protocol stack are implemented at the headend.
Since the upstream scheduler is implemented
at the RMD, the requests from the CMs are processed for the 
the grants at the RMD to coordinate the upstream transmissions 
over the cable link reducing the request-to-grant delay.
Therefore, the CM request-to-grant delay is independent of the distance of 
RN from the headend.
\paragraph{Advantages}
R-MACPHY produces significantly lower delay when compared to R-PHY especially
the CIN distance is very large. ULL applications can be readily supported as
the upstream scheduler is implemented very close the CM reducing the request-to-grant
delay. 
\paragraph{Disadvantages}
RMD device has to implement the scheduler on the remote node increasing the 
CAPEX and OPEX of the remote node. As remote node implements more 
functions of the CMTS protocol stack in a distributed fashion,
the flexibility of the network reduces as
compared to centralized architectures, such as R-PHY.

\subsection{Functional PHY-Split}
\begin{figure}[t!] \centering
	\includegraphics[width=5in]{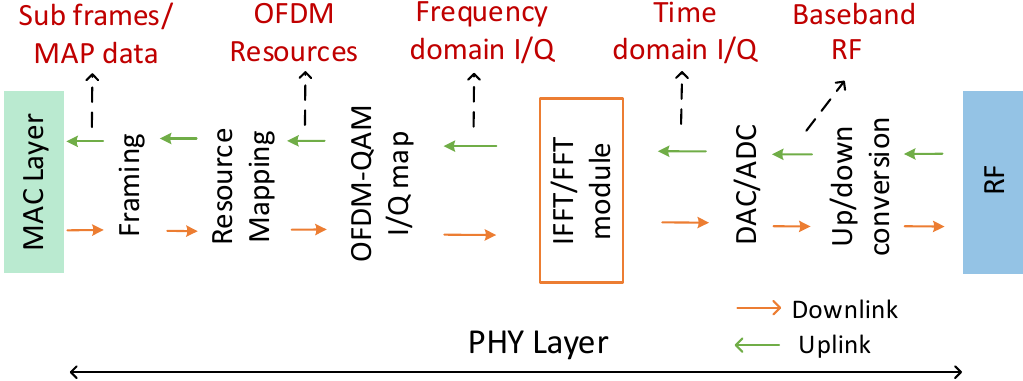}
	\caption{PHY-split architecture.} 
	\label{fig_split_phy}
\end{figure} 

The maximization of the “cloudification” of CMTS DOCSIS and 
LTE functions results in the
increased flexibility and reduced CAPEX 
and OPEX for the Multiple System Operators (MSOs) and 
cellular operators. The cloud implementation of DOCSIS and LTE functions 
can be easily achieved at the same physical location, remote data center or cloud.
Both LTE and DOCSIS 3.1 also share similar transceiver PHY characteristics for 
the OFDM implementation that can be exploited to design 
the simultaneous support for LTE and DOCSIS over the HFC network which is critical to
the 5G type of applications, such as health monitoring and security \cite{veeranna2016hardware}. 
The general overview of the physical layer for LTE and 
DOCSIS is shown in the Fig.~\ref{fig_split_phy}. In the downstream direction,
the data from MAC layer is
processed to PHY frames and mapped to OFDM resource locations which is then 
converted to frequency domain QAM I/Q symbols (see~\ref{sec:freq:iq}) based on the 
modulation and coding schemes. The
QAM I/Q symbols are then IFFT transformed to 
get the complex time domain samples. 
This time domain samples (see~\ref{sec:time:iq}) 
are then converted 
to analog RF signal for the transmission over the cable link. To note, the processing of I/Q 
information before the IFFT/FFT module is specific to DOCSIS and LTE 
protocol whereas the processing of 
I/Q after the IFFT/FFT module is relatively same for both DOCSIS and LTE.
Thus, we can separate the functions based on the IFFT/FFT module such 
that IFFT/FFT is implemented at the remote node to support the
the architectures and mechanisms to simultaneously 
operate LTE and DOCSIS over the HFC network. 
To the best of our knowledge, there exists no prior research 
for simultaneously supporting LTE and cable so as to efficiently 
utilize the optical fiber (fronthaul) 
resources of the already installed HFC plant.

\section{\MakeUppercase{Proposed Cross-Functional Split Interaction}} 
\label{sec:split:inter}
\begin{figure}[t!] \centering
	\includegraphics[width=4.5in]{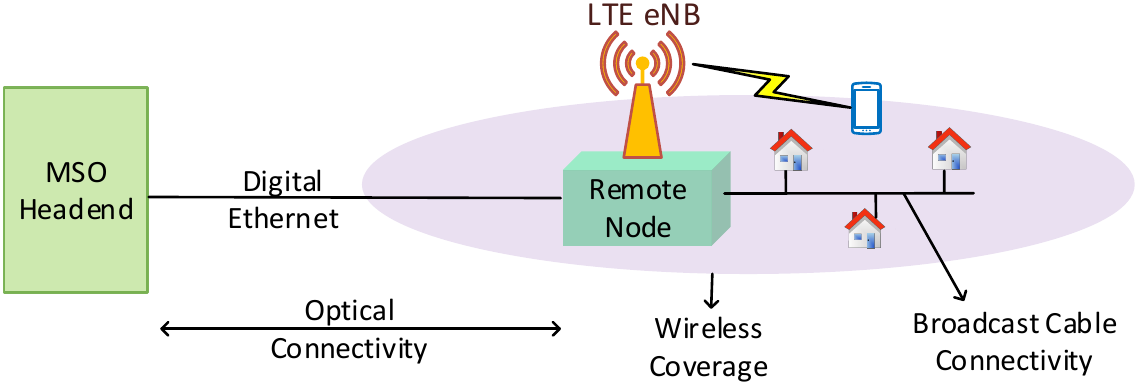}
	\caption{Deployment of the LTE eNB RRU at the Remote Node (RN).}
	\label{fig_LTE_cable}
\end{figure} 
The digital optical remote node (ORN) in the DCCAP architecture 
is deployed closer to the CM (users). Close proximity of the 
ORN to the residential 
subscribers would help in establishing the wireless LTE 
connectivity by deploying
an LTE eNB RRU at the ORN site. As show in Fig.~\ref{fig_LTE_cable} 
installing an RRU at the ORN site 
is advantageous to the operators due to the close proximity of the users. 
With the establishment of LTE connectivity by the MSOs users can be 
wirelessly connected to the MSO's core network 
for the Internet connectivity increasing the service capabilities of the MSOs.
In addition, support of LTE eNB RRU at the ORN also reuses the 
existing HFC infrastructure reducing the CAPEX and OPEX for the MSOs in 
providing the additional services of LTE.

Traditionally, the CRAN and DCCAP architectures split the 
functions linearly based on the protocol stack 
(i.e., MAC, PHY etc.), 
and implements the split parts at the ORN and cloud.
Additional to protocol stack split, our proposed mechanism 
enables cross-split interaction through signalling
to facilitate the caching and prefetching (see Sec~\ref{sec:caching}) 
of redundant information between the functional splits. More
specifically, cross-functional split interaction involves 
the communication of control information between
baseband unit and radio unit.
\subsection{Proposed Shared Remote-FFT (R-FFT) Node}
\begin{figure}[t!] \centering
	\includegraphics[width=5in]{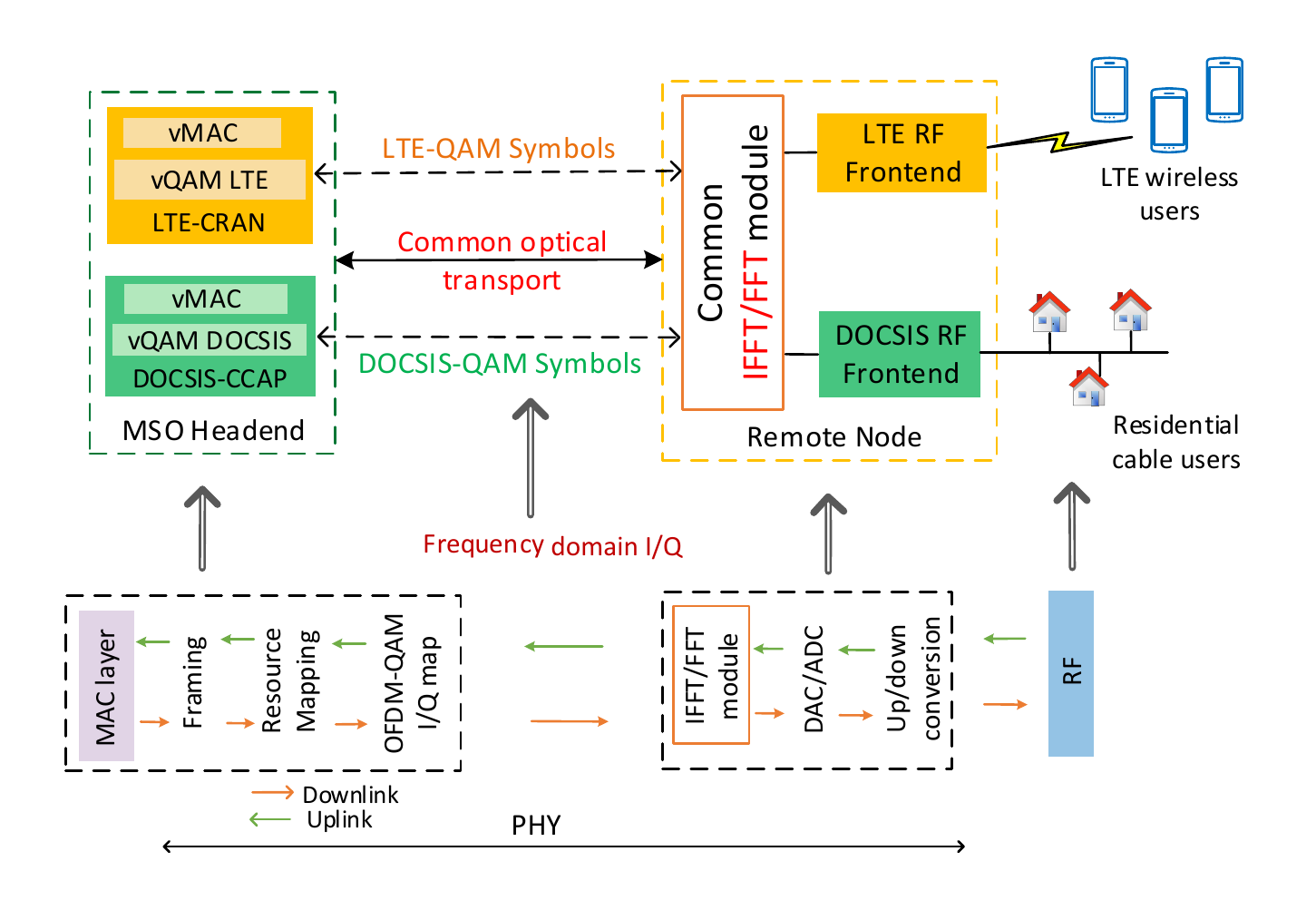}
	\caption{Remote FFT (R-FFT) implements the FFT module at the remote node.}
	\label{fig_rfft}
\end{figure} 

In the uplink direction, the proposed remote node R-FFT converts the 
incoming DOCSIS RF signal from the CM to an encapsulated data bits format 
that can be transported over digital fiber link for additional processing and
onward forwarding at the headend. In a similar way, for the downlink, 
RF signals are generated from the incoming formatted data bits and 
sent out on the RF link to the CMs. For LTE, an eNB can use a wide 
range of licensed spectrum with a single largest carrier component 
of 20 MHz, however the bandwidth can be extended further by carrier 
aggregation technique resulting in larger effective bandwidth. 
The R-FFT node effectively converts the upstream LTE RF signal 
from the wireless users and transports the signal digitally to the 
BBU/CRAN via the digital fiber link. In the downlink direction, 
the R-FFT node converts the digital information to an LTE RF 
signal that can be transmitted wirelessly to the users.

A Remote-DAC/ADC (R-DAC/ADC) would require peak rate 
transmissions \cite{rdac}. Considering a system with a downlink 
bandwidth of 100 MHz and a 1024 QAM modulation format, 
which requires 12 bits per sample, results in a total bit rate over the fiber of (100 MHz) x (12) x (Frequency sampling rate). 
For a frequency sampling rate of 2.5 times (slightly greater than the Nyquist rate),
 the total required bit rate is 3 Gbps. In contrast, 
an equivalent R-PHY system requires a data rate of (100 MHz) x (10 Bits/Hz) x
(90~\%) =  900~Mbps; whereby, 100 MHz is for the resource allocation 
across the entire bandwidth, 10 Bits/Hz is for the 1024 QAM, and 90~\% 
is for the effective channel with a 10~\% guard band.
This simple example comparison indicates a threefold increase of the 
data rate in the R-DAC/ADC system compared to the R-PHY system. 
Note that the R-DAC/ADC system requires a constant peak 
data rate to transport the I/Q samples, regardless of the 
amount of user data present in the RF signal. 
In contrast, similar to R-PHY/MAC, in R-FFT system, 
the data rate depends on the 
amount of user data. In the above example data rate calculation, 
we have considered the maximum allocation 
of the user data in the R-PHY system. 

Our motivation is to address the increased fiber data rate 
through a balanced split among the functions within the PHY node, 
while keeping the simplicity of the R-DAC/ADC and the CRAN systems. 
The existing R-DAC/ADC node requires some digital circuitry, 
such as a CPU, for the DAC and ADC control. Therefore, we believe 
that the FFT/IFFT implementation is not a significant additional 
burden for the remote node~\cite{jang2016study}. 
Figure~\ref{fig_rfft} shows the implementation of vMAC 
for both LTE and DOCSIS at the Headend and 
a common FFT module at the remote node. 
The advantages of proposed FFT implementation at the remote node include: 
\begin{itemize}
	\item[i)] flexible deployment support for LTE and DOCSIS
	\item[ii)] requires lower data rate $(R_i^F)$ to transport frequency 
	domain I/Q as compared to time-domain I/Q $(R_i^B)$.
	\item[iii)] data tones carrying no information are zero valued 
	in the frequency I/Q samples effectively resulting a 
	lower date-rate over the fiber channel in both LTE and DOCSIS with the
	chance for statistical multiplexing, and
	\item[iv)] possible caching of repetitive frequency QAM I/Q samples, 
	such as Reference Signals (RS) and pilot tones. 
\end{itemize}

To emphasize the important data rate aspect, we note that in the 
proposed shared R-FFT, the data rate required over the fiber channel is 
directly proportional to the user traffic. We believe this is an 
important characteristic of the FFT functional-split whereby we 
can achieve multiplexing gains by combining multiple R-FFT nodes or 
by the enabling the concurrent support of LTE as illustrated in 
Fig.~\ref{fig_LTE_cable}. With the similar characteristics of DOCSIS,
the LTE user traffic also translates to proportional data rate 
over the fiber channel. In addition, the proposed mechanism enables 
the complex signal processing of the PHY layer to be implemented 
at the headend. Examples of the signal processing operations include channel
estimation, equalization, and signal recovery, which can be implemented with
general-purpose hardware and software. In addition, the processing of 
digital bits, such as the LDPC, which is necessary for the forward error
correction, can also be implemented at the headend. Thus, the proposed approach
reduces the cost of the remote nodes and increases the flexibility of changing 
the operational technologies. The software implementations at the headend can be
easily upgraded while retaining the R-FFT node hardware since the node hardware
consists only of common platform hardware, such as elementary DAC/ADC and 
FFT/IFFT components. Thus, the proposed approach eases the change/upgrade of
technologies. That is, the R-FFT node has minimal impact on technology 
advancements because the blocks within a remote node are elementary 
or independent of most technology advancements. 

\subsection{Proposed Remote Caching and Prefetching}
\label{sec:caching}
\begin{figure}[t!] \centering
	\includegraphics[width=5in]{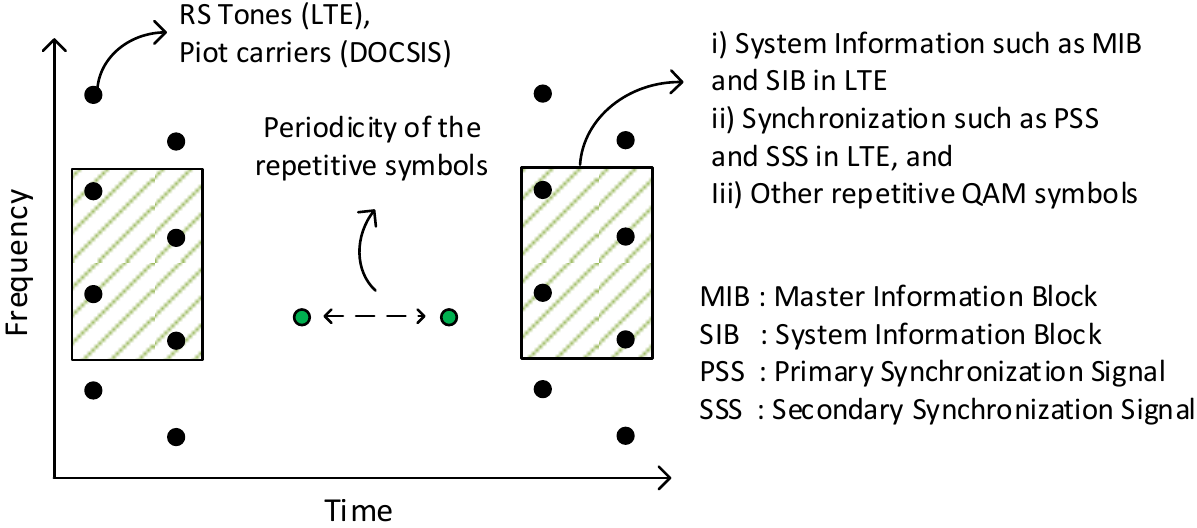}
	\caption{Caching at the remote node.}	
	\label{fig_caching}
\end{figure}
In order to further reduce the bandwidth in addition to functional split process, 
several techniques, such as I/Q compression~\cite{nanba2013new, nieman2013time,guo2012cpri, joung2013base} 
can be employed. In contrast, we propose resource element (time and frequency slot) 
allocation based remote caching. If some part of the information is regularly and 
repeatedly sent over the interface, a higher (orchestration, in case of SDN) 
level of the signaling process can coordinate caching mechanisms. For example, 
there is no need to transmit the downlink I/Q samples of the pilot tones as
they would remain constant in the DOCSIS. Figure~\ref{fig_caching} illustrates the overview 
of repetitive QAM symbols in the LTE and DOCSIS. There is a process of 
up-sampling and zero padding before and after the FFT/IFFT block, 
depending on the FFT size and the effective subcarriers, which can 
potentially be controlled and cached through the headend (depending on 
the implementation). Zero padded QAM symbols as well as certain reserved resource elements 
can be excluded from the 
transmissions while signaling the changes 
to the cache management at the R-FFT node.
As compared to downstream, upstream information must be entirely 
transported to the headend to process all the signal component 
received by the R-FFT receiver. 
\subsubsection{LTE Networks}
\paragraph{Reference Signal (RS) Tones Caching}
RS tones are the pilot subcarriers which are embedded throughout the operational 
bandwidth of the wireless system for the channel estimation to equalize the 
impairments of received wireless signal. Disturbances to the 
wireless signal are more prominent compared to propagation of signal 
in the wired channel and therefore
the RS tones are added in close proximity with 
each other to accurately estimate the channel based 
on the channel characteristics, such as coherence-time and coherence-bandwidth. 
For a single antenna, the RS tones are typically spaced 6 subcarriers apart 
in frequency such that 8 RS tones exists in a single subframe, i.e, 14 OFDM symbols
and single Resource Block (RB), i.e., 12 subcarriers of LTE. Thus, the overhead
due to RS tones in the LTE resource grid is
\begin{equation}
\text{RS Overhead} = \frac{8}{12 \times 14 } = 4.7~\%.
\label{eq:rs}
\end{equation}
Therefore, approximately 4.7~\% of I/Q transmissions over the 
digital fiber can be saved from the RS tones caching at the 
remote node regardless of the system bandwidth.  

\paragraph{PHY Broadcast Channel (PBCH) Caching}
PHY Broadcast Channel (PBCH) carries the Master Information Block (MIB)
which is broadcasted continuously by the
eNB regardless of the user connectivity.
MIB includes the basic information about the LTE system 
such as the system bandwidth and control information specific to LTE channel.
PBCH/MIB always uses central
6~RBs (i.e, 72~subcarriers) for the duration of 
4~OFDM symbols to broadcast the MIB data.
PBCH space in the resource grid is inclusive of the 
RS tones used in the calculation of Eq.~\ref{eq:rs} and therefore
needs to subtracting while calculating the overhead for MIB transmission.
PBCH/MIB occurs once every 40~ms and there exists 4~redundant versions of 
MIB which will be broadcasted with the offset of 10~ms. Thus, 
essentially PBCH/MIB occurs once in every 10~ms (radio frame). The 
PBCH/MIB overhead for the entire LTE system is
\begin{equation}
\text{PBCH Over.} = \frac{6 \times 12 \times 4 - (8 \times 6)}{1200 \times 14 \times 10} = 0.142~\%.
\label{eq:mib}
\end{equation}
Although 0.142~\% is less, relative overhead of PBCH/MIB data 
increases significantly for lower bandwidth LTE system. Moreover, our evaluation of the
PBCH overhead considers the full RBs i.e., 1200 subcarriers, 14 OFDM symbols and 10 subframes
in the LTE. In a real system the allocation broadly varies user density i.e, 
generally only part of the RBs will be used increasing the overhead effects. 
For example, for a 1.4~MHz system (used in an IoT type of applications) 
with full RBs, the overhead jumps to 
\begin{equation}
\text{PBCH Over.}_{1.4\text{MHz}} = \frac{6 \times 12 \times 4 - (8 \times 6)}{72 \times 14 \times 10} = 2.3~\%.
\label{eq:mib:14}
\end{equation}
Therefore, up to 2.3~\% I/Q transmissions over the 
digital fiber can be saved from the PBCH/MIB caching at the remote node.

\paragraph{Synchronization Channel Caching}
Synchronization Channel (SCH) consists of Primary Synchronization Sequence (PSS)
and Secondary Synchronization Sequence (SSS) which is broadcasted continuously
by the eNB regardless of the user connectivity. PSS and SSS are the special
sequence which helps in the cell synchronization of wireless users by identifying
the physical cell ID and the frame boundaries of the LTE resource grid.
PSS/SSS occurs every 5~ms (twice per radio frame) and uses central 6~RBs over 2 
OFDM symbols. Similar to Eq.~\ref{eq:mib}~and~\ref{eq:mib:14}, the overhead due to
PSS/SSS in 20~MHz and 1.4~MHz system is 
\begin{equation}
\begin{split}
\text{SCH Over.} = \frac{6 \times 12 \times 4}{1200 \times 14 \times 10} = 0.171~\%. \;\;\;\:\\
\text{SCH Over.}_{1.4\text{MHz}} 
= \frac{6 \times 12 \times 4}{72 \times 14 \times 10} = 2.8~\%. \;\;\;\;\;\;\;
\end{split}
\label{eq:sch}
\end{equation}

\paragraph{System Information Block (SIB) Caching}
In a similar way, the caching mechanism can also be extended 
to the System Information Blocks (SIBs) broadcast messages 
which is through the PHY Downlink Shared Channel (PDSCH) of the LTE.
There are 13 different types of SIBs, from SIB1 to SIB13. 
SIB1 and SIB2 are mandatory broadcast messages while the transmission of other SIBs 
depends on the relation between the 
serving and neighbor cell configurations. On a typical deployment, 
SIB3 to SIB9 are configured and can be combined in a single message block
for the resource block allocation. Typical configuration of the 
RB allocation type schedules the SIB over 8 RBs across
14~OFDM symbols in time (1~subframe) are used for the transmission of SIB1 and SIB2, 
and with the effective periodicity (with redundancy version transmission) 
of 2 radio frames (20~ms). The overhead and caching gain from the 
of SIB1 and SIB2 transmissions while subtracting the corresponding 
RS tones overhead $8 \times 8$, i.e., 8 tones per RB for 8 RBs is
\begin{equation}
\begin{split}
\text{SIB Over.} = \frac{8 \times 12 \times 14 - (8 \times 8)}{1200 \times 14 \times 20} = 0.381~\%. \;\:\\
\text{SIB Over.}_{1.4\text{MHz}} 
= \frac{8 \times 12 \times 14 - (8 \times 8)}{72 \times 14 \times 20} = 6.3~\%. \;\;\;\;\;
\end{split}
\label{eq:sib}
\end{equation}
Caching of higher order SIBs i.e., from SIB3 to SIB 9 
can achieve further savings, however, the resource allocation 
and periodicity can widely vary to accurately estimate the overhead.

The stationary resource elements across the time domain, such as 
SIB information is allowed to change over a larger time scale in terms of 
hours and days, at such times the cached elements are refreshed or 
re-cached through cache management and signalling procedures, see Sec.~\ref{sec:cache:mgt}.
 
\subsubsection{Cable Networks}
In the DOCSIS~3.1 downstream pilot subcarriers are modulated 
by the CMTS with a predefined modulation pattern which is known to all the
CMs to allow the interoperability.
Two types of pilot patterns are defined in the DOCSIS~3.1 for OFDM time
frequency grid allocations, i) continuous, and ii) scattered. 
In the continuous pilot pattern, the pilot tones with a predefined modulation 
occur at fixed frequencies in every symbol across time. 
Whereas, in the scattered pilot pattern, the pilot tones are sweeped to 
to occur at each frequency locations but at different symbols across time.
The scattered pilot pattern has a periodicity of 128 OFDM symbols along the time dimension such that the pattern repeat for the next cycle. 
Scattered pilots assist in the channel estimation for the
equalization process of demodulation. In a typical deployment~\cite{cabledeploy}, 
operational bandwidth is 192~MHz corresponding to the FFT size of 8K
with the 25~kHz subcarrier spacing. The total number
of subcarriers in a 192~MHz system is 7680 out of which 
there are 80 guard band subcarriers, 88 continuous pilot subcarriers and 
60 scattered pilot subcarriers. Therefore, the overhead due to redundant 
subcarriers in the DOCSIS system which can be cached at the remote node is
\begin{equation}
\text{Cable Over.} = \frac{80 + 88 + 60}{7680} = 2.9~\%.
\label{eq:cable}
\end{equation}  

\subsection{Memory Requirements for Caching}
Caching of frequency domain I/Q symbols in the OFDM 
comes at the cost of caching memory implementation at the remote node.
Each I/Q symbol that needs to be cached is a complex number
with real and imaginary part. For the purpose of evaluation, we consider 
each part of the complex number represented by 10 bits resulting
in 20 bits in total for each frequency domain QAM symbol. The caching of
RS tones in the LTE results in 4.7~\% of the savings in the 
fronthaul transmissions as show in the Eq.~\ref{eq:rs}. In the duration of 
each OFDM symbol, 2 RS tones exists for every 12~subcarriers. For the 
20~MHz system, there would be 200 RS tones. Therefore,
total memory required to cache the data
of QAM symbols corresponding
to the RS tones is 
\begin{equation}
\centering
\text{RS Tones Mem.} = (2 \times 100) \times 2 \times 10 = 4000~\text{bits}.
\label{eq:rs:mem}
\end{equation}  
Similarly, caching of PBCH, SCH and SIB data requires 
\begin{equation}
\begin{split}
\text{PBCH Mem.} = (4\times12\times6 - (8 \times 6)) \times 2 \times 10 \;\;\;\;\;\;\; \\
= 4800~\text{bits}, \;\;\;\;\;\;\;\;\;\;\;\;\;\;\;\;\;\;\;\;\;\;\;\;
\;\;\;\;\;\;\;\;\;\;\;\;\;\;\;\;\: \\
\text{SCH Mem.} = (6 \times 12 \times 4) \times 2 \times 10 = 5760~\text{bits},\;\;\: \\
\text{SIB Mem.} = (8 \times 12 \times 14-(8 \times 8)) 
\times 2 \times 10 \;\;\;\;\; \\ =5760~\text{bits}.
\;\;\;\;\;\;\;\;\;\;\;\;\;\;\;\;\;\;\;\;\;\;\;\;
\;\;\;\;\;\;\;\;\;\;\;\;\;\;\;\;
\end{split}
\label{eq:pbch:mem}
\end{equation}
Whereas, for the DOCSIS, the the memory requirements for the
caching of continuous and scattered pilots is  
\begin{equation}
\centering
\text{Pilot Tones Mem.} = (80+88+60) \times 2 \times 10 = 4560~\text{bits}.
\label{eq:pilot:mem}
\end{equation}
Thus, based on Eqs.~\ref{eq:rs}-\ref{eq:cable}, total  
savings of approx. 7~\% to 18~\% can be achieved in the
fronthaul transmissions when the full resource allocation over the
entire bandwidth is considered in both LTE and DOCSIS.
For lower allocations i.e, when there 
exists lesser user data, the caching process can achieve much 
larger relative benefits. 
In an extreme case, when the user data is not present, all the cell 
specific broadcast data information can be cached at the 
remote node such that the fronthaul transmissions 
can be seized reducing the power consumption.
The total memory for the caching required at the remote node based 
on the Eqs.~\ref{eq:pbch:mem}-\ref{eq:pilot:mem} is approximately 19120~bits.
The implementation of cache at the remote node 
which is lesser than the size of 25~KB is relatively very simple 
and of no practical burden to the existing remote node. Therefore, 
we believe the savings of more than 7~\% with almost negligible implementation
burden is a significant benefit in the CRAN functional PHY-split. 
 
\subsubsection{Signalling and Cache Management}
\label{sec:cache:mgt}
Signalling mechanism facilitates the cache management processes. 
Cache management operation involves, 
i) transporting the caching information to the remote nodes,
ii) updating the cached information at the remote nodes with new 
information, and
iii) establishing the rules for prefetching of cached resource 
elements at the remote node.
Signalling agents at the headend/cloud
and remote node coordinate with each other 
through a separate (i.e., non-I/Q transport)
logical connection between headend/cloud and remote node.
Some of the information which has been cached may change over time but 
at a larger timescale compared to the transmission of 
I/Q from headend to the cloud. The signalling overhead which arises 
from the cache management is negligible because of the timescale of
signalling operations. Headend/cloud can notify the remote node 
with the changes 
through signalling to keep the information up to date at the remote node.
Prefetching of the cached content has to be precisely executed 
with accurate placement of the subcarriers information in the particular 
time and frequency locations to an IFFT/FFT computational module. 
 
\subsection{Transport Networks and Protocols}
\subsubsection{Transport Networks}
An R-FFT node can be connected to headend through variety of transport 
network solutions over the optical fiber connectivity as well as dedicated 
mmWave links. The fundamental requirement is to support the CRAN and CCAP 
demands while virtualizing the QAM functions at the headend. 
Some of the technologies that can be considered are:
\paragraph{Dark Fiber} A fiber network that is deployed in excess of 
the existing requirement is referred to as a dark fiber network. 
Typically, dark fiber resources are abundant but with limited accessibility. 
If a headend can be connected to an R-FFT node using dark fiber, a large data 
rate with low latency connections can be established, supporting safe 
and reliable transport of required I/Q samples.
\paragraph{Optical Wavelength Division Multiplexing (WDM)} WDM establishes 
end to end connectivity with the dedicated wavelength aggregated to meet 
the user demands. WDM with the potential to reach large data rates and low 
latencies can easily support the connectivity of the R-FFT with the headend.
\paragraph{Passive Optical Network (PON)} A PON network typically provides 
high data rates in the Point to Multipoint Ethernet services for residential 
and Fiber to The Premises (FTTP) type of applications. If PON technology is 
used to establish the connectivity, careful latency limitations must be evaluated. 
Several interfaces, such as CPRI, exists to support the 
connections over the PON transport network.
\paragraph{Wireless} A standard and dedicated microwave or millimeter-wave 
wireless links can support large data rate and low latency connection between 
two nodes. However, wireless links are limited by the range and 
sometimes require Line of Sight (LoS) operation.

\subsubsection{Protocols}
Protocol is required to coordinate the transmissions of I/Q data over the
transport network. A strict latency requirement for the CRAN and DCCAP 
architectures limits the choice of generic protocols over 
Ethernet. Some of the fronthaul protocol that exists for the 
transport of information between headend/cloud and radio node 
are:

\paragraph{Radio over Fiber (RoF)}
Radio over fiber (RoF) transports the radio frequency signal 
over an optical fiber link by converting the electrically modulated signals
to optical signal. RoF signals are not converted in frequency but superimposed
over optical signals to achieve the benefits of optical transmissions,
such as reduced sensitivity to noise and interference
RoF signals are optically distributed to end nodes where the
optical signals are directly converted to electrical signal with
minimal processing reducing the cost of the remote node. 
The downside of the RoF is the analog transmission of the optical 
signal which attenuate over distance over the fiber as compared to 
the digital data over the fiber.

\paragraph{Common Public Radio Interface (CPRI)}
Common Public Radio Interface (CPRI) \cite{CPRI} defines the protocol
to transport the digitized I/Q data through the encapsulated CPRI frames. 
As compared to RoF, CPRI provides more reliable end-to-end connection between
headend/cloud and the remote node. Dedicated TDM channels can be established 
to supports multiple logical connections supporting different air interfaces. 
The downsides of the CPRI are the strict timing and synchronization 
requirement as well as support for only the fixed functional split to 
transport time domain I/Q samples. To overcome the fixed functional split
in the eCPRI is being currently developed to support a new functional 
split with ten folds of decrease in the data rate requirement over the 
fronthaul link. In comparison to eCPRI and CPRI protocols, 
our proposal is an enhancement to 
the base protocols where the repeated I/Q can be cached when frequency 
I/Q samples are transported over the fronthaul link.

\paragraph{Open Base Station Architecture Initiative (OBSAI)}
Open Base Station Architecture Initiative (OBSAI)~\cite{obsai} is 
similar to CPRI in which the digitized time domain I/Q samples are 
transported over fronthaul interface.
In contrast to CPRI, OBSAI interface is an IP based connection.
A digital interface which is based on an IP logical connection 
can be implemented over any generic Ethernet link providing a flexible
connectivity between headend/cloud to remote node.

\paragraph{External PHY Interfaces}
Downstream External PHY Interface (DEPI)~\cite{depi} and Upstream External PHY Interface
(UEPI)~\cite{uepi} enable the common transport mechanisms between 
R-PHY and CCAP core. DEPI
and UEPI are based on the Layer 2 Tunneling Protocol version 3 (L2TPv3).
The L2TPv3 transparently transports the Layer 2 protocols 
over a Layer 3 network creating the psedudowires (logical connections).
For the R-MACPHY, since only the MAC payload is required to 
be transported to the headend for processing of upper layer
data, strict requirements for the latency as in the case of R-PHY 
can be relaxed, allowing
any generic tunneling protocol to be used between headend and R-MACPHY.

\section{\MakeUppercase{Proposed Common Platform for LTE and Cable Networks}}
\label{sec:common:infra}
\begin{figure*}[t!] \centering
	\includegraphics[width=\textwidth]{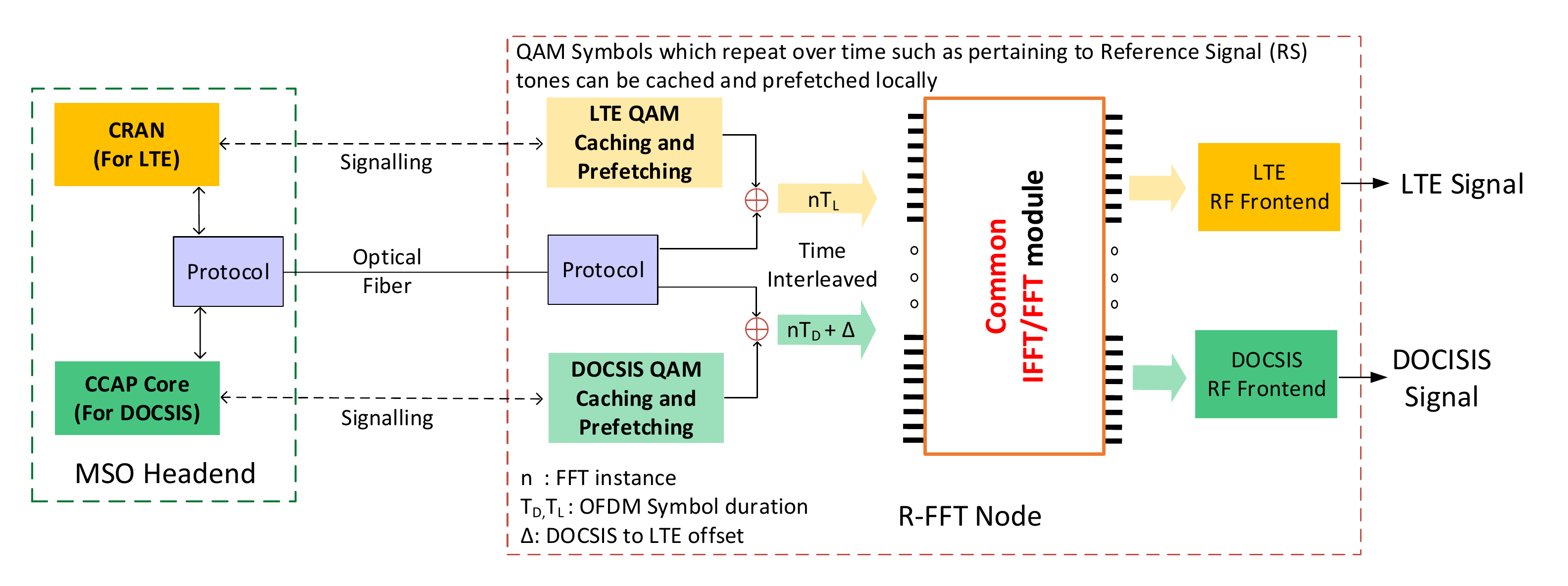}
	\caption{Reuse of infrastructure.}
	\label{fig_common_fft}
\end{figure*}
The last hop connectivity to the user devices is typically 
enabled by wireless technologies. Although, any wireless interface, 
such as Bluetooth, ZigBee, or WiMax, can provide the wireless 
last hop connectivity, WiFi is a very promising technology. 
WiFi has proven to be simple and efficient for managing large 
numbers of connections as well as to serve large
data rates to connected nodes. Also, importantly, WiFi has the capability
of delivering high data rates (up to Gbps) in the unlicensed bands
over a reasonable coverage area (up to 100 m), and in indoor 
environments. In contrast, the LTE wireless interface, which is widely used 
for commercial cellular communications in licensed bands, is capable of 
delivering up to several hundreds of Mbps over larger geographical areas 
(up to several tens of miles). Several efforts are underway to integrate 
these two wireless technologies so as to enable even higher data rates 
on the order of several Gbps. LTE-U and LTE-LAA [6] are newly developed wireless 
interfaces where a primary LTE link opportunistically utilizes the 
unlicensed WiFi bands. The next generation personal computers 
are expected to support LTE connectivity in conjunction with 
Ethernet and WLAN. 
However, to meet the future demands 
to support concurrent connectivity, deployment and management of
independent LTE and WiFi technologies by the same operator 
incurs large expenses to the operators. We envision a solution
where LTE base station can be established at the 
remote node in the DCCAP architecture.
To our knowledge, the
implementation of FFT at the remote node to simultaneously support
cable and LTE infrastructure is novel to our proposal. 

\subsection{Common IFFT/FFT for LTE and DOCSIS}
The protocol of LTE and DOCSIS share the same physical properties as
they depend on OFDM for the physical layer modulation technique.
Implementation of the OFDM is dependent on the design of 
FFT computations~\cite{he1998designing}.
The property by which both LTE and DOCSIS require 
same IFFT/FFT operations to be 
performed for each OFDM modulation and demodulation can be exploited
to reuse the existing infrastructure provided such mechanism 
can sustained on the computing hardware.
Thus, the main motivation of the FFT implementation at the remote node 
is to bring out  the common platform at the remote 
nodes while the transmission formats are 
flexibly realized at the headend for heterogeneous 
protocols based on OFDM.
Figure~\ref{fig_common_fft} describe the internals of the
R-FFT architecture simultaneously supporting cable and LTE. 
Generally, in the downstream direction, an IFFT operation is performed 
once every OFDM symbol duration. The LTE OFDM symbol duration is 
approximately 72 µs, and for DOCSIS, the OFDM symbol duration 
is 84.13 µs. However, the actual time to compute IFFT can span 
from few microseconds to several tens of microsecond. 
Considering that there exists a large portion of idle 
time durations in the IFFT module during the FFT computation, 
we can interleave the I/Q input in time such that same 
IFFT/FFT module can be used for multiple technologies. 
By reusing the IFFT/FFT computing structures we can reduce 
the complexity of the hardware, be more power efficient, 
and reduce the cost of R-FFT node. 

\subsection{FFT Computations Interleaving Timing Discussion}
\begin{figure}[t!] \centering
	\includegraphics[width=5in]{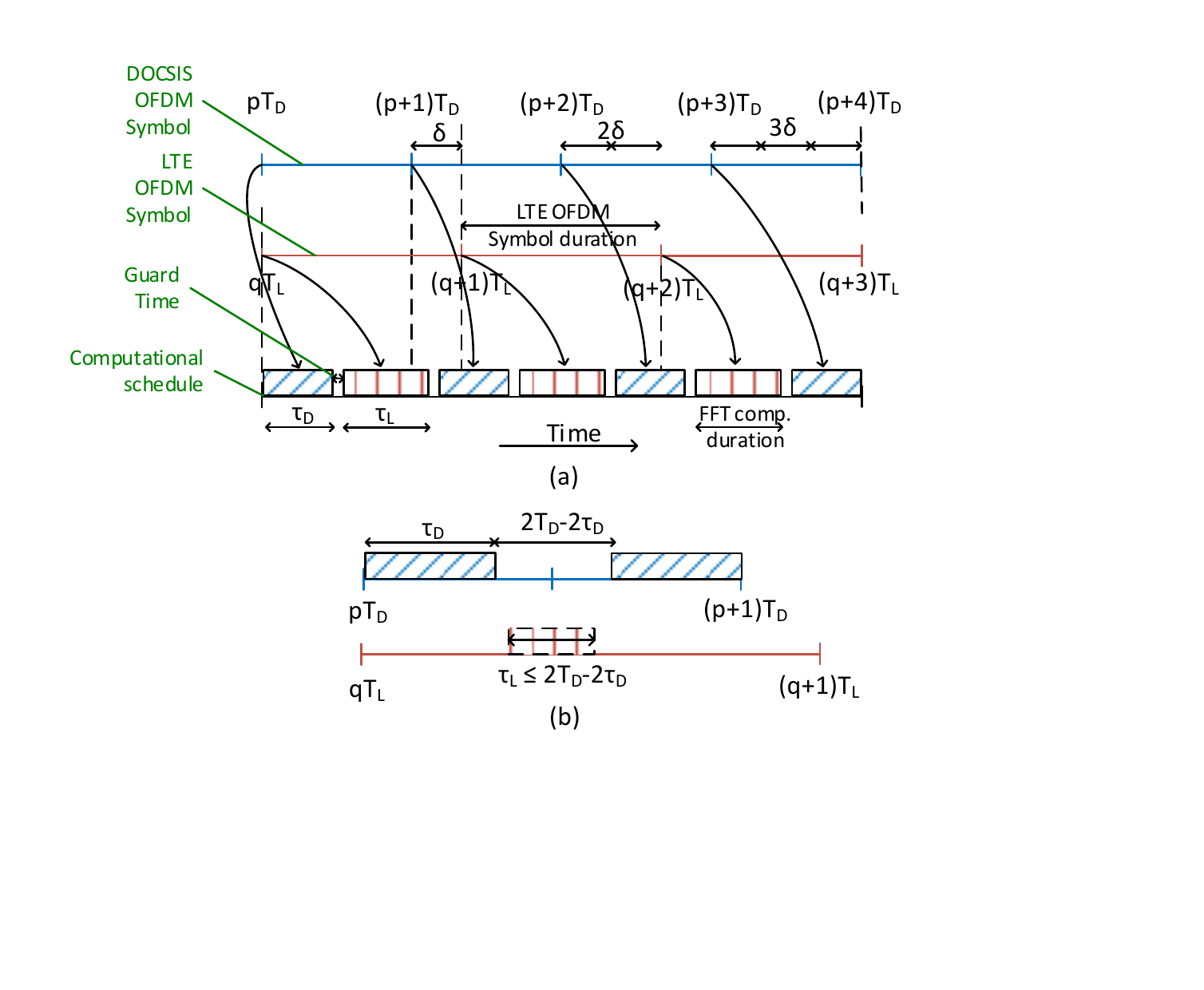}
	\caption{The IFFT/FFT interleaving.}
	\label{fig_timing}
\end{figure}
In this section we preset the timing schedule analysis 
for the interleaving of IFFT/FFT computations on 
a single computing resource. Figure~\ref{fig_timing} shows the
basic timing diagram to schedule the FFT computations on the
computing resource. $T_D$ and $T_L$ are the OFDM symbol durations of 
DOCSIS and LTE respectively. Similarly, $\tau_D$ and $\tau_L$
are the durations of FFT computations for DOCSIS and LTE.
While $p$ and $q$ are the timing indexes for the frame of reference
for DOCSIS and LTE independent periodic events. 
WE also assume that $T_L$ and $T_D$ start at the time index p
and q without any offset as show seen in the Fig.~\ref{fig_timing}

For illustration, in this article we assume the OFDM
symbol duration of LTE is greater then the symbol duration of 
DOCSIS, i.e., $T_L > T_D$, Consequently, the duration
of FFT computations depends on the FFT size where we assume 
$\tau_D > \tau_L$, as DOCSIS has the larger FFT size compared to LTE. 
Also, it is most likely that heterogenous technologies operate with 
different OFDM symbol times. Thus, 
we define the difference in the OFDM symbol duration as 
$\delta= T_L - T_D$. 

\paragraph{Feasibility Discussion}
Suppose if there is no offset between the cycles 
as shown in the start of timing schedule in Fig.~\ref{fig_timing}
and $k/l$ denotes the fraction of OFDM symbol durations, 
where, $k$ and $l$ if exists are the minimum positive integers,
such that $T_L/T_D=k/l$,
for every $k$ cycles of $T_D$ there would be $l$ cycles of $T_L$. 
That is, when two periodic signal are overlapped, 
the difference of periodicities $T_L$ and $T_D$, $\delta$, 
at each cycle turns out to be integer multiple of the 
previous cycle. 
For example, in the Fig.~\ref{fig_timing} at the
first cycle we define $\delta = T_L - T_D$, for the second cycle
the difference is $2T_L-2T_D = 2\delta$ and similarly for the third
cycle the difference is $3\delta$. Since $T_L > T_D$ From 
the fundamental principles we can evaluate that after  
$k=\lceil T_L/\delta \rceil = \lceil T_L/(T_L-T_D) \rceil $ 
cycles of $T_D$, the overlapping behavior repeats. Also,
the  For example,
in the Fig.~\ref{fig_timing}, for every $k=4$, which corresponds to
4 cycles of $T_D$ and 3 cycles of $T_L$, the behavior repeats.
Therefore, if the stability is ensured for $k$ cycles of $T_D$ or
$l$ cycles of $T_L$, the system is stable and feasible.

Thus, considering the larger OFDM symbol duration $T_L$
as the reference, where we have $l$ cycles of $T_L$ for the secondary
periodicity of combined $k$ and $l$ cycles. For the system 
to be stable, the total 
FFT computation durations from combined LTE and DOCSIS 
i.e., $l\tau_L + k\tau_D$ should not exceed 
the duration of $l$ cycle of $T_L$ i.e.,
\begin{equation}
\centering
l\tau_L + k\tau_D \leq lT_L.
\label{eq:stability}
\end{equation}  
We know that $k$ cycles of $T_D$ is equal to the $l$ cycles of $T_L$,
i.e., $kT_D = lT_L$ and therefore, $k=\frac{lT_L}{T_D}$. Substituting $k$
in the Eq.~\ref{eq:stability} we get,
\begin{equation}
\centering
\begin{split}
\left(\frac{lT_L}{T_D}\right) \tau_D+l\tau_L\leq lT_L \\
\frac{T_L \tau_D}{T_D} + \tau_L \leq T_L\\
\tau_D T_L + \tau_L T_D \leq T_L T_D.
\label{eq:stability2}
\end{split}
\end{equation}
In addition to Eq.~\ref{eq:stability2}, as shown in the Fig.~\ref{fig_timing}
we need to ensure that
FFT computation  durations are sufficiently small to support the 
interleaving. That is, for interleaving $\tau_{\max(T_L, T_D)}$ on
the periodic cycles of $\min(T_L, T_D)$, we need at least the space of  
\begin{equation}
\centering
\tau_{\max(T_L, T_D)} \leq 2\min(T_L, T_D)-2(\tau_{\min(T_L, T_D)}).
\label{eq:stability3}
\end{equation} 

\paragraph{Constraints}
The end of each OFDM symbol duration marks the
trigger point for the FFT computing scheduling request. 
Our fundamental analysis is based on
the fact that IFFT/FFT computing takes much less time 
compared to the actual OFDM symbol duration, especially
as the computing hardwares are becoming more 
advanced~\cite{yeh2003high, arun2016design}.
As this factor is dependent on the hardware we impose 
following constraints to design the interleaving scheduling
procedure of two heterogeneous technologies on the
same hardware, i) the FFT computing
durations should not exceed their OFDM symbol duration, and
ii) the FFT computations must be finished before the start of
next OFDM symbol, i.e.,
\begin{equation}
\centering
\begin{split}
\tau_L < T_L, \;\;\;\;\;\;\;\;\;\;\:\\
\tau_D < T_D, \;\;\;\;\;\;\;\;\;\;\:\\
p\tau_L < (p+1)T_L,\\
q\tau_D < (q+1)T_D.
\end{split}
\label{eq:comp.constr}
\end{equation}
Thus, in addition to Eq.~\ref{eq:stability2}, the constraints presented in
the Eq.~\ref{eq:comp.constr} must be satisfied in the process interleaving
of computations to have no adverse effects on the operational technologies.

\paragraph{Effect of Guard Time}
Guard time separates two consecutive computations to compensate for the 
load and read time of the I/Q to the FFT structure while supporting the 
coexistence of two technologies on the same computation module.
We define $\theta$ as the constant guard time which is applied 
to each scheduling set of the FFT computation. Based on Eq.~\ref{eq:stability2},
the impact of guard time can be evaluated as
\begin{equation}
\centering
(\tau_D+\theta)T_L + (\tau_L+\theta)T_D \leq T_DT_L,
\label{eq:guardtime}
\end{equation}
which can be written as,
\begin{equation}
\centering
\tau_DT_L + \tau_LT_D \leq T_DT_L - \theta(T_L+T_D).
\label{eq:guardtime2}
\end{equation}

\begin{algorithm}[t!]
	\caption{Caching and FFT Computation Procedure}
	\label{algo:interleaving}
	\SetKwInOut{Input}{input}
	\SetKwInOut{Output}{output}
	\nonl {\bfseries{1. CRAN/Headend}} \newline
	\nl (a) Identify cachable I/Q samples. \newline
	    (b) Create Caching rules. \newline
	    (c) Signal the rules and data for caching.	  
	\newline
	\nonl \If{Cached I/Q samples requires updating}{
		\nonl  Signal remote node for cache renew or flush.
		\nonl }
 \nonl {\bfseries{2. Remote Node}}   \newline
\nonl\ForEach{OFDM Symbol in $T_D$ and $T_L$}{		
  \nonl          \If{Caching is enabled}{
  \nonl				  Read cache and I/Q mapping\; 
  \nonl                    Insert I/Q cache-read to recieved I/Q\;}     
  \nonl	    	\If{FFT module is free}{
  \nonl				  Schedule I/Q for FFT based on Eqs.~\ref{eq:stability}-\ref{eq:comp.constr}\;} 
  \nonl          \Else{
  \nonl	Schedule at completion of current execution\; 
  \nonl }  \nonl}	   	    
\end{algorithm}

\paragraph{Implementation}
\label{sec:tau}
Sharing of remote node infrastructure for multiple technologies can 
be in both upstream and downstream, as the computations are 
performed independent of each other 
even for wireless full-duplex communications. 
The FFT computation duration 
$\tau$ can be aggregate of multiple instances of OFDM symbol, for example, in
the case of carrier aggregation in LTE (or channel bonding in DOCSIS), 
there would be an OFDM symbol 
for each carrier component resulting in $\tau_L=\tau_1+\tau_2\dots\tau_C$, 
where $C$ is the number of carrier components. Similarly, computations resulting 
from multiple LTE eNBs at a single node can be aggregated and abstracted to 
a single $\tau_L$ resulting in a lager $\tau_L$ can be very 
close the symbol duration $T_L$. Proposed approach can also be 
easily extended to more than two technologies which depend 
on FFT computations by sharing the the remote node.
 
\section{\MakeUppercase{Performance Evaluation}}
\label{sec:perf:eval}
In this section we present the performance evaluation of the R-FFT node where
the infrastructure of a remote node is 
shared for two independent LTE and DOCSIS technologies.
\begin{figure*}[t]
	\begin{tabular}{cc}
		\includegraphics[width=2.8in]{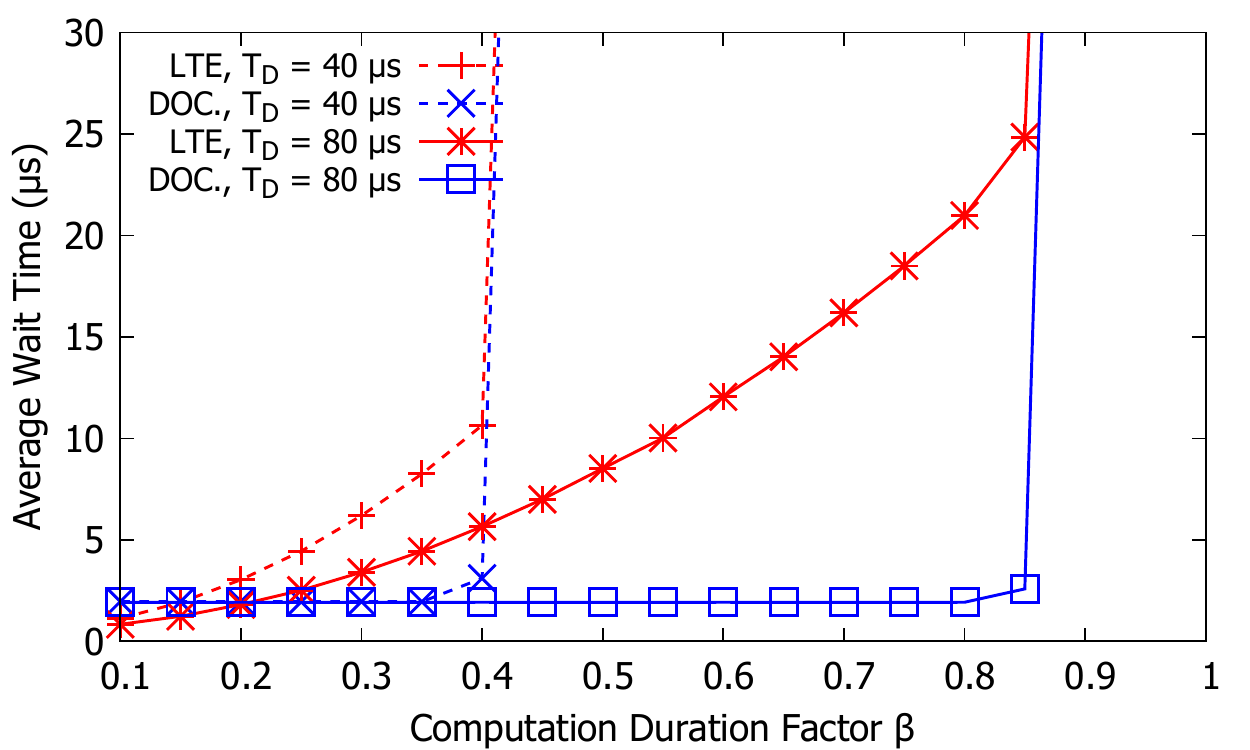} \vspace{-0.1cm}
		&\includegraphics[width=2.8in]{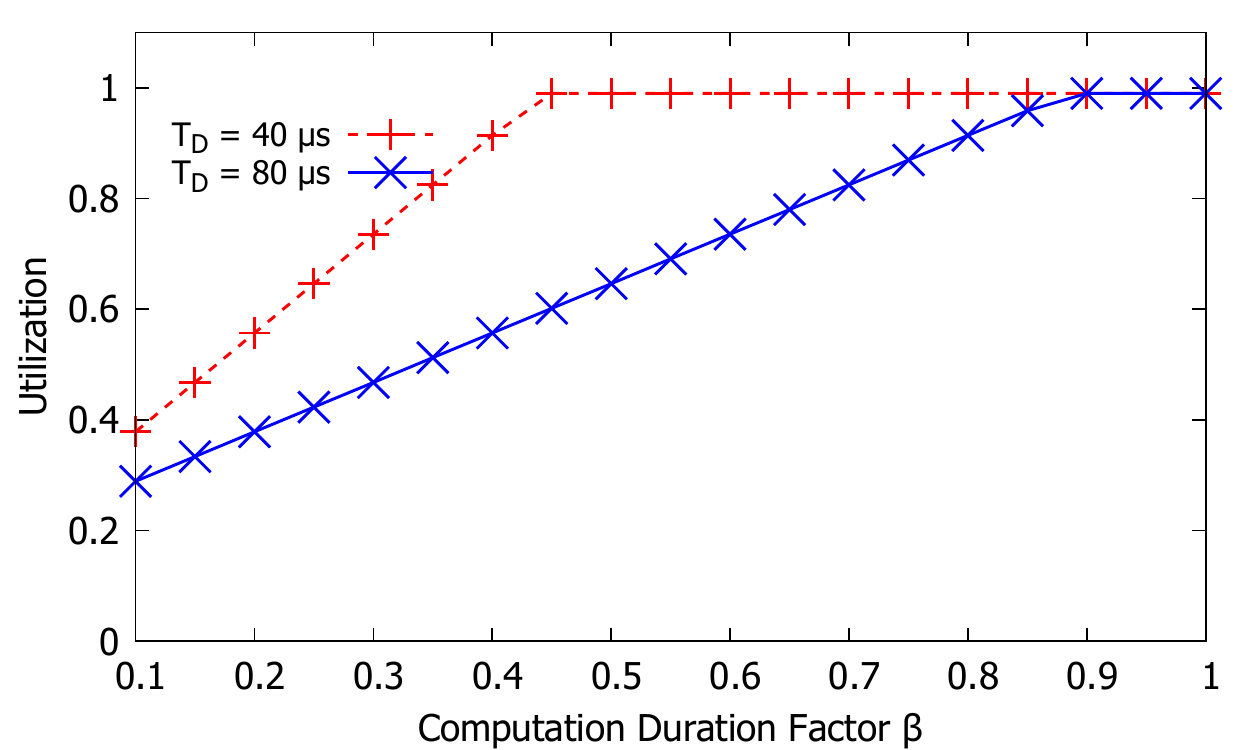} \vspace{-0.1cm}\\
		\vspace{+0cm}
		\scriptsize (a) $T_L=71.4~\mu\textit{s, } \tau_L=14.28~\mu s\text{, and } \theta=0~\mu\textit{s}$
		&\scriptsize (b) $T_L=71.4~\mu\textit{s, } \tau_L=14.28~\mu s\text{, and } \theta=0~\mu\textit{s}$\\
		\includegraphics[width=2.8in]{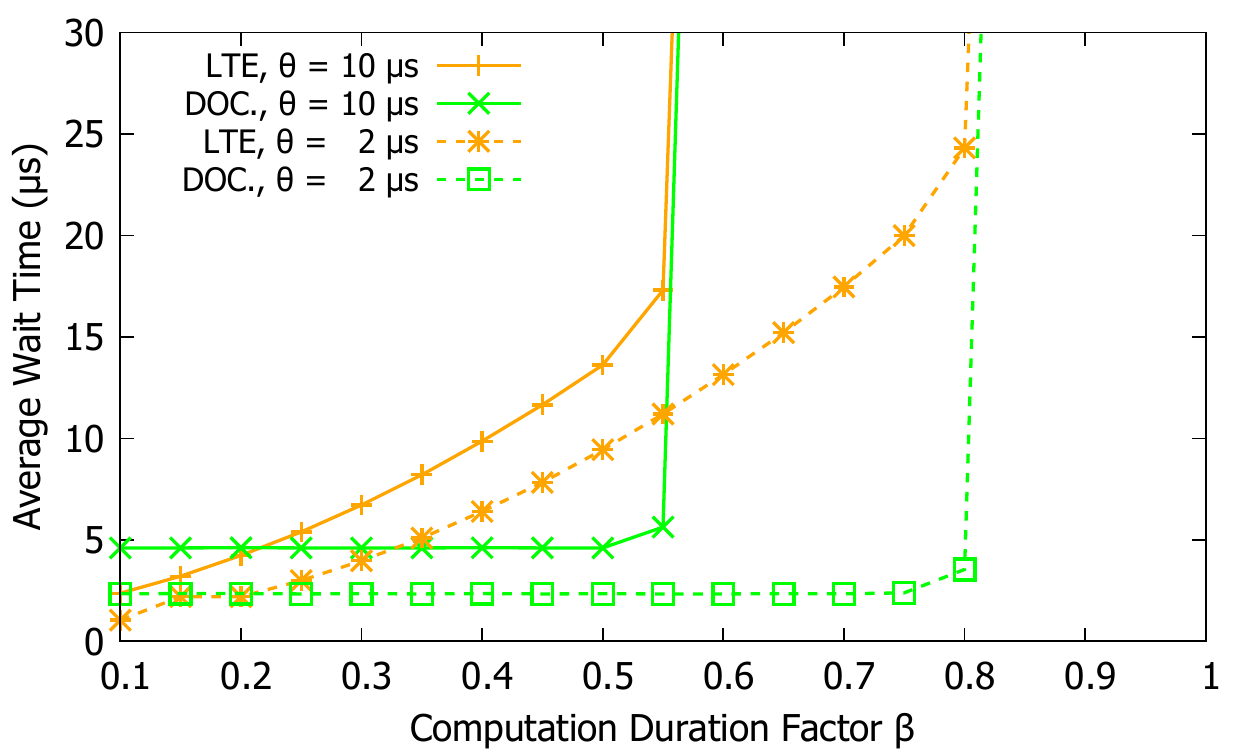} \vspace{-0.1cm}
		&\includegraphics[width=2.8in]{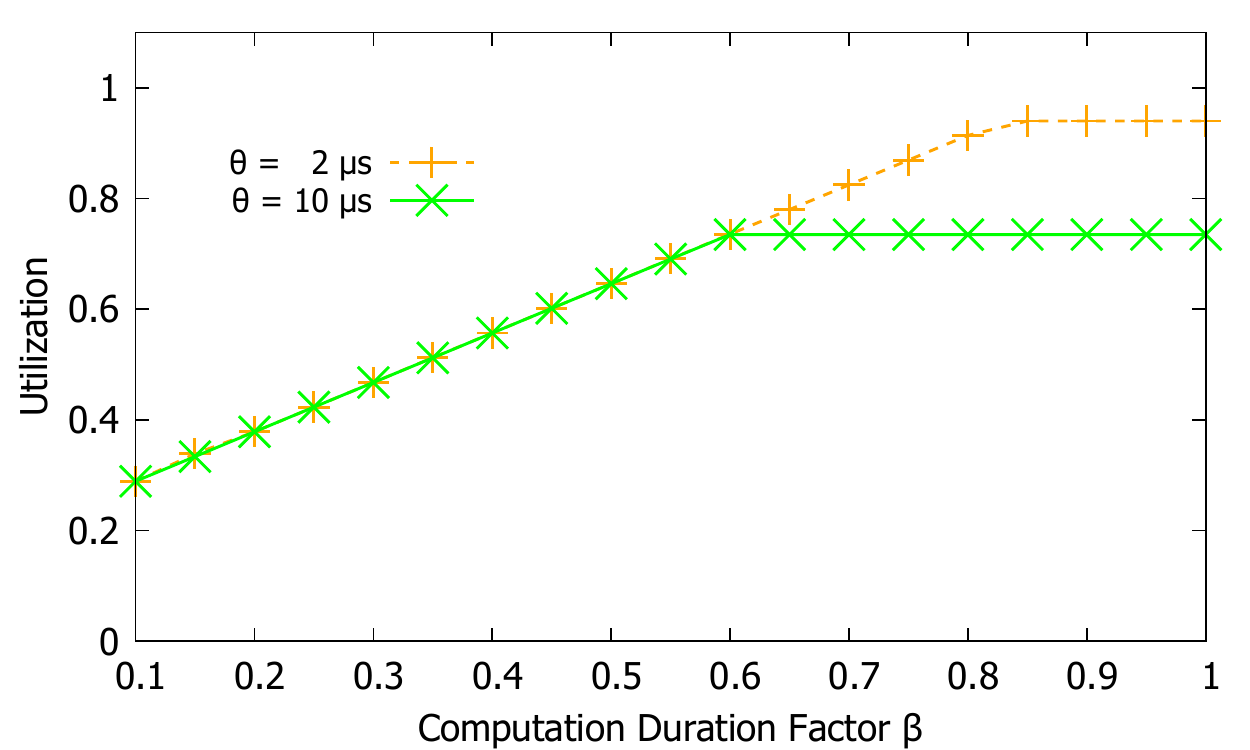} \vspace{-0.1cm}\\
		\scriptsize (c) $T_L=71.4~\mu\textit{s, } \tau_L=14.28~\mu s\text{, and } T_D=80~\mu\textit{s}$ 
		&\scriptsize (d) $T_L=71.4~\mu\textit{s, } \tau_L=14.28~\mu s\text{, and } T_D=80~\mu\textit{s}$ \\
	\end{tabular}
	\caption{Simulation results for 
		FFT interleaving procedure.} 
	\label{fig_tau_theta}
\end{figure*}

\subsection{FFT Module Sharing}
In the initial evaluation, we verify the proposed FFT interleaving procedure. 
we have implemented the FFT interleaving mechanism 
as a discreet event simulation framework in OMNET++.
A LTE and DOCSIS OFDM symbols are generated every 
$T_L$ and $T_D$ durations respectively. For LTE with normal cyclic prefix
the OFDM symbol duration $T_L$ is 71.4~$\mu$s, while the DOCSIS 
symbol duration can be either 40 or 80~$\mu$s. 
For each OFDM symbol arrival at the radio node, 
FFT computation is queued by the scheduler. Typically, the FFT size
for the LTE is 2K, whereas the FFT size for the DOCSIS can be 4K or 8K
based on the symbol duration (i.e., 40 or 80~$\mu$s).
Computation duration $\tau$ for the FFT computation 
can vary based on the number of carrier
components and number of radio nodes supported by the 
remote node as described
in Sec.~\ref{sec:tau}. 
Interleaving procedure should be aware of the computation duration for
each technology to ensure the stable operation of infrastructure sharing. 
To evaluate the system for wide range of FFT computation duration,
we define a computation duration factor
\begin{equation}
\centering
\beta=\frac{\tau_L}{\tau_D}.  
\end{equation}
We choose the FFT computation duration $\tau_L$ to be 20~\% of 
the $T_L$, which is $0.20 \times 71.432~\mu\textit{s} = 14.2832~\mu\textit{s}$ 
and the computation factor $\beta$ is varied 
from $0$ to $1$ with an effective sweep of 
the FFT computation duration $\tau_D$ (i.e., $\min(T_L,T_D)$) 
from $0$ to $T_L$, which is effectively $0 \leq \tau_D \leq T_L$. 
We primarily evaluate the FFT sharing mechanism with 
two performance metrics, average wait time and utilization.
Average wait time is the average time required to
schedule the FFT computation after the arrival of an OFDM symbol.
Whereas, the utilization parameter is defined as the ratio of 
total computation time to total elapsed time.

Figure.~\ref{fig_tau_theta} presents the performance evaluation 
of proposed FFT sharing mechanism. 
Throughout the evaluation, the OFDM symbol duration 
of LTE is kept constant $T_L=71.4~\mu$s.
Figure.~\ref{fig_tau_theta}(a) 
shows the average wait time of LTE and DOCSIS as a function of 
computation duration factor $\beta$ for different values 
of DOCSIS OFDM symbol duration, $T_D=40\mu$s and $T_D=80\mu$s.
We can observe from the Fig.~\ref{fig_tau_theta}(a), the 
average wait time of LTE increases linearly as the computation factor 
$\beta$ is increased which corresponds to increase
in the FFT computation duration of DOCSIS $\tau_D$. This shows that 
the computation duration of one technology has the direct impact on 
the average wait time of another as they compete with each other
for the computational resource. We also observe that the average
wait time of both LTE and DOCSIS 
for $T_D = 40~\mu$s linearly increases up to $\beta=0.45$ and 
overshoots to a very large indicating the instability of the system.
This is because, for the FFT computation sharing mechanism to be 
stable, the Eq.~\ref{eq:stability2} must be satisfied, i.e, 
$\tau_D \leq (T_L T_D - \tau_L T_D)/T_L$, which is, 
$\tau_D \leq (40 \times 71.4 - 14.28 \times 40)/71.4 = 32~\mu\textit{s}$. 
When $\beta=0.45$, the corresponding 
$\tau_D = 0.45 \times T_L = 0.45 \times 71.4~\mu\textit{s} = 32.13~\mu\textit{s}$,
surpassing the stability limit based on Eq.~\ref{eq:stability2}, which is $>32~\mu\textit{s}$. 
This behavior can also be observed in Fig.~\ref{fig_tau_theta}(b) 
which plots the utilization as a function of
computation factor $\beta$. While the utilization of FFT computation module 
increases linear with $\beta$, the system 
approaches nearly $100\%$ utilization when $\beta>0.45$ for $T_D=40~\mu\textit{s}$ 
reaching the stability limit of the system. 
Similarly, when the DOCSIS OFDM symbol duration $T_D$ is
changed from $40~\mu\textit{s}$ to $80~\mu\textit{s}$, the 
behavior of average wait time is similar to $T_D = 40~\mu\textit{s}$, but the
system becomes unstable when $\beta > 0.9$. When $\beta$ is $0.9$, the value 
of $\tau_D$ is $0.9 \times T_L = 0.9 \times 71.4~\mu\textit{s} = 64.26~\mu\textit{s}$,
which is slightly  greater than stability limit of 
$\tau_D \leq (80 \times 71.4 - 14.28 \times 80)/71.4$ or $\tau_D \leq 64~\mu\textit{s}$. 
Consequently, we can observe from Fig.~\ref{fig_tau_theta}(b), the utilization
of system approaches to $100\%$ for $\beta\geq0.9$ for $T_D=80~\mu\textit{s}$.
   
Fig.~\ref{fig_tau_theta} (a) and (b) corresponds to the evaluation 
when guard time $\theta=0~\mu\textit{s}$ for different values of
DOCSIS OFDM symbol duration $T_D=40$ and $80~\mu\textit{s}$. In contrast, 
Fig.~\ref{fig_tau_theta} (c) and (d) evaluates the system for different 
guard times $\theta=2$ and $10~\mu\textit{s}$ for $T_D=80~\mu\textit{s}$. 
Fig.~\ref{fig_tau_theta} (c) plots the average wait time as a function of
computation factor $\beta$, whereas, Fig.~\ref{fig_tau_theta} (c) plots the
system utilization as a function of $\beta$. Fig.~\ref{fig_tau_theta} (c) 
shows the same behavior as in Fig.~\ref{fig_tau_theta} (a), where the average 
wait time for LTE increases linearly as the $\beta$ is increased. 
The scheduling of $\tau_D$ experiences 
a constant delay of $2.3~\mu\textit{s}$ until the stability limit as 
the computation duration of LTE is retained constant in the simulation
$\tau_L=14.28~\mu\textit{s}$. 
In contrast to Fig.~\ref{fig_tau_theta} (a), 
system reaches stability limit for lower values of $\beta$ in (c) 
and this behavior pronounced as the guard time is increased. 
More specifically,
when $\theta$ is increased from 0 (see Fig.~\ref{fig_tau_theta} (a) for $T_D = 80 ~\mu\textit{s}$) to $2~\mu\textit{s}$, the system becomes 
unstable for $\beta=0.85$ as compared to $\beta=0.9$. 
Based on Eq.~\ref{eq:guardtime2}, for a system with guard time to be 
stable, $\tau_DT_L + \tau_LT_D \leq T_DT_L - \theta(T_L+T_D)$ or
$\tau_L \leq (T_DT_L - \theta(T_L+T_D)-\tau_LT_D) / T_L$.
When $\beta=0.85$ which corresponds to $\tau_D=60.69~\mu\textit{s}$, with $\theta=2~\mu\textit{s}$, the stability
condition evaluates to be $tau_D \leq 59.75$, therefore, the system is
unstable for $\beta=0.85$ indicated by the very 
large in the average wait time for $\beta>0.85$.
Similarly, for $\beta=0.6$ which corresponds to $\tau_D=42.84~\mu\textit{s}$, with $\theta=10~\mu\textit{s}$, the stability
condition evaluates to be $\tau_D \leq 42.79$, therefore, the system is
unstable for $\beta=0.6$.
Additionally, the effect of guard time has the direct impact on the
utilization of the system. We see from Fig.~\ref{fig_tau_theta} (d), the maximum
achievable utilization of the system is only up to 
94~\% when guard time $\theta=2~\mu\textit{s}$ for $\beta>0.8369$. 
As value of $theta$ is increased
to $10~\mu\textit{s}$, the maximum achievable utilization is
reduced to 73.5\% for $\beta>0.59$. 
The saturation in the utilization also
indicates the instability of the system, where the average wait time
grows to a very large value for $\beta>0.8369$ and $\beta>0.59$,
when guard time $\theta=2$ and $2~\mu\textit{s}$ respectively. 

\subsection{End-to-End Delay Evaluation}
\begin{figure*}[t]
	\begin{tabular}{cc}
		\includegraphics[width=2.9in]{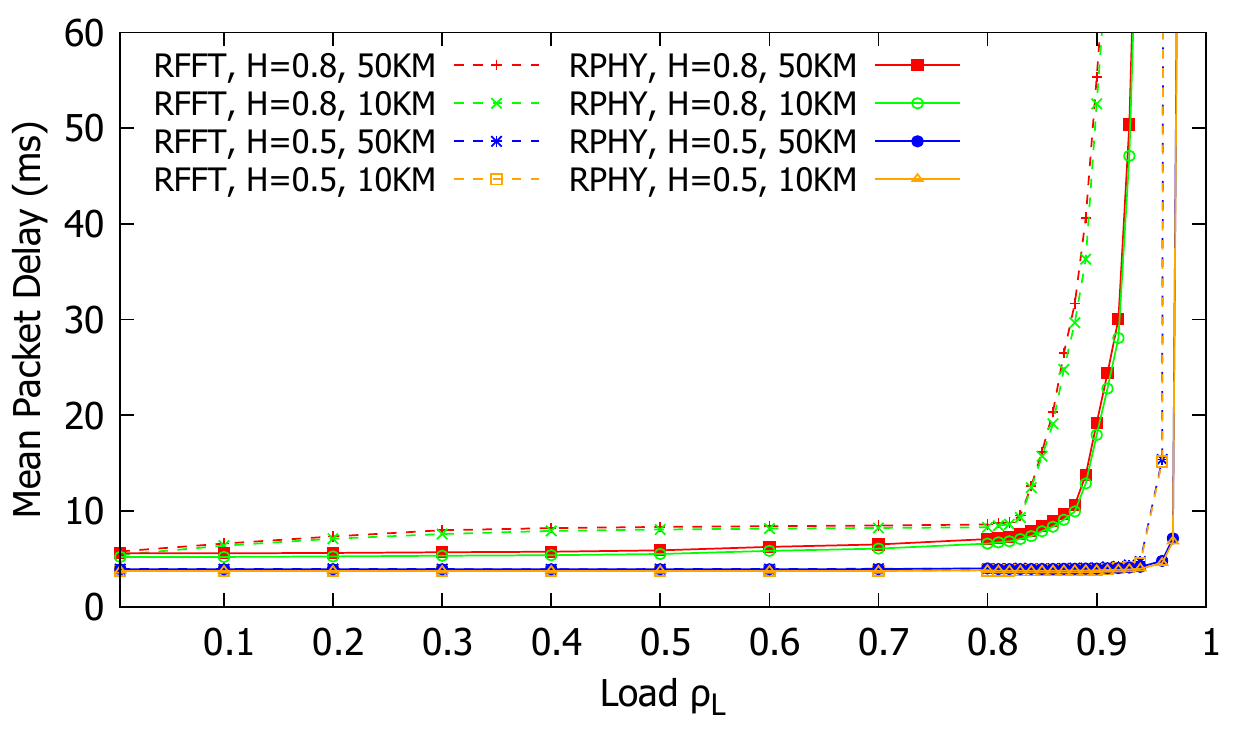} \vspace{-0.1cm}
		&\includegraphics[width=2.9in]{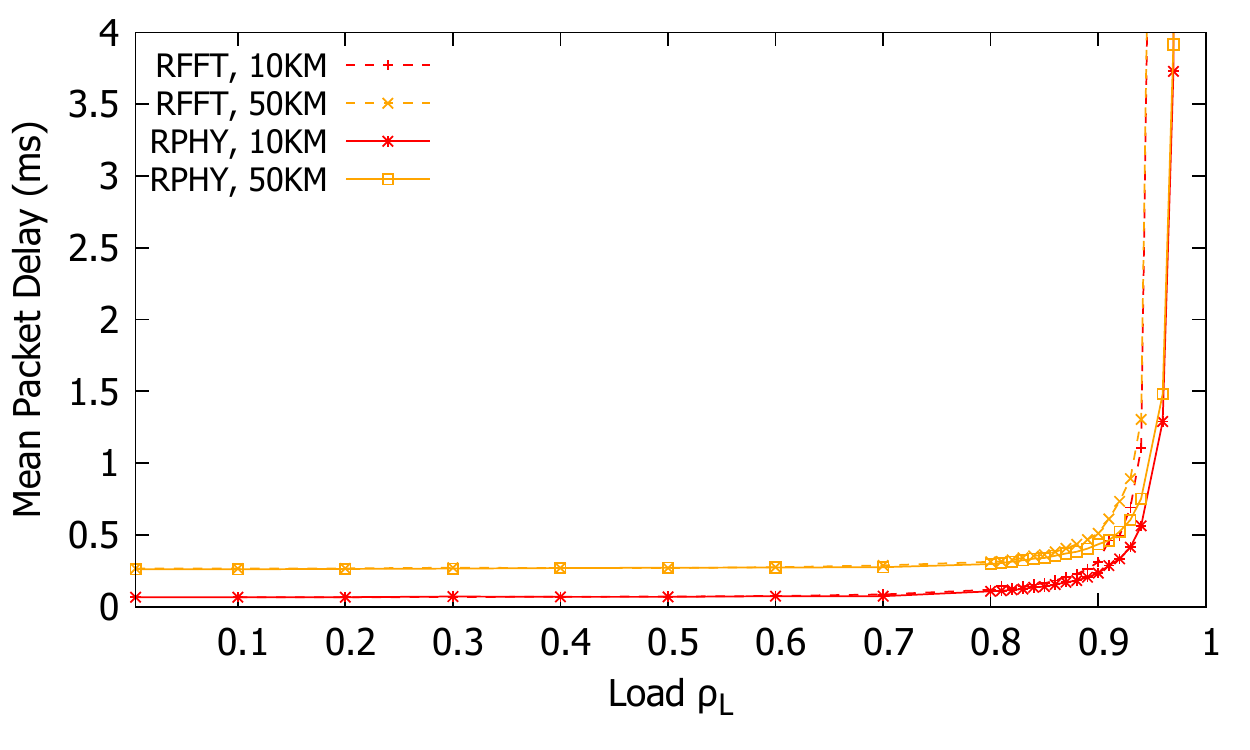} \vspace{-0.1cm}\\
		\vspace{+0cm}
		\scriptsize (a) DOCSIS Delay (end-to-end), $\rho_c=0.2$
		&\scriptsize (b) LTE Delay (fronthaul), $\rho_c=0.2$, $H=0.5$\\
		\includegraphics[width=2.9in]{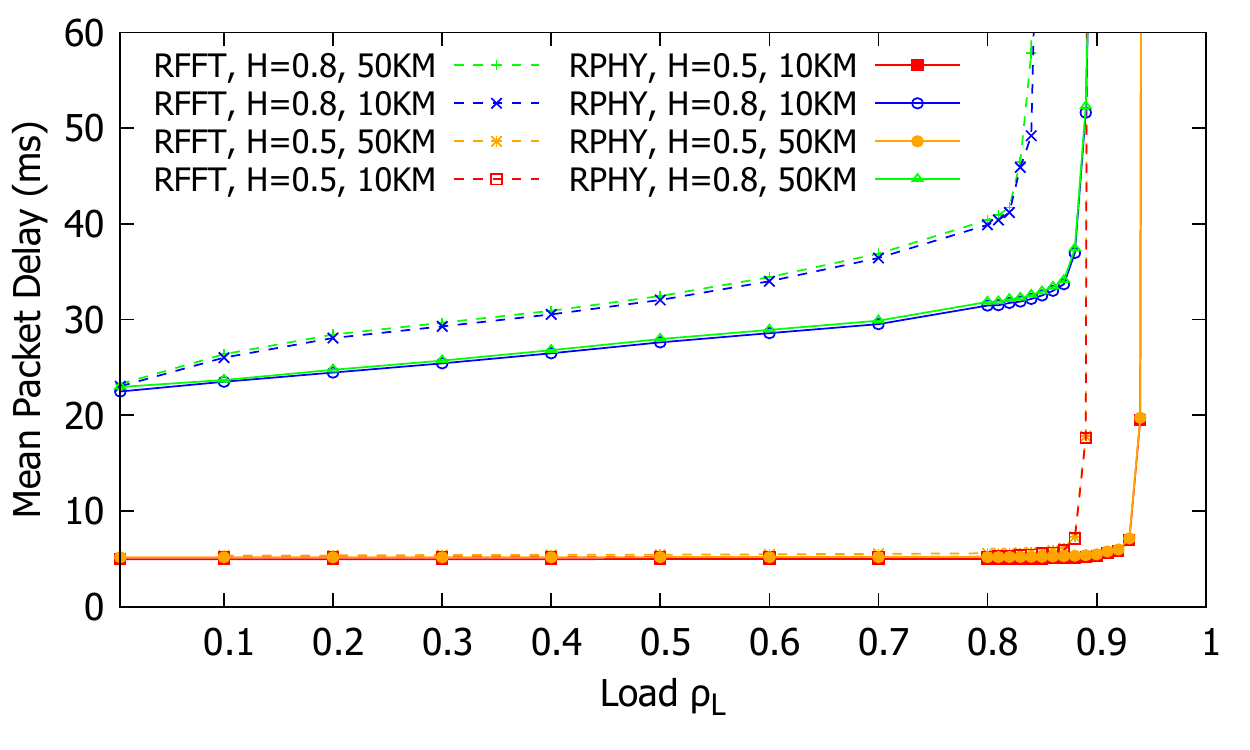} \vspace{-0.1cm}
		&\includegraphics[width=2.9in]{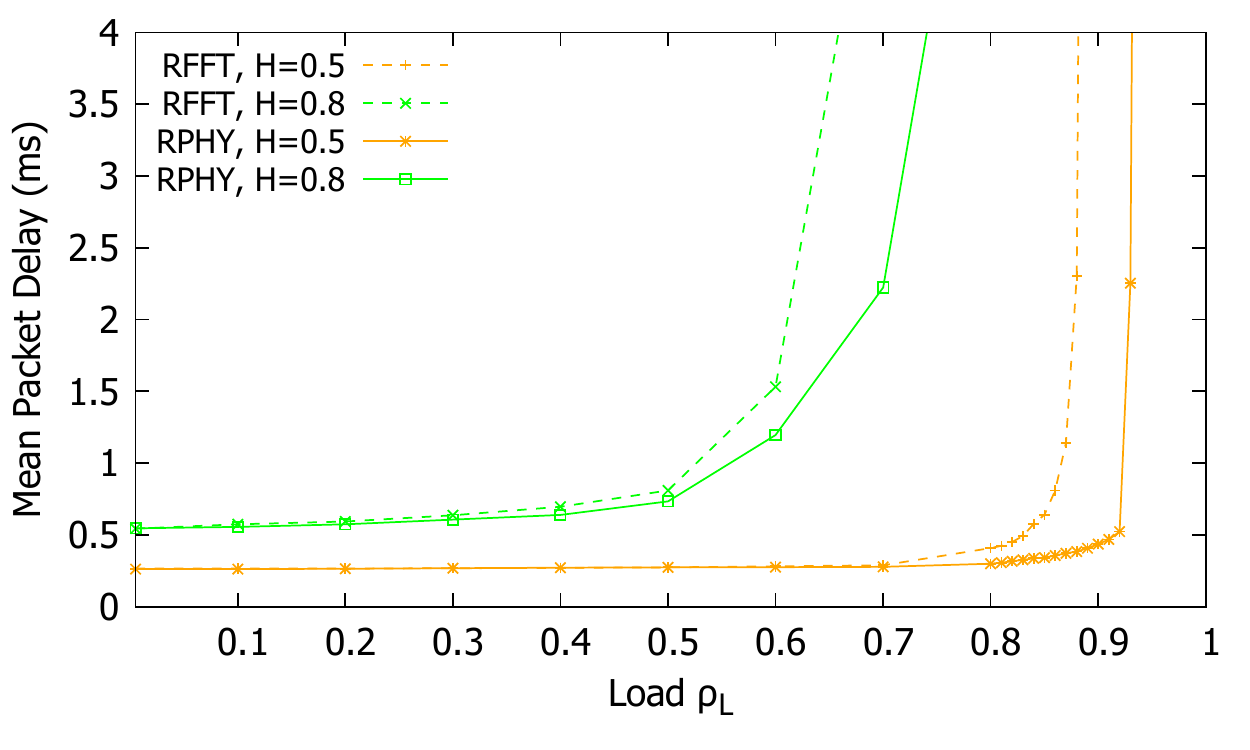} \vspace{-0.1cm}\\
		\scriptsize (c) DOCSIS Delay (end-to-end), $\rho_c=0.6$
		&\scriptsize (d) LTE Delay (fronthaul), $\rho_c=0.6$, $d=50$~km\\
	\end{tabular}
	\caption{Simulation results for RFFT and RPHY.}
	\label{fig_rfft_perf}
\end{figure*}
\begin{figure}[t!]
	\centering
	\includegraphics[width=3.5in]{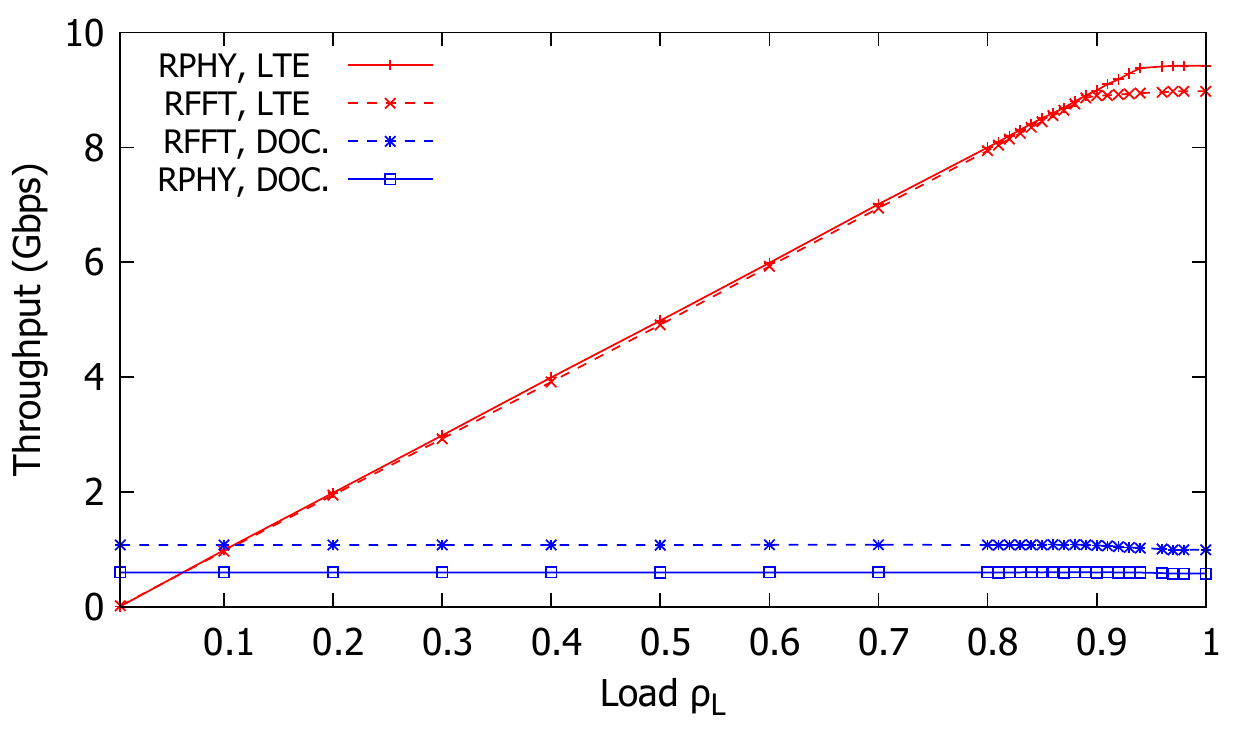} 
	\vspace{-0.3cm}
	\caption{Average DOCSIS (cable) and LTE throughput.}
	\label{fig_rfft_tput}
\end{figure}
Subsequently, in this evaluation, we study the delay characteristics of R-FFT
deployment in comparison to existing R-PHY deployments. In addition, 
we also present the impact of cable traffic on the LTE fronthaul link when
the same cable infrastructure is shared between LTE and DOCSIS.
We developed a simulation framework in a discreet event 
simulator OMNET++ to implement the DCCAP cable architecture of 
the HFC network. A remote node (RPHY or RFFT node) is connected 
to the headend through an optical connectivity separated by a 
distance $d$ and data rate of $R_o=10$~Gbps. 
200~Cable Modems (CMs) are connected to the remote node through
a analog broadcast cable connectivity. We assume the DOCSIS~3.1 protocol
to coordinate the cable transmissions in the cable broadcast medium 
with the bidirectional cable data rate $R_c=1$~Gbps.
We implemented a Double Phase Polling (DPP) protocol\cite{cho2009dou}
for the scheduling of 200~CMs over the cable segment. In case of 
RPHY, DOCSIS PHY frames are digitized and transported over Upstream 
External PHY Interface (UEPI) where the REQ packets were prioritized. 
Whereas for the RFFT, the upstream cable 
data is converted to frequency I/Q and transported as a generic UDP packet.
FFT size of $4K$ which corresponds to $T_D = 40~\mu s$ and QAM size of $12$~bits,
and the code rate of $9/10$ are used in the 
process of converting data to frequency domain I/Qs.
Each complex number representing an I/Q symbol
is digitized with $20$~bits.
We assume the deployment of an LTE RRH at the site of 
remote node in conjunction to the operation of RFFT or RPHY 
node sharing the optical connectivity 
of data rate $R_o$ to the headend where the BBU CRAN is implemented.
We note that our study is focused on deployment of LTE networks
over the HFC cable infrastructures. Therefore, we implement and
evaluate the performance of a cable network when an
LTE fronthaul traffic is simultaneously injected to the shared optical link.
We have modeled a typical FIFO queue at the remote node to froward the 
LTE packets to CRAN/BBU. 
We model the self similar traffic generation 
at each CM with varying burst levels controlled 
by the hurst parameter $H$ with the average packet size of $472$~KB. 
The hurst parameter $H=0.5$ corresponds to the Poisson traffic, and the
burstiness increases as the value $H$ is increased.

Figure~\ref{fig_rfft_perf} presents the performance evaluation of the
RFFT and RPHY network in presence of the LTE fronthaul traffic. 
In Fig.~\ref{fig_rfft_perf}(a) we show the mean packet delay of the DOCSIS
as a function of LTE fronthaul traffic intensity $\rho_L$ which corresponds 
to the LTE data rate $R_L = \rho_L \times R_o$ for different optical distance 
$d$ and traffic levels burstiness $H$, with fixed cable 
traffic intensity of $\rho_c = 0.2$ which corresponds to the 
traffic of $\rho_c \times R_c = 0.2 \times 1$~Gbps $=200$~Mbps. 
Delay is considered for an end-to-end connection of the cable 
connectivity i.e., from the CM to headend.
From Fig~\ref{fig_rfft_perf}(a) we observe for the 
shorter optical distance which is typically deployed for an LTE
fronthaul link has the minimal impact on the cable infrastructure for both
RFFT and RPHY deployments. More specifically, 
for $H=0.5$ in RFFT (and RPHY) for distance $10$~km and $50$~km 
we observe the delay difference of roughly $0.2$~ms 
(delay of $3.73$~ms for $10$~km RFFT $H=0.5$ and 
the delay of $3.93$~ms for $50$~km RFFT $H=0.5$)
when the distance is changed from $10$~km to $50$~km. 
This difference in the delay is from the propagation
delay in the optical fiber which is approximately 
$(d_2-d_1)/c = 40~\text{km}/2\times10^8~\text{m/s}=0.2$~ms.  

We also observe a minute difference in the
delay between RPHY and RFFT implementation at the remote node for 
the Poisson traffic, but a larger difference for higher burstiness
traffic. For example, at $\rho_L=0.5$ 
we observe the mean packet
delay values of $3.75$~ms and $3.73$~ms for 
RFFT and RPHY respectively.
The difference in the delay of nearly $0.02$~ms 
is due to the average transmission delay of 
the cable data which is inflated by the process of 
I/Q conversion and digitization. For the 
code rate of $9/10$ and the QAM size of $12$~bits 
the cable data would be inflated by 
factor of $(10/9) \times (1/12) \times 2 \times 10=1.85$. Thus,
the RFFT packets undergo an additional delay 
at the remote node for the 
complete transmission of data to headend. 
However, this effect is pronounced for $H=0.8$
because of the traffic with the higher 
burstiness. 
The behavior of the DOCSIS delay for RPHY and RFFT 
remains constant until $\rho_L=0.96$ for $H=0.5$ and $\rho_L=0.8$ 
for $H=0.8$, as
the optical link is not congested. 
However for $\rho_L>0.8$,
the effective throughput over the optical link 
reaches more than the capacity $R_o$ crossing 
the stability limit of the system. For e.g., 
the instantaneous cable traffic of RFFT on the 
optical link for $H=0.5$ and 
$\rho_c=0.2$ would be approximately 
$200~$Mbps$\times 1.85=370$~Mbps or $0.37$~Gbps.
Thus, when $\rho_L>0.963$ the effective
traffic would be more than $R_o$ 
resulting in the very large value of delay due 
to instability of the system. 

As seen in Fig.~\ref{fig_rfft_perf}(b) which plots the 
mean packet delay for the LTE fronthaul traffic for RFFT and RPHY 
for different optical distances of $10$ and $10$~km. LTE fronthaul delay
the directly impacted by deployed cable technology when the traffic from
LTE and cable are aggregated at the remote node. 
RFFT induces more delay on the LTE fronthaul traffic compared to the RPHY, 
which is mainly resulting from the increase in the effective data rate
over the optical link. Also, we can observe the 
effect of stability limit of the optical link which is 
in both Fig.~\ref{fig_rfft_perf}(a) and (b) [see for e.g., 
the overshoot of the delay 
when $\rho_L=0.96$ for RFFT, $H=0.5$ and $d=50$~km 
at Fig.~\ref{fig_rfft_perf}(a) and (b)]. 
Therefore, we can infer that RFFT and RPHY has the similar 
delay characteristics for cable traffic as well as 
the LTE fronthaul traffic.    

To further study effect of cable traffic load, in 
Fig.~\ref{fig_rfft_perf}(c) we evaluate the behavior of mean packet delay
for DOCSIS when cable load is $\rho_c = 0.6$. Fig.~\ref{fig_rfft_perf}(c) plots the
mean packet delay as the function of LTE fronthaul traffic $\rho_L$. 
Similar to Fig.~\ref{fig_rfft_perf}(a), we observe a narrow difference in the delay 
performance between RFFT and RPHY for the Poisson traffic ($H=0.5$). But, for the 
self similar traffic with $H=0.8$, we observe the difference of more 
than $10$~ms between RFFT and RPHY which is partly due to the 
temporary cable link saturation resulting from the 
instantaneous load reaching more than the cable link 
capacity $R_i$ $\rho_c$ and $20$~\% which is reserved 
for cable maintenance. This difference is even 
further pronounced by the 
combined effect from the data inflation at the remote node. 
As the LTE fronthaul traffic $\rho_L$ approach $0.8$ and $0.94$ 
for RFFT and RPHY respectively, the mean packet delay packet 
increases to a very large value resulted by the
optical link saturation. For $\rho_c=0.6$, the effective rate over the 
optical link is $\rho_L + (1.85 \times \rho_c)$, therefore, 
when $\rho_c = 0.6$, the optical link is saturated for 
$(1 -1.85 \times 0.6) = 0.89$ which results in the 
large value of delay at $\rho_L = 0.8$ and $0.94$ for 
RFFT and RPHY with $H=0.5$. The throughput behavior 
is verified in Fig.~\ref{fig_rfft_tput} which plots the 
cable and LTE throughput as a function of LTE traffic 
intensity $\rho_L$. We can compare Fig.~\ref{fig_rfft_perf}(c) and 
Fig.~\ref{fig_rfft_tput} to see that delay and the average
throughput reaching stability at $\rho_L = 0.8$ and $0.94$ for 
RFFT and RPHY with $H=0.5$. On the other hand, when the distance is increased from $10$~km 
to $50$~km, we see the pronounced effect in 
delay difference between the RPHY and RFFT 
in Fig.~\ref{fig_rfft_perf}(c). 
This effect is because of the 
combined impact of propagation delay and transmission delay 
due to data inflation of RFFT on the DOCSIS scheduler.

To understand the effect of 
traffic burstiness along with the cable load
on the LTE fronthaul traffic,
we show the mean packet delay as a function of LTE fronthaul 
traffic $\rho_L$ in Fig.~\ref{fig_rfft_perf}(d). 
We can relate the results in 
Fig.~\ref{fig_rfft_perf}(c) to Fig.~\ref{fig_rfft_perf}(d) 
as such the traffic with higher burstiness $H=0.8$ 
increases delay of both cable and LTE fronthaul traffic. 
Additionally, the large fronthaul delay values at  $\rho_L = 0.8$ and $0.94$
indicates the stability limits for the Poisson traffic ($H=0.5$) 
for RPHY and RFFT. However,
the fronthaul delay is more impacted the higher burstiness ($H=0.5$) 
as compared to the Poission traffic ($H=0.8$) in the RFFT deployment.
For example, even when the fronthaul load is $\rho_L=0.5$ 
(i.e, below the stability limit), 
the fronthaul delay is almost increased
by $4$ times from $0.27~$ to $2.04~$ms for $H=0.8$ in the RFFT deployments. 
Therefore, if the functional split of LTE requires a dedicated 
QoS requirements, then a resource allocation 
mechanism of optical link may be employed for the dedicated assignments. 

\section{\MakeUppercase{Conclusions}}
We have presented a mechanism for Time-Frequency resource elements (QAM) caching along with the 
FFT sharing for LTE and DOCSIS coexistence. We have comprehensively evaluated the FFT 
interleaving procedure through the simulation evaluation.
Additionally we have also provided the 
performance comparison between the RFFT and RPHY remote nodes while 
coexisting with LTE networks.
Our study  numerically evaluate and show that 
RFFT performance is very close to RPHY. As 
RFFT achieves more flexibility and more 
scalability RFFT can be preferred for deployment. Additionally we also 
discuss the impact of functional split cable traffic on 
the LTE fronthaul traffic.

	\chapter{\MakeUppercase{Future Work Outline: The SDN-LayBack: An SDN-Based Layered Backhaul
	Architecture for Dense Wireless Networks}}

\section{\MakeUppercase{Introduction}} 

\subsection{Need for a New Backhaul Architecture for Wireless Networks}

\begin{figure}[t]
    \centering
    \includegraphics[width=1\textwidth]{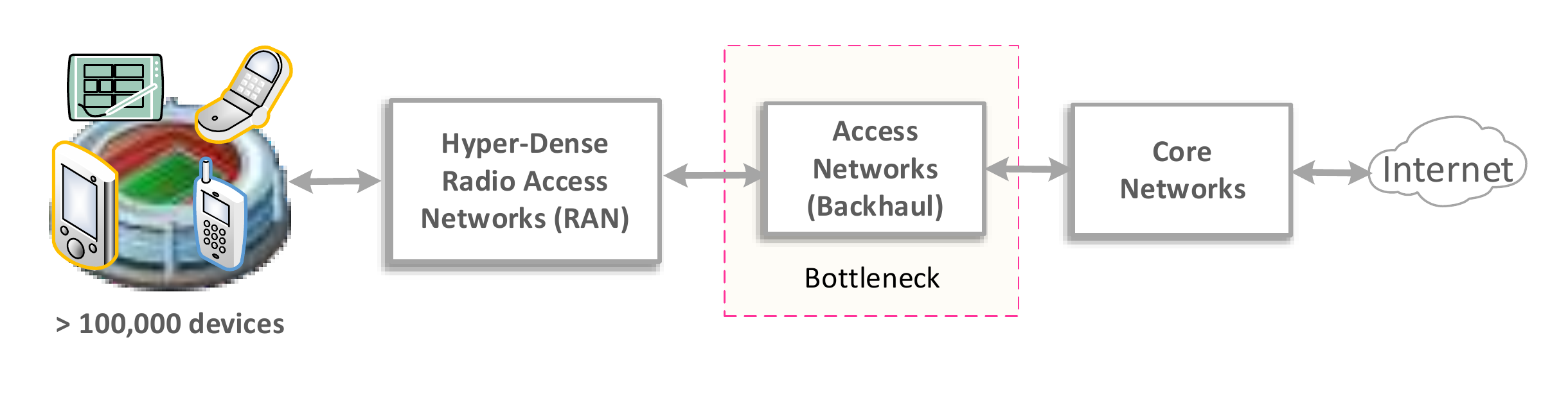}
    \caption{Bottleneck in the wireless Internet access chain.}
    \label{Bottleneck}
\end{figure}

A plethora of wireless devices running a wide range
of applications connect to radio access networks (RANs), as illustrated in Figure~\ref{Bottleneck}.
RANs provide end device to radio node (base station) connectivity.
The radio nodes are connected to the core networks by technology-specific backhaul access
networks, such as LTE or WiMax backhaul access networks.
Present day hand-held wireless end devices have very high
processing and memory capabilities, supporting various
applications, including resource demanding ones such as live 4K video streaming.
Advanced RANs, such as LTE-Advanced,
support up to several hundred Mb/s in downstream
by exploiting a range of physical wireless layer techniques, such as
multi-carrier aggregation, opportunistic utilization of unlicensed
bands, and millimeter wave technologies. 
A dense infrastructure of small cells with appropriate interference
management is a critical technique for advanced RANs~\cite{nak2013tre}.
At the same time, the core networks already employ high-capacity optical links
providing abundant transmission capacity.

However, the backhaul access networks have emerged as a critical
bottleneck in wireless Internet
access~\cite{and2014ove,ghi2015rev,rat2010fem,wan2015bac}.
Today's backhaul access networks typically require
highly-priced bulky network equipment with proprietary locked-in control
software~\cite{ber2015sof}.
They also need to be highly reliable and have high
availability.
It is therefore very difficult to
add new network functionalities, reconfigure operational states,
or upgrade hardware as technology advances.
These aspects have combined to stifle progress in backhaul access
networks, causing them to become a critical bottleneck that can stall
the progress of wireless networking and undermine the advances in the
complementary RANs and core networks. It can therefore be concluded that there is a need for innovative approaches, such as a new architecture,
to enable technological progress in terms of backhaul access networks.

\subsection{SDN-LayBack - an Advanced Wireless Network Backhaul Architecture}
This paper proposes \emph{SDN-LayBack, a fundamentally novel backhaul architecture
for wireless backhaul access networks based on software defined networking
(SDN)}~\cite{Bradai,moh2015con}.
In contrast to other recently proposed backhaul architectures, such as
CROWD \cite{seb2015dyn}, 
iJOIN \cite{Dongyao},
U-WN \cite{Shengli}, and Xhaul~\cite{oli2015xha}, 
our SDN-LayBack architecture
consistently decouples the wireless RAN technology
(such as LTE or WiFi) from the backhaul access network.
In addition, our SDN-LayBack architecture flexibly accommodates
highly heterogeneous RANs,
ranging from sparse cell deployments in rural areas to
extremely high density cells in crowded stadiums.
Further, to the best of our knowledge, SDN-LayBack is the first
architecture to simultaneously  provide a new clean-slate platform for
cellular backhaul and support the coexistence with current cellular access backhaul technologies, such as LTE (4G/3G),
WCDMA (3G), and GSM (2G).
Previous clean-slate architecture proposals,
such as the one proposed by Ameigeiras et al.~\cite{Ameigeiras}, do \textit{not}
support existing cellular backhaul technologies, and
hence have limited functionalities. SDN-LayBack is layered on top of the existing backhaul networks and flexibly supports a wide range of RANs through a network of SDN switches and a hierarchical SDN controller pool.

This paper also introduces \emph{an innovative four-step intra-LayBack handover protocol} within a given
gateway in the SDN-LayBack architecture. Evaluations indicate 60~\%
reductions of the signalling load in comparison with conventional LTE handover.
Finally, the article discusses future research directions within the
SDN-LayBack architecture framework, focusing on
interference mitigation, device-to-device (D2D) communication, and video streaming~\cite{seeling2012video, seema2011towards, chikkerur2011objective,
van2009implications,van2008traffic, seeling2004network, kangasharju2002distributing, fitzek2001mpeg}.

\section{\MakeUppercase{SDN-LayBack: a Novel SDN-Based Layered Backhaul Architecture}}  \label{arch:sec}

\begin{figure}[t!]
    \centering
    \includegraphics[width=1\textwidth]{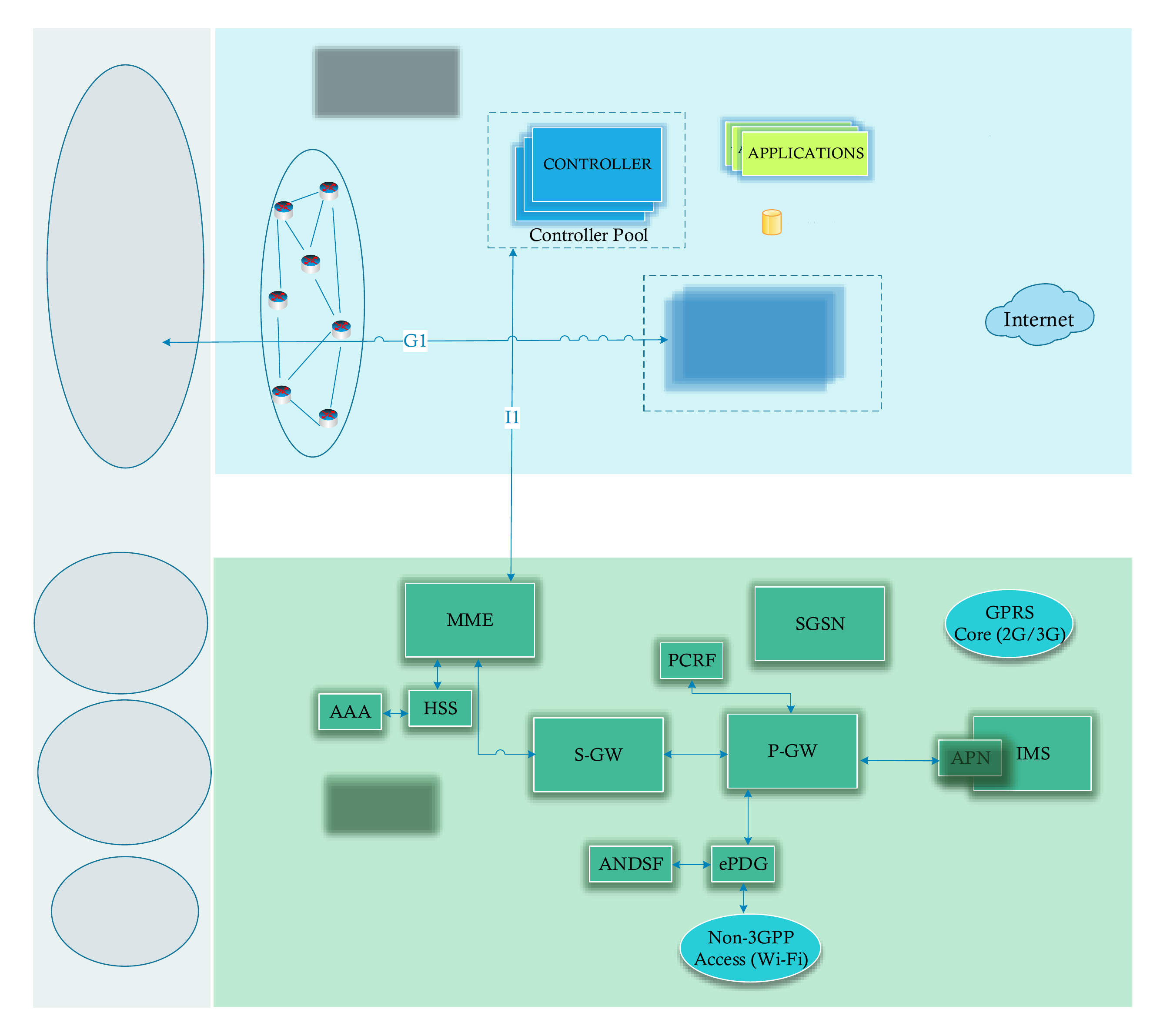}
    \caption{SDN-LayBack architecture.}
    \label{LA:fig}
\end{figure}

\subsection{Overview} 
As already mentioned, the proposed SDN-LayBack is a novel SDN-based layered backhaul access network
architecture which can be layered on top of an existing backhaul access network (e.g., SAE-3GPP-LTE) in an evolutionary manner. Fig.~\ref{LA:fig} illustrates SDN-LayBack's deployment on top of the classic LTE architecture.

Conventional backhaul access networks include technology-specific
functional blocks for different RAN technologies.  For instance, in the LTE architecture, macro cells are
served by an LTE Serving-Gateway (S-GW), while pico/femto cells can be
served by an Home-eNodeB-Gateway (HeNB-GW) (or S-GW), as illustrated in
the bottom left of Fig.~\ref{LA:fig}.
RAN and backhaul are thus coupled in conventional wireless networks, e.g., a
WiMax radio node cannot be used within an LTE architecture. Only with
an additional dedicated network entity, an evolved packet data
gateway (ePDG), can a WiFi radio node be connected to LTE backhaul
with IPSec tunneling, as illustrated in the bottom middle of
Fig.~\ref{LA:fig}.

In contrast, our proposed SDN-LayBack architecture decouples the RAN from
the backhaul, as illustrated in Fig.~\ref{LA:fig}.
The figure also indicates SDN-LayBack's four architectural layers:
\emph{RAN Layer}, \emph{SDN Backhaul Layer}, \emph{Core Networks Layer},
 and \emph{Evolutionary Internetworking Layer}.

In the SDN-LayBack architecture, heterogeneous RAN technologies, such as LTE
eNBs, pico/femto LTE, WiFi, and WiMax, can be flexibly
deployed and accommodated in the \emph{RAN Layer}. The \emph{RAN Layer} interacts
with the \emph{SDN-LayBack Backhaul Layer} through the network of flexible SDN (OpenFlow, OF)
switches and onwards through the programmable gateway pool to the Internet with the \emph{Core
Networks Layer}.  The data plane of the SDN-LayBack backhaul network consists of
the network of SDN switches and the gateway pool. This data plane is
centrally controlled by the SDN controller pool and the applications (with corresponding databases). SDN-LayBack interfaces with the classic backhaul access network architecture through \emph{the Evolutionary Internetworking Layer}.

\subsection{Outline of Layers in SDN-LayBack Architecture} 
This section outlines the key components and functionalities of the
proposed LayBack architecture and contrasts the LayBack architecture with previous designs. 

\subsubsection{Radio Access Network (RAN) Layer} 

\paragraph{Wireless End Devices}  

Mobile wireless end devices are heterogeneous and have a wide range of
requirements. Providing reasonable quality-oriented services to
every device on the network is a key challenge to the
wireless network design.
The proposed SDN-LayBack architecture takes a unique approach to provide
requirement-specific network connectivity to every device that is
connected to the SDN-LayBack network.
Future devices will likely be
highly application-specific, such as a speed sensor on a race track
or a health monitoring biosensor. 
The SDN-LayBack architecture is designed to be adaptive
to the different environments, such as an air port terminal, public park, or
university/school. This enables SDN-LayBack to
support a wide range of device requirements, such as
real-time D2D video streaming at a large sporting event or music concert.

\paragraph{Radio Nodes}  
Radio nodes, such as the evolved NodeB (eNB) in LTE or an
access point (AP) in WiFi, provide RAN services to
the end devices. Aside from LTE and WiFi, there exists a wide range of
wireless access technologies (and protocols), including
WiMax, Zig-Bee, Bluetooth,
and near field communication (NFC).
These wireless protocols have unique advantages and serve unique purposes.
A fluidly flexible backhaul that homogeneously supports
multiple different types of radio nodes does not yet exist, but is
highly desirable.
Therefore, our SDN-LayBack architecture is
fundamentally designed to work with multiple RAN and communication technologies
by isolating the RAN layer from the backhaul network.

In existing backhaul access networks, a given
RAN technology requires a corresponding specific backhaul. For instance, the
LTE radio access network can only operate with LTE
backhaul network entities. This restriction of
specific RAN technologies to specific backhaul
technologies limits the usage of other RAN technologies on
LTE backhaul networks and vice-versa.
In addition, femto cells are expected to operate
on multiple RAN technologies~\cite{Bennis}. 
To address this restrictive structure of present RAN-backhaul
inter-networking, we propose our SDN-LayBack architecture to flexibly
support multiple types of radio nodes, including software defined
radios integrated with SDN enabled radio nodes.  
In an interesting approach, the reconfigurable antennas 
\cite{saeed2012new, costantine2013new, saeed2015flexible, 
	saeed2016reflection, saeed2016radiation, costantine2014mimo, saeed2016inkjet} can be also 
be further extended SDN to achieve more flexibility in the RAN design.

Moreover, the system modeling for the RAN can be borrowed from the 
control literature. Techniques to design the systems for multiple specifications have been extensively discussed in 
\cite{puttannaiah_acc_2016,puttannaiah_acc_2015,puttannaiah_cdc_2015,puttannaiah_echols_gnc_2015,puttannaiah_2013_msthesis}. 
Optimization problems can be formulated that address multiple
objectives subject to constraints, by exploiting convex optimization ideas.
Problems on a set of nonlinear hybrid systems, involving mixed-integer
problems, can be solved using convexification
\cite{puttannaiah_nandola_ecc_2013}. 
High fidelity numerical simulators and system identification techniques can
be used to efficiently solve optimization problems \cite{puttannaiah_cartagena_acc_2016} using SDN.

\subsubsection{SDN Backhaul Layer}  
The SDN backhaul layer encompasses the SDN switches (with hypervisor), a
controller pool, and a gateway pool, as well as the applications and corresponding databases.
Next, the main aspects of the SDN switches, controller pool, applications, and gateways are briefly outlined in this section.

\paragraph{SDN (OpenFlow) Switches} 
SDN capable switches are today typically based on the OpenFlow (OF)
protocol 
and are therefore sometimes referred to as
OpenFlow switches or OF switches.
We present SDN-LayBack in the context of OpenFlow switches, but emphasize that
the SDN-LayBack architecture applies to any type of SDN-capable switch.
OF-switches are capable of a wide range of functions, such as forwarding a
packet to any port, duplicating a packet on multiple ports, modifying the
content inside a packet, or dropping the packet.
The switches can be connected in a mesh to allow disjoint flow paths
for load balancing.

\paragraph{Controller Pool} 
The SDN-LayBack controller pool configures the entire OF-switch fabric
(network of OF switches) by adding flow table entries on every switch.
The controller pool is also able to dynamically configure and program
the gateway pool functions, such as queuing policies, caching,
buffering, charging, as well as QoS \cite{guck2016function, fitzek2002providing} and IP assignment for the end devices.
The SDN-LayBack controller pool coordinates with existing (legacy)
cellular network entities, such as the mobility management entity (MME) and
other 3GPP entities, 
to coexist with present architectures.
Connections from the SDN-LayBack controller to the 3GPP entities can be
provided by extending the existing tunneling interfaces at the 3GPP entities.
We define extensions of the tunneling interfaces as a part of
the \emph{Evolutionary Internetworking Layer} in Fig.~\ref{LA:fig}.
Thus, the proposed SDN-LayBack architecture can enable communication
between, the new
flexible RAN architecture in the top left of Fig.~\ref{LA:fig} as well as
legacy technology-specific RAN and backhaul networks, see bottom
of Fig.~\ref{LA:fig}. 

\paragraph{Applications} 
Applications are programs executed by the controller for
delegating instructions to individual SDN switches. The controller
applications realize all the network functions required for Internet
connectivity, such as authentication and
policy enforcement, through the switch flows.
Radio nodes require RAN-specific interfaces for their operation
(e.g., the X2 interface in LTE), 
We realize these interfaces through controller applications on the network of
SDN-switches. Therefore, for LTE in SDN-LayBack, eNBs are enabled with
inter-radio node X2 interfaces along with other required interfaces.
Cellular networks have many network functions, such as the
Automatic Network Discovery and Selection Function
(ANDSF) 
of 3GPP. As an example, to replace ANDSF
in LayBack, a dedicated controller application will be responsible for
delegating network access policies to the devices.
Databases assist the SDN controller applications in their operations. 

\paragraph{Gateways} 
Gateway functions are programmed by the SDN controller
to perform multiple network functionalities in order
to simultaneously establish connectivity to heterogeneous RANs, such as
LTE, Wi-Fi, and Wi-Max.
For example, the gateway in SDN-LayBack functions as
both S-GW and P-GW for an LTE radio node.

\subsection{SDN-LayBack Hierarchical Micro-Architectures}
\label{ma:sec}
\begin{figure}[!t]
	\centering
	\includegraphics[width=1\textwidth]{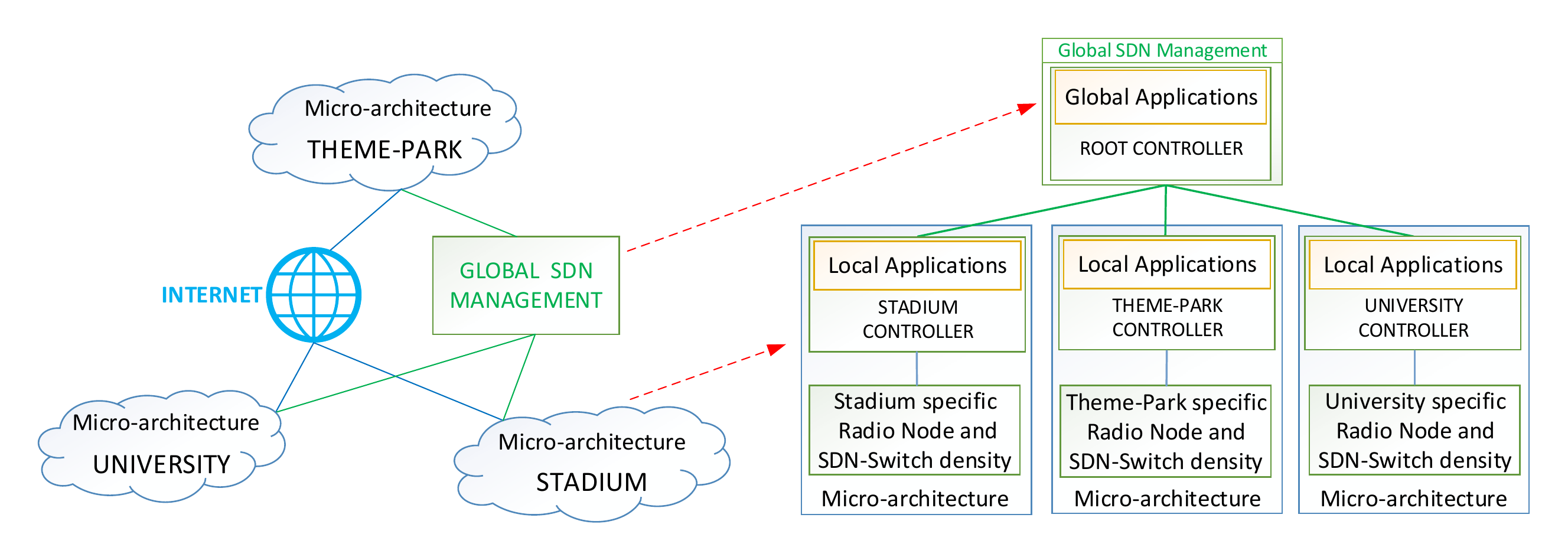}
	\caption{Environment-specific micro-architectures and controller
		hierarchy.}
	\label{fig:microarchitectures}
\end{figure}

The SDN-LayBack architecture can provide a wide range of
heterogeneous services to localized regions through
hierarchical SDN-LayBack micro-architectures.
A micro-architecture hierarchy consists of a global (root) controller
and multiple environment-specific local controllers located in the
respective environments, as illustrated in Fig.~\ref{fig:microarchitectures}.
The local controllers can tailor applications for a specific
environment, e.g., a park with a sparse user population that may be
located adjacent to a stadium with a very high user density.
Accordingly, micro-architectures allow to
classify applications specific to an environment
as local applications and more common applications as global applications.
For example, WiFi offloading and adaptive video streaming
can be considered as local
applications, whereas applications that serve a larger purpose, such as
interference management, can be considered as global applications.

\section{\MakeUppercase{Intra-LayBack Handover: A Radically Simplified Handover Protocol}}
\begin{figure}[t]
	\centering
	\includegraphics[width=1\textwidth]{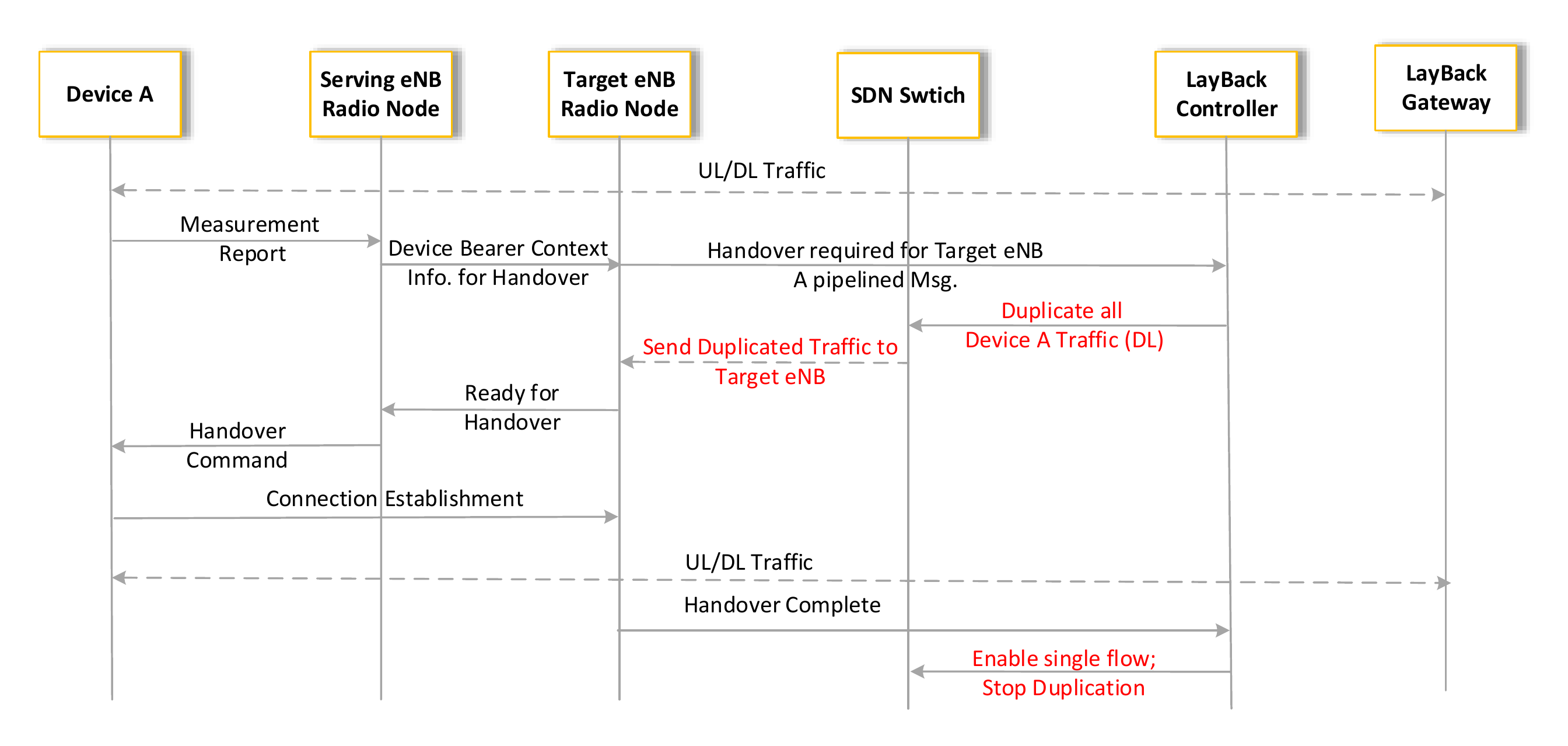}
\caption{Proposed simplified intra-LayBack handover protocol.}
	\label{fig:handover}
\end{figure}

\subsection{Motivation: Signalling Load Due to Frequent Handovers}
The LTE protocol has generally higher signalling overhead
compared to legacy 3G protocols.
A recent forecast from Oracle Corp.~\cite{OracleSignalling}
expects a large increase of the global LTE signalling traffic, which is
predicted to surpass the total global IP traffic growth by 2019.
The increase of signalling traffic can result in
service interruptions to customers and financial losses to cellular operators.
Therefore, new architectures
with simplified signalling mechanisms are necessary to tackle the
growth of signalling in the cellular networks.

Signalling in the cellular backhaul is required for a wide range of
actions, including initial attach and handovers.
Initial attach procedures are completed only when a device requests
connectivity for the very first time. The initial attach procedure can result in
large signalling overhead; however, it is executed only once.
Handovers are initiated when a device measures and reports
poor signal quality to the network.
Depending on the configuration, the neighboring cell signal
quality along with the quality of serving cell signal may be
considered for handover decisions.
Handovers require large
amounts of signalling in the backhaul and occur very frequently.
Especially in ultra dense networks (UDNs), where the cell
range is reduced to relatively small areas on the order of several
meters (e.g., 20~m in femto cells), handovers are very frequent.
Even slow movements of devices
can cause device moves between multiple cells, resulting in handovers.
Hence, especially in UDNs, handovers are the source of serious bottlenecks in the backhaul entities.

\subsection{Proposed 4-Step Intra-LayBack Handover} 
We propose \emph{Intra Lay-Back Handover}, as the first mechanism
to exploit the traffic duplication
property of SDN to radically simplify the handover procedure in
cellular backhaul.
In comparison to X2 based LTE handover \textit{within} the same S-GW,
our proposed method for handover within the same LayBack gateway
(Intra-LayBack Handover) achieves:
1) 100\% reduction of the signalling load at the gateway,
i.e., completely eliminates the handover signalling load on the LTE
gateways,
2) 69\% reduction of signalling load compared to the LTE MME, and
3) 60\% reduction in the overall signalling cost.

We employ four steps in the handover protocol,
as illustrated in Figure~\ref{fig:handover}.

\emph{Step 1 Handover Preparation}

When a measurement report satisfying
the handover conditions arrives at the serving eNB (SeNB), the device
bearer context information is forwarded to the target eNB (TeNB) and a
message requesting handover is sent from the SeNB to the SDN-LayBack controller.

\emph{Step 2 Enable Traffic Duplication}

LayBack controllers enable the traffic
duplication at the SDN switch through an OpenFlow
message for all the downlink traffic related to the device waiting to
be handed over. The SDN switch continuously changes the headers (tunnel
IDs) such that traffic related to the device reaches at TeNB. The bearer
information pertaining to the device which has already been
received by the TeNB from the SeNB is used to receive the duplicated traffic
at the TeNB. Once the duplicated traffic related the device is
received at the TeNB, a message is sent to the SeNB indicating the readiness
for the handover. The SeNB then sends a handover command to the device
(i.e., RRC reconfiguration in LTE).

\emph{Step 3 Perform Wireless Handover}

Upon receiving the handover command, the device
breaks the current wireless link (a hard handover)
and establishes a new wireless connection to the TeNB (through reserved
RACH preambles in LTE) \cite{tyagi2014impact, tya2015con, gur2017hyb, vilgelm2017latmapa}. 
By then, duplicated traffic arrived prior to wireless
connection reestablishment is waiting to be forwarded to the device in the
new connection. Downlink traffic is then forwarded to
the device as soon as the wireless connection is established.

\emph{Step 4 Stop Traffic Duplication}

When the device traffic
flow path through the new wireless connection
is successful in both the uplink and downlink, then the
TeNB sends a handover complete message to the LayBack controller.
The controller sends
an OpenFlow message to the SDN switch to stop the duplication by
disabling the device traffic flow to the SeNB.

Other handover types, such as inter-LayBack gateway
handover between different LayBack gateways, can be accomplished
through extensions of the presented intra-LayBack handover.

\begin{figure}[t]
	\centering
	\includegraphics[width=1\textwidth]{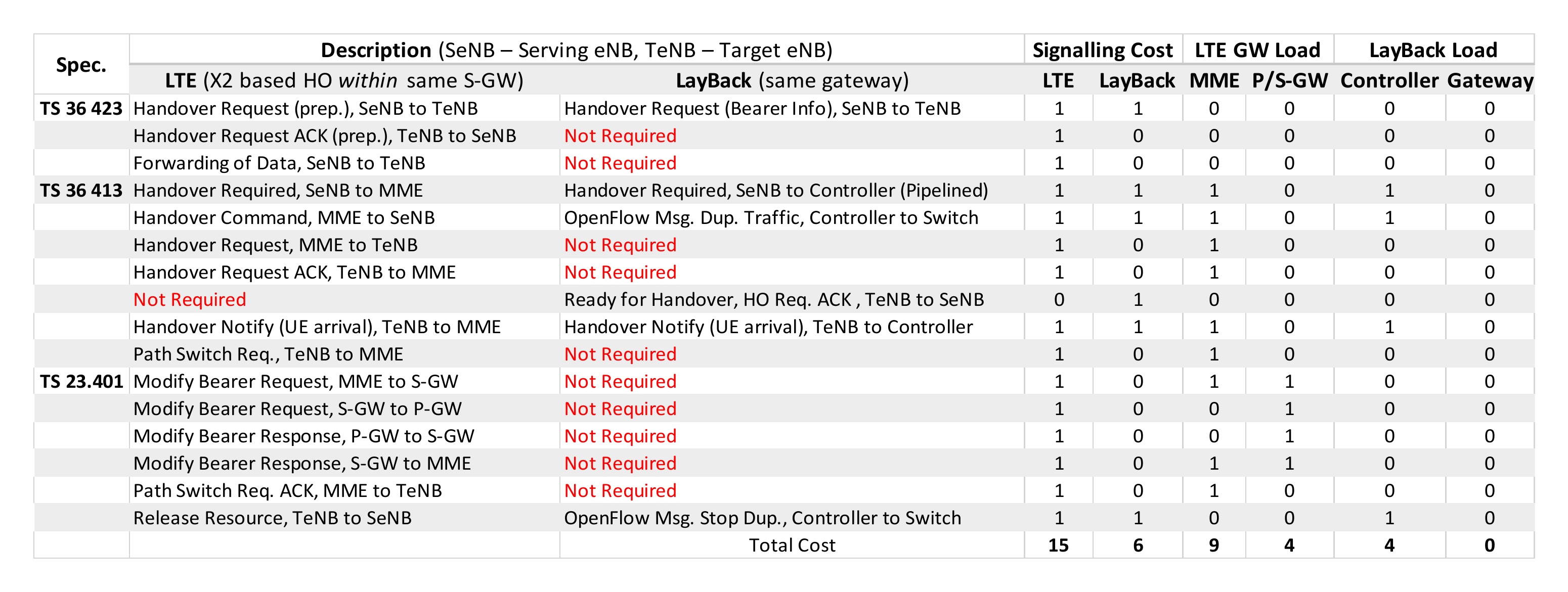}
	\caption{Handover cost and load comparison in LayBack and LTE backhaul}
	\label{tab:handover}
\end{figure}
\subsection{Signalling Overhead Evaluation} 
Table~\ref{tab:handover} characterizes the amount of required
signalling messages
for X2-based handovers in LTE and for intra-LayBack handovers
within the same gateway.
We observe from Table~\ref{tab:handover} that the overall signalling
cost is reduced from 15 messages in LTE to 6 messages in the proposed
intra-LayBack handover, a 60~\% reduction of signalling overhead.
We note that we count an OpenFlow message sent from the controller to an SDN
switch as a signalling message in this evaluation.
We further observe from Table~\ref{tab:handover} that the load
of processing 9 handover signalling messages at the MME in LTE is reduced
to processing 4 handover signalling messages at the controller in LayBack.
Moreover, the P/S-GWs in LTE need to process 4 signalling messages for
a handover, whereas the LayBack gateway is completely oblivious to the
handover mechanism. That is, the LayBack handover completely eliminates the
handover signalling load on the gateway.

The LTE handover changes the bearers (UE context) on the gateway,
requiring overall a high handover signalling load, whereby large
signalling loads have to be processed at the LTE MME and P/S-GW.
In contrast, the LayBack handover reduces the overall signalling load
and employs the SDN switches to change the packet flow path.
Essentially, the LayBack handover moves the handover burden from
the  gateway to the network SDN switches.
The SDN switches can share the handover burden, avoiding bottlenecks.

\section{\MakeUppercase{Summary and Future Directions}}  
\subsection{Summary of SDN-LayBack Architecture and Handover Protocol}

We have introduced SDN-LayBack,
a novel layered backhaul architecture for wireless networks.
SDN-LayBack consistently decouples the radio access networks (RANs)
from the backhaul, permitting the flexible backhauling
of highly heterogeneous RANs.
Within the SDN-LayBack architectural framework, we have introduced
a first protocol mechanism, namely a 4-step handover protocol for
handovers within the scope of an SDN-LayBack gateway.
The proposed
LayBack handover mechanism relieves the signalling bottleneck at the
backhaul gateways (e.g., LTE S-GW and P-GW)
by (a) traffic duplication at SDN switches, and
(b) bearer information forwarding from the currently serving eNB to the
target eNB.
Moreover, the LayBack handover reduces the signalling load on the
backhaul control entities (e.g., MME) by (a) reducing the required
signalling messages, and (b) moving some of the signalling messages
from mobility control entities to the SDN controller.
Our comparison of the handover signalling load indicates that
LayBack achieves a 60~\% reduction of the overall handover signalling load
compared to the conventional
LTE handover. Moreover, LayBack completely avoids handover processing at
the gateway and distributes the handover processing load over a network
of SDN switches, avoiding processing bottlenecks.

\subsection{Future Direction: Interference Management Protocol} 

There are many exciting avenues for future protocol developments and
evaluations within the SDN-LayBack architectural framework.
Since interference is one of the main limitations for the deployment
of small cells, an important future research direction is to
exploit the capabilities of the SDN-LayBack architecture for
interference mitigation protocols.
We can exploit the central control (with a global holistic
perspective) in SDN-LayBack for coordinating
the long-term wireless configurations, such as center
frequency, bandwidth, and time-sharing patterns,
as well as protocol-specific configurations (e.g., random access channel
parameters in LTE).
We emphasize that the goal of SDN-LayBack interference management
would \textit{not} be to manage the physical
resources, such as sub-frame power control in LTE, at the micro level.
Instead, SDN-LayBack interference management assigns high-level configurations
in a coordinated manner to the radio nodes so as to mitigate interference.

\subsection{Future Direction: D2D Communication Protocol}  
Device-to-Device (D2D) communication will likely play an important role in
the future of real-time content sharing and is another promising
direction for SDN-LayBack protocol development.
The LayBack architecture can be exploited for D2D communication so as
to avoid flooding on the access network.
More specifically, the SDN controller can
maintain a subscriber (location) data base to coordinate the connection
establishment and routing between D2D communication partners.

\subsection{Future Direction: Video Streaming Protocol}  
Another direction is to develop and evaluate collaborative video streaming
protocols within the micro-architecture framework
(Section~\ref{ma:sec}) of SDN-LayBack.
For densely populated areas,
e.g., airport terminals and stadiums, it is highly likely that
multiple RAN technologies, e.g., LTE and WiFi, are deployed,
and multiple simultaneous (unicast) video streams are being
transmitted to (or from) devices within the coverage area of the
considered micro-architecture, which is controlled by a local controller.
A LayBack video streaming protocol can exploit the multiple ongoing video
streams and multiple RANs to collaboratively trade off the video traffic
characteristics (e.g., traffic bursts from complex video scenes)
with RAN transmission conditions to achieve improved video experience
across an ensemble of video streams. In addition to existing techniques
such \cite{suresh2007superresolution, suresh2007robust, suresh2006robust, suresh2006discontinuity, aravind2012multispinning, suresh2014hog} can be applied to the SDN LayBack framework.

%


\clearpage
\newpage
\addtocontents{toc}{}
\SingleSpace
\bibliographystyle{IEEEtran}
\bibliography{refs2}

\clearpage
\addtocontents{toc}{\vspace*{-.32in}\noindent APPENDIX\par}
\DoubleSpace
\begin{appendices}
\renewcommand{\thesection}{\Alph{section}}
\end{appendices}
\end{document}